\documentstyle[epsf]{article}
\fussy
\newcommand{\AmS}{{\protect\the\textfont2
  A\kern-.1667em\lower.5ex\hbox{M}\kern-.125emS}}
\def\be{\ifnum \count1=0 $$ \else \begin{equation}\fi}
\def\ee{\ifnum\count1=0 $$ \else \end{equation}\fi}
\def\ele(#1){\ifnum\count1=0 \eqno({\bf #1}) $$ \else 
\label{#1}\end{equation}\fi}
\def\req(#1){\ifnum\count1=0 {\bf #1}\else \ref{#1}\fi}

\def\bea(#1){\ifnum \count1=0   $$ \begin{array}{#1}
\else \begin{equation} \begin{array}{#1} \fi}

\def\eea{\ifnum \count1=0 \end{array} $$
\else  \end{array}\end{equation}\fi}

\def\elea(#1){\ifnum \count1=0 \end{array}\label{#1}\eqno({\bf #1}) $$
\else\end{array}\label{#1}\end{equation}\fi}

\def\cit(#1){
\ifnum\count1=0 {\bf #1} \cite{#1} \else 
\cite{#1}\fi}

\def\bibit(#1){\ifnum\count1=0 \bibitem{#1} [#1    ] \else \bibitem{#1}\fi}
\def\ds{\displaystyle}

\def\p{\partial}

\makeatletter
\@addtoreset{equation}{section}
\makeatother

\newcommand{\foot}{\footnote}
\newcommand{\ZZ}{\hbox{Z\hspace{-3pt}Z}}
\newcommand{\semi}{\mathbin{\hbox{\hskip2pt\vrule height 5.0pt depth -.3pt
 width .25pt \hskip-1.6pt$\times$}}}
\newcommand{\IN}{\hbox{I\hspace{-2pt}N}}
\newcommand{\IR}{\hbox{I\hspace{-2pt}R}}
\newcommand{\IC}{\,\hbox{I\hspace{-6pt}C}}
\newcommand{\IP}{\hbox{I\hspace{-2pt}P}}
\newcommand{\IH}{\hbox{I\hspace{-2pt}H}}

\newcommand{\ra}{\rightarrow}
\newcommand{\cA}{{\cal A}}
\newcommand{\cB}{{\cal B}}

\newcommand{\cF}{{\cal F}}
\newcommand{\cO}{{\cal O}}

\newcommand{\bra}{{\langle}}
\newcommand{\ket}{{\rangle}}
\newcommand{\C}{{\bf C}}
\newcommand{\bP}{{\bf P}}

\newcommand{\ttx}{{\theta_{\tilde x}}}
\newcommand{\tty}{{\theta_{\tilde y}}}

\newcommand\mabs{\noindent $\bullet$ }
\newcommand\dd{ {\rm d} }

\def\npb#1(#2)#3{{ Nucl. Phys. }{B#1} (#2) #3}
\def\plb#1(#2)#3{{ Phys. Lett. }{#1B} (#2) #3}
\def\pla#1(#2)#3{{ Phys. Lett. }{#1A} (#2) #3}
\def\prl#1(#2)#3{{ Phys. Rev. Lett. }{#1} (#2) #3}
\def\mpla#1(#2)#3{{ Mod. Phys. Lett. }{A#1} (#2) #3}
\def\ijmpa#1(#2)#3{{ Int. J. Mod. Phys. }{A#1} (#2) #3}
\def\cmp#1(#2)#3{{ Comm. Math. Phys. }{#1} (#2) #3}
\def\cqg#1(#2)#3{{ Class. Quantum Grav. }{#1} (#2) #3}
\def\jmp#1(#2)#3{{ J. Math. Phys. }{#1} (#2) #3}
\def\anp#1(#2)#3{{ Ann. Phys. }{#1} (#2) #3}
\def\prd#1(#2)#3{{ Phys. Rev. } {D{#1}} (#2) #3}
\def\ptp#1(#2)#3{{ Progr. Theor. Phys. }{#1} (#2) #3}
\def\aom#1(#2)#3{{ Ann. Math. }{#1} (#2) #3}

\newcommand\figinsert[4]
{\begin{figure}[htb]
\newcommand\figsize{#3}
\epsfysize\figsize
\centerline{\epsffile{#4}}
\caption{\leftskip 1pc \rightskip 1pc
\baselineskip=10pt plus 2pt minus 0pt {\sl {#2}}}
\label{fig:#1}
\vskip -.3cm
\end{figure}}
\newcommand\figref[1]{Fig.\ref{fig:#1}}

\begin{document}

\hyphenation{author another created financial paper re-commend-ed}

\title{On the Geometry behind $N=2$ Supersymmetric Effective Actions in 
Four Dimensions.\thanks{An extended version of Lectures presented at 
the Trieste Summer school 1996 and the 33rd Karpacz school 
on String dualities 1997. Partly supported by the Edison fund.}
       		}
\author{A.\ Klemm 
\\
Enrico Fermi Institute, 
University of Chicago, 
5640 S. Ellis Avenue, \\
Chicago IL 60637, USA}

\maketitle

\begin{abstract}
An introduction to Seiberg-Witten theory and its relation 
to theories which include gravity.
\end{abstract}


\section{Introduction}

In the last years it has become clear that consistency requirements 
restrict the non-perturbative properties of supersymmetric theories 
much more then previously thought. In fact it turned out that such 
theories cannot be consistently ``defined'' without referring
to their non-perturbative structure.

The prototypical examples are the $N=2$ $SU(2)$ supersymmetric 
Yang-Mills theories of Seiberg and Witten \cit(swI). 
Self-consistency seems to require a duality to be at work, 
which interchanges an electrical and a 
magnetic description of the same low energy theory. 
A short introduction into this duality 
will follow in section (\ref{sduality}). 
The set of states, which are elementary in one description, are 
solitonic in the other. Depending on the scale one of the 
descriptions is preferred because its coupling is weak. In particular  
the description of the effective $SU(2)$ gauge theory 
can be replaced in its strongly coupled infrared regime by a 
magnetic $U(1)$ gauge theory, which couples weakly to massless 
magnetic monopoles. Vice versa, if one starts at low scales with 
the weakly coupled magnetic $U(1)$ theory one gratefully notices 
that it can be replaced by the asymptotically free $SU(2)$  theory 
before it hits its Landau pole in the ultraviolet. 
These theories are probably the first examples of globally 
consistent nontrivial continuum quantum field theories 
in four dimensions. A review of these theories is given in 
section (\ref{neq2}). 

What is known about these theories, namely the exact masses of the 
BPS states and the exact gauge coupling, is so far not derived from 
a first principle high energy formulation but rather from some knowledge 
of the symmetries of the microscopic theory and global consistency 
conditions of the low energy Wilsonian action, as defined in section 
(\ref{defaction}). The reconstruction from consistency 
requirements is subject of section (\ref{reconstruction}), it leads in 
particular to an uniformization problem, whose solution is 
discussed in subsection (\ref{uniformisation}). 
It seems rather difficult to go beyond these results without 
a deeper understanding the microscopic theory.
  
The BPS states are the lightest states, which carry electric or magnetic 
charge. Their mass is proportional to a topological central term in the 
supersymmetry algebra, see Appendix A. 
The BPS masses and the gauge coupling have a 
remarkable geometrical interpretation, as described in section 
(\ref{geompict}). In particular for $SU(2)$ theories there is 
an auxiliary elliptic curve,
in real coordinates a surface, whose {\sl volume} gives the gauge 
coupling and whose period integrals give the masses. For higher rank 
groups special Riemann surfaces can be constructed, which encode these 
informations in a similar way. The discussion of these auxiliary 
surfaces is subject of section (\ref{curves}).

To include gravity one has essentially to replace the Riemann surfaces
by a suitable Calabi-Yau threefold and to consider the effective 
action of the ten dimensional type II ``string'' theory  compactification 
on the Calabi-Yau manifold. Basic properties of Calabi-manifolds 
are summarized in section (\ref{calabiyau}). 
As in the pure gauge theory case one can allow for considerable
ignorance of the details of the microscopic theory, which 
describes gravity, and can nevertheless obtain certain 
properties of the effective theory exactly. 
Since the periods are proportional to the masses and vanish at the 
degeneration points of the manifold, the question about possible 
light spectra in the effective action becomes a question about 
the possible degeneration, or in other words, an issue of singularity theory. 
What happens at the possible degeneration sheds, on the other hand, 
light on the microscopic theory. This story is familiar from the type 
II/heterotic string duality\cit(huto) in six dimensions. 
The singularities a $K3$ can acquire are the classified
$ADE$ surface singularities, see sect. \ref{localmirror}, and lead 
for the type IIa theory by wrapping of two-branes to precisely the 
massless non-abelian gauge bosons which are required to match the 
gauge symmetry enhancements of the heterotic string on $T4$ 
\cit(wittencomments)\cit(aspinwallenhance). 
The possible singularities of Calabi-Yau threefolds are far richer and 
lead not only to gauge theories with or without matter but also 
to exotic limits of $N=2$ theories in four dimensions 
which fit a microscopic description in terms of non critical 
string theories.       

The perturbative string sector of the type II theory 
is less complete then the magnetic or the electric field theory 
formulation of Seiberg-Witten, it contains neither electrically nor 
magnetically charged states. The welcome flip side of this coin 
is that the magnetically and electrically charged states, which
are solitonic, appear more symmetrically. 
Both types can be understood as wrappings of the $D$-branes of the Type II 
theory around supersymmetric cycles  
of the Calabi-Yau manifold, see section (\ref{branes}). 
In the type IIb theory solitonic states 
arise by wrapping three branes around sypersymmetric  Calabi-Yau 
three-cycles. They can lead to solitonic hypermultiplets, 
which were interpreted as extremal black holes in \cit(strominger), or to 
solitonic vector multiplets\cit(klmvw). In the appropriate double 
scaling limit, which decouples gravity, $M_{pl}\rightarrow \infty$, 
and the string effects, $\alpha'\rightarrow 0$, \cit(kklmv) these solitons 
are identified with the Seiberg-Witten monopole and non-abelian gauge 
bosons respectively\cit(klmvw).

Mirror symmetry maps type IIb theory to type IIa theory, the odd branes 
to even branes and the odd supersymmetric cycles to even supersymmetric 
cycles. In the type IIa theory the non-abelian gauge bosons can now be 
understood as two-branes wrapped around non-isolated 
supersymmetric  two-cycles, which are in the geometrical phase of 
the CY manifold nothing else then non-isolated holomorphic 
curves. One can easily ``geometrical engineer'' configurations of 
such holomorphic curves, which will lead in the analogous scaling 
limit \cit(kklmv) to prescribed gauge groups, also with controllable 
matter content\cit(kkv). Using local mirror symmetry
\cit(kkv) it is possible to rederive in this way the Seiberg-Witten 
effective theory description by the Riemann surfaces from our 
present understanding of the non-perturbative sector of the type 
IIa string alone. 

One very important aspect of the embedding of the 
$N=2$ field theory into the type II theory on CY manifolds 
is that the field theory coupling constant is realized 
in the type II description as a particular  geometrical 
modulus. The strong-weak coupling duality is accordingly realized 
as a symmetry which acts geometrically on the CY moduli. 
In fact all properties of the non-perturbative field theory can be 
related to geometrical  properties of the CY manifold, 
e.g. the space-time instanton contributions of Seiberg-Witten 
are related to invariants of rational curves embedded in the 
Calabi-Yau manifold, etc.\cit(kkv).    
   
In the type IIb theory the solitonic states originate most symmetrically, 
namely from the wrapping of three-branes. We do not have really a  
non-perturbative formulation of the type II theory yet. One can try to keep 
the advantages of the symmetric appearance of the solitons and yet simplify 
the situation by modifying the above limit to just decouple gravity 
\cit(klmvw). As reviewed in \cit(lerche) this gives rise to non-critical 
string theories in six dimensions and a quite intuitive picture for the 
solitonic states as $D$-strings wound around the cycles of the Riemann 
surfaces. At this moment we do not understand these non-critical string 
theories well enough to infer properties which go much beyond what can 
be learned from the geometry of the singularities, rather at the moment the 
geometrical picture wins and predicts some basic features of these 
yet illusive theories.

An important conceptual and technical tool in the analysis is mirror
symmetry. Aspects of this will be discussed it section 
(\ref{mirrorsymmetry}). This includes a discussion of some properties 
of the relevant branes sec. (\ref{branes}), special K\"ahler geometry 
sec. (\ref{specialkaehler}), the deformation of the complexified K\"ahler 
structure sec. (\ref{complexkaehler}) with special emphasis on the
point of large K\"ahler structure sec. (\ref{largekaehler}) as well as 
the main technical tool, the deformation theory of the complex 
structure sec. (\ref{complexstructure}). The duality between the heterotic
string and the type II string is shortly discussed in sec. 
(\ref{stringduality}).

We find it very useful to present in some detail an example where all
the concepts presented in these lectures come together, the so called 
$(ST)$-model, which corresponds to a simple $K3$ fibration 
Calabi-Yau, sec. (\ref{example}). In principle it should be possible to 
try to understand this example first and go backwards in the text 
when more background material is needed.

\section{Electric-magnetic duality and BPS-states}
\label{sduality}
Before we turn to the $N=2$ case we shall give a short review of 
the concept of $S$-duality in field theory and in particular 
in $N=4$ supersymmetric theories. This is in order to introduce some 
concepts, where they are realized in the simplest way, and to prepare 
for the more complicated situation in $N=2$ theories. 
There exist already highly recommendable reviews 
\cit(olive),\cit(harveyrep) on the subject, so we will be very 
brief here.        

The semi-classical mass bound saturated for the 
Prasad-Sommer\-field-Bo\-go\-mol'nyi (BPS) states in a pure $SU(2)$ gauge 
theory\footnote{Strictly speaking 
a $SU(2)/Z_2\simeq SO(3)$, as a rotation by $2\pi $ is trivial in the 
adjoint.} 
without matter and a Higgs in the adjoint with potential $V={\lambda\over 4}
(\phi^a \phi^a-v)^2$ is given by
\be
M\ge   |v(n_e + \tau n_m)|.
\ele(bps) 
Here $n_e$ and $n_m$ are integral electric and magnetic charge quantum 
numbers of the state and $\tau$ is a combination of the gauge coupling and 
the $\theta$-angle\footnote{Historically the $\theta$-angle was 
considered in this context only later \cit(wittentheta)\cit(cardy).}  
\be
\tau={\theta\over 2 \pi}+{4 \pi i\over g^2}\ . 
\ele(tau) 
The mass of the elementary purely electrically charged 
$W^\pm$ bosons is of course just given by the Higgseffect, as 
reproduced by (\ref{bps}). 
The purely magnetic states are {\sl solitonic}, the simplest 
being the t'Hooft-Polyakov monopole configuration. The semiclassical 
mass bound for these configurations has been derived in \cit(PSB) 
(see \cit(colemann) for a review). 

There is a very remarkable fractional linear symmetry (see below, why only 
integer shifts $\tau\rightarrow \tau+a$, $a\in \ZZ$ are considered) 
in these formulas 
\bea(rl)
{\bf S}: & \left\{
\begin{array}{rl}
n_e&\rightarrow   n_m \cr 
n_m&\rightarrow - n_e  \cr 
\tau& \rightarrow - {1\over \tau} \cr
v&\rightarrow v \tau 
\end{array}
\right. \\ [4 mm]
{\bf T}: & \left\{
\begin{array}{rl}
n_e&\rightarrow  n_e - n_m  \cr
n_m&\rightarrow  n_m  \cr    
\tau& \rightarrow \tau + 1  \cr
v&\rightarrow   v 
\end{array}
\right.
\elea(sdual)
which generate a $PSL(2,\ZZ)$ action 
\be
\tau\mapsto {A\tau +B\over C \tau + D}
\ele(frac)
on $\tau$ with $A,B,C,D\in \ZZ$ and $AD-BC=1$ as well as 
an $SL(2,\ZZ)$ action on the electro-magnetic `charge' vector 
$$\left(\matrix{n_m\cr n_e}\right)\mapsto 
\left(\matrix{D&-C\cr -B& A}\right)\left(\matrix{n_m\cr n_e}\right)$$
The S-action\footnote{Often called so im mathematics books, which might 
be the reason for the name $S$-duality. Another possible origin of the
name is that the dilaton modulus on which the symmetry acts in 
string theory is also called $S$. }, 
which exchanges in particular 
the $W^\pm$ bosons with t'Hooft-Polyakov monopoles and inverts the gauge 
coupling (for $\theta=0$), was conjectured by Montonen and 
Olive \cit(mo) to be in {\sl some sense } a symmetry in the 
full quantum theory. It is obvious that a naive microscopic 
realization cannot possibly work in a normal quantum field theory:

\mabs Because the gauge bosons and the monopoles are not in 
the same Lorentz group representation 

\mabs In the quantum theory the coupling ${4 \pi i\over g^2(\mu)}$ 
will have different scale dependence in the original theory 
and its dual, which make a simple interpretation
of the inversion symmetry impossible. 

\mabs  The semiclassical 
analysis of the mass relied on the assumption that $V=0$ for the BPS 
configuration \cit(PSB), which will be invalidated by quantum corrections.

All these objections evaporate however in a theory with $N=4$ 
global supersymmetry, the maximal possible in four dimensions. 
Only here the gauge bosons and the monopoles are both 
in the same susy multiplet; the {\sl ultrashort} $N=4$ multiplet. 
The coupling does not run in $N=4$ theories as the beta function 
is exactly zero, in fact the full theory is believed not only to be scale 
invariant but actually {\sl conformal}. Finally the potential in the 
supersymmetric quantum theory is $V\equiv 0$. 

As for the validity of (\ref{bps}) in the quantum theory, 
it was shown in \cit(ow) that the supersymmetry algebra 
gets central extensions in the presence of non-trivial 
vacuum configurations. A simple analysis of the representation 
theory of supersymmetry algebras with $N$ (even) supersymmetry 
generators in the presence of central extensions 
$Z_i$, see app. A and  Ch. II of \cit(wessbaggerbook), shows that the mass of 
all multiplets is bounded by $M\ge |Z_i|$, $i=1,\ldots, (N/2)$ 
and the multiplets whose mass 
is actually $M=|Z_i|$, $\forall\, i$, 
consist of  $2^N$ degrees of freedom, while the generic ones, with the
minimal spin difference in the multiplet ($N/2$), have $2^{2N}$ 
degrees of freedom. The central 
charge  as calculated in \cit(ow) from the anti-commutator of the
super-charges in non-trivial vacuum configurations reproduces the 
semiclassical formula 
(\ref{bps}), i.e. $Z=v(n_e+\tau n_m)$.  If the existence of BPS 
saturated states is established by a calculation in the 
semiclassical regime the supersymmetry algebra protects these 
states from wandering off  the bound, neither by perturbative 
quantum effects nor by non-perturbative effects, as long as the 
supersymmetry is unbroken.    

Because the $\theta$-angle appears in front of the topological 
term in the Lagrangian
\bea(rl) 
{\cal L}&=-{1\over 4 g^2} F^{\mu\nu} F_{\mu\nu} -
{\theta \over 32 \pi^2} F^{\mu\nu}  {^*}F_{\mu\nu} \\ [ 2 mm]
&=-{1 \over 32 \pi} {\rm Im}\biggl(\tau 
(F^{\mu\nu}+i {^*}F^{\mu\nu})(F_{\mu\nu}+i {^*}F_{\mu\nu})\biggr)
\elea(theta)
only $\theta$-shifts by an integer $\tau\rightarrow 
\tau+1$ alter the classical action 
by a multiple of $2 \pi$ and hence the weight factor in the path integral by
an irrelevant phase shift. In this context it is important to note 
that an $n$-instanton effect in this normalization will be 
weighted by $e^{2\pi i n\tau}$.
This is the reason for the integrality 
condition which leads to $SL(2,\ZZ)$ as duality group. 
At quantum level it is again only the $N=4$ theory which allows 
for the definition of a microscopic $\theta$-angle. 

The electric and magnetic charges as defined from the Noether current 
of a dyon with quantum numbers $(n_m,n_e)$ are $q =  e n_e- {\theta e \over 2 \pi}n_m$ and $p={4 \pi\over e}n_m$. For consistent 
quantization, pairs of dyons $(p,q)$ $(p', q')$ must satisfy the Dirac-Zwanziger quantization condition
\bea(rl) 
q p' - q'p&=4 \pi(n_e n'_m- n'_e n_m)\cr 
          &= 2\pi n,\ \ n\in \ZZ \cr          
\elea(dz)
actually here with $n/2\in\ZZ$, which is good as we want later 
to include quarks which have half-integer charges in this 
units. The condition (\ref{dz}) generalizes immediately to theories 
with $r$ electric charges and $r$ magnetic charges, where it requires
integrality\foot{In units where the ``quark'' charge is $1$.} 
of the sympletic form $\vec n_e \vec n'_m - \vec n'_e \vec n_m = n$.
Quite generally one can argue\cit(apsw) that a theory  containing 
simultaneously massless states for which (\ref{dz}) does not vanish 
is conformal (and does not admit a local Lagrangian description).

\figinsert{lattice}
{Charges of dyons, which fulfill the Dirac-Zwanziger quantization
condition, lie on a lattice $\Lambda$ spanned by $e\tau$ and 
$e$ in the complex plane ($e$ is set to one). }
{1.8truein}{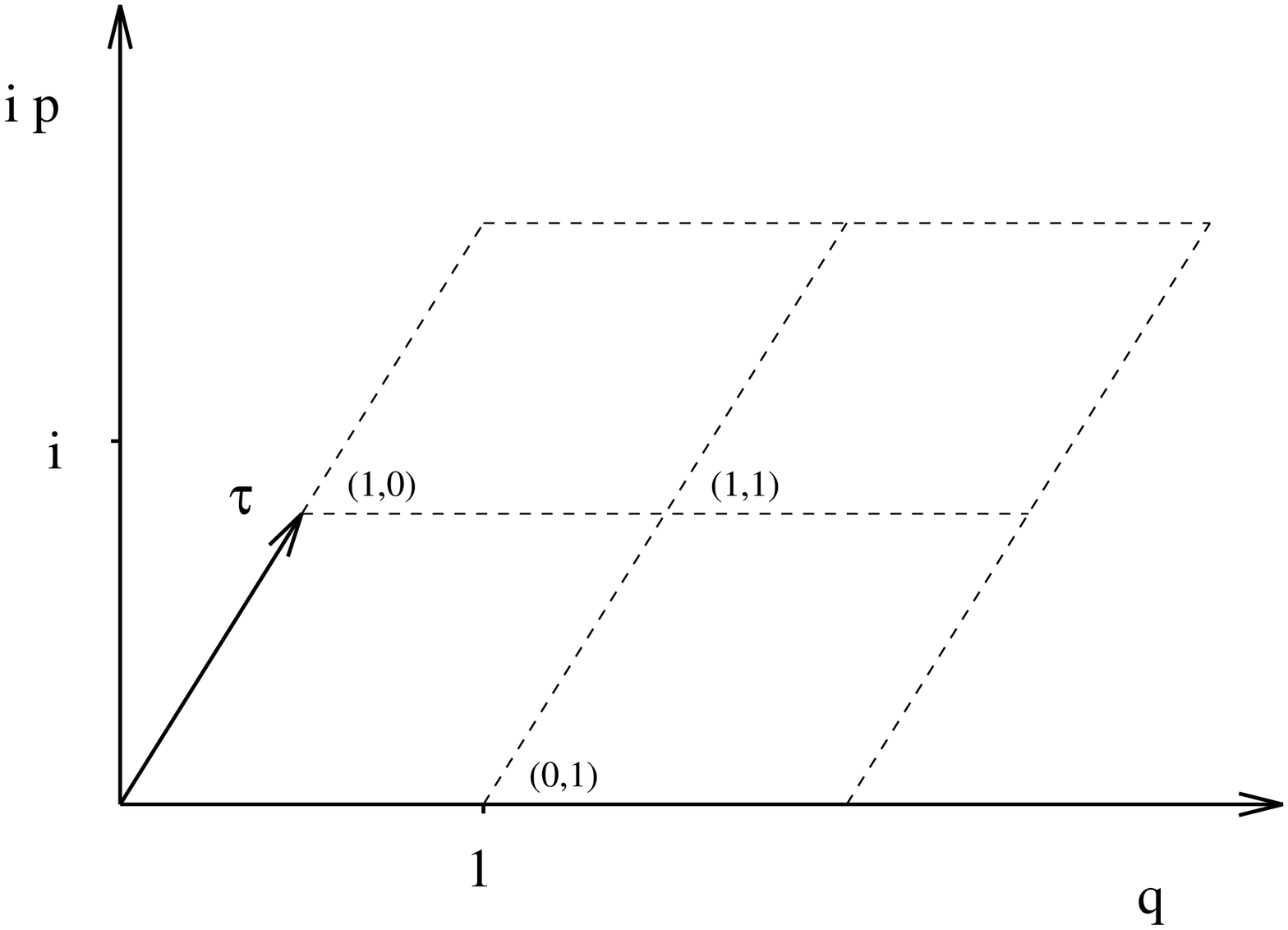}

Under the mild assumption that $W^\pm$-bosons exist in an $N=4$ 
theory as stable BPS states and $\tau$ is a generic complex number, i.e. the
lattice $\Lambda$ in figure 1 is not degenerate, 
$S$-duality makes very non-trivial predictions: The existence of 
the dyonic BPS states in the  $SL(2,\ZZ)$ orbit. 
These must be stable if their decay into other BPS states is forbidden 
by mass and charge conservation. 
By (\ref{bps}) and the assumption that $\tau$ is generic, 
stability means simply that $(n_e,n_m)$ must be co-prime integers. 
Some of these predicted stable multimonopole configurations\footnote{See 
also \cit(harveyrep).} have been found in \cit(sen) for the 
(broken) $SU(2)$, which has triggered renewed interest into 
$S$-duality.
For $SU(3)$ analogous predictions were checked in \cit(gl). 
The relevant discrete duality group for the latter case is 
$SP(4,\ZZ)$. 

A more genuine strong coupling test was performed in 
twisted $N=4$ theories on various manifolds \cit(vwneq4).

\section{$N=2$ supersymmetric Yang-Mills theory}
\label{neq2}
From the last section it is clear that the $S$-duality cannot 
be realized as a symmetry in the microscopic $N=2$ $SU(2)$ Yang-Mills 
gauge theory. As it turned out from the analysis of Seiberg and Witten, 
a subgroup of $SL(2,\ZZ)$ (or $SP(2\times {\rm rank} (G), \ZZ)$ for 
general gauge group $G$) is realized in this case on the {\sl abelian} gauge 
fields of the {\sl effective action}. 

This symmetry and an intriguing combination of microscopic and macroscopic 
arguments makes it possible to determine the terms up to two derivatives in 
the effective actions {\sl exactly}. In fact $N=2$ supersymmetry implies 
that the functions, which determine the $N=2$ effective action to this 
order, are holomorphic and, as one might expect, they are closely related to 
automorphic forms of the relevant symmetry groups, which are in the simplest 
cases subgroups of $SL(2,\ZZ)$. Reviews on the subject can be 
found in \cit(bilal)\cit(lerche)\cit(alvarezrev).

The theory of automorphic functions of subgroups of $SL(2,\ZZ)$ is 
an old and extremely well studied mathematical subject, so that once 
the group is known the functions can be quickly related (in many 
different ways) to well known ones. This theory has been 
also used in \cit(fmrss)\cit(nahm) to clarify some assumption made in 
\cit(swI)\cit(swII). We will review some aspects of this approach.
However the theory of automorphic function becomes more difficult and 
less studied for the multi parameter cases involving subgroups of   
$SP(2\times {\rm rank} (G), \ZZ)$.

Therefore we focus in section (\ref{geompict}) 
mainly on a closely related approach, which
identifies the electro-magnetic charge lattice $\Lambda$ 
with the {\sl integral homology lattice} $H^1(X,\ZZ)$ 
of an auxiliary Riemann surface $X$. In this approach the bilinear form
(\ref{dz}) will be identified with the intersection form on $X$ 
and the relevant functions, which determine the effective action, 
can be obtained from period integrals of the Riemann surface.

If one includes gravity, which has to be done by embedding 
the supergravity in string theory (at least according to our 
present understanding), the invariance groups will be still  discrete 
subgroups of (at least) $SP(2\times ({\rm rank}(G) + 2),\ZZ)$, where the 
extension by $2$ comes from the dilaton and the graviphoton multiplets 
respectively. However, while the physical quantities are here in general 
not related in a simple way to the developing  map 
(see section (\ref{uniformisation}))
of the discrete group \cit(candelasetal), 
they are still related in a simple way to the periods 
of a CY threefold, which is of course not auxiliary, but part of 
space-time. In the point particle limit the Riemann surfaces can also be 
understood as part of the space-time geometry \cit(kklmv)\cit(klmvw)
\cit(kkv).  

Let us summarize first the properties of $N=2$ theories which become 
important for the discussion.

\subsection{Definition of the $N=2$ Wilsonian action}
\label{defaction}

\mabs {\sl BPS short multiplets:} In the $N=2$ theory the monopoles 
and the matter are in short $N=2$ hyper multiplets $Q$ 
with maximal spin ${1\over 2}$, see (\ref{hypermult}), and the gauge 
bosons are in short $N=2$ vector multiplets $\Phi$ with maximal 
spin $1$, see (\ref{vectormult}).

\mabs {\sl Perturbative corrections:} Perturbative corrections are 
present, but due to the non-renormalization properties of $N=2$ theories
\cit(sv)\cit(nsvz), extremely simple. 
In particular the perturbative part of the scale dependence 
of $g$ comes only from wave function renormalization at one loop 
and is given  by
\bea(rl)
\mu {d\over d \mu} g & =\beta(g), \quad {\rm with} \\ [ 2 mm]
\beta(g)&= -{g^3\over 16 \pi^2}
\sum\left({11\over 3} C^R_{g.b.}-{2\over 3} 
C^R_{f}-{1\over 6} C^R_{s}\right)=:-{g^3\over 16 \pi^2}\kappa
\elea(beta) 
Here $C^R$ is the quadratic Casimir invariant in the 
representation\footnote{For $U(1)$ one has $C_{f}={1\over 2} 
q_{f}^2$ and $C_{s}={1\over 2} q_{s}^2$ with $q$ the $U(1)$ charge.} $R$: 
$C^R \delta_{ij}:={\rm Tr}(T_i T_j)$ and the sum is 
over gauge bosons, Weyl (or Majorana) fermions and real scalars 
in the loop. For $SU(2)$ one has $C^{adj}=2$ and $C^{fund}={1\over 2}$, 
so from (\ref{vectormult},\ref{hypermult}) we see that $\Phi^{adj}$ 
contributes $4$, $Q^{adj}$ contributes $-4$ and $Q^{fund}$ contributes 
$-1$ to $\kappa$. In general
\be
\kappa=2 N_c-N_f\ .
\ele(kappa)
For $SU(2)$ that leaves us with the following 
possibilities 

1.) $\beta(g)=0$: That is the case for 
$\Phi^{adj}$ plus one $Q^{adj}$, the field content of an $N=4$ 
ultrashort multiplet. Another possibility is one $\Phi^{adj}$ plus 
four $Q^{fund}$; this leads to another scale invariant theory 
with a differently realized $SL(2,\ZZ)$ invariance \cit(swII). 

2.) $\beta(g)<0$: the number $N_f$  of $Q^{fund}$ is less 
then four: that leads to asymptotic free field theory which we will
mainly discuss in this chapter, following \cit(swI),\cit(swII). 

3.) $\beta(g)>0$: there are various possible field contents. 
This possibility cannot be realized consistently as  gauge 
field theory with only global susy. 
However it can be realized as a field theory limit of string 
theory \cit(km)\cit(kmp). That signals the fact
that inclusion of gravity is essential for the consistency of 
the theory.

\mabs {\sl The Coulomb branch:} In the pure gauge 
theory the complete scalar potential comes 
from the $D$-terms:
\be
V(\phi)\ =\ {1\over g^2}{\rm Tr}[\phi,\phi^\dagger]^2\ ,
\ele(dpot)
where $\phi=\phi_i T_i$ are the scalar components of 
$\Phi^{adj}$. There is family of lowest energy 
configurations, $V(\phi)\equiv 0$, parameterized by vacuum
expectation values $a_i$ of the $\phi_i$ in the direction of a 
Cartan-subalgebra of the gauge group, as for those field
configurations the commutator in (\ref{dpot}) vanishes.  
E.g. for $SU(2)$ the flat direction can be  parameterized by 
$\phi=a \sigma_3 $, where $\sigma_3$ is the third 
Pauli-matrix. If $a\ne 0$ the $SU(2)$ breaks to $U(1)$ 
and the $W^\pm$ vector multiplets become massive with 
$M = \sqrt{2} |a|$. This corresponds to spontaneous 
generation of a central charge. Similarly if the field $A$ 
couples to an hypermultiplet the latter becomes
massive as a short multiplet with $M=\sqrt{2} |a|$.
Generally one refers to the parameters which parameterize the 
possible vacua as {\sl moduli} and the branch of the moduli space, 
which correspond to vev's of scalars in the vector multiplets as 
{\sl Coulomb} branch. As is 
clear from the $D$-term potential, the rank of the gauge group
will not be broken on the Coulomb-branch. We will see later on 
that the $N=2$ vector moduli space has a rigid special 
K\"ahlerstructure.   

\mabs {\sl The Higgs branch:} For $N=4$ theories 
the Coulomb branch is the only branch of 
the moduli space. For $N=2$ theories with $r$ hyper multiplets, one 
can have a gauge invariant superpotential, which can be written
in terms of the chiral $N=1$ super multiplets, defined below 
(\ref{vectormult},\ref{hypermult}) as                                 
\be 
W=\sum_{i=1}^{r} \tilde Q_i \Psi Q^i + m_i \tilde Q_i Q^i,
\ele(superpot)   
with suitable summation over the color indices to make this a 
singlet. Flat directions can emerge in the scalar potential if
at least two masses $m_i$ are equal. If the scalar of a charged
short hyper multiplet gets a vacuum expectation value the gauge group
is broken to a group of {\sl lower} rank, the corresponding
gauge bosons absorb the degrees of freedoms of the short hyper multiplet
and become heavy as {\sl long} vector multiplets with 3 d.o.f in a 
heavy vector, 4 Weyl fermions and  5 real scalars. In this way one 
gets rid of pairs of BPS states.  The branch of the moduli space 
parameterized by the hyper multiplet vev's is called the {\sl Higgs} 
branch. An essential point is that the scalars of the vector 
multiplets do not affect the kinetic terms of the hyper moduli space 
and vice versa. This follows from the absence of the corresponding 
couplings in the general $N=2$ effective actions \cit(specialkaehler) 
(see in particular 4th ref.). As one can treat the bare masses 
and the scale as vector moduli vevs the Higgs branch receives neither 
scale nor mass dependent corrections \cit(apsw). 
It maintains its classical {\sl hyperk\"ahler} structure. 
For example for quark hyper multiplets 
in the fundamental of $SU(2)$ flat directions 
emerge for $N_f>1$, $m_i=0$ and for $a=0$. 
For $N_f=2$ these are two copies of $\IC^2/Z_2$ touching 
each other and the Coulomb-branch at the origin \cit(swII) .

Of course for higher rank gauge groups we can 
have in general mixed branches, where the maximal gauge 
group is broken to a non-abelian subgroup by hyper multiplet 
vev's, which in turn has a Coulomb-branch 
parameterized by the vev's of its abelian gauge fields etc.

\mabs {\sl The central charges:} 
The BPS formula (\ref{bps}) is still protected by the $N=2$ 
supersymmetry algebra, in fact the analysis in \cit(ow) was carried out
for $N=2$, but the central charge term  becomes now scale dependent. 
One includes the bare masses $m_i$ of the quark hyper multiplets in
 $Z$ to reproduce the  BPS mass for the short quark hyper multiplets, 
so the BPS and central charge formulas for $SU(2)$ read  
\bea(rl)
M&\ge \sqrt{2}\, |Z|\\ [ 3 mm]
Z&= n_e a + n_m a_D +s^i {m_i}
\elea(bpsmass)
where, contrary to $m_i$,  $a$ and $a_D$ are scale dependent 
functions. $s_i$ are charges of global $SO(2)$'s carried only by 
the quarks of the i'th flavor. That is, the fixed lattice $\Lambda$
spanned by $(e \tau,e)$ in \figref{lattice} is be replaced (for $m_i=0$)
by a scale dependent lattice spanned by $(a_D(u),a(u))$. In particular 
this lattice can degenerate, which reduces the number of stable 
dyons drastically, see sect. \ref{consistencychecks}.  

\mabs {\sl Formal integration of the high energy modes and effective action:} 

For $a>0$ the charged sector develops a  mass gap\footnote{For simplicity 
we set $m_i=0$ in the following, otherwise 
we have to assume that $a \gg m_i$.}. 
At least formally one can integrate out the high energy 
modes $\phi_{high}$ of the charged states including the 
$W^\pm$ vector multiplets and the quark hyper multiplets, which have 
mass proportional to $a$. That leaves us with the effective action 
$H_{eff}(a,\phi_{low})$ of an abelian supersymmetric gauge 
theory without matter. More precisely the {\sl Wilsonian} 
effective action $H_{eff}(a,\phi_{low})$ is defined by summing over 
all high energy modes above some infrared cutoff scale, which 
is set to be equal to $a$ \cit(sv)
\be
\exp\lbrack -H_{eff}(a,\phi_{low})\rbrack = \ds{\sum_{\phi_{high}\atop E > a} 
\exp\lbrack -{H_{micro}}(\phi_{low},\phi_{high})\rbrack}\  .
\ele(eff) 
 Again formally the result can be expanded 
in terms of the slowly fluctuating fields $\phi_{low}$ and its derivatives 
$H_{eff}(a,\phi_{low})=\int d^4 x  \lbrack m(a,\phi_{low}) 
+ f(a,\phi_{low}) $ $(\partial \phi_{low})^2 + 
g(a,\phi_{low}) (\partial \phi_{low})^4\ldots\rbrack$. 
The Seiberg-Witten Wilsonian action differs slightly from the 
usual definition in that only  the charged high energy modes are
integrated out.

\figinsert{loop}
{Wave function renormalization of the effective coupling.}
{0.8 truein}{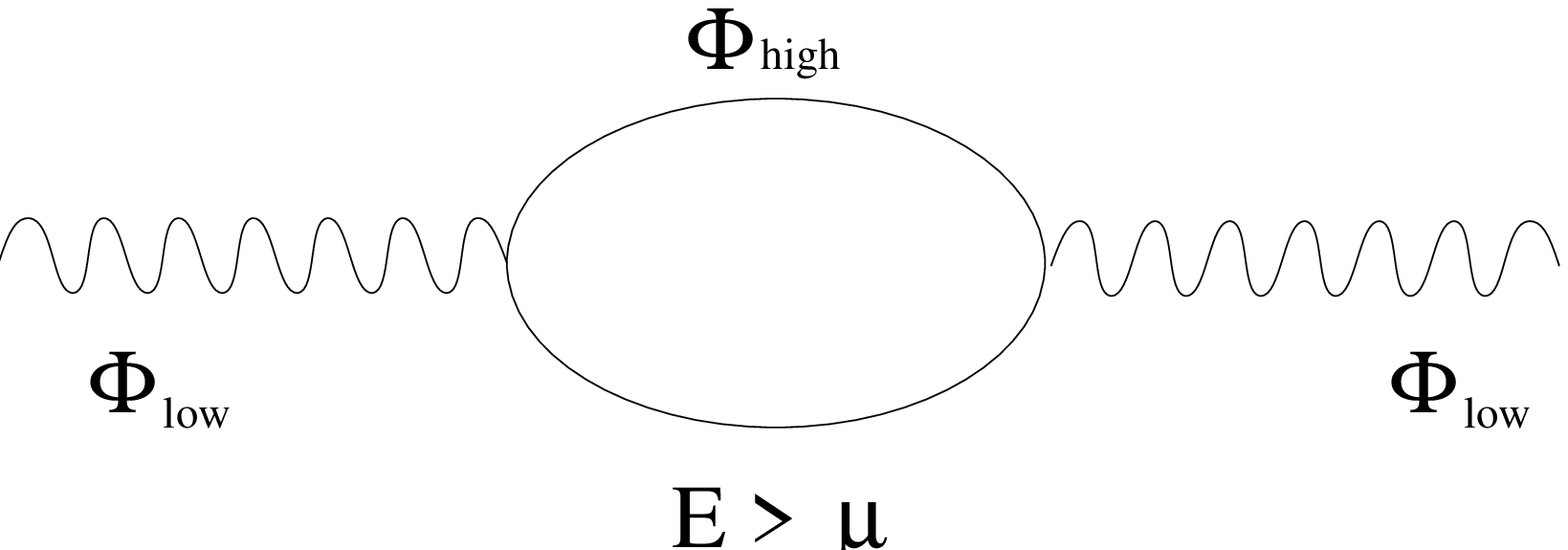}
In the  Wilsonian action the dependence of the 
{\sl effective Wilsonian} coupling constant on the scale $a$ 
due to (one-)loop effects can be determined for 
$N=2$ theories\cit(sv)\foot{It is explained 
in\cit(ahm) (comp. \cit(sv)) 
how this holomorphic coupling constant is related to the one particle 
irreducible\cit(nsvz). In particular for N=2 and up to two 
derivatives the Wilsonian action coincides with the 
1PI effective action. For an explicite derivation of the 
Wilsonian action in the non-abelian case see \cit(dgr).} 
as follows. Above the scale $a$ one includes the $W^\pm$ and the 
quarks as light degrees of freedom in the one-loop wave 
function renormalization and the coupling runs with 
the scale according to (\ref{beta}) for the microscopic 
$SU(2)$ gauge theory. Below the scale $a$ the above mentioned 
degrees of freedom freeze out and, as the $\beta$ function 
of the low energy $U(1)$ gauge theory without matter is zero, 
the coupling becomes constant. As non-perturbative effects 
are weighted with (\ref{nonpert}) this perturbative picture above 
is a good approximation for $g_{eff}$ as long as 
$\mu,a\gg\Lambda$, where $\Lambda$ is defined as $\Lambda_{QCD}$
for the microscopic theory. It does not make sense at all for  
$\mu,a\approx \Lambda$. As we will see the electric $U(1)$ gauge 
theory without matter is not the relevant effective theory in this region.

\figinsert{renorm}
{One-loop running of the effective coupling.}
{1.6truein}{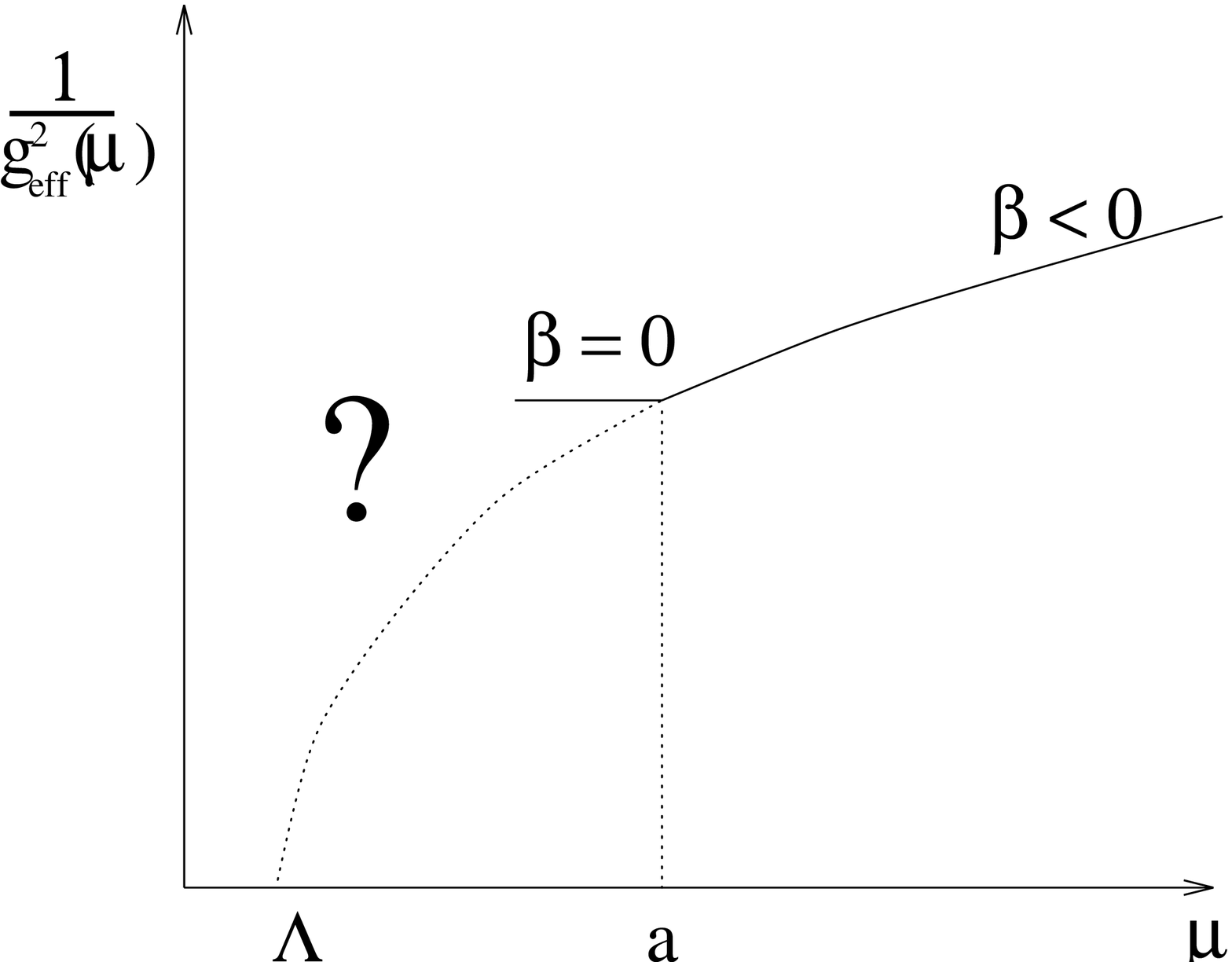}
It is of course extremely difficult to actually carry 
out the step (\ref{eff}).   
However $N=2$ supersymmetry provides integrability conditions, known
as rigid special K\"ahler structure \cit(rkaehler),
which allow to express the terms up to two derivatives 
in the low energy effective action through the holomorphic
prepotential ${\cal F}({A})$ 
\cit(sv),\cit(seibergachtacht),\cit(seibergneundrei), with
$\Phi=:{A} \sigma_3$ 
\be
{\cal L}_{eff}={1 \over 4 \pi}{\rm Im} \biggl[\int d^4 \theta 
{{\partial {\cal F}(A)\over \partial {A}}}
\bar { A}+\int d^2 \theta {1\over 2} 
{\partial^2 {\cal F}({A})\over \partial {A}^2} W^\alpha W_\alpha\biggr] 
\ele(effaction)
Here ${A}$ is ``photon'' multiplet,    
$W_{\alpha}:=-{1 \over 4} {\bar D}^2 D_\alpha V$ is the abelian  
field strength as derived from the $N=1$ photon vector multiplet 
$V$ (\ref{vectormult})
and  
\be 
K( A, {\bar{ A}}):={i\over 2}\left({{\partial \bar {\cal F}\over 
\partial {\bar A}}{ A}}-{{\partial  {\cal F}\over 
\partial {A}}{ \bar A}}\right)
\ele(rkpot)
is the real K\"ahler potential for the metric on the field 
space, which is hence a K\"ahler manifold. 

With the identification  
\be 
\tau({A}):=  {\partial^2 {\cal F}({A})\over \partial {A}^2} \ . 
\ele(gaugecoupling)
the bosonic pure gauge parts read as in (\ref{theta}), however with field 
depended effective coupling constant and $\theta$-angle, see
\cit(alvarezrev) for the full action. 
In particular in lowest order in derivatives of the effective action
$\tau(a)={\theta\over 2 \pi}+{4\pi i\over g^2}$, 
will be parameterized by the vev of the adjoint Higgs and positivity of
the kinetic term requires ${\rm Im}(\tau(a))>0$.  
${\cal F}$ is a holomorphic section of a line 
bundle over the Coulomb branch of the moduli space, whose 
reconstruction from monodromy data and global consistency 
requirements will be the main task in the remainder of this section.

\subsection{Reconstruction of the Wilsonian action}
\label{reconstruction}
\mabs {\sl The perturbative part} of ${\cal F}$ is obtained by 
first integrating (\ref{beta}) with $\mu:=a$ as 
explained, which yields $\tau={i\kappa\over 2 \pi} 
\log\left(a\over \Lambda\right)+c$ and then integrating 
(\ref{gaugecoupling})
\be
{\cal F}_{1-loop}= {1 \over 2} \tau_{cl}a^2+ {\kappa i\over 8 \pi } 
\log\left(a^2\over \Lambda^2\right) a^2\ .
\ele(oneloop) 
Here $\tau_{cl}$ is the bare value and as discussed for $SU(2)$ with
$N_f$ quarks, $\kappa=2 N_c - N_f$. 

\mabs {\sl Non-perturbative effects}:
 
The $n$-instanton contribution to $\tau$ will be weighted by 
$\exp{2 \pi i n \tau}$ (comp. (\ref{theta})) and according 
to the perturbative running of the coupling constant 
(\ref{beta}) this can be rewritten in leading order as
\be   
e^{2\pi i n \tau}=\left({\Lambda\over a}\right)^{\kappa n}.
\ele(nonpert)
Considering the zero-modes in an instanton background one learns that 
these contributions are not forbidden \cit(das)\cit(seibergachtacht). 
So one expects them generically to be non-zero and the 
non-perturbative contribution to ${\cal F}$ can be formally written as
\be
{\cal F}={\cal F}_{1-loop}+{a^2\over 2 \pi i}\sum_{n=1}^\infty {\cal F}_n 
\left(\Lambda \over a\right)^{\kappa n} . 
\ele(finfinit) 
As we will see in a moment the solution of Seiberg and Witten contains 
the exact information about all instanton coefficients ${\cal F}_n$ and 
therefore the exact non-perturbative gauge coupling 
(\ref{gaugecoupling}).

\mabs {\sl An apparent inconsistency:} It is instructive 
to realize that (\ref{finfinit}) cannot be a description of 
the theory everywhere on the Coulomb branch. Simply because
the metric ${\rm Im}(\tau(a))$ cannot be bounded from below, as
the Hessian ${\rm det}(\partial_{a_i} \partial_{a_j} {\rm Im}\tau(a))
\le 0$ ($a=a_1+i\ a_2$), as follows immediately from the 
Couchy-Riemann equations for the 
{\sl holomorphic} function $\tau(a)$. In a microscopic theory 
that would be disastrous, here it just means that we will enter 
regions in the moduli space where the degrees of freedom of the
effective action (\ref{effaction}) are not any more the relevant 
ones.   

\mabs {\sl Global symmetries of the moduli space:}
From the above it is clear that $a$ cannot be a good global 
variable on the moduli space. As a matter of fact it is not 
even locally in the semiclassical limit $a\rightarrow \infty$ a good 
variable, because the Weyl-reflection acting $a\rightarrow -a$ is 
part of the gauge symmetry, so that $a$ covers 
the physically inequivalent theories {\sl twice}. 
A good global variable should approximate in that limit Weyl-invariant 
quantities, here for ${\rm SU}(2)$, $u\propto {\rm Tr} (\phi^2)={1\over 2} a^2$, so one defines
the expectation value of the Weyl-invariant quantity in the full 
quantum theory
\be
u={\rm Tr} (\bra  \phi^2 \ket)
\ele(udef) 
as global variable of the moduli space. The choice of 
Weyl-invariant parameters for other groups is explained below 
\ref{sym} .

$N=2$ extended supersymmetry has an $U(2)_R$ global 
symmetry rotating the supercharges \cit(wessbaggerbook). 
The symmetry is often split into $SU(2)_R\times U(1)_R/Z_2$ to adapt 
for its action on the $N=1$ field content. 
It is easy to see that the  $U(1)_R$ symmetry  is a chiral symmetry 
\cit(swI). Due to the chiral anomaly 
$$\partial_\mu J_5^\mu=-{2 \kappa \over 32 \pi^2} F ^* F,$$ 
with $\kappa=2 N_c-N_f$ as before\footnote{The one-loop beta function 
and the chiral anomaly are in a ``multiplet of anomalies'' 
as explained in \cit(seibergachtacht). This fact relates the argument 
here to the argument leading to the $(\Lambda/a)^{\kappa n}$ 
non-perturbative terms in (\ref{nonpert}), which likewise break 
the global $U(1)$.}, 
one gets a change of 
the Lagrangian under the $U(1)_R$ 
rotation by $e^{2 \pi i \alpha}$, which is 
\be 
\delta {\cal L}_{eff} = - \alpha  {2 \kappa \over 32 \pi^2} F ^* F\ . 
\ele(shift)
That implies, compare (\ref{theta}), that the $U(1)_R$ is broken to  
$Z_{2 \kappa}$ \cit(seibergachtacht). 
The later acts on $\phi$ as $\phi\rightarrow 
e^{2 i \pi n/(2 N_c-N_f)} \phi$, $n\in \ZZ$. In particular 
for pure $SU(2)$ this means that there is an action
\be 
SU(2):\ N_f=0:\ \ZZ_2:  \ \ u \rightarrow -u . 
\ele(symmetryi)  
In principle one should keep the above philosophy and 
introduce in view of (\ref{symmetryi}) now $z=u^2$ as parameter 
labeling the vacua, which are inequivalent under global symmetries. 
In\cit(swI) this is not done, because the singularities in the 
$u$-plane have a somewhat easier physical interpretation, 
as we will see below. However there is a slight catch here, 
namely that the monodromy group on the $u$-plane will 
{\sl not} generate the full quantum symmetries of the theory, 
they will miss of course (\ref{symmetryi}). 

For $SU(2)$ with matter there is a very important additional 
symmetry. It stems from the fact that in $SU(2)$ the quarks $Q$ and
anti quarks $\tilde Q$ are in the same representation and  
(\ref{superpot}) allows for an $O(2 N_F)$ action on $(Q,\tilde Q)$ 
\cit(swII). The $\ZZ_2$ {\sl parity} in $O(2 N_f)$ 
\be
Q_1\leftrightarrow \tilde Q_1
\ele(halfrot) 
is also anomalous and the anomaly 
is such that it cancels the half rotations in (\ref{shift}). 
The anomaly free $Z_{2\kappa}$ is therefore
in the presence of quarks enlarged to an $Z_{4\kappa}$. To 
summarize one has 
\bea(rl) 
N_f&=1:\ \ZZ_3:  \ \ u \rightarrow \exp{2\pi i\over 3} u ,\cr 
N_f&=2:\ \ZZ_2:  \ \ u \rightarrow - u , \cr
N_f&=3:\ { \rm no \ symmetry\ on\ } u . 
\elea(symmetryii)
Alternatively one can analyze the instanton zero modes in the 
presence of matter as in \cit(das), which shows that non-trivial 
configuration in (\ref{theta}) occur only for even instanton 
numbers and therefore half theta shifts are allowed. 
We will come back to the symmetry considerations in section 
\ref{dyonsymmetries}.

\mabs {\sl Duality symmetry:} The physically most relevant question
is: What are the light BPS states in the effective action 
in regions where (\ref{finfinit}) ceases to be the right 
low energy description and how many phases will the theory 
have ? The answer to this questions is presently not given 
by a first principle analysis but by  minimal assumptions and  
a posteriori consistency checks.
 
We will make here a pragmatic choice of assumptions, which we 
consider natural\footnote{They can be chosen weaker at the expense of 
some additional argumentation, see e.g. \cit(fmrss).}:  
$u$ is the modulus of the theory. That can actually be justified from
the $N=2$ Ward-identities \cit(howewest), see also section 
(\ref{consistencychecks}). The $u$-plane is compactified, by a 
one point compactification to an $\IP^1$. The effective action in 
every other region of the moduli space can be described by a 
{\sl local} Lagrangian, which is related by a $SL(2,\ZZ)$ duality 
transformation to the description at infinity, see below. 
Finally to pin down the number of phases, we will make the assumption 
that no BPS state acquires an infinite mass, apart from the 
semiclassical region at infinity, inside the $u$-plane \cit(fmrss). 

To justify the duality assumption consider the  
bosonic piece of the $N=2$ 
Lagrangian\footnote{The inclusion of the fermionic part in 
this duality transformation is straightforward.} (\ref{theta}) 
\bea(rl) 
S&= -{1\over 32 \pi} {\rm Im}\biggl[\int \tau(a) (F+i {^*}F)^2\biggr] 
\\ [ 3 mm]
& =-{1\over 16 \pi} {\rm Im}\biggl[\int \tau(a) (FF+i {^*}FF)\biggr] 
\elea(thetaII)
and enforce the Bianchi identity $dF=0$ by a Lagrange multiplier 
 field $A_{D\mu}$. The term, which is to be integrated over 
to enforce $dF=0$, can be also interpreted as the {\sl local} coupling of a 
{\sl dual} gauge field $A_{D_\mu}$ to a magnetic monopole with charge 
normalization 
\be
\epsilon^{0\mu\nu\rho}\partial_\mu F_{\nu\rho}=
8\pi \delta^{(3)}(x) \ .
\ele(chargenorm)
It is suitably rewritten as
\bea(rl)
&{1\over 8 \pi}\int A_{D\mu} \epsilon^{\mu\nu\rho\sigma}
\partial_\nu F_{\rho\sigma}={1\over 8 \pi}\int {^*}F_D F\\ [ 2 mm]
&\ \ ={1\over 16 \pi}{\rm Re}
\biggr[\int ({^*}F_D-i F_D)(F+i{^*}F)\biggl]\ ,
\elea(lagrange)
such that one can perform a Gaussian integration over  $F$ 
after  adding (\ref{lagrange}) to (\ref{thetaII}). 
This leads to the dual action
\be 
S=-{1\over 32 \pi} {\rm Im}\biggl[\int {-1\over \tau(a)} 
(F_D+i {^*}F_D)^2\biggr]\ .
\ele(thetaIII)
By the general structure of $N=2$ 
supersymmetry the dual action must be expressible by a holomorphic
prepotential ${\cal F_D(A_D)}$ as in (\ref{effaction}) plus a 
$U(1)_{\rm mag}$ invariant superpotential (\ref{superpot}).
Note in particular that the mass of a short 
hyper multiplet containing a magnetic monopole 
depends according to (\ref{superpot}) in the dual local Lagrangian 
description on the vev $a_D$ of the scalar in the dual gauge vector 
multiplet $A_D$ which contains the gauge potential of a dual 
$U(1)_{\rm mag}$, that is $M=\sqrt{2} |a_D|$ after the obvious 
identification in agreement with (\ref{bpsmass}). 
The vacuum 
expectation values $a$ and $a_D$ are of course not independent but
will both depend on $u$. Comparing the expression in front of the 
kinetic terms in (\ref{thetaIII}) and (\ref{effaction}) in terms of 
$\cF_D(A_D)$ 
\bea(rl)
\ds{-{1\over \tau(A)}}&=-\left[ \p (\p_A {\cal F}(A))\over \p_{A}\right]^{-1}
\\ [ 2 mm]
& \  \  =\left[ \p (\p_{A_D} {\cal F}_D(A_D))\over \p_{A_D}\right]
=\tau_D(A_D)
\elea(tnaedt)
one learns that one has to identify $A_D=\p_A {\cal F}(A)$ and 
$A=-\p_{A_D}{\cal F}_D(A_D)$. This can be used to express the metric
in field space $(ds)^2={\rm Im}\, \lbrack \tau \rbrack {\rm d} a 
{\rm d} {\bar a}$ 
as
\bea(rl)
(ds)^2&={\rm Im} \left[\p^2 {\cal F} \over \p^2 a \right] 
{\rm d} a {\rm d} {\bar a}=
{\rm Im}\, {\rm d} a_D\, {\rm d}{\bar a}
\\ [ 2 mm]
      &=-{i\over 2}({\rm  d} a_D {\rm d} \bar a -
     {\rm d} a {\rm d} {\bar a}_D) \\ [ 2 mm]
      &=\ds{-{i\over 2}\left( {{\rm d} a_D\over {\rm d} u} 
         {{\rm d} \bar a\over {\rm d} \bar u} - 
        {{\rm d} a\over {\rm d} u} {{\rm d} {\bar a}_D\over {\rm d} 
       \bar u}\right) {\rm d}u \, {\rm d} \bar u }
\elea(metricinvariance)
in an obviously $SL(2,\IR)$ invariant way. We know from (\ref{nonpert})
that the $a\rightarrow a + s a_D$ shift invariance will be broken. At
worst, if all instantons numbers are present, to discrete shifts 
$s\in \ZZ$ hence $SL(2,\IR)$ to $SL(2,\ZZ)$. On 
\be
\tau(u)=\ds{ \left({ {\rm d} a_D\over {\rm d} u}\right)\over 
\left({{\rm d}a \over {\rm d}u}\right)}
\ele(taudef) 
the $SL(2,\ZZ)$ will act then as $PSL(2,\ZZ)$. 

Let us summarize the general linear symmetry, which can be realized 
on the abelian gauge fields $\vec V=(\vec a_D(\vec u),\vec a (\vec u))^t$
and the global charge vector $\vec s$ of a $r={\rm rank}(G)$ gauge 
group with $N_f$ flavors. As we have discussed this symmetry must be an 
invariance of the BPS mass  formula and the metric of the 
abelian gauge fields 
$$\begin{array}{rl} 
M&=\ds{\left|\sum_{i=1}^r (n^i_e a^i + n^i_m a^i_D) +\sum_{j=1}^{N_f} 
s_j m_j\right|}\\ [3 mm]
ds^2&=\ds{-{i\over 2}\sum_{i=1}^r({\rm  d} a^i_D {\rm d} \bar a^i -
     {\rm d} a^i {\rm d} {\bar a}^i_D)}\  .
\end{array}
$$ 
Again in a non-trivial instanton background one can have only discrete
shifts. The symmetry is therefore expected to be 
$({\bf M}, H)\in Sp(2 r, \ZZ) \semi \ZZ^{N_f}$ and 
acts on fields $\vec V $ and quantum numbers  
$\vec Q:=(\vec n_m,\vec n_e)^t$, $\vec s$ and $\vec m$ as 
\bea(rl) 
\vec V&\rightarrow {\bf M} \vec V+ H \vec m, \quad \vec Q\rightarrow 
({\bf M}^{-1})^t \vec Q  \\ [2 mm]
\vec s& \rightarrow \vec s - H \vec Q \ .
\elea(dualitygroup)
In the case of vanishing bare masses one can in addition rotate 
$\vec a$, $\vec a_D$ simultaneously by a phase. 
For reasons discussed below (\ref{shift}) this $U(1)$-symmetry is 
closely related to the shift symmetry and likewise broken 
to a discrete group by the chiral anomaly. 

What subgroups of this general invariance are finally realized in the 
theory will depend technically speaking on the monodromies of 
$\vec V:=(a_D(u),a(u))^t$ induced by the local physics, see next paragraph.
More conceptual one can directly try to address the question what 
non-perturbative states can be present in the spectrum, 
see section (\ref{dyonsymmetries})

\mabs {\sl The Riemann-Hilbert problem :}

Let $u_i$ the putative singularities: for 
${\rm SU}(2)$ we have then a flat holomorphic 
$SL(2,\ZZ)$-bundle $\vec V\rightarrow \{\IP^1 \setminus 
\{u_1, \ldots ,u_{s} \}\}={\cal K}$ over the 
Coulomb branch have to specify a particular section $\vec V(u)$, 
which will determine the effective action up two derivatives 
everywhere in ${\cal K}$. Such a section is uniquely determined by 

\noindent a.) the monodromies of $\vec V(u)$ around the $u_i$ and  

\noindent b.) the values of  $\vec V(u)$ at the $u_i$ \cit(yoshida).

As it is always helpful to understand the local physics let us 
first discuss, which effective local Lagrangian leads to which 
monodromies. 
We have assembled the information to discussed the 
monodromy of $\vec V$ at $u\propto \infty$. From 
(\ref{oneloop}) and (\ref{udef}) one sees that the monodromy relevant 
non-analytic piece is $a_D$ and $a$ is
\be
\vec V:=\left(\matrix{&a_D(u)\cr &a(u)}\right)\propto \left(
\matrix{{i\kappa \sqrt{2 u}\over 4 \pi}\log(u/\Lambda^2)
\cr \sqrt{2 u}}\right),
\ele(leadinf)
leading for pure $SU(2)$ to a monodromy matrix\foot{For the 
monodromies to be in $SL(2,\ZZ)$ one needs for $N_f\neq 0$ a 
different charge normalisation, cff. section \ref{ne2ne4conventions}.} 
\be 
M^\infty=\left(\matrix{-1& \phantom{-} 2 \cr \phantom{-} 0&-1}\right)
\ele(matinfinity)
which transforms $\vec V\rightarrow M^{\infty} \vec V$, 
if we take $u$ around the singularity at 
$u_0=\infty$ clockwise.

Next we investigate the possibility suggested by duality that we have 
a magnetic $U(1)$ coupled locally to a monopole (or more 
generally a dyon of charge $(n_m,n_e)$), which becomes massless $a_D=0$ 
at $u=u_0$. Because of (\ref{symmetryi},\ref{symmetryii}) 
there must be a physical equivalent situation at $u=-u_0$ etc. Again the 
theory has a mass gap for $a_D\ge 0$ and we use $a_D$ as the scale 
parameter of the effective action. Especially the determination of the 
perturbative running of the effective coupling $\tau_D(a_D)$ follows 
the same logic as explained above \figref{renorm}. The difference is that, 
because of the opposite sign of the $\beta$-function (\ref{beta}), 
the theory becomes now weakly coupled for $a_D=0$, while 
perturbative -- the theory will have a Landau-pole -- and 
non-perturbative effects become relevant for large ${\rm Im}[ {a_D}]$. 
Near $u=u_0$ the function $a_D(u)$ is analytic, i.e. in leading order 
$a_D\propto c(u-u_0)$.
Also similarly	 as near infinity one can easily see that the non-perturbative 
corrections will give an analytic contribution of type 
$\left(a_D\over \Lambda\right)^n$. Integrating (\ref{beta}) for 
the dual magnetic $U(1)_{mag}$ with a monopole of charge 
$1$ according to (\ref{chargenorm}) (compare footnote below 
(\ref{beta} )) one has
$${\p^2 {\cal F}_D\over \p^2 a_D}=\tau_D(a_D)\propto 
-{i\over \pi} \log( a_D)\ .$$
From $a(u)=-{\p {\cal F}_D\over \p a_D}$ we learn that the 
monodromy relevant piece of $\vec V$ near $u \propto u_0=\Lambda^2$ 
is
\be
\vec V\propto \left(
\matrix{c_0(u-u_0)
\cr {i\over \pi} c_0(u-u_0)\log(u-u_0)+ a_V}\right)\ ,
\ele(monmonopole)
which leads upon counter-clockwise analytic continuation around
$u\propto \Lambda^2$ to the monodromy matrix
\be 
M^{\Lambda^2}_{(1,0)}=\left(\matrix{&\phantom{-} 1& 0 \cr &-2& 1}\right)\ .
\ele(matmonopole)
The non-zero constant $a_V$ is of course not relevant for the 
monodromy, but its presence, established a posteriori from the 
explicit solutions (\ref{ada}) $a_V={4\over \pi}$, 
is very important as otherwise $a_D(u_0)=a(u_0)=0$, which would 
imply that electrically and magnetically charged states would 
become simultaneously massless. A conformal point\cit(apsw) at $u_0$ 
would contradict the selfconsistency of the solution.

Consider now a dyon of charge $(n_m,n_e)$, which becomes massless at 
a point $\tilde u$ in the  moduli space, i.e.  $\tilde 
a_D(\tilde u):=n_m a_D(\tilde u)+n_e a(\tilde u)=0$ 
and let $\tilde a(\tilde u):=k a_D(\tilde u)+l a(\tilde 
u)$ be the ``photon'', which couples locally 
to that dyon. Invariance of the metric (\ref{metricinvariance}),
means that $\vec {\tilde V}=({\tilde a}_D,{\tilde a})^t=
C\vec V$, with $C\in SL(2,\ZZ)$. 
By the one-loop analysis the monodromy relevant terms of 
$\vec {\tilde V}$ near $u_0$ look exactly as in (\ref{monmonopole}) and 
the counter-clock-wise analytic continuation around $\tilde u$ will 
lead to a 
monodromy $\tilde M$ on $\vec {\tilde V}$  as in 
(\ref{matmonopole}). Transforming 
this back\footnote{We do not have to determine $k,l$ actually, the 
knowledge $\det C=1$ is enough.} to the old basis $V$ we get the 
general dyon monodromy $M_{(n_m,n_e)}:=C^{-1}\tilde M C$ 
\be 
M_{(n_m,n_e)}=\left(\matrix{1+ 2n_m n_e&2 n_e^2 \cr -2n_m^2&
1-2n_m n_e}\right)\ .
\ele(genmonodromy)

There is a consistency requirement on the choice of the $2\, r$ 
monopoles, which can be seen as follows. Chose now a generic base point 
$u_b$ and draw a counter-clock wise loop starting and ending on 
$u_b$ around each singular point, where a monopole become massless. 
Define the label $i$ in $u_i$ 
$i=1,\ldots 2\, r$ by the order a counter-wise rotating ray from 
$u_b$ would hit them. The combination of these paths can be deformed 
to a big loop around
all singularities $u_i$ and since we are on a $\IP^1$ sphere it can be 
slipped over to a loop that encloses clockwise the singularity at 
$u=\infty$, hence we get a compatibility condition for these monodromies 
\be 
M_\infty=M_{u_{2 r}}\ldots M_{u_1} \ .
\ele(cons)  

Suppose now we knew that $2\, r$ dyons become massless at some 
points $u_i$ in $\IP^1$ symmetric under (\ref{symmetryi}) and consistent
with (\ref{cons}). This provides us with the data mentioned at the
beginning of this section and allows us to reconstruct $\vec V(u)$.
Clearly if $u$ is a label for the vacuum the physics should not 
depend on way we have reached a particular point in the $u$-plane. 
The physical invariance group $\Gamma$ must therefore contain 
the subgroup of the modular group $\Gamma_M\subset SL(2,\ZZ)$, 
which is generated by the monodromies $M^\infty$ and 
$M^{u_i}_{(n^i_m,n^i_e)}$. Also we know that it has to be 
augmented by the symmetries (\ref{symmetryi},\ref{symmetryii}).

\subsubsection{The uniformisation problem} 
\label{uniformisation}

Let us recast the problem posed above in a very well studied and more 
intuitive form. Fixing the monodromies also means fixing the local 
branching behavior of $\tau(u)$. Clearly this map will be vastly 
multivalued. For instance from (\ref{leadinf},\ref{taudef}) follows 
$\tau(u)\sim {i\over \pi} \log(u)$ at infinity and the monodromy around 
infinity identifies then $\tau \sim \tau+2 n$. Physically that is 
very reasonable because that corresponds just to the shift of theta by an 
(even) integer, which is irrelevant in view of (\ref{theta}). 
Can we reconstruct  $\tau(u)$ with ${\rm Im}(\tau)>0$ from its local 
branching data,  knowing that it is $SL(2,\IR)$ multivalued with action 
as in (\ref{frac}) and holomorphic away from the branch-points ? 
The above question is known as the uniformisation problem and the  
answer was given in detail at the end of the last century, see 
\cit(kleinfricke) for classical and \cit(lehner) 
for more recent reviews. In fact this classical theory answers also 
two essential physical questions: What are the admissible combinations 
of massless dyons and what is the range of the gauge coupling in 
the truly inequivalent physical theories. 

The latter question is answered by construction a fundamental 
region $F$ for $\Gamma$ as action on $\tau$ in the upper half-plane 
$\IH^+$, i.e. $F=\IH^+/\Gamma$. The essential facts about the
fundamental region we need are summarized in Appendix B.

\mabs {\sl The developing map:}
Specifying the fundamental region $F$ is tantamount to specifying 
$\Gamma$ up to conjugation and $\tau(u)$ is given by the developing 
or Fuchsian mapping $\tau:\IH^+\rightarrow F$ \cit(kleinfricke). 
From the local properties of the developing 
map encoded in $F$ and the prescribed mapping to the corners it was 
shown by H.\ A.\ Schwarz, (see \cit(kleinfricke),\cit(nehari)\cit(yoshida) 
for reviews) that it fullfils the so called Schwarzian 
differential equation, which is really in the heart of the theory 
\be
\{\tau, u\}=2 Q
\ele(unif) 
where the  $SL(2,\IC)$ invariant Schwarzian derivative 
is defined by 
$$\{\tau,u\}:={\tau'''\over  \tau'}-{3\over 2} 
\left(\tau''\over\tau'\right)^2,$$ 
with $'={\rm d}/{\rm d} u$ and 
$$2 Q:=\sum_{i=1}^{n} {1\over 2} {{1-\alpha_i^2}\over (u-u_i)^2}+
{\beta_i\over u-u_i}+\gamma \ . $$

The real $\alpha_i$ are the inner angles of the fundamental region $F$
of $\Gamma$, the real $\beta_i$ are also fixed by $F$ or by the 
asymptotic of $\tau$ at the $u_i$ and $\infty$. 
Up to an $SL(2,\IC)$ transformation $F$ is specified by $3 n$ 
parameters, namely the radii and the positions of the centers of the arcs
(see. Appendix B). 
After removing the $SL(2,\IC)$ invariance $3n-6$ real parameters are left. 
In (\ref{unif}) we count $3n+1$ real parameters ($u_i,\alpha_i,\beta_i$) 
and $\gamma$. But $3$ real parameters can be removed by an $SL(2,\IR)$ 
transformation which allows to put three points $u_i$ on a fixed 
position on the real axis. 

Note furthermore that $\{ \tau(u),u\}\sim 1/u^4$ for $u\rightarrow \infty $ 
if $\tau$  is regular at $u=\infty$, that is  
if $\tau=\sum_{i=0}^\infty c_i u^{-i}$. Comparing this with 
the Laurent expansion of (\ref{unif}) fixes another four parameters. 
Similar if $\tau$ is not regular at $\infty$, i.e. $u_i=\infty$ then 
either $\tau\sim u^{-\alpha_i}\times {\rm reg}$ if $\alpha_i>0$ 
or $\tau\sim {\rm log} u$ if $\alpha_i=0$. In both cases 
$\{\tau,u\}\sim {1\over 2}(1-\alpha^2_i)u^{-2}$ which
removes likewise $4$ parameters.

That (\ref{unif}) describes indeed the developing map can be seen 
as follows: first note that (\ref{unif}) is $SL(2,\IC)$ invariant thanks
to the special properties of the Schwarzian derivative and 
then check that $\tau$ has the right local properties i.e. 
$\tau\sim \log (u)$ is local solution near $u_i$ 
with $\alpha_i=0$ and similar $\tau\sim u^{p/n_i}$ is a 
local solution near $u_i$ for finite angles $\alpha_i=2 \pi /n_i$.
Using the property $\{ \tau, u\} =-\{u,\tau,\}/ 
\left({\dd^2 u\over \dd^2 \tau}\right)$
we can write the differential equation for the slightly more 
difficult inverse problem to determine $u(\tau)$  
\be
\{\tau, u\}=- 2 Q \left(\dd^2 u\over \dd^2 \tau\right)\ .
\ele(infunif)
  
It is easy to verify the essential fact that the non-linear 
equation (\ref{unif}) is solved by ratios of solutions 
$\tau={\varpi_1\over \varpi_2}$ of the following linear differential equation
\be
\varpi'' + Q \varpi =0. 
\ele(lineq)
It is clear that if one is only interested in  $\tau(u)$, there is 
an ambiguity in the association of the linear differential equation 
(\ref{lineq}), because we can multiply $\varpi_1,\varpi_2$ by an entire 
function $g(u)$. This ambiguity in the entire function has to be used 
to obtain from $\tau={\varpi_1\over \varpi_2}$ 
via (\ref{taudef}) the functions 
$a_D(u),a(u)$ with the right leading behavior (\ref{leadinf}) and 
(\ref{monmonopole}) as
\be
{{\rm d}\over {\rm d} u}\vec V(u)= g(u) 
\left(\matrix{\varpi_D(u)&\cr \varpi(u)& }\right)=:\vec \varpi (u)\ .
\ele(diffbzg)

A short look on the local indicial problem\footnote{A good reference on 
ordinary differential equations is\cit(ince).} of (\ref{lineq}) with ansatz 
$\varpi_i=(u-u_i)^r\times {\rm reg}$ at $u_i$ , i.e. 
$r(r+1)+1-(1-\alpha_i^2)/4=0$, shows that we get two power series 
solutions $[x^{r_1}(c_0+c_1x\ldots ),x^{r_2}(c_0+c_1x\ldots )]$  
with $r_i={1\over 2}(1\pm \alpha_i)$ iff $\alpha_i\neq 0$ and iff 
the indices degenerate for $\alpha_i=0$ the local solutions are of the 
form $[\sqrt{u-u_i},\sqrt{u-u_i}\log(u-u_i)]$.

The authors \cit(fmrss) consider a $U(1)$ section $f(u)$ defined 
such that $\vec V = f' \vec \varpi -f \vec \varpi' =: W(f,\vec \varpi)$. 
By (\ref{lineq}) it follows that $\vec V'=(f''+Qf)\vec \varpi$, 
hence $g(u)=(f''+Qf)$. 
Comparing now the  local behavior of $\varpi_i$ with (\ref{leadinf}) we see 
that $f$ has to have a simple pole at infinity and 
from (\ref{monmonopole}) we see that $f$ has to have a zero of 
order ${1\over 2}$ at every point, were a dyon mass comes down. 
Since $f$ is an entire function 
it's pole orders and zero orders have to add up to zero. Hence
if one does not allow for further poles of $f$ at points were $\vec \varpi$ 
are regular we cannot accommodate more then {\sl two} dyon singularities. 
Poles of $f$ at points were the $\vec \varpi$ are regular would lead to 
poles in the BPS masses as follows from 
$\vec V = W(f,\vec \varpi)$. Such an argument 
appears\footnote{In fact the argumentation in \cit(fmrss) 
does not require the assumption of specific monodromies inside 
$SL(2,\IC)$  beside
the one at infinity \cit(fmrss).} in \cit(fmrss) and shows that at least if 
we want to avoid the appearance of infinitely heavy particles inside 
the $u$-plane we have to have precisely two light dyons. We can 
therefore restrict in ({\ref{unif}) to $n=3$ and in (\ref{cons}) 
to $r=1$.

\figinsert{fundamental}
{The strip of width 2 above the two largest arcs is the
fundamental region of monodromy group $\Gamma(2)$ as found by 
the method of isometric cycles described in app. {\bf B}. 
Its area is by (\ref{area}) is $A=2 \pi$ hence six 
times the one of $SL(2,\ZZ)$.
Because of the identification (\ref{symmetryi}) the fundamental region 
of the {\sl quantum symmetry\ } group of pure $SU(2)$ is given by the 
hatched region, which corresponds to the group $\Gamma_0(2)$, 
the subgroup of $SL(2,\ZZ)$ with $C=0\ {\rm mod} \ 2$. The marked 
point at $\tau_0=-3/2+i/2$ is the $Z_2$ orbifold point of 
$\Gamma_0(2)$,  hence by (\ref{area}) $A=\pi$. In particular the 
identification by the $T$ generator (\ref{sdual}) 
$\tau\rightarrow \tau + 1$ is realized in 
the $N=2$ theory, while the $S$ generator is {\sl not} realized.}
{2.7truein}{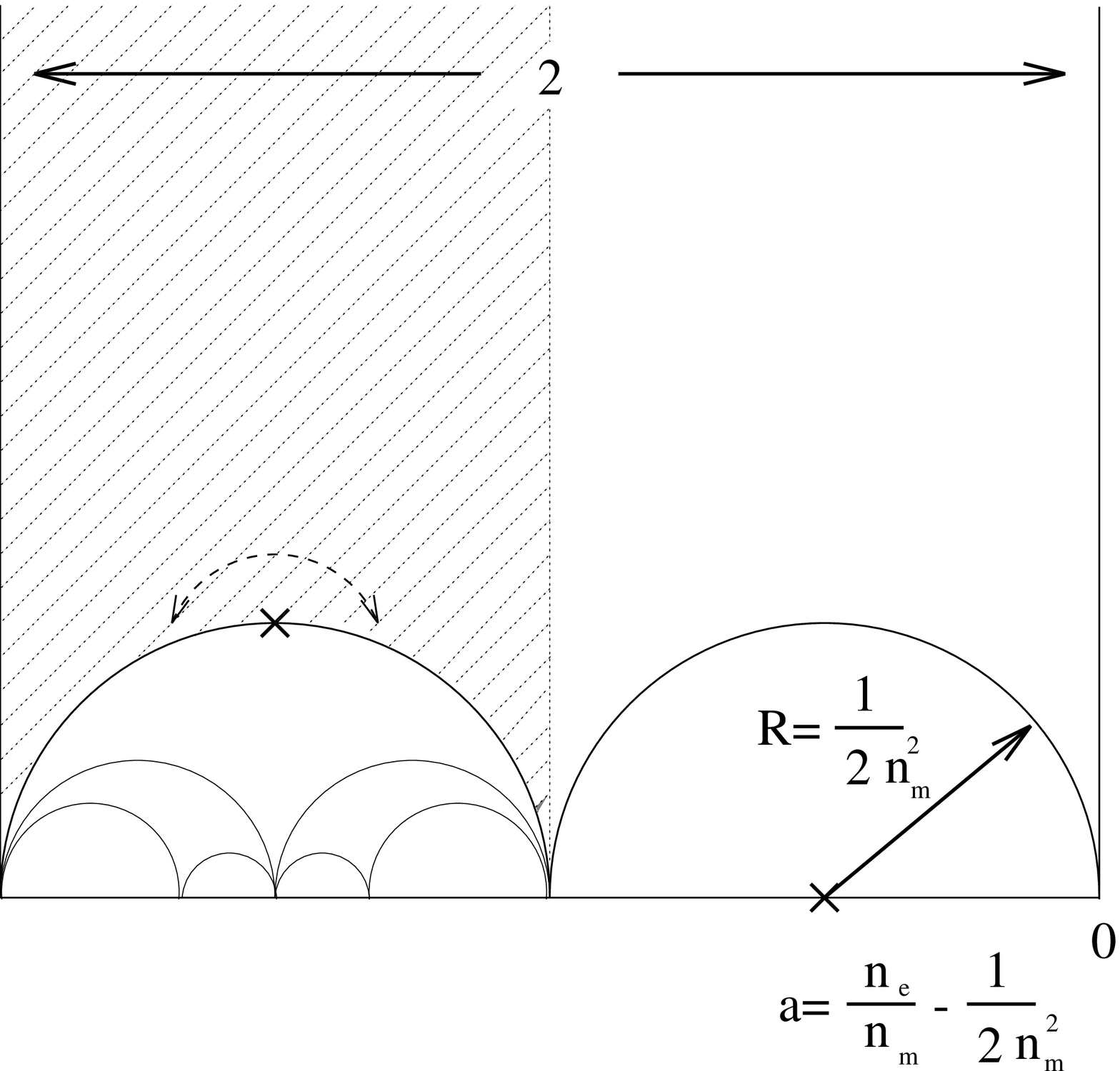}

\mabs {\sl The solutions:}

Now if $n=3$ one can choose from the 10 redundant parameters in $Q$ 
as the free parameters in (\ref{unif}) the angles $\alpha_i$ 
and the uniformization problem is {\sl solved} by the Schwartz-triangle 
functions, which are ratios of hyper geometric functions see e.g. 
\cit(erdelyii). This cases were completely studied in the last century, 
for general discrete subgroups of $SL(2,\IR)$. 
Especially if $\alpha_1=\alpha_2=\alpha_3=0$ as for our three necessarily 
parabolic elements the subgroup 
$\Gamma$ is uniquely determined, if the boundary conditions
are obeyed. For pure ${\rm SU}(2)$ it is given by the index $6$ 
congruence subgroup denoted by $\Gamma(2)$, which is defined as in 
$$\Gamma(N)=\left\{ \left. 
\left(\matrix{A & B \cr C & D}\right)\in SL(2,\ZZ)\right|\matrix{
&\!\!\!!\!\!  B,C=0 \, {\rm mod} \, N \cr 
& \!\!\!\!\!  A,D=1 \, {\rm mod} \, N } \right\} .$$
Alternatively one can argue that the pairs of massless dyons, 
which satisfy (\ref{cons}), are precisely 
$M^{\Lambda^2}_{(1,n)}$ $M^{-\Lambda^2}_{(1,n-1)}$$=M^{\infty}$. Any two of these 
matrices ge\-ne\-ra\-te\footnote{The fact that the pairs labled by $n$ are
equivalent, was referred to as dyon democracy in \cit(swI).} 
$\Gamma(2)$. The choice of 
$n$ corresponds to the $\theta$ shift symmetry, that is different 
choices correspond to the same physics, so we may chose $n=0$.

The linear differential equation (\ref{lineq}) is equivalent, 
in the sense explained below, to the hyper geometric equation 
${\cal L}f=0$ with
\be
{\cal L}= z(1-z) {\dd^2\over \dd^2 z} +[c-(a+b+1) z]
{\dd\over \dd z}-ab
\ele(hyp)
with parameters 
\be 
a={1\over 2}(1+\alpha_\infty-\alpha_0-\alpha_1),\quad
b={1\over 2}(1-\alpha_\infty-\alpha_0-\alpha_1),\quad
c=1-\alpha_0\ .
\ele(paramidentification)
Here by the indices on the $\alpha$ we indicate the associated 
singularities $z_i$ in (\ref{unif}), which have been fixed to be 
$0,1,\infty$. This differs from the choice we made in the $u$-plane
$-\Lambda^2,\Lambda^2,\infty$. To check the parameter identification 
we note that a second order linear differential equation
\be 
\varpi''+p \varpi' + q\varpi =0
\ele(genlin) 
can be brought to the form (\ref{lineq}) by substitution of 
$f(z)=g(z)\varpi(z)$ with $g(z)=\exp-{1\over 2}\int^z p dz'$. 
No matter how we write (\ref{genlin}) by choosing a particular 
$g(z)$ the invariant quantity on which the definition of $\tau$ 
depends is 
\be 
Q=q-{p'\over 2}-{p^2\over 4} \ . 
\ele(inv) 
Using this definition of $Q$  it is easy to check from 
(\ref{unif}) with $\beta_i=0$ and (\ref{hyp}) the parameter 
identification (\ref{paramidentification}) for the triangle groups.

From the physics point of view there is distinguished form of 
(\ref{genlin}) namely the one for which $g(u)$ in (\ref{diffbzg}) 
is constant. As is turns out the hypergeometric equation with 
$a=b={1\over 2}$ $c=1$ is itself the  preferred form. 
This can be easily seen by putting the singularities in (\ref{hyp}) 
from $z=0,1$ to $u=\pm \Lambda^2$ by the substitution $z={1-u\over 2}$
which transforms it to ${\cal L}\vec \varpi=0$ with
\be 
{\cal L} =\p_u^2-{2 u \p_u\over \Lambda^4- u^2}-
{1\over 4 (\Lambda^4-u^2)}
\ele(periodeqi)
Now we can chose solutions $\varpi_D,\varpi$, 
which lead to the correct leading behavior (\ref{leadinf}), 
(\ref{monmonopole}) with constant $g$.
Hence we can commute ${\cal L}\p_u$ to $\p_u {\widehat {\cal L}}$ so that 
$\vec V(u)$ is determined (the argument is up to additive constant, 
which has to be set to zero) by ${\widehat {\cal L}}\vec V=0$ with 
\be 
{\widehat {\cal L}} =\p_u^2-{1\over 4 (\Lambda^4-u^2)}
\ele(periodeqii)
This equation can be brought also in the hypergeometric form 
(\ref{hyp}) with $(a,b,c)=(-{1\over 4},-{1\over 4},{1\over 2})$ 
by substituting $\alpha:=u^2/\Lambda^4$. 
Hence we get compact formulas for the 
physical $a_D(u),a(u)$, which determine the masses of the BPS states
\bea(rl)
a_D(\alpha)  &= \ds{{i \Lambda\over 4} (\alpha-1)\,  _2F_1\left({3\over 4},
{3\over 4},2;1-\alpha\right)}\\ [ 3 mm] 
a(\alpha)&=\ds{{\sqrt{2}\Lambda} \alpha^{1\over 4} \ 
_2F_1\left(-{1\over 4},{1\over 4},1;{1\over \alpha} \right)\ .}
\elea(ada)

\subsection{N=2 versus N=4 conventions }
\label{ne2ne4conventions} 
There exist two conventions of charge normalizations
in the literature. In the $N=4$ conventions the smallest 
occuring electric charge, that of the $W^+$ boson, is set to one.
Since one can add matter to $N=2$ theories the smallest charge 
is now that of the quarks and is set often to one in the $N=2$ 
conventions. There is no change in the magnetic charge units however. 
E.g. in $SU(2)$ the effect is that the $W$-bosons have charge $|2|$ in 
the $N=2$ units and to keep (\ref{bpsmass}) one has to transform 
$(a_D,a)\mapsto (a_D,a/2)$, $\tau \mapsto 2 \tau$ and conjugate 
the monodromies by 
\be 
M\mapsto C^{-1}MC\ \ {\rm with}\ \ C=\left(\matrix{1&0\cr 0&2}\right) \ .
\ele(scale)
For pure $SU(2)$ the group $\Gamma_M$ generated by the monodromies 
in the new conventions becomes $\Gamma^0(4)$, which are the
$SL(2,\ZZ)$ transformations with $B=0 \ {\rm mod } \ 4$, instead 
of $\Gamma(2)$. Because of (\ref{symmetryi}) the full quantum symmetry 
is in this case $\Gamma^0(2)$.  
As the $T$ shift of (\ref{sdual}) is $\tau\mapsto \tau+2$ in the
new conventions the $\Gamma^0(2)$ is {\sl not} the canonical $\Gamma^0(2)$ 
subgroup of the  $SL(2,\ZZ)$ S-duality group  we 
started with. It is conjugated to the more canonical $\Gamma_0(2)$  
by a (not physical) $S$ duality, in general 
$\Gamma^0(N)=S \Gamma_0(N) S^{-1}$. 
It is however the canonical subgroup $\Gamma^0(2)$ subgroup of the 
$SL(2,\ZZ)$ 
found for the other conformal theory with $N_f=4$ flavors see sect.
(\ref{dyonsymmetries}).
The fundamental regions for $\Gamma^0(2)\subset \Gamma^0(4)$ 
looks exactly as the fundamental regions $\Gamma_0(2)\subset \Gamma(2)$ 
depicted in the \figref{fundamental} except that the whole 
figure is scaled such that the indicated width becomes $4$. 
Counterintuitively the scaling does not affect the hyperbolic 
areas as it is clear from formula (\ref{area}), so the index of the 
groups in ${\rm SL}(2,\ZZ)$ does not change. 
Similar the fundamental region for $\Gamma_0(4)$ looks like the 
$\Gamma(2)$ area in \figref{fundamental} when scaled to width $1$. 

\subsection{The symmetries on the dyon spectrum.}
\label{dyonsymmetries}

Let us investigate, purely from symmetry considerations, what 
states could be there. To discuss that it is useful to adopt 
the $N=2$ conventions and to think the groups $\Gamma\in {\rm SL}(2,\ZZ)$ 
as canonical subgroups of the $N_f=4$  ${\rm SL}(2,\ZZ)$. Bare masses of the 
quarks are set to zero in the following.  

For $N_f=0$, there are no dyons with the smallest charge unit, so the 
possible states are $(n_m, 2 n_e)$ and the subgroup leaving them 
invariant is $\Gamma^0(2)$.

For $N_f>0$ the dyons can be labeled by their 
$2 N_f$-fermion zero modes, which form after quantization a 
${\rm Spin}(2 N_f)$ representation 
\cit(swII). The elementary hyper multiplets transform in the 
vector representation of $SO(2 N_f)$, while monopoles (dyons) are in 
the spinor representation and as was pointed out in \cit(swII)
they are in the different conjugacy classes $s$ or $c$ depending
of whether they carry in addition even or odd electric charge.

For $N_f=2$ ${\rm Spin}(4)={\rm SU}(2)\times {\rm SU} (2)$ with
center $Z_2\times Z_2$. That means that the transformation  
properties of a state $(n_e,n_m)$ w.r.t. the center must 
given by the $Z_2$ charges $((n_e+n_m)\ {\rm mod}\ 2, 
n_e \ {\rm mod}\  2)$. Especially states with 
$(n_m,n_e)=(2 k+1,l)$ are spinor classes and should transform 
among themselves i.e. the $C$ in the $SL(2,\ZZ)$ transformation 
must be even, while $B$ can be $1$, this mixed $s$ and $c$ 
classes, but that is allowed because the corresponding outer
isomorphism is realized in ${\rm SO}(4)$. That implies that the 
${\rm SL}(2,\ZZ)$ 
is broken to $\Gamma_0(2)$.

For $N_f=3$ the center of ${\rm Spin}(6)=SU(4)$ is $Z_4$ and
since vectors have charge $2$ the $Z_4$ must act as 
$\exp {2 \pi i\over 4} (n_m+2 n_e)$ on dyons. In particular
the vectors have charges $(n_m,n_e)=(4k,l)$ the spinors
$(4k+1,l)$, $(4 k+3,l)$ and the scalars $( 4 m+2,l)$, which
means that $c=0\ {\rm mod} \ 4$ i.e.  ${\rm SL}(2,\ZZ)$ must be 
broken to $\Gamma_0(4)$.

For $N_f=4$ the relevant ${\rm Spin}(8)$ has center $Z_2\times Z_2$ 
but the $Z_2$ charges are $o:(0,0)$, $v:=(0,1)$, $s:(1,0)$ and 
$c:(1,1)$. The $Z_2$ charges of dyons must be $(n_m \ {\rm mod}\ 
2 , n_e \ {\rm mod} \ 2)$. Now the minimal shifts $b=1$, $c=1$ of
${\rm SL}(2,\ZZ)$ permute the $v,s,c$ classes but that could still be a 
valid symmetry as ${\rm Spin}(8)$ allows for an outer automorphism
called triality symmetry, which in fact permutes this classes by  
$S_3$. We can define a homomorphisms  $h:SL(2,\ZZ)\rightarrow S_3$ 
by modding the matrix entries $A,B,C,D$ by $2$, so that the total 
symmetry group can be the semi direct product 
${\rm Spin}(8)\semi {\rm SL}(2,\ZZ)$.

If we accept these subgroups, we get the generating 
monodromies and the associated massless particles and 
solutions without further effort. We can read off generating
monodromies from the fundamental region as they are the ones 
which conjugate the arcs of $F$ in pairs. Alternatively we may consider 
dyons with the smallest electric and magnetic charges, which generate 
according to (\ref{cons}) the symmetry $\Gamma_M$, which is
up to the discrete symmetry (\ref{symmetryi},\ref{symmetryii}) the 
quantum symmetry $\Gamma$. Note that in $N=2$ conventions one has 
to rescale $M_{(n_m,n_e)}$ of (\ref{genmonodromy}) we call the 
rescaled monodromy $\tilde M_{(n_e,n_m)}:=M_{({n_m \over \sqrt{2}},
{n_e\over \sqrt{2}})}$. Let us summarize the quantum symmetry 
groups $\Gamma$,
the monodromy groups and defining generators corresponding to the 
shortest massless dyon states for the 
cases in turn
$$\begin{array}{rl}
N_f=0:&  \Gamma^0(2):\  \Gamma^0(4): 
          \tilde M_{(1,0)} \tilde M_{(1,-2)}=T_0\\ [ 3 mm]
N_f=2:&  \Gamma_0(2):\  \Gamma(2):\ M_{(1,0)} M_{(1,-1)}=T_2 \\ [ 3 mm]
N_f=3:&  \Gamma_0(4):\  \Gamma_0(4):\  \tilde M_{(2,0)}\tilde M_{(2,-1)}=T_3,
\end{array}
$$
Here the $(2,0)$ is expected according to section \ref{sduality} not to 
be a stable monopole, but at best a bound state a threshold.  
$T_{N_f}$ is the semiclassical monodromy due to the $\beta$-function 
logarithm and the Weyl-reflection 
it is $T_{N_f}:=- (T^{N_f-4})$. $\tau(u)$ will be given by the Schwarz 
triangle function with appropriate boundary conditions 
and $a_D(u),a(u)$ can again be very simply obtained from 
solutions of hypergeometric functions. The situation for 
$N_f=4$ is in some sense the simplest as $\tau$ will
not depend on $u$. 

For $N_f=1$ we have (at least) four singularities because of the 
$Z_3$ symmetry and cannot expect such an extremely easy relation 
to the triangle functions of a subgroup of $SL(2,\ZZ)$. From the 
double scaling limit of the $N_f=4$ theory see (\ref{masscurves}) 
and the Lefshetz monodromy (compare the discussion in \ref{hyperelliptic})
one finds that the three monodromies are 
associated to the following massless particles 
$\tilde M_{(1,0)}\tilde M_{(1-1)}\tilde M_{(1,-2)}=T_1$.

\subsection{Consistency checks:}
\label{consistencychecks}
\mabs {\sl Consistency checks from instanton coefficients :}
This explicit solutions can of course be used to calculate 
${\cal F}$ everywhere in the moduli space. For instance 
the first coefficients in (\ref{finfinit}) for pure ${\rm SU}(2)$ 
are given by
$$
\vbox{\offinterlineskip\tabskip=0pt
\halign{\strut\vrule#&
\hfil~~$#$~~&
\hfil~~$#$~~&
\hfil~~$#$~~&
\hfil~~$#$~~&
\hfil~~$#$~~&
\hfil~~$#$~~&
\hfil~~$#$~~&
\vrule#\cr
\noalign{\hrule}
&n  &1 & 2 &  3& 4   & 5   & 6     & \cr
&&&&&&&& \cr
&{\cal F}_n&
\ds{1\over 2^2}&
\ds{5\over 2^{7}}&
\ds{3\over 2^{7}}&
\ds{1469\over 2^{16}}&
\ds{4471\over 2^{15} \cdot 5} &
\ds{40397\over 2^{20}}  &\cr
&&&&&&&& \cr
\noalign{\hrule}}
\hrule}
$$

The function 
\bea(rl)
{\cal G}:=&{i \pi \over 2} \int (a_D \dd a - a \dd a_D)\cr 
         =& i\pi ({\cal F}-{1\over 2} a \dot \cF)\ ,
\elea(calg)
with $\dot {}:={\dd\over \dd a}$, is obviously modular invariant. 
It is easy to see that ${\cal G}$ behaves at the cusps like $u$ 
since it is modular it must be therefore that\footnote{One may also 
use (\ref{periodeqii}) to check  that 
${\dd^2\over \dd^2 u}{\cal G}\equiv 0$. Vice versa it must be
that e.q.(\ref{periodeqii}) is of the form ${\cal L}=\p_u^2-1/p_i(u)$ 
also for $N_f=1,2,3,4$, which is true compare (\ref{pfeqs}).}  
${\cal G}= u+const.$ and 
the constant is zero as one can see from the 
vanishing of ${\cal G}$ at $u=0$. 
It was later shown in \cit(howewest) from the $N=2$ Ward-identities 
that ${\cal G}={\rm Tr} (\bra  \phi^2 \ket)$, which justifies the 
assumption that ${\rm Tr} (\bra  \phi^2 \ket)$ is the good variable
in the moduli space. Note furthermore that, because of 
${\cal F}=a^2 f(a /\Lambda)$ and using  ${\cal G}=u$ we get 
\be 
\Lambda {\dd \over \dd \Lambda} F=-{i\over \pi} u.
\ele(trivial)  
Now transforming the dependent variable in (\ref{periodeqii}) 
from $u\rightarrow a(u)$ and using ${\cal G}=u$ and the 
fact that $a(u), a_D$ is a solution one 
gets \cit(matone) a differential equation for ${\cal G}$
\be 
(1-{\cal G}^2)\ddot {\cal G}+{1\over 4} a \dot {\cal G}^3=0
\ele(insteq)  
and the same equation for the analogous defined ${\cal G}_D$.
From (\ref{insteq}) one can derive a recursion relation for 
the instantons coefficients, which can be found in \cit(matone). 

Eq. (\ref{insteq}) governs the non-perturbative effects in
the strong and the weak coupling region, it should in principle 
be understandable directly from explicit non-perturbative 
calculations. At least ratios of the instantons coefficients 
have been successfully compared for $SU(2)$ with $N_f<4$ in 
\cit(instchecks) by a very tedious direct calculation.
This is  a certainly very encouraging independent check for 
the solutions,

\mabs {\sl Consistency checks from the dyon spectrum} 

The consistency checks on the dyon spectrum for $N_f=4$ are
quite similar to the $N=4$ case. $\tau$ does not depend on $u$ and 
once $\tau$ is generically fixed the lattice spanned by, say normalized
vectors, $1$ and $\tau$ is non degenerate and one has to check that 
bound states of the  stable dyons for which $(n_e,n_m)$ is coprime 
exist, since they must be present in theory as they occur in the 
${\rm SL}(2,\ZZ)$ orbit (on which the ${\rm Spin}(8)$ representations
are mixed) of e.g. the stable $(0,1)$ electron. 

For $N_f<4$ the lattice, spanned by $a_D(u)$ and $a(u)$, is $u$ dependent
and degenerates at a subspace $K:=\{p\in {\cal K}|{\rm Im}(k(p))=0\}$ 
in the moduli space, where we define $k(u):=a_D(u)/a(u)$. $K$ is called curve 
of {\sl marginal stability}. It was known for $N_f=0$ 
that beside the elementary electrically charged particles only 
configurations for the monodromy generating states and their 
$\theta$-shifted companions, i.e $(1,n)$ 
($N=4$ conventions) exist semiclassically. 
That turns out to be a general feature and the nontrivial prediction 
concerning new dyons are the existence of the $(2,n)$ bound states 
with $n$ odd  in the  $N_f=3$ theory \cit(swII), which were found 
in fact later \cit(ssz).

On the other hand it is an internal consistency check that the 
monodromy generating dyons are the {\sl only} magnetically charged 
states in the spectrum at semiclassical infinity. 
For that to work the topology of 
the set $K$ must be such that the existing dyons and 
elementary particles {\sl cannot} be transformed by a monodromy loop in 
the $u$-plane into unwanted states and continued to the semiclassical 
region without encountering a point on $K$, where all unwanted states 
can decay.  From (\ref{bpsmass})\footnote{We consider $m_i=0$} it is 
clear that that $k(u_s)={a_D(u_s)\over a(u_s)}=-{n_e\over n_m}$
is rational at the singularity $u_s$ due to the massless dyon $(n_m,n_e)$. 
So by construction $K$ contains these singularities. 
If $K$ exists outside the singular set it must be there a 
continuous codimension one subspace. That is easy to see, because 
$a, a_D$ are holomorphic, so $k$ is a harmonic function outside the 
singularities and $\dd k=0$ would imply that the Hessian vanishes, but
the Hessian of $k$ is proportional to ${\rm Im}\, \tau$ and can vanish only 
at the singularities, compare \figref{fundamental}. 
Now because of the $a_V$ term in 
(\ref{monmonopole}) $\dd k\neq 0$ also at the singularities, so that 
$K=0$ is a smooth real codimension one curve everywhere. 
Note again the importance of the $a_V\neq 0$ constant, if it 
were not present the ${\rm Im }\, k=0$ subset would be 
stuck at the singularities as the ${\rm Im}\, \tau=0$ ``curve'' in 
fact is.   

Let us discuss e.g. the situation for ${\rm SU}(2)$ with $N_f=0$ in the
$N=4$ conventions. The up to the $T$-shift closed path in the complex 
$k$ plane runs from $k(u=-\Lambda^2)=-1$, at which the 
$(1,1)$ dyon is massless, along the real axis to $k(u=\Lambda^2)=0$, 
at which the $(1,0)$ state is massless, to the $k(u=-\Lambda^2)=1$, 
at which the $(1,-1)$ state is massless, must therefore have a smooth 
pre-image in the $u$-plane. Hence $K$ has the right topology\footnote{From 
the explicit expressions e.g. (\ref{ada}) for $N_f=0$ it turns out that 
$K$ looks roughly like a symmetric ellipse with apheliae at 
$u=\pm \Lambda^2$ and periheliae at 
$u\approx \pm 0.86\Lambda^2 i$.}. Note that the values of $k(u)$
at $u=\pm \Lambda^2$ depend really on the strip $R_\infty$ we have 
chosen in \figref{fundamental}.

Besides the aspect that certain states can decay on $K$ there is a 
second important aspect related to the existence of $K$, which has 
been used \cit(febi) to construct the spectrum 
in the weak and the strong coupling region. 
No other states than the dyons $(1,n)$, which are responsible for 
the monodromies should become massless at $K$.
More precisely the existence of a stable dyon with coprime integers 
$(n_m,n_e)\neq(1,n)$ in a region of the moduli space, which can be connected 
(without crossing $K$) to a point on $K$ on which it would become 
massless is forbidden. It would lead 
to an additional singularity incompatible with the actual solutions.  
Since ${\rm Re}\, k(u)$ takes a continuous
set of real values this restricts the possible stable dyons drastically.
The argument is facilitated by working with the parameter which
labels truly inequivalent vacua, namely  $\alpha=u^2/\Lambda^4$. 
This parameterization identifies the $(1,0)$ monopole singularity 
at $u=\Lambda^2$ and the 
$(1,-1)$ dyon singularity at $u=-\Lambda^2$, as it should, and creates 
an $\ZZ_2$ singularity at the origin, which corresponds to the $\ZZ_2$
fixpoint in fundamental region of $\Gamma_0(2)$.  The monodromies 
can easily worked out from the solution (\ref{ada}). As expected they
contain phases $\rho$ with $\rho^4=1$ corresponding to the $\ZZ_4$-action on 
$\vec V$. The action on $\tau$ is generated by
$M^\infty=\left(\matrix{-1&1\cr 0&-1}\right)$, 
$M^1=\left(\matrix{1&0\cr-2&1}\right)$ and
$M^0=\left(\matrix{1&1\cr-2&-1}\right)$ with $M^0M^1=M^\infty$
and on $(n_m,n_e)^t$ it acts by $(n_m,n_e)^t\mapsto 
(M^{-1})^t (n_m,n_e)^t$. 
  
\figinsert{mod}
{The vector moduli space of $N_f=0$ ${\rm SU}(2)$ theory is parameterized by 
$\alpha=u^2/\Lambda^4$ and compactified to a sphere. $K$ divides the 
sphere into the strong coupling region containing $\alpha=0$ and the 
weak coupling region containing $\alpha=\infty$. On $K$ ${\rm Re}\ 
k(\alpha)$ runs continuously from $0$ at $\alpha=1$ to $1$ at $\alpha=1$. 
Note that there are branch cuts running along $\overline{0,1}$ and 
along $\overline {1,\infty}$. } 
{1.4truein}{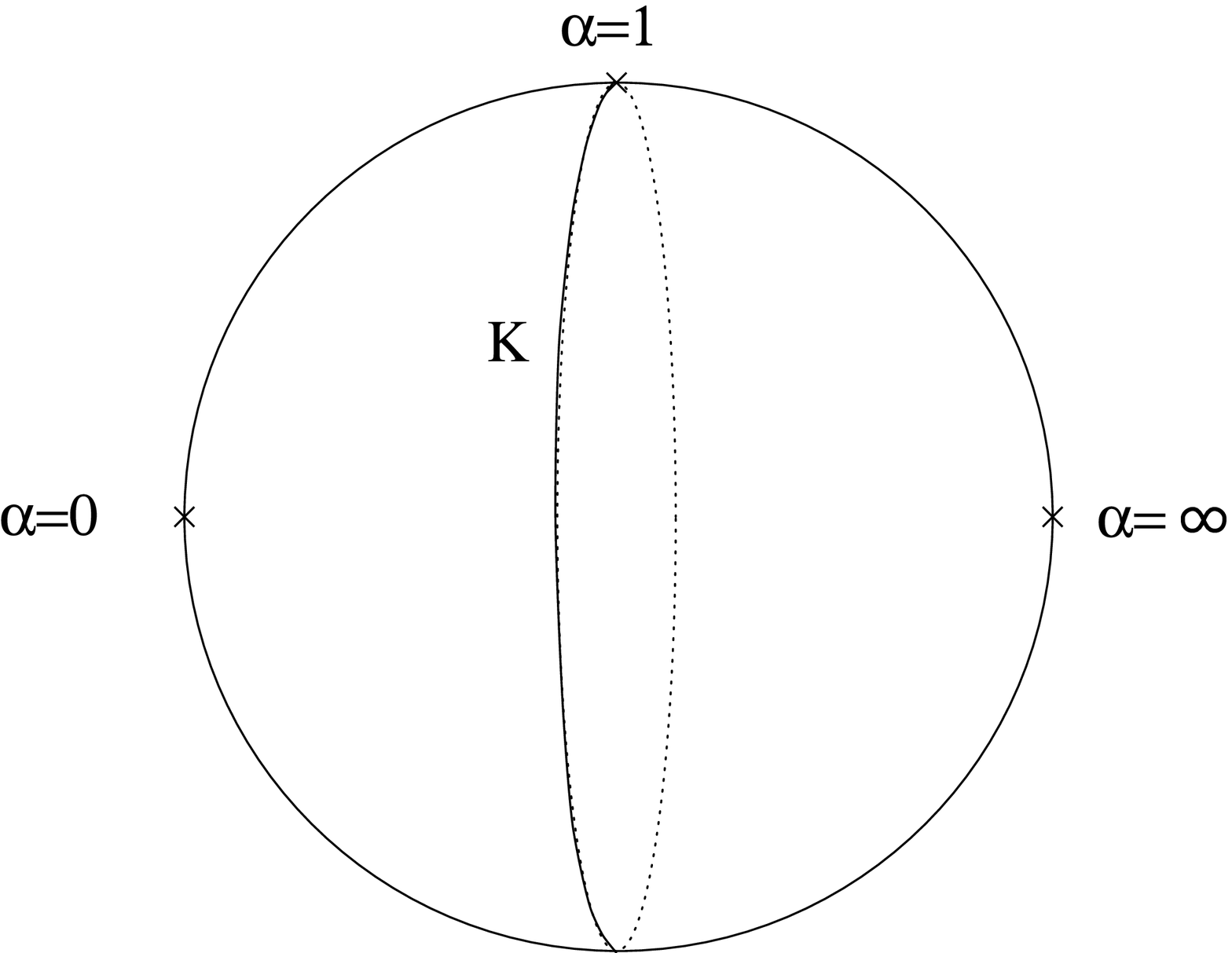}

From \figref{mod} we see that the $\alpha$-sphere is divided
by $K$ into two regions. 

\mabs {\sl The weak coupling spectrum:}
The region at infinity is governed by the $M^\infty$ monodromy. 
By this monodromy a stable state with $|n_m|>1$ can be converted 
into one for which $-n_e/n_m\in [0,1]$ hence all these states are forbidden. 
Beside the $(1,n)$ dyons and their antiparticles the $W^\pm$ bosons 
with $(0,\pm 1)$ are of course also allowed as they are stable under 
the $M^\infty$ action and do not become massless at $K$. 

\mabs {\sl The strong coupling spectrum:}
Now we ask again for the states $(n_m,n_e)$ for which $-n_e/n_m$ 
{\sl cannot} be brought by the $M^0$ action into the interval 
$[0,1]$. The conclusion is that for our choice of the strip $R_\infty$ 
this strong coupling spectrum consists only of the massless monopole 
$(1,0)$ and its antiparticle.

\mabs This picture predicts in particular at $K$  the decay of the
charged vector multiplet of the $W^\pm $ $(0,\pm 1)$ 
boson into two hyper multiplets of magnetic 
monopoles $(-1,\pm 1)$ and $(1,0)$, which is possible from
mass by charge and conservation, if $a_D/a\in [0,1]$. Note that
the $(-1,\pm 1)$ state  is identified by the $(M^0)^\pm$ monodromy 
with the $(-1,0)$ state.

Finally convincing evidence for the consistency of Seiberg-Witten 
solutions come from the connectedness of these theories via limits 
in the quarks masses. Starting from massive $N_F=4$ one gets 
indeed every other theory, by sending part of the quark masses 
to infinity, see the discussion above (\ref{masscurves}). Such 
arguments apply also to the higher rank groups and can be best 
discussed in the geometrical picture to which we turn now.

\section{The geometrical picture :}
\label{geompict}
\subsection{General ideas}

We have seen in (\ref{periodeqi},\ref{periodeqii}) that there were 
differential equations completely adapted the problem of finding
the exact BPS masses and the exact gauge coupling. Were
do this equations come from ? In context of the uniformization 
problem it was already observed by \cit(kleinfricke) that e.g. 
(\ref{periodeqi}) is the Picard-Fuchs equation fulfilled by the 
period integrals of a specially parameterization family of an 
elliptic curves ${\cal E}(u)$. In particular the solutions 
$\varpi_D$ and $\varpi$, which solve the uniformization problem 
\be
\tau(u)={\varpi_D(u)\over \varpi(u)}\ 
\ele(taudefii)
correspond to the integrals of the holomorphic differential $\omega$
(\ref{holform}) over homology cycles which generate $H^1({\cal E},\ZZ)$, i.e.
\be 
(\varpi_D(u),\varpi(u))=(\oint_B\omega ,\oint_A \omega)\ . 
\ele(periods)
That is not very surprising as the maximal discontinuous 
reparametrization  group of the torus is ${\rm SL}(2,\ZZ)$ and  if we 
insist to stay within a parameterization family, which obeys some 
additional finite symmetries the ${\rm SL}(2,\ZZ)$ will broken down 
to a subgroup of finite index in ${\rm SL}(2,\ZZ)$ just like e.g. 
$\Gamma(2)$. 
Moreover if one finds a form $\lambda$ such that 
the integrals $\oint_{\gamma}\lambda$ are well defined for
$C\in H^1({\cal E},\ZZ)$ and 
\be 
\p_u\lambda=\omega + {\rm exact\  form }
\ele(lamdef)
then we find from (\ref{taudef})
\be 
(a_D(u),a(u))=(\oint_{B} \lambda,\oint_{A} \lambda)\  .
\ele(adatorus)
From the above requirements it is clear that 

\noindent a.) $\lambda$ is a 
meromorphic form, as the holomorphic form is unique, but with 

\noindent b.) vanishing residues as otherwise the integral would 
depend on the path. 

In the cases with non-zero masses  condition b.) is too strong. In
fact the shift by $H$ in (\ref{dualitygroup}) has the explanation
that one picks up a contribution from the residue, if the cycle 
defining $a_D(u),a(u)$ undergoes a Lefshetz monodromy.  

\figinsert{phi}
{A scetch of the identification of the electro-magnetic charge lattice
with the  independent cycles of the torus }
{1.1truein}{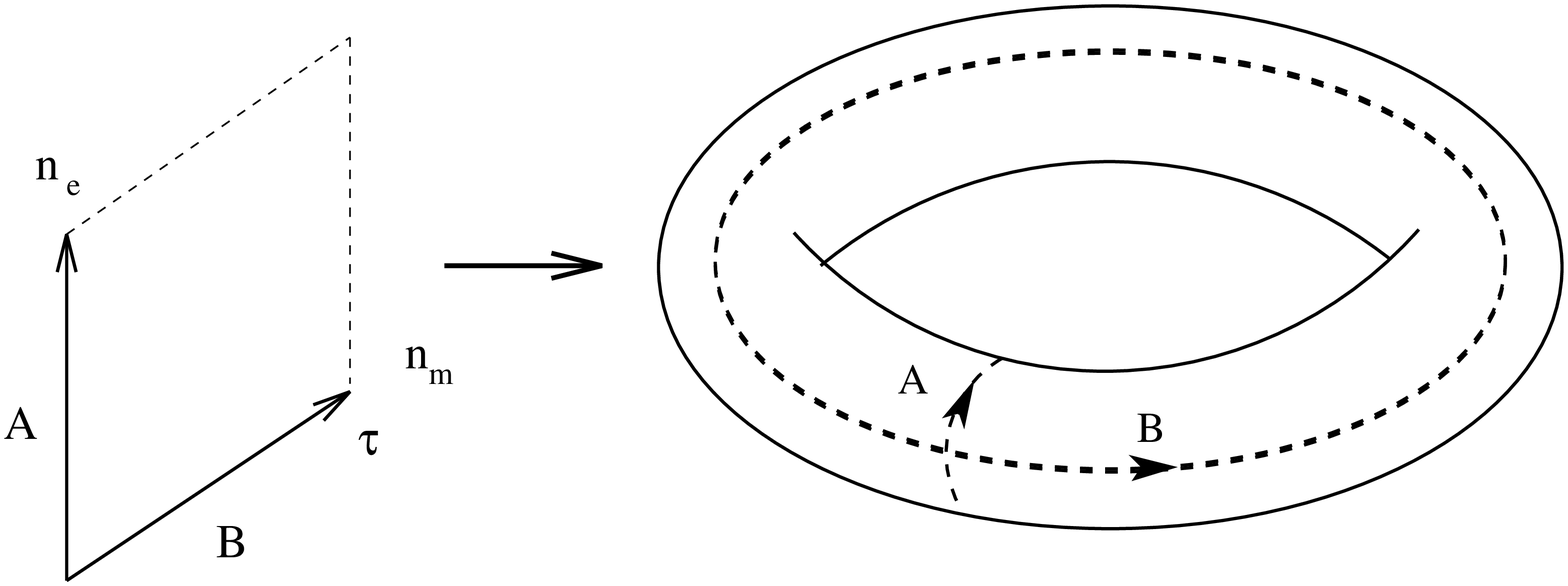}

Eq. (\ref{adatorus}) defines an identification of the  
electro-magnetic charge lattice $\Lambda$ \figref{lattice} with the lattice of 
integral homology $H_1({\cal E},\ZZ)$.
\be 
\phi:\Lambda \rightarrow H^1({\cal E},\ZZ)\ .
\ele(id) 
By the symmetries of the problem the identification can be made 
in various ways, e.g. for the scale 
invariant theories (\ref{id}) is actually up to ${\rm SL}(2,\ZZ)$ reflecting 
electro-magnetic duality.  For the scale dependent families 
${\cal E}(u)$ the ambiguity is reduced to a subgroup of 
$SL(2,\ZZ)$. That comes essentially because we have to identify
the particles, which become massless, with the vanishing 
cycles of the family. Another choice which was made  
in (\ref{adatorus}) was the orientation  of the cycles. 
Reversing globally the orientation correspond to the exchange of 
particles and antiparticles.

\mabs {\sl Positivity of the metric and Riemann bilinear relations:}

A very nice feature of this geometric interpretation is that  
${\rm Im}(\tau)$ is the normalized {\sl volume} of the torus, 
so positivity of the metric is guaranteed by construction. 
Let us see how this is derived and how it generalizes to 
guarantee positivity of the metric (\ref{metricinvariance}) 
as obtained from the periods 
of a general Riemann surface $X$. One a even dimensional 
manifold of real dimension ${\rm dim}=2 r$ with $r$ odd and  
$2 k={\rm rank}(H^r(X,\ZZ))$ 
one can always chose a symplectic basis $A^i$, $B_i$ $i=1,\ldots,k$ 
of $H_r(X,\ZZ)$ i.e. with the intersection pairing 
\bea(rl) 
A^i\cap A^j&=B_i\cap B_j=0\cr 
A^i\cap B_j&=(-)^r B_j\cap A^i = \delta^i_j \ .
\elea(intersectg) 
For our Riemann surfaces and the threefold CY we discuss later this 
choice is up to ${\rm SP}(2 k,\ZZ)$. If $r$ is even there will be a 
nontrivial signature associated to the bilinear pairing
(\ref{intersectg}), which we will calculate in section \ref{genprop} .

By Poincar\'e duality we can also chose a {\sl topological} basis for 
$H^r(X,\ZZ)$  $\alpha_i$, $\beta^i$ $i=1,\ldots, k$ 
with the following properties
\bea(rl) 
\int_X \alpha_i\wedge \alpha_j&=\int_X \beta^i\wedge \beta^j=0
\\ [ 3 mm] 
\int_X\alpha_i\wedge \beta^j&=(-)^r \int_X \beta^j\wedge \alpha_i=\delta^j_i
\\  [3 mm]
\int_{A^j} \beta^i&=\int_{B_j} \alpha_i=0
\\ [3 mm] 
\int_{A^j} \alpha_i&=\delta^{j}_i ,\ \ \int_{B_j} \beta^i=\delta^i_j
\ .
\elea(poinc) 
The topological basis we fix according to a given choice of topological 
cycles and it will not be holomorphic w.r.t. to the complex structure, 
which varies with the moduli. Using  the moduli dependent 
basis of {\sl holomorphic } forms $(1,0)$ on a Riemann surface: 
$\omega_i$ $i=1,\ldots,k$ (\ref{holoforms}) and the $(k,2 k)$ period matrix 
\be
({\bf W_D}_{ji},{\bf W}_{ji}):=\left(\int_{B_i}
 \omega_j,\int_{A^i} \omega_j\right)
\ele(periodmatrix)
the definition of $\tau$ from (\ref{taudefii},\ref{periods}) can be  
generalized  to ${\bf T}={\bf W}^{-1}{\bf W_D}$. 
In fact first for $g=1$ we get with (\ref{holform})
$i\int_X \omega\wedge \bar \omega=i\int_X dz\wedge d {\bar z}=2 
\int_X dx\wedge dy=2\, {\rm vol}(X)$ and on the other hand by 
developing $\omega$ and $\bar \omega $ in the basis $\alpha, \beta$,
i.e. $\omega = \varpi_D \beta + \varpi \alpha$,  we get
$i\int_X \omega\wedge \bar \omega=i  \varpi[\bar \tau-\tau]\bar \varpi=2 
|\varpi|^2
{\rm Im}(\tau)$, hence ${\rm Im}(\tau)>0$. By considering all bilinear
pairings between $\omega_i$ and $\bar \omega_j$ as well as between 
$\omega_i$ and $\omega_j$  one gets a straightforward generalization 
to higher genus \cit(griffithharris) which yields the first and second 
Riemann bilinear relation:
\bea(rl)
{\bf T}-{\bf T}^t&=0\cr
{\rm Im} [{\bf T}]&>0\ . 
\elea(riemannbil)
Also the identification (\ref{id}) generalized immediately
to higher rank lattices $\Lambda$ and general Riemann surfaces $X$. 
The electro-magnetic charge lattice $\Lambda$ with $H_1(X,\ZZ)$ maps 
the generalization of the symplectic bilinear form (\ref{dz}) to the 
intersection form (\ref{intersect}). 
Once the choice (\ref{intersect}) has made 
states with only electric charge quantum numbers will be identified 
with one sort of cycles, say the $A$-cycles, and purely magnetically 
charged must then be identified with the $B$ cycles. 
Now, if we have a special parametrisation family with $k$ deformations
$u_{i+1}$ $i=1,\ldots, k$ and have identified a meromorphic form 
$\lambda$ with $\omega_i=\p_{u_{i+1}} \lambda+ {\rm exact \ form}$ and 
$a_D^i=\oint_{B_i}\lambda$, $a^i=\oint_{A_i}\lambda$ then the 
positivity of the metric ${\rm Im}[{\bf T}_{ij}]={\rm Im}[\p_{a_i} a^j_D]$ 
in every direction in field space is guaranteed from 
${\rm Im}[{\bf T}]>0$, while the integration condition for the 
existence of ${\cal F}$ with $a_D^i=\p_{a_i}{\cal F}$ is 
${\bf T}-{\bf T}^t=0$ !

\mabs {\sl Lefshetz formula and one loop $\beta$-function:}

This makes  Riemann surfaces candidates whose periods can describe 
the effective action of theories with higher rank  gauge groups. 
The task is then to find parameterization families of Riemann surfaces, 
which have the right discrete symmetries and give the right monodromies. 
The monodromies of algebraic varieties on the middle homology 
$H_r(X,\ZZ)$ are determined by the cycles which shrink to zero volume 
at the singular degenerations of the variety. 
These are called the {\sl vanishing cycles}. More precisely the Lefshetz 
formula gives the monodromy action on an arbitrary cycle 
$C\in H_r(X,\ZZ)$ along a path in the moduli space around a {\sl complex}
codimension one locus, where one cycle $V\in H_r(X,\ZZ)$ vanishes as
\be 
M_V:C\mapsto C +(-1)^{(r+1)(r+2)/2} n (C\cap V) V\ ,
\ele(lefshetz) 
Here the integer $n$ depends on the local parameterization of the
singularity by the moduli, which is forced to us from the 
family (see below). In singularity theory the parameterization 
is chosen such that $n=1$. 

In the simplest cases the vanishing cycle has the topology of 
an $S^r$ and its self intersection number is 
$(V\cap V)=(-1)^{r (r-1)/2}(1+(-1)^r)$ (\cit(arnold), Lemma 1.4). 
In particular for $r$ odd, we get then a physical interpretation 
of (\ref{lefshetz}) as a shift from the one-loop $\beta$-function as
in the discussion below. For $r$ even on the other hand (\ref{lefshetz}) 
is a reflection in $H_r(X)$ on the hyperplane perpendicular 
to $V\in H_r(X)$, which is just a Weyl-reflection, if the 
intersection pairing is proportional to the Cartan matrix as 
in the zero dimensional example (\ref{cartanintersection}) and on
$K3$ surfaces \cit(bpv).

Up to phase factors, which can come from the forms 
$\lambda$ (or $\omega$) eq. (\ref{lefshetz}) describes 
also the monodromies on the period vectors $\vec V$ (or $\vec \varpi$). 
In view of the map (\ref{id}) one might ask, which states are mapped 
to these vanishing cycles and what is the local physics associated 
to the monodromy. For the most generic singularity, where precisely 
one cycle shrinks to zero as above, the answer is simple. 
The period $a_V:=\oint_V \lambda $ is proportional to the mass of the 
light {\sl charged } particle  $\Phi_V$ (and its antiparticle) which sets the 
infrared cut off in (\ref{eff}).
The ratio between the mass of this particle $\Phi_V$ and its magnetic 
(or electrical) dual particle will be zero at the singularity. 
After a basis transformation which diagonalizes 
${\bf T}$ we calculate the gauge coupling of the gauge boson(s), 
which couple locally to $\Phi_V$. From (\ref{beta}) we get 
$\tau_{V}\propto \kappa\log(a_V)+{ holomorphic}$ and  
the period dual to $a_V$ will be therefore 
$a_D^V=\oint_{V_D}\lambda \propto \kappa a_V \log(a_V)+
holomorphic$. This gives rise in the new basis rise to a shift, which 
corresponds in the old basis to (\ref{lefshetz}). 
If $\lambda$ is regular at the degeneration, as it turns out to be the 
case for singularities due to magnetically charged states, 
the mass of the particle will  actually go to zero. 
The Lefshetz theorem is quite useful to make consistency 
checks on the curves for the higher rank gauge groups \cit(klty).

In type IIB string theory an analogous picture arises, when a 
single cycle in the middle homology of a CY shrinks 
to a point \cit(strominger)\cit(gms). The wrapping of a $D$-3-brane 
around the vanishing 3-cycle leads to an object which looks from
the four dimensional point of view like a black-hole. By (\ref{blackmass}) 
its mass is proportional to the volume of  the vanishing cycle. In 
particular at the degeneration point $a_V=0$ this particle cannot
be integrated out, but has to be included in the Wilsonian supergravity 
action, just as in the rigid case the magnetic monopole. Very similarly it  
produces  an one-loop $\beta$ function logarithm in the coupling 
of the dual gauge field, which gives rise to the shift in 
(\ref{lefshetz}).

\mabs {\sl Variants of the idea:} 
The columns of the period matrix (\ref{periodmatrix}) 
define a lattice $\Lambda_X$ from which the Jacobian variety of the genus
$g$  Riemann 
surface is constructed as ${\cal J}(X)=\IC^g/\Lambda_X$. There 
is a natural generalization in which one imposes the condition 
(\ref{riemannbil}) on arbitrary non degenerate rank $r$ lattices 
$\Lambda^a$. The quotients $\IC^r/\Lambda^a$ with this restriction 
are known as {\sl abelian varieties}\footnote{The Riemann bilinear
relation ensure that these tori can be embedded into a projective 
space. To find the embedding, i.e. the problem discussed for the two torus
in the next section, is an interesting and hard problem 
\cit(mumford).}\cit(griffithharris). 
If the abelian variety has complex dimension greater then two, it is 
not necessarily the Jacobian variety of a Riemann surface. 
In physics context mainly Jacobian and Prym varieties occur. 
In the later cases the Riemann
surface admits an automorphism, so that periods get identified and 
the abelian variety is defined from the quotient of the period 
lattice. In fact in this way one can define infinitely many Riemann
surfaces of different genus, which describe the same gauge group. 
Cases which have no geometrical interpretation from a Riemann surface 
seem rare, comp. sect. (\ref{quantummoduli}).  

While there were probably no consistent $N=2$ theories in 4d before
the work of Seiberg-Witten, one can satisfy at least the basic 
consistency requirements with a Riemann surface, which admits a 
differential form $\lambda$ and gives rise to structure, like
in (\ref{adatorus}). General theorems about the degeneration 
of the periods integrals  imply that 
there are always local coordinates in the moduli space so that the
the periods degenerate no worse then with a logarithmic singularity at
the discriminate such that the effective action can always be 
determined (comp. section (\ref{curves})). This may lead to the discovery 
of interesting exotic $N=2$ theories in four dimensions.

\subsection{The curves for ${\rm SU}(2)$.}
\label{curves}
We will now discuss examples of Riemann surfaces, which correspond to 
gauge groups (with matter). I.e. the necessary discrete symmetries are 
realized and the periods have the prescribed physical monodromies and 
the right asymptotic behavior. For $SL(2,\ZZ)$ and the subgroups 
$\Gamma^0(N)$ and $\Gamma_0(N)$ the corresponding families of elliptic 
have been partly constructed  long time ago in the context of the 
uniformization problem. Their periods are hence, 
comp. sect. (\ref{uniformisation}), related to hypergeometric 
functions of type $_2F_1$. 
It is clear by the geometric ansatz and completely 
compatible with physics that the periods will always be Fuchsian 
functions\cit(deligne)\cit(yoshida). 
E.g. for pure $SU(3)$ the holomorphic periods where found \cit(klt) 
to fulfill Appells\cit(appell)\cit(erdelyii) Vol. I 
$F_4({1\over 3},{1\over 3},{2\over 3},{1\over 2},4 {u_2^3\over 27\Lambda^6}, 
{u_3^2\over \Lambda^6})$ system and the $a^i$, $a_D^i$ periods 
fullfil Appells $F_4({1\over 6},{1\over 6},{1\over 3},
{1\over 2},4 {u_2^3\over 27\Lambda^6}, 
{u_3^2\over \Lambda^6})$ system (see (\ref{sym},\ref{suncurves}) 
for the definition of $u_i$), but in contrast to the ${\rm SL}(2,\ZZ)$ 
case the functions are in general not known.

\subsubsection{\sl Algebraic form of the torus:} 
\label{algtor}
Let us shortly review how the algebraic description of 
the torus arises \cit(lang)\cit(serre). 
Define the torus in the standard 
form $T=\IH/\Lambda$, were $\Lambda$ is spanned by
the {\sl periods} $\varpi_1$ and $\varpi_2$.  
I.e. $T$ is the fundamental cell of $\Lambda$ identified 
(orientation preserving) on opposite sides. 
We might normalize the periods such that the lattice is 
spanned by $\pi \tau$ with $Im(\tau)>0$ and $\pi$. 
Now we want to map $T$ into a set given by an algebraic constraint. 
To do this one needs first well defined functions on $T$, 
i.e. $f(z)=f(z+\pi)=f(z+\pi\tau)$ for $z\in \IH$. 
The  Weierstrass function
\be
\wp(z,\tau)={1\over z^2}+\sum_{\omega\in \Lambda\setminus \{0\}}
\left({1\over (z-\omega)^2} -{1\over \omega^2}\right)
\ele(weierf)
has this property. It is easy to see that it converges in 
$T$, but has poles on the lattice sites. Moreover $\wp$ 
fulfills the differential equation 
\be 
\left({\dd \over \dd z}
\wp\right)^2=4 \wp^3-g_2\wp-g_3
\ele(weierdiff)
with $g_2(\tau)=60\sum_{\omega \in \Lambda\setminus \{0 \} }{1\over 
\omega^4}=
2/3^2 E_4(\tau)$ and\foot{Note that because of the normalization of the lattice $\Lambda$ 
we have $G_k=\pi^{2 k}\sum_{\omega \in \Lambda \setminus \{0\}}{1\over \omega^{2k}}$, with
$G_k$ as in \cit(lang).}
$g_3(\tau)=140\sum_{\omega \in \Lambda\setminus 
\{ 0 \} }{1\over \omega^6}=(2/3)^3 E_6(\tau)$. 
By identifying $x=\wp(z),y={\dd \over \dd z}\wp(z)$ every point 
$z$ in $T$ is mapped to a point on the algebraic constraint in $\IC^2$ 
\be 
y^2=4 x^3 - g_2 x - g_3 \ . 
\ele(weierstrass) 
This is true apart from the lattice points on $T$, which are
mapped to infinity in the $x,y$-plane. This must be rectified 
by compactifying the latter to an $\IP^2$. As it is clear from 
the construction the holomorphic differential 
$\omega:=\dd z$ can be written as
\be 
\ds { \dd z = {\dd x\over y}}
\ele(holform) 
and gives by integration over the cycles just the 
normalized period vector $(\pi,\pi \tau)$. 

$E_4$ and $E_6$ are known as Eisenstein series, which are normalized so that 
they have a nice $q:=\exp(2 \pi i \tau)$ expansion
\bea(rl)
E_4(\tau)&=1+240\sum_{n=1}^\infty{n^3 q^n\over 1- q^n}\\ [ 3 mm]
E_6(\tau)&=1-504\sum_{n=1}^\infty{n^5 q^n\over 1- q^n\, .}
\eea
They are the, up to multiplication, unique automorphic 
(or modular) functions of weight $4$ and $6$, i.e. 
$E_{2k}({A\tau +B\over C \tau+D}) = (C\tau+B)^{-2k} E_{2 k}(\tau)$, 
which are holomorphic in the whole upper half-plane $\IH$. 
Every modular function with this holomorphicity 
property and weight $2 k$ can be written as degree $2 k$ weighted 
polynom in $E_4$ and $E_6$ (or $g_2$ and $g_3$ of course)\footnote{These
facts appear in any review on elliptic functions see e.g. 
\cit(serre),\cit(lang).}. 
$E_4$ has simple zero at $\tau=i$ and $E_6$ at $\tau=\exp(2\pi i/3)$. 
The value at infinity is $g_2(i\infty)={120 \over \pi^4}\zeta(4)=2^2/3$
and $g_3(i\infty)={280\over \pi^6}\zeta(6)=(2/3)^3$. So the 
combination with lowest modular weight, which has a simple zero at infinity 
is $g_2^3-27 g_3^2=2^{12} \eta^{24}$ proportional to the 24th power of the 
Dedekind $\eta$-function, which has  product representation $\eta:=q^{1\over 24}\prod_{n=1}^\infty (1-q^n)$.
The $j(\tau)$ function is the unique modular invariant function with a 
simple pole at infinity
\be
J(\tau)={g_2^3\over \Delta}\  {\rm where } \  
\Delta=g_2^3-27 g_3^2 
\ele(jinvariant)
is the discriminant of the elliptic curve\footnote{See appendix C and 
below.}. 
Up to a factor $1728$ it has an integral expansion
\be
j(\tau)=1728 J(\tau)={E_4^3\over \eta^{24}}={1\over q} + 744 + 196884 q+21493760 q^2 + \ldots
\ele(jfunct)
In the following we will see $j$ frequently as a function of the 
specific parameter $u$ of the parameterization family ${\cal E}(u)$.
In this case the identification  $j(u)=j(\tau)$ will give us an 
invariant characterization of the family ${\cal E}(u)$! 
A similar theory for the modular functions of higher genus
Riemann surfaces is discussed e.g. in \cit(Igusa).

\mabs {\sl Universal Picard-Fuchs equation:}

The integrals 
$$
\varpi_C=\oint_C \omega =\oint {\dd x\over \sqrt{4 x^3-g_2(u) x-g_3(u)}},
$$ 
with contours $C$ as in \figref{sliced}, called {\sl elliptic} integrals; 
they are not elementary. Instead of direct integration, which can
be done only after expanding the integrand, one can derive a 
differential equation for them. This is done by deriving a differential
operator with the property ${\cal L}(u) {\dd x\over y} = 
{\p f\over \p x} \dd x $. As $C$ is closed ${\cal L}(u) \varpi_C=0$ and
since there are only two independent solutions corresponding to the two
independent integrals over $A$ and $B$
cycles a second order ${\cal L}(u)$ must exist.  Such differential 
equations are called Picard-Fuchs equations. We explain in 
Appendix \ref{c} two ways how the Picard-Fuchs equations can be derived.

To appreciate the r\^ole of the $j$-invariant and link
the discussion here to the one in section (\ref{reconstruction}), 
note that every elliptic curve with an arbitrary parameterization 
$s$ can be brought in the form (\ref{weierstrass}), see footnote \ref{jfoot},  
and the Picard-Fuchs equation can then be written in the useful 
universal form
\bea(rl) 
&\varpi ''+p\varpi'+q \varpi =0\quad {\rm with} \\ [ 3 mm]
p&=-\log'({3\over 2 \Delta}(2 g_2 g_3'-3 g_2' g_3)), \\ [ 2 mm]
q&={1\over 12} (p \log'\Delta + \log''\Delta )-{1\over 16} 
(g_2 (g_2^2)'-12 (g_3^2)') \ ,
\elea(upf) 
where $'={\dd\over \dd s}$.

If one now changes the coordinate $s\rightarrow J=j/1728$ one gets 
an universal Picard-Fuchs equation for the rescaled periods 
$\Omega=\sqrt{g_2\over g_3} \varpi$ depending only on $J$, 
see e.g. \cit(klryI) and an universal expression for
\be
Q=\left({3\over 16 (1-J)^2}+{2\over 9 J^2}+{23\over 144 J 
(1-J)}\right)\ ,
\ele(qj)
which we recognize after a short calculation as the $Q$ appearing 
in the Schwar\-zian differential equation (\ref{unif}) for 
$\alpha_0={1\over 3}$, $\alpha_1={1\over 2}$ and $\alpha_\infty=0$. 
I.e. $j(\tau)$ is the inverse of the developing map for $SL(2,\ZZ)$
itself.

\mabs {\sl The $N=4$ and $N=2$ $N_f=4$ curves:} 
The tori for the scale invariant theories are expected from the 
sections (\ref{sduality}) and (\ref{dyonsymmetries}) to exhibit exact
${\rm SL}(2,\ZZ)$ invariance. Therefore they should parameterized
by the $u$ independent parameter $\tau$, that is (\ref{weierstrass}) 
with $g_2(\tau)$ and $g_3(\tau)$ is in principle the correct form. 
In view of (\ref{bps}, \ref{bpsmass}) $a_D,a$ depends however on $u$
\bea(rl)
a_D=& \tau a \cr
a=&\left\{ \begin{array}{rl} 
           {1\over 2 }\sqrt{2u}& {\rm for } \ N_f=4 
           \cr     
           \sqrt{2u}& {\rm for } \ N=4 
          \end{array}\right.
\eea
Because of (\ref{diffbzg}) this implies that $\vec \varpi=
{\cal N} \sqrt{2/u}(\tau,1)^t$ with ${\cal N}=1/4$ for $N_f=4$ and 
${\cal N}=1/2$ for $N=4$. We can rescale $dz \rightarrow 
{\cal N}\sqrt{2/u} dz$ to get that. For later comparisons in scaling 
limits one wants to work always with the standard $(1,0)$ form $\dd z=\dd x/y$.
So one  rescales in addition $x\rightarrow x/ u$ and 
$y \rightarrow {y u^{-{3\over 2}}/2}$. This leads to an $u$ 
dependent form of the curve 
\be 
y^2=x^3-{1\over 4} g_2(\tau) x u^2-{1\over 4} g_3(\tau) u^3,
\ele(conformcurve)
while the $(1,0)$ form is transformed back to the standard one. 
The left hand side of (\ref{conformcurve}) can be factorized  
$y^2=\prod_{i=1}^3(x-e_i(\tau)u)$, where the zeros are 
given by the Jacobian Theta functions \cit(serre)  
$e_1(\tau)-e_2(\tau)=\theta_3^4(\tau)$, 
$e_3(\tau)-e_2(\tau)=\theta_2^4(\tau)$,
$e_1(\tau)-e_3(\tau)=\theta_4^4(\tau)$. 

\mabs {\sl The $N=2$, $N_f \le 4$ curves:}

It was explained in \cit(swII) how to use the global ${\rm SO}(8)$ 
symmetry acting on the quarks to incorporate the bare masses into 
(\ref{conformcurve}). An alternative derivation using the constraint
on the residua of $\lambda$ form the inhomogeneous transformation
law in (\ref{dualitygroup}) was also given in \cit(swII). 
A particular nice representation of the corresponding 
curve\footnote{And generalizations of this curve to other 
gauge groups.} was found in \cit(arsh) 
\be
y^2=(x^2-u)^2-4 h (h+1)\prod_{i=1}^4(x-m_i- 2 h \mu) 
\ele(mastercurve)
with $h={\theta_2^4\over \theta_4^3-\theta_2^4}$, 
$\mu={1\over N_f}\sum_{i=1}^4 m_i$ and 
$$\lambda={x-2 h \mu\over 2 \pi i} \dd {\rm log}
\left(x^2-u-y\over x^2-u+y\right)
\ . $$
The curves for the asymptotic free $N_f<4$ theories can be obtained 
from  (\ref{mastercurve}) by considering the double scaling limit in which
$M\rightarrow \infty$ and $ \tau \rightarrow i\infty$ such that 
$\Lambda^{4-N_f}:=64\sqrt{q} M^{4-N_f}$ defines the finite scale of 
the $N_f<4$ theory. The leading terms of $\theta^4$ functions are 
$\theta_2^4=16 q^{1/2}+\cO(q^{3/2})$,
$\theta_3^4=1+8 q^{1/2}+\cO(q)$,
$\theta_4^4=1-8 q^{1/2}+\cO(q)$. With this one gets for the 
$N_f=0,\ldots,3$ cases the curves
\be
y^2=(x^2-u)^2-\Lambda^{4-N_f} \prod_{i=1}^{N_f}(x+m_i)
\ele(masscurves)
To show the equivalence of these curves with the ones in\cit(swII) we
check that the $j$-invariants\footnote{\label{jfoot} 
A curve which is given by a quartic 
constraint $y^2=\prod_{i=1}^4 (x-e_i)=ax^4+4bx^3+6c x^2+4 d x+e$ 
is converted to a cubic form $y^2=\prod_{i=1}^3(x-\tilde e_i)=
Ax^3 +3B x^2 + 3 C x +D$ by mapping one of the zeros $e_i$ to 
infinity, see e.g. \cit(erdelyii), and vice versa. 
For convenience we note that from the cubic we get 
$g_2=3 2^{2/3}(B^2-AC)$ and
$g_3=3 A B C-A^2D-2B^3$, while from the quartic we get 
$g_2=a e-4 b d+3 c^2$ and $g_3=a c e+2 b c d - a d^2-c^3- e b^2$.}  
for $N_f=0,1$ are identical with ones of the corresponding 
curves in \cit(swII). For the others a 
shift in the origin of $u$ is required e.g. for  $N_f=2$ 
$u\rightarrow u-\Lambda^2/8$. Because of the absence of the 
a symmetry in the $u$-plane there is an unfixed shift in $u$ for the 
$N_f=3$ curve. The last reference in \cit(instchecks) suggest
another choice of the shift then \cit(swII).  
E.g. with (\ref{upf}) or the formalism in appendix \ref{c} one can
easily obtain the Picard-Fuchs equations for $\oint \omega$ and 
$\oint \lambda$ and the equivalent of (\ref{insteq}). Explicit 
expressions for periods and prepotential appear in the literature, 
see e.g. \cit(iy)\cit(om) and with emphasis on the modular properties 
\cit(nahm)\cit(bs). E.g. for the cases with vanishing bare mass 
the Picard-Fuchs equations  for ${\cal L}_{N_f}\oint \lambda=0$ are  
\bea(rl)
{\cal L}_{N_f}&=\ds {{\dd^2\over \dd^2 u}-{1\over p_{N_f}(u)}, \ 
{\rm for } \ N_f=0,\ldots, 3}\\ [3 mm]
p_0&=4(u^2-\Lambda_0^4),\quad p_1=4 u^2+{27\Lambda_1^6 \over 64u},
\\ [ 3 mm]
p_2&=4 (u^2- {\Lambda_2^4\over 64}),\quad p_3=4u(u- {\Lambda_3^2\over 64})\ .
\elea(pfeqs)
and the first few coefficients of the prepotentials are also 
calculated in \cit(iy).

Theories with massive matter have very interesting 
singularities, where electric and magnetic charged states become 
simultaneously massless. As discussed in \cit(apsw) this leads to 
conformal theories. The different conformal fixed points in 4d 
can be classified \cit(apsw).

\subsection{Hyperelliptic curves and application of the Lefshetz formula}
\label{hyperelliptic}
One may recast the equation (\ref{weierstrass}) in the form
\be 
y^2(x,u)=p(x,u)=\prod_{i=1}^4 (x-e_i(u))\, .
\ele(genell)
This defines the torus as a double covering of the $x$-plane, 
which is compactified to $\IP^1$ and has branch cuts along 
$\overline{e_1,e_2}$ and $\overline{e_3,e_4}$. The $A$- and 
$B$-cycle are defined in \figref{sliced} .

\figinsert{sliced}
{Integration contours along the $A$ and $B$-cycle in
the double covered $x$-plane. The plane in front and the plane behind 
are glued along the upper and the lower banks of the cuts. Both 
planes will be compactified to a $\IP^1$.} 
{1truein}{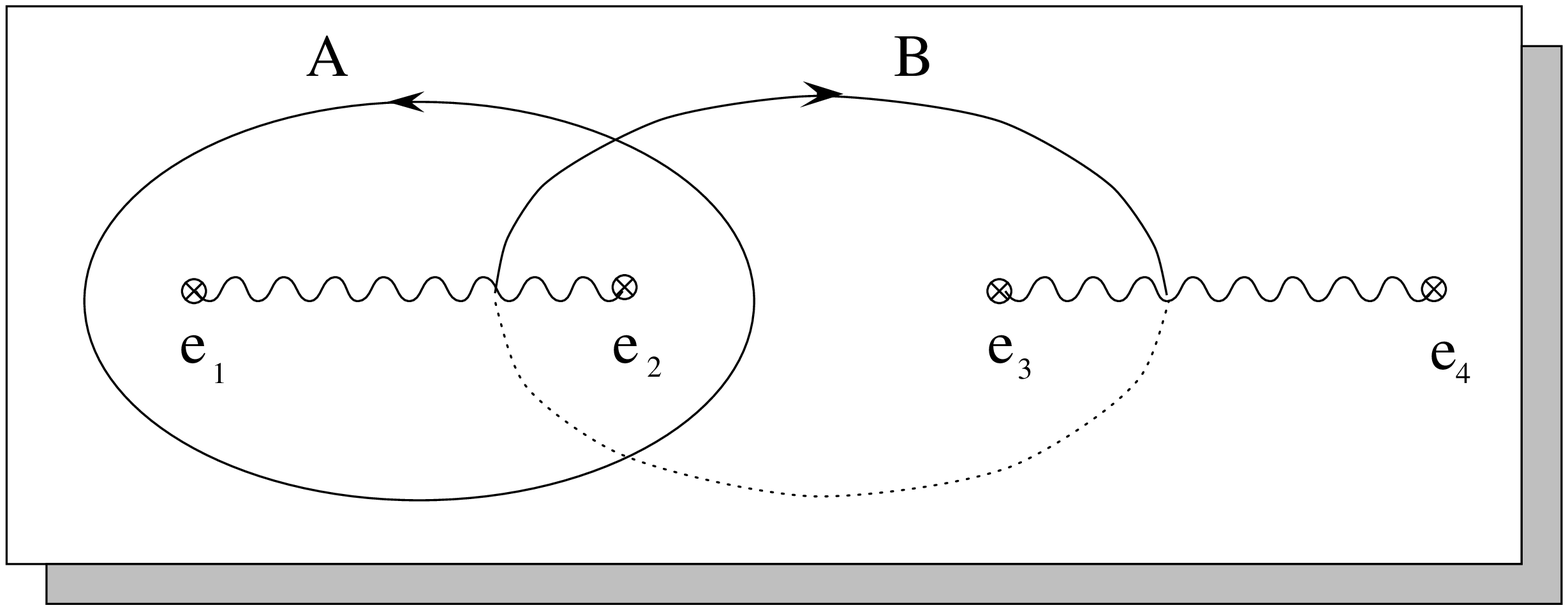}

It is clear by \figref{sliced} that the construction (\ref{genell})
can be generalized to Riemann surfaces with not genus one but 
$r$ holes. If $p(x,u)$ has degree $2(r+1)$, there will be 
$r+1$ cuts $\overline {e_i,e_{i+1}}$ $i=1,3,\ldots,2r+1$. 
Such genus $r$ curves are known as hyperellitic curves. They 
have $2r-1$ independent parameter namely the $2r+2$ locations 
of the zeros minus the three parameter of the invariance 
$SL(2,\IC)$ invariance group of the $\IP^1$ on which $x$ is 
compactified. A general $g>1$ Riemann surface has by Riemanns count 
$3g-3$ parameter, see e.g. \cit(griffithharris) section 2.3. 
To obtain the ${\rm SU}(n)$ curves we have to define special
parameterization families of genus $g=n-1$ with $n-1$ parameters.

\mabs {\sl The discriminant:}
The Riemann surface becomes singular when the roots $e_i(u)$ collide or 
differently said when one (or more) one-cycle(s) vanish.
The codimension one locus in the moduli space where this 
happens is called {\sl discriminant} and defined as the 
zero locus of
\be 
\Delta(u)=\prod_{i<j} (e_i(u)-e_j(u))^2\ .
\ele(disci)
It is essentially that we chose a compactification of the 
moduli space. For instance in the $SU(2)$ case if we do not 
compactify the $u$-plane to a $\IP^1$ we would miss semiclassical 
singularity, which is at infinity in the $u$-plane. 

Precisely at the points in the moduli space where $\Delta(u)=0$ 
the dimension of the normal space to the constraint 
$P:=y^2-p(x,u)$ is not minimal and an equivalent way of defining 
the discriminant is therefore 
as the locus in the moduli space, where the homogenized constraint 
$P=y^2z-4x^3+g_2(u)x z^2+g_3 z^3$ fails to be 
{\sl transversal} in $\IP^2$. That is at points where
$P=0$ and  $dP=(\p P/\p x)\dd x +(\p P/\p y)\dd y + 
(\p P/\p z)\dd z=0$ have common solutions in $\IP^2$ i.e. 
for  $(x_0:y_0:z_0)\ne (0:0:0)$. The corresponding locus is known 
as resultant of the equations $\p P/\p x_i=0$, $P=0$ and can be 
easily calculated, without determining the roots of course, 
see appendix \ref{c}. That yields in the $(x,y)$ patch for 
(\ref{weierstrass}) $\Delta(u)$ as defined in (\ref{jinvariant}).
This definition of the discriminant generalizes immediately
to hypersurfaces of arbitrary dimension.

\mabs {\sl Application of the Lefshetz formula:}
In the $u$-plane the singularities of the family of tori
occur just at points. These points are called stable if only 
two of the branch points come
together. The monodromy at a stable branch is very simple to 
describe by the  Lefshetz formula. We define a reference 
point $u_0$ and consider a closed counter clockwise loop $G$ 
in the $u$-plane encircling the singular point $u_V$ at 
which a cycle $V$ vanishes. The  monodromy on a cycle 
$C\in H^1({\cal E},\ZZ)$ then given by (\ref{lefshetz}). 

The angle $n \pi$  corresponds to the relative movement of the branch 
points defining the vanishing cycle around each other if we complete the 
loop $G$ in the $u$-plane.  The factor $n$ can be either 
determined from the local form of $p(x,u)\propto (x-e_+)(x-e_-)(x^2-u^n)$ 
at the singularity or equivalently from the leading behavior of 
the discriminant 
\be 
\Delta(u)=(u-u_V)^n+O((u-u_V)^{n+1}).
\ele(nfac) 
In the present case one can proof (\ref{lefshetz}) directly by 
graphically studying the deformations of the contours, 
or the leading parts of the  integrals at the degeneration, 
the general proof uses the latter approach and can be 
found in \cit(arnold).

Let us consider this for the simple example of the stable 
$SU(2)$ curve 
\be
y^2=(x-\Lambda^2)(x+\Lambda^2)(x-u),
\ee 
which has obviously the same $j$-function (\ref{gamiij}) as our 
weighted representation (\ref{poli}). The discriminant is 
by (\ref{jinvariant}) and footnote (\ref{jfoot}) 
$\Delta=4 (\Lambda^2)^2 (u-\Lambda^2)^2(u+\Lambda^2)^2$, 
where the factor $\Lambda^4$
corresponds to the singularity at $u\sim \infty$, we consider 
$(\Lambda^2:u)$ as homogeneous  variables of $\IP^1$. All degenerations
are stable and we may identify in \figref{sliced} the $x$-plane
with the $u$-plane that is $e_1=-\Lambda^2$, $e_2=\Lambda^2$, $e_3=u_0$ 
and $e_4=\infty$. Now if  $u=e_3$ loops around $e_2=\Lambda^2$ 
the $B$ cycle vanishes $V=B$ and $(B,A)^t\mapsto 
\left(\matrix{1&0\cr -2&1}\right)(B,A)^t$. According to the
map from the charge lattice $\Lambda$ to $H^1({\cal E},\ZZ)$ 
we have in general $V=n_m B + n_e A$ and the corresponding 
monodromies on $\vec V$ or  $\vec \varpi$ become 
exactly (\ref{genmonodromy}). The monodromy around infinity is 
$T^{-2}$ on the cycles, but there will we a sign change from the 
continuation of the forms, hence we reproduce also $M^{\infty}$.   
Note that the $2$ in $M^\infty$ comes by (\ref{nfac}) 
from the leading  behaviour $(\Lambda^2)^2$ of $\Delta$ at 
infinity . 

\mabs {\sl The general degeneration of the periods.}

The discriminant in multi moduli cases will be a, in general  
singular, algebraic variety of codimension one in the moduli space,
which can have many components. 
For instance if more then two zeros of $p$ collide at a point 
in the moduli space then transversality fails for the 
discriminate as subspace of the moduli space itself $\dd \Delta=0=\Delta$. 
It was shown by Hironaka \cit(hironaka) in a much more general context, 
which is also relevant to the moduli space of CY manifolds, 
that such singularities can be always, but not uniquely, resolved by
quadratic transforms (compare sect. \ref{swtII}), such that the 
discriminante components become normal 
crossing divisors. This procedure is important to get variables $z_i$ 
in which the solutions of the Picard-Fuchs equation have only Fuchsian 
singularities\cit(deligne)\cit(yoshida). 
I.e. around the normal crossing divisors at $z_i=0$
the solutions can always be locally expanded as $z_1^{p_1/q_1}
\ldots z_r^{p_r/q_r} \sum_{\vec n,\vec k}
\log^{k_1}(z_1)\ldots \log^{k_r}(z_r)  c_{\vec n,\vec k} z^{\vec n} +holomorphic $, where $p_i,q_i \in \ZZ$, 
$k\in \IN_0$ and $\sum_{i=1}^r k_i\le {\rm dim}(X)$  
after a suitable resolution procedure, comp. \cit(arnoldII) Chap II.3.8. 
For this procedure we discuss an explicit example in (\ref{swtII}).

\subsection{The ${\rm SU}(n)$ curves}

\subsubsection{The classical moduli space}
As for ${\rm SU}(2)$ the flat directions of (\ref{dpot}) 
will be parameterized for any gauge group by the fields in the 
Cartan sub-algebra. For ${\rm SU}(n)$ we may choose for 
the moment $\phi=\sum_{k=1}^{n-1} a_k H_k$ with $H_k=E_{k,k}-
E_{k+1,k+1}$, $(E_{k,l})_{i,j}=\delta_{ik}\delta_{jl}$ 
as coordinates of the classical moduli. For generic $a_i$ 
the gauge group will be broken to the maximal torus $U^{n-1}(1)$.
If the some eigenvalues $e_i(a)$ of $\phi$ coincide $SU(n)$ is 
only broken to a bigger subgroup $H\subset SU(n)$, e.g. in case of two 
eigenvalues to ${\rm SU}(2)\times {\rm U}^{n-2}(1)$. 
As in the $SU(2)$ case the $a_i$ parameterize a multicover of 
the physical moduli space consisting of orbits under the 
Weyl-group. The Weyl-group acts by conjugation on $\phi$ 
therefore the following characteristic polynomials are Weyl-invariant 
\bea(rl) 
F_{A_{r}}(x,\vec u)&={\rm det}\lbrack x- \phi\rbrack=
                \prod_{i=1}^{n} (x-e_i(a))\\ [ 2 mm]            
                &=\ds{x^n-\sum_{l=1}^{n} u_{l}(a) x^{n-l}}\\ [2 mm] 
\elea(sym)
and so are their coefficients, which are the symmetric polynomials 
in the $e_i$: $u_k(a)=(-1)^{k+1} \sum_{j_1<\ldots <j_k}$ $ e_{j_1} 
\ldots e_{j_k}$, see e.g. \cit(langalg). These expressions can be used
as the Weyl-invariant parameters. They are up to signs the Chern 
classes $c_i(\phi)$ of $\phi$, a definition we will need later
\bea(rl) 
{\rm det}\lbrack x- \phi\rbrack &=\sum_{i=0}^n(-)^i  x^{n-i} c_i(\phi)\\ [2 mm]
              &=x^n {\rm det}(1-{\phi\over x})=x^n e^{{\rm Tr} 
            \log (1-\phi/x )}\\ [2 mm] 
              &=x^n\exp\left(-\sum_{k=1}^{\infty} 
               {{\rm Tr}(\phi^k)\over x^k k}\right)\ .
\elea(symchern)
Due to the tracelessness of $\phi$ the first Chern class vanishes.
Under the global non-anomalous $Z_{2N_c}$ discussed  
above (\ref{symmetryi}) the $u_k$ transform with charge $k$. 

Following\cit(arnold) we call $F_{A_r}(x,\vec  u)$ the miniversal 
deformation of the $A_r$ singularity and 
$W^z_u=\{x\in \IC:F(x,\vec u)=z, ||x||<\epsilon \}$ 
its {\sl level set}. It is in our case zero dimensional and 
we can apply Lefshetz formula to its ``middle'' homology. E.g. the
the zero level  set $W^0_{u_0}$  of $F_{A_r}(x,0,\ldots,0,1)=x^{r+1}-1$
are points in the $x$-plane, for our choice of $\vec u_0$ the unit roots 
$e_k=\exp(2 \pi i(k-1)/(r+1))$, $k=1,\ldots,r+1$ with $\sum_{k=1}^{r+1} e_k=0$.
A basis of vanishing cycles which correspond to the simple roots $\alpha_k$ 
of $A_r$ is $V_k=e_k-e_{k+1}$ $k=1,\ldots,r$. The non-vanishing 
intersections  are
\bea(rl)
V_i\cap V_i&=2\\ [ 2 mm]
V_i\cap V_{i+1}&=-1 \ ,  
\elea(cartanintersection) 
i.e. the Cartan matrix of $A_r$ and the Lefshetz 
formula  $M_{V_k}: X\mapsto X-( X \cap V_k) V_k$ is identified 
with the Weyl-reflections on the simple roots 
$S_{\alpha_k}: x \mapsto x - 2 {\bra x,\alpha_k\ket \over \bra 
\alpha_k \alpha_k\ket}\alpha_k$, which generate the Weyl-group.
The mass of the gauge boson $W_{\alpha_k}$ due to the Higgs effect 
in the Coulomb branch is 
\be 
M=|Z_{\alpha_k}|, \ {\rm with } \ Z_{\alpha_k}=e_k-e_{k+1}=:\vec n^k_e \vec a .
\ee
and corresponds precisely to the distance of the points $e_i$ in the
$x$-plane, see \figref{su6}. We may label the zeros of $F_{A_k}$ by 
$e_{\lambda_i}$ with $\vec \lambda_i \vec a:=e_i$. 
The $\vec \alpha_k=\vec \lambda_k-\vec \lambda_{k+1}$ become then the 
root vectors in the Dynkin basis. 

The {\sl level bifurcation set} is the discriminant of zero level set 
$W_u^0$ and since $\Delta_0=\prod_{i<j}(e_i-e_j)^2$ it gives the loci of
classical enhancement of the gauge group, where the mass 
gauge bosons $W_\alpha$ with $\alpha $ a positive root vanishes. 
E.g. for 
\bea(rl)
{\rm SU}(2):&  \Delta_0=u_2 \\ [ 2 mm]
{\rm SU}(3):&  \Delta_0=4 u_2^3-27 u_3^2\ . 
\eea    

\figinsert{su6}
{Level set and vanishing cycles for $A_r$. 
All lines correspond to vanishing cycles associated with non-abelian
gauge bosons. The solid lines represent the simple roots. 
We may chose a orientation for the other cycles such that they are 
associated with positive roots. By orientation reversal one gets 
then the anti-particles.}
{2.1 truein}{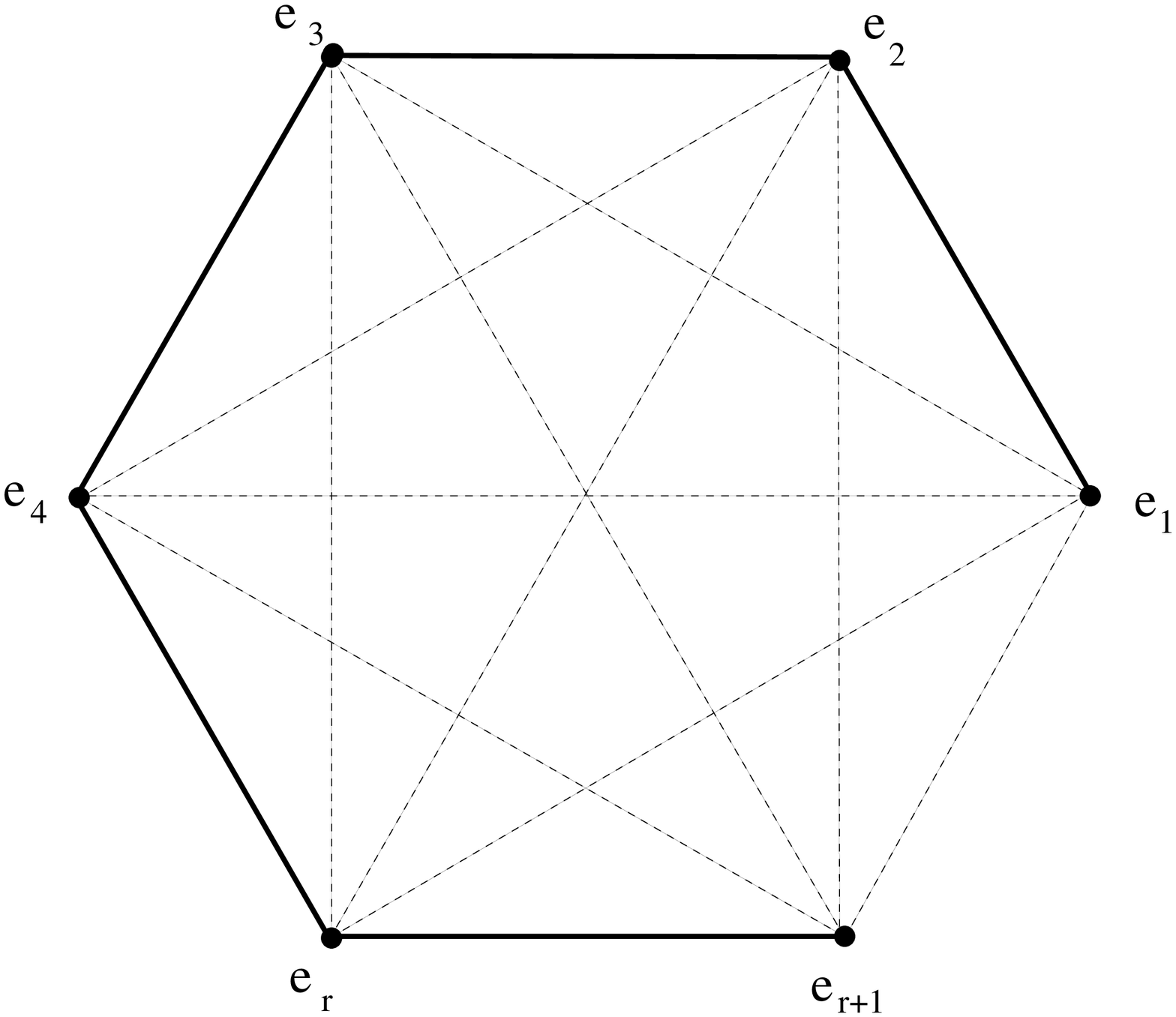}

\subsubsection{The quantum moduli space}
\label{quantummoduli}

How the quantum moduli space arises from the 
classical moduli space needs to be understood in 
this framework essentially just for $SU(2)$ the 
generalization is then almost immediately. The classical moduli space
with it's singularities is drawn in \figref{split}. The line
connecting the roots $e_1$ and $e_2$ of  $F_{A_1}(x,1)=x^2-1$ 
in the upper picture of \figref{split} corresponds to the vanishing 
cycle of the $W^+$-boson. We now want to describe a procedure which
replaces this vanishing cycle of the gauge boson with the vanishing
cycle of a  magnetic monopole and a dyon. We consider first a 
a deformation of $F_{A_1}$ namely 
$F^{\Lambda^2}_{A_1}(x,u)=(x^2-u)+\Lambda^2$. 
The zero level set of  $F^{\Lambda^2}_{A_1}(x,u)$ and in particular the
vanishing cycle is smoothly deformed by turning on 
$\Lambda^2\approx i\epsilon$ 
to run between $e_1^+$ and $e_1^+$ , as shown in the second raw of 
\figref{split}. For the 
$F^{-\Lambda^2}(x,u)$ deformation with $\Lambda^2=-i\epsilon$ 
the same applies and the image of the
classical  vanishing cycle runs between $e_1^-$ and $e_2^-$. That implies 
by continuity that the singularity 
$p=F^{\Lambda^2}(x,u)F^{-\Lambda^2}(x,u)=(x^2-u)^2-\Lambda^4 $
has two vanishing cycles $V^{-}$ and $V^{+}$ in the finite $u$-plane. 
If we consider the hyperelliptic curve $y^2=p$ and choose 
the cuts and the homology basis as in the last picture in \figref{split} 
then wee see from (\ref{id}) (compare \figref{phi}) immediately that 
the vanishing cycles correspond to the magnetic monopole 
$(n_m,n_e)=(1,0)$ and the  dyon $(n_m,n_e)=(1,-2)$.  
Using e.g. footnote (\ref{jfoot}) one calculates  
$\Delta=(2 \Lambda)^8(u-\Lambda^2)(u+\Lambda^2)$ and 
the Lefshetz formula with $n=1$ gives for the monopole 
monodromy $M_{(1,0)}=\left(\matrix{1&0\cr -1&1}\right)$
and for the dyon $M_{(1,-2)}=\left(\matrix{-1&4\cr -1 & 3}\right)$.
Furthermore we have $M^\infty=\left(\matrix{-1&4\cr 0 &-1}\right)=
M_{(1,0)}M_{(1,-2)}$, which establishes this 
curve as the $\Gamma_0(4)$, i.e. the ${\rm SU}(2)$ in $N=2$ 
conventions.

\figinsert{split}
{Splitting of classical level-set and vanishing cycle for ${\rm SU}(2)$.}
{2.1 truein}{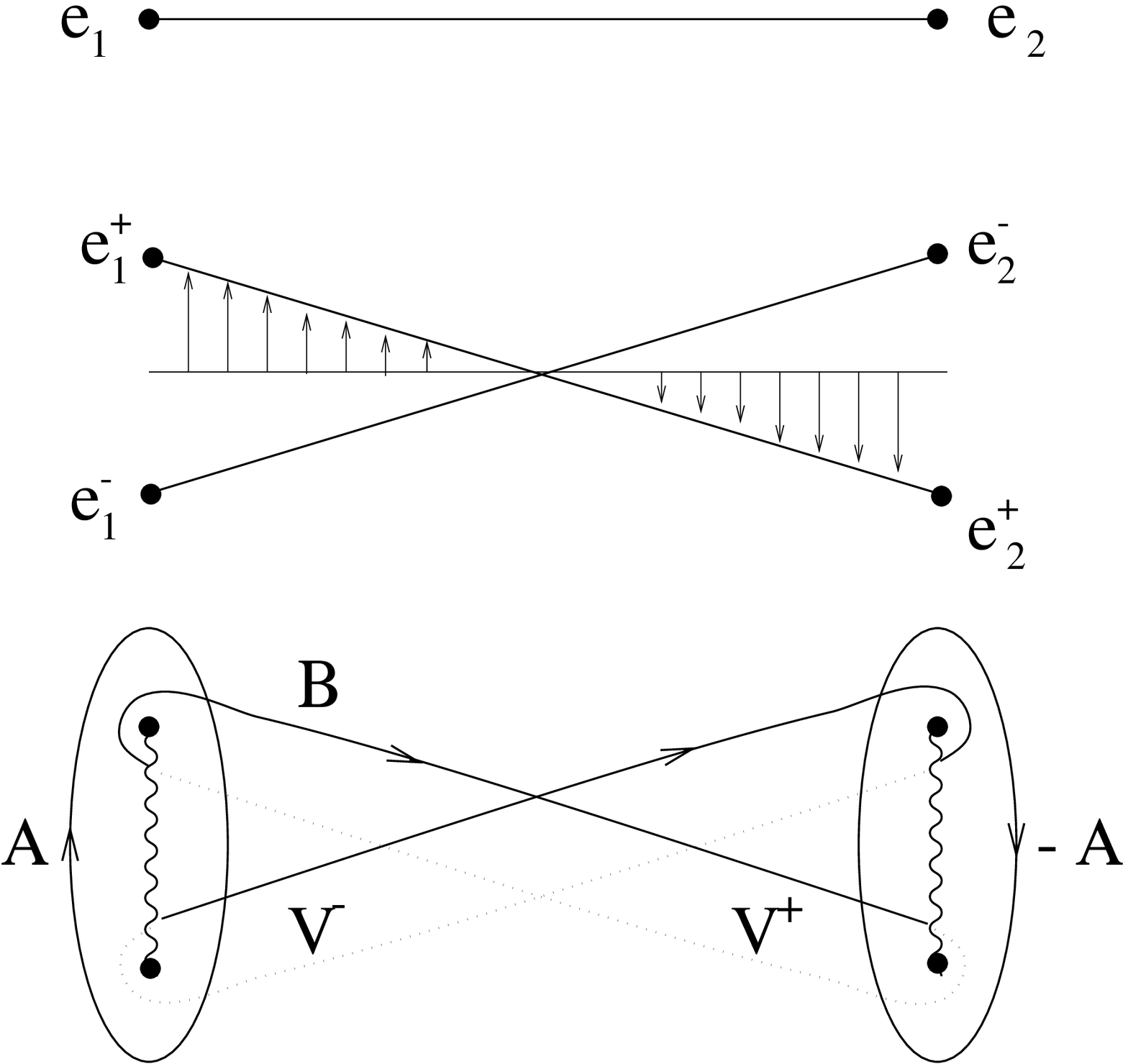}

In general the genus $g=r=n-1$ hyperelliptic curves 
\cit(klty)\cit(af) 
\be 
y^2= (F_{A_r}(x,\vec u))^2-\Lambda^{2n},
\ele(suncurves) 
seem to give a consistent  description of the non-perturbative 
effective action for the ${\rm SU}(n)$ theories. As for the ${\rm
SU}(2)$ the classical level-set and the classical vanishing 
cycles \figref{su6} 
will be doubled. Just as in \figref{split} for small 
$\Lambda^n=\pm i \epsilon$  the $+$ copy will be rotated slightly 
clockwise and the $-$ copy counter clockwise, such that each 
$W_{\alpha}$ with $\alpha>0$ will split into two dyons. As in
\figref{split} we can take for the basis of the $A$-cycles clockwise
contours around $\overline {e^+_{\vec \lambda_i},e^-_{\vec \lambda_i}}$, 
$i=1,\ldots,r$. They are then by definition purely electric 
$(\vec n_m,\vec n_e)=(\vec 0, \vec \lambda_i)$.  For the purely magnetic 
$B$-cycles we can take the vanishing cycles in the $+$ copy of the 
classical level-set, which are associated with the simple roots, 
they have charges $(\vec \alpha_i,\vec 0)$ $i=1,\ldots,r$. 
The charges of the other 
vanishing cycles follows be expanding them in the above described base. 
These are all vanishing cycles which occur at finite values of the
$u_i$. The factorization of the discriminate  
\be
\Delta=\prod_{i<j}
(e^+_{\lambda_i}-e^+_{\lambda_j})^2 
(e^-_{\lambda_i}-e^-_{\lambda_j})^2 
\ee
reflects this fact. By the parameterization  it is clear
that one can degenerate the curve such that an arbitrary combination 
of $+$ roots {\sl or} arbitrary combination of  $-$ roots come together. 
Similar as in the massive ${\rm SU}(2)$ case 
(comp. end of sec. (\ref{algtor})), mutually non-local dyons can 
become simultaneously massless for pure $SU(n)$ with $n>2$ at 
the points where the corresponding combination of $+$ or $-$ roots 
come together, e.g. for ${\rm SU}(3)$ if all $+$ or $-$ roots 
coincide \cit(agdo).
They are non-local in the sense that their mutual symplectic 
form (\ref{dz}) does not vanish.    

The curves (\ref{suncurves}) have many 
consistency properties built in per construction. 
Most notably in the classical level set one can push $k$
zeros $e_i$ off to infinity and reducing thereby $A_r$ singularity to an
$A_{r-k}$ singularity. This carries over for the curves (\ref{suncurves}) 
and allows e.g. to recover the ${\rm SU}(2)$ from the 
corresponding limits of the ${\rm SU}(3)$ curve. Furthermore the 
semi-classical monodromies,  which follow from the perturbative 
one-loop prepotential\foot{Here $C$ is the Cartan matrix.}
\be 
\cF={1\over 2} \tau (a^t C a)+{i\over 4 \pi i}\sum_{\alpha>0} 
Z_\alpha \log \left[ Z_\alpha^2\over \Lambda^2\right]
\ele(sunpert)
are automatically reproduced. The effective action can be evaluated
using e.g. the choice of the meromorphic form  
\be
\lambda={1\over 2 \sqrt{2} \pi}\p_x F_{A_r}(x,\vec u){x\dd x\over y}+
{\rm exact \ forms} \ ,
\ee
which gives upon derivation $\p_{u_{i+1}}\lambda = \omega_i + {\rm exact \ form}$
with 
\be 
\omega_i={x^{g-i-1} \dd x \over y}, \ {\rm with}\ i=1,\ldots,g
\ele(holoforms)
a basis of holomorphic $(1,0)$-forms. For later use we note finally that
useful Laurent representation for the curves, which is given by 
the reparameterization $y\rightarrow z+F_G$ followed by division by $z$.
\be
z+{\Lambda^{2{\rm cox}(G)}\over z}+2 F_{G}(x,\vec u)=0 \ .
\ele(martinecform) 
This form appears  with the Seiberg-Witten differential 
\be
{\lambda}_{SW}={1\over 2 \sqrt{2} \pi} x(z,u) {\dd z\over z}
\ele(lambdalaurant)
 naturally in the relation of the Seiberg-Witten 
result to integrable models \cit(integrable)
and string theory and is particular in the generalization of 
(\ref{suncurves}) to $ADE$-gauge groups \cit(mw).

As was mentioned there are curves of different genera, but with 
special additional symmetries, which describe the same effective 
action. E.g. for the simply-laced gauge groups $G$ one can consider
the characteristic polynomial in every representation of $G$.
$F^{\cal R}_G(x,\vec u)={\rm det}_{\cal R}\lbrack x-\phi \rbrack $ 
and gets a representation of the curve by shifting the highest 
Chern class $c_h$ as in $F^{\cal R}_G(x,\vec c, c_h+z+
{\Lambda^h\over z})=0$. Here $h$ is dual Coxeter number \cit(mw).       
Non-simply laced Lie groups can be obtained if the monodromy
in $x$ generates beside the Weyl-group an outer automorphism. E.g.
the $G_2$ representation can be understood as an $D_4$ representation
where the triality automorphism is part of the monodromy group and 
therefore identifies the three symmetric roots.

\subsubsection{The solutions for $SU(3)$}

For $SU(3)$ the Picard-Fuchs differential operators 
for the periods $a^1_D,a^2_D,a^1,a^2$ were derived in \cit(klt)
using the methods indicated in app. C 
\bea(rl)
{\cal L}_1&=(27 \Lambda^6- 4 u^3 - 27 v^2)\partial_u^2-
12 u^2 v \partial_u\partial_v-3 uv \partial_v-u\\ 
{\cal L}_2&=(27 \Lambda^6- 4 u^3 - 27 v^2)\partial_v^2-
36 u v \partial_u\partial_v-9 v \partial_v-3 \ .
\elea(picop)
As a consequence of (\ref{picop}) also the simple operator
$(u\partial_v^2 - 3 \partial_u^2)$ vanishes on the solutions.
Introducing the variables $\tilde \alpha={4 u_2^3\over 27 \Lambda^6}$ 
and $\tilde \beta={u_3^2\over \Lambda^6}$ (\ref{picop}) can be identified 
with Appell's system
$F_4({1\over 6},{1\over 6},{1\over 3},
{1\over 2}; \tilde \alpha, \tilde \beta)$, see  
\cit(appell) and \cit(erdelyii) Vol. I. 
The discriminante is found essentially 
by the method described in app. C and reads in the 
$\alpha, \beta,\gamma$ variables 
\be
\Delta=\alpha\beta\gamma(\alpha^2+\beta^2+\gamma^2-
2(\alpha\beta+\beta\gamma+\alpha\gamma))\ ,
\ele(sudreidis)
where we compactified the moduli space to a $\IP^2$, which has
homogeneous variables $(\alpha:\beta:\gamma)$ with 
$\gamma=27 \Lambda^6$ and $\tilde \alpha=\alpha/\gamma$, 
$\tilde \beta=\beta/\gamma$. The $\gamma$  factor
of the discriminante was actually detected by the 
analysing the singularities of the differential 
equations (\ref{picop}) at infinity.

\figinsert{su3dis}
{Quantum discriminante for $SU(3)$ in the $F_4$ parametrisation. 
The semiclassical regions
are at the $\alpha= \gamma=0$ and $\beta= \gamma$ locus. 
The magnetic dual semiclassical regions 
are at $Q_1$. At $Q_2$ the Riemann surface develops a cusp and
mutually non-local states, for which the Dirac-Zwanziger 
product does not vanish, become simultaneously massless. This
conformal point was analysed in detail by Argyres and Douglas. }
{2.1 truein}{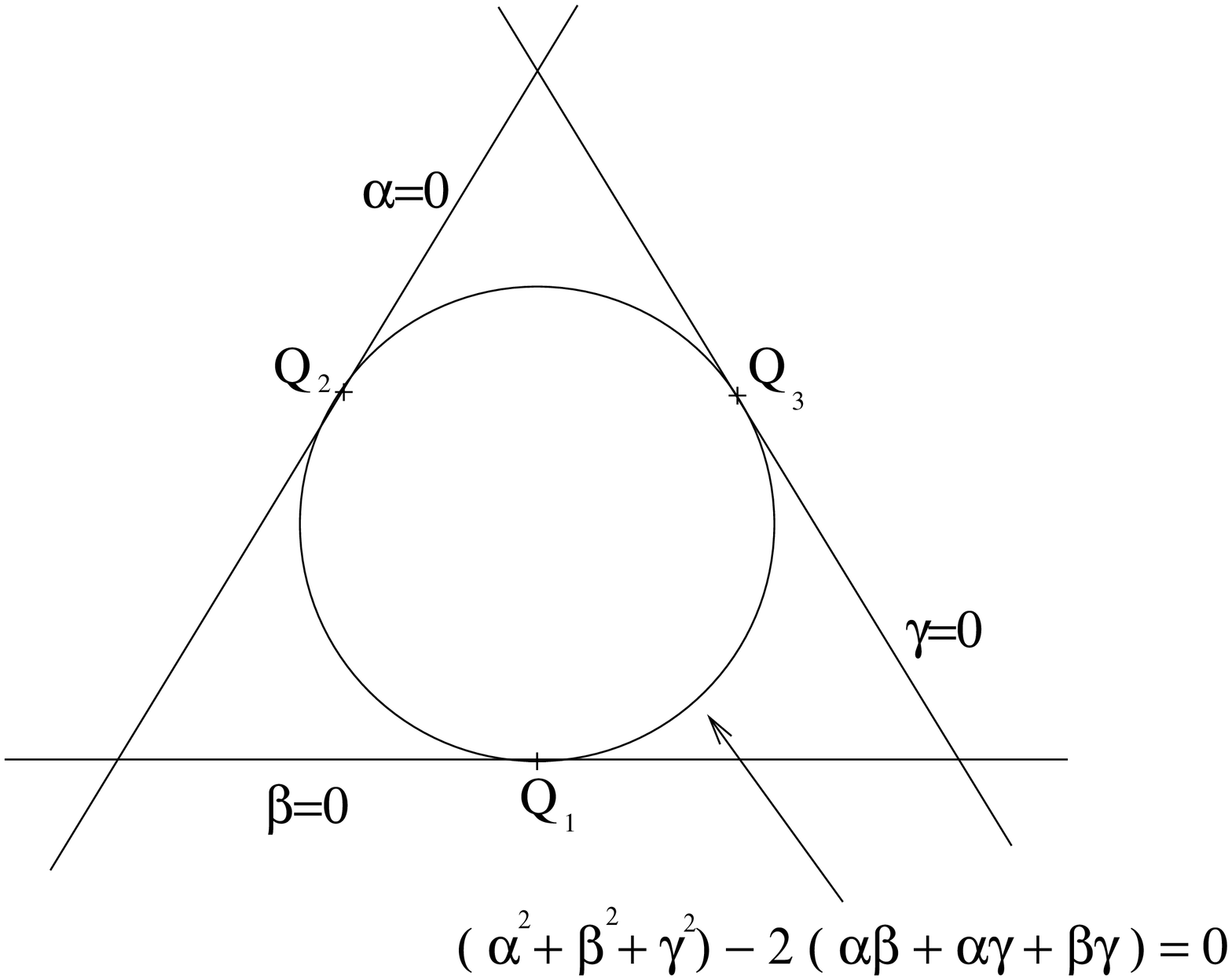}

It is technically a non-trivial task to analytically continue 
the solutions to all regions in the moduli space, which were 
solved in \cit(klt) by computing the leading terms of the integrals 
(\ref{adatorus}) directly. 
The expansions of ${\cal F}, {\cal F}_D$ in the semiclassical 
regions $\alpha=\gamma=0$, $\beta=\gamma=0$  and at the magnetic
dual region around $Q_1$ can be found in \cit(klt). The conformal 
point $Q_2$ was analysed in detail in \cit(agdo). In fact all tangencies
at $Q_1,Q_2,Q_3$ can be treated completly analogous to the discussion,
which can be found in section (\ref{example}) around \figref{blowi}. 

For the $SU(3)$ matter case see\cit(eft). 
For a fairly complete discussion of the Picard-Fuchs 
systems of $N=2$ theories we refer to \cit(imns)\cit(alis).

\section{Calabi-Yau manifolds}
\label{calabiyau}
In this chapter we will summarize some facts about the cohomology 
and the geometry of Calabi-Yau manifolds. Reviews motivated from string 
theory about this subject can be found in 
\cit(candelaslectures)\cit(hubschbook)\cit(hkt)\cit(greenerev).

\subsection{General properties}
\label{genprop}
By definition these are compact K\"ahler manifolds with 
vanishing first Chern class. The vanishing of the first Chern 
class implies by the theorem of Yau \cit(yau) that there exists\footnote{
More precisely given a K\"ahlerform $J$ one can find a Ricci-flat
metric $G_{i,\bar \jmath}$ such that $J^\prime=i G_{i,\bar \jmath}
d z^i d \bar z^{\bar \jmath}$ is in the same cohomology class as $J$.
The existence of the metric is inferred from existence of
a solution to the Monge-Ampere equation. There is no nontrivial case
in which this metric is actually known.}  a Ricci flat metric on the 
CY manifold. The converse is trivial,
since the first Chern class is represented by the Ricci 
two form $R_{i\bar j} d z^i d {\bar z}^{\bar j}$ and one essential
property of the Chern classes is their independence of the 
actual choice of the K\"ahler metric. The holonomy group of a 
generic K\"ahler manifold of complex dimension
$d$ is $U(d)$. On Ricci flat manifolds it is inside 
$SU(d)$. This can be easily seen as it is the trace part 
of the Riemann tensor which generates the $U(1)$ part of $U(d)$. 

We will use the term CY manifold for a Ricci flat 
K\"ahler manifold whose holonomy generates all of $SU(d)$. 
In three complex dimensions this rules out the complex 
three dimensional torus with trivial holonomy as well 
as the product of a complex one dimensional torus times the $K3$ surface 
with holonomy $SU(2)$, which have four and two covariantly 
constant spinor fields, respectively. This  leads for the compactification 
of the heterotic string to $N=4$ and $N=2$ supersymmetry in 
four dimensions, while compactification on a CY manifold
leads to the phenomenologically preferred $N=1$ supersymmetry. 
The $2d$ dimensional (co)tangent vectors split into the 
$d\oplus\bar d$ representation of $SU(d)$. Especially the
holomorphic $(d,0)$ forms which are completely antisymmetric in their
indices  transform therefore as $SU(d)$ singlets, 
i.e. they are covariantly constant and in fact non vanishing. 
The converse holds also: A K\"ahler manifold with a non vanishing 
covariant constant holomorphic $(d,0)$ form has to 
have a trivial $U(1)$ holonomy part and from that a vanishing 
Ricci-tensor, i.e. a vanishing first Chern class. 

Equivalence classes $[\omega]$ ($\omega\sim \omega'+\bar \partial \lambda$) of 
forms $\omega$ with $(p,q)$ index structure, i.e. in local 
coordinates written as  $\omega=\omega_{i_1,\dots, i_p,\bar 
\jmath_1,\ldots,\bar \jmath_q} dz^{i_1}\! \ldots\!  
dz^{i_p}$ $d\bar{z}^{\bar \jmath_1}\ldots d{\bar z}^{\bar \jmath_q}$, 
which are $\bar \p$ close ($\bar \omega=0$), generate the 
Dolbeault cohomology groups $H^{p,q}(X)$. Canonical representatives 
are the harmonic forms, which are annihilated by 
the $\bar \p$ Laplacian $\Delta_{\bar \p}= {\bar \p}
{\bar \p}^*+{\bar \p}^*{\bar \p}$. The relation to the 
more conventional De Rham cohomology groups $H^k(X)$ in which $\bar \p$ is 
replaced by the ordinary exterior derivative $d$ is given by
the  basic result in Hodge theory, which states that the 
cohomology groups on a K\"ahler manifold have a decomposition
(see e.g. \cit(griffithharris))
$$H^k(X,\IC)=\oplus_{p+q=k} H^{p,q}(X).$$
The rank of these cohomology groups are known as Hodge-numbers
and denoted by $h^{p,q}(X):={\rm rank}\ H^{p,q}(X)$. For the   

There are isomorphisms (a) and (b) among the  cohomology groups for 
all K\"ahler manifolds (see  e.g. \cit(griffithharris)):
\begin{itemize}
\item{(a)} The Hodge *-duality (a version of Poincar\'e $*$-duality 
for K\"ahler manifolds, which respects the Hodge decomposition)
implies $H^{p,q}(X)\simeq H^{d-p,d-q}(X)$.
\item{(b)} Complex conjugation $H^{p,q}(X)\simeq H^{q,p}(X)$.
\item{(c)} For CY manifolds we have in addition due to the 
possibility to contract with the projective unique holomorphic $(N,0)$-form 
the so called holomorphic duality $H^{p,0}(X)\simeq H^{d-p,0}(X)$.
\end{itemize}

Now a $(1,0)$-form transforms in the $d$ 
representation of $SU(d)$ and is 
therefore not covariantly constant. The existence of such a form 
on a Ricci-flat manifold  would contradict 
Bochners Theorem stating that the $(r,0)$-form 
$\omega$ is covariantly constant if 
\be 
\Theta=R^n_m \omega_{n,i_2,\ldots i_r} 
\omega^{m,i_2,\ldots i_r}+
{r-1\over 2} R_{n,m}^{p,q}\omega_{p,q,i_3\ldots i_r} 
\omega^{m,n,i_3,\ldots i_r}
\ee
is positive semi definite. Therefore $h^{1,0}=0$ and from 
this it is clear that there will be no Killing vector fields and
hence no continuous isometries on a CY manifold. 
In general one can show $h^{r,0}=0$ for $0<r<d$ for 
CY $N$-folds; for threefold this follows e.g. 
from the symmetries (c,b,a).  

If  we arrange the Hodge numbers of a CY three-fold in the
Hodge square  
\be 
\matrix{1&      0  &  \ldots      &          0&         1\cr    
          0&h^{d-1,1}&  \ldots      &h^{d-1,d-1}&         0\cr
     \vdots&   \vdots&              & \vdots    & \vdots   \cr      
          0&h^{1,1}  &  \ldots      &h^{1,d-1}  &         0\cr
          1&      0  &  \ldots      &          0&         1}
\ele(hodgesquare)
then 
(a) is the rotation symmetry by $\pi$ around the center of 
(\ref{hodgesquare}), (b) the reflection symmetry on the (SW)-(NE) 
diagonal.

The Hirzebruch-Riemann-Roch theorem for vector bundles 
$W$ over the space $X$ gives useful identities between 
the  dimensions of $H^{p,q}(X)$ and the Chern classes 
$c_i(T_X)\in H^{2k}(X,\IC)$ \cit(hirzebruch). The latter are symmetrical
polynomials in  the matrix valued curvature 2-form  
$\Theta=R^{k}_{l i \bar j} \dd z^i \wedge \dd \bar z^{\bar i} $
of the corresponding bundle.  Here the bundle is the tangent
bundle and $R^k_{l i \bar j}$ is the usual complex 
curvature tensor. The explicit
expression follow from (\ref{symchern}) by identifying 
$\phi$ with ${i\over 2 \pi}\, \Theta$, e.g. $c_1(T_X)={i\over 2 \pi}
{\rm tr} \Theta$, $c_2(T_X)={1\over 2 \cdot 4 \pi^2} \left({\rm Tr}\, 
\Theta \wedge \Theta- {\rm Tr}\, \Theta \wedge {\rm Tr}\, \Theta\right)$, 
compare e.g. \cit(egh). With the definitions $\chi(X,W)=
\sum_{i=0}^d (-1)^i {\rm dim} H^i(X,W)$ and
$c_0,\ldots,c_n$, Chern classes of the tangent bundle of $X$ (often simply
called the Chern classes of $X$) and $d_0,\ldots,d_r$ Chern classes of 
the vector bundle $W$ one has 
\be
\chi(X,W)=\int_X \kappa_d\left[\sum_{i=1}^{{\rm rank}(W)} e^{\delta_i} 
\prod_{i=1}^d {\gamma_i\over 1- e^{-\gamma_i}}\right],
\ele(hirz)
were $\kappa_n[ ]$ means taking the coefficient of the n'th 
homogeneous form degree, the $\gamma_i$ and $\delta_i$ are formal 
roots of the the Chern characters $c(T_X):=\sum_{i=0}^d c_i(T_X) s^i=
\prod_{i=1}^d (1-\gamma_i)$ and
$c(W):=\sum_{i=0}^q d_i t^i=\prod_{i=1}^q(1-\delta_i )$. 

We want
to use that for the alternating sums over the columns 
in (\ref{hodgesquare}), the so called arithmetic genera 
$\chi_q=\sum_p (-1)^p {\rm dim}\ H^p (X,\Omega^q)$. 
One way of evaluating the 
right-hand side of (\ref{hirz}) is to express the formal roots via 
symmetric polynomials in terms of the Chern classes, but it
is simpler and more instructive to take from (\ref{hirz}) 
only the message that $\chi_q$ depend on the Chern classes of $X$ in 
an universal manner for all complex manifolds and evaluate 
(\ref{hirz}) for easy cases, e.g. all possible products of $\IP^{n_i}$.
 
This calculation will use $h^{i,j}(\IP^n)=\delta^{i,j}$, $i,j=1,\ldots, n$, 
$c(T\IP^n)=(1+J)^{n+1}$, the Whitney product formula 
(see e.g. chapter IV of \cit(Bott)), 
the K\"unneth product formula $H^p(X \times Y)=\sum_{k+l=p}
H^k(X)\otimes H^l(X)$ \cit(Bott) and the fact that $\int_{\IP^n} J^n=1$. 
In two dimensions for the ``products'' 
$\IP^2$ and $\IP^1\times \IP^1$ this yields
\footnote{The only CY 
one-fold is the complex torus with $\chi=0$.}straightforwardly:  
\bea(rl)
\chi_0&={\ds {1\over 12}\int_X (c_1^2 +c_2)}, \\ [3 mm] 
\chi_1&={\ds {1\over 6}\int_X (c_1^2 -5 c_2)} 
\elea(chernII)
Using the Gauss-Bonnet interpretation of the Euler-number $\chi$  
as integral over the top Chern form $\chi=\int_X c_2$, it follows 
immediately from (\ref{hodgesquare}), (\ref{chernII}) that the 
Euler-number of any CY two-fold  is $24$ and $h^{1,1}=20$. 
Remarkably this harmonize perfectly with the anomaly 
cancellation condition of six-dimensional $N=2$ 
supergravity \cit(seibergKIII). In fact the only topological type 
of CY two-folds is given by the famous $K3$-surfaces. For a 
physics oriented review see\cit(aspinwallkiii).    

Unfortunately for  CY three-folds ($c_1=0$)
\bea(rl)
\chi_0&={\ds {1\over 24}\int_X c_1 c_2}, \\ [ 3 mm] 
\chi_1&={\ds {1\over 24}\int_X (c_1 c_2 -12 c_3)}
\elea(chernIII)
the first equation is trivially fulfilled ($c_1=0$) and 
we get from (\ref{chernIII}) just the fact that the Euler number is 
divisible by two 
\be 
\chi=\int_X c_3= 2(h^{2,1}-h^{1,1}),
\ee
which follows of course also from the arithmetic definition of the 
Euler number $\chi=\sum_p (-1)^p \chi_p=
\sum_{p,q} (-1)^{p+q}h^{p,q}$ and the symmetries of the Hodge square.

For four-folds, which might become relevant to describe 
the non-perturba\-tive behavior of the $N=1$ string theories
we have the relations
\bea(rl)
\chi_0&={1\over 720}\int_X (c_1 c_3-c_4 +3 c_2^2 +4 c_1^2 c_2-c_1^4)\\ [2 mm]
\chi_1&={1\over 180}\int_X(3c_2^2-31 c_4-14 c_1 c_3+4c_1^2 c_2-c_1^4)\\ [2 mm]
\chi_2&={1\over 120}\int_X ( 79 c_4- 19 c_1 c_3+ 3 c_2^2+4 c_1^2 c_2-c_1^4).
\elea(chernIV)
For CY manifolds in the sense above we have $c_1(T_X)=0$, 
$\chi_0=2$. Using this in (\ref{chernIV}) and the symmetries (a)-(c)
of the Hodge diamond implies\footnote{Beside this
it implies $\int_X c_2^2$ is even. It also seems
that $c_2^2\geq0$, indicating that $\chi\ge -1440$.} the following 
relation among the Hodge numbers  
\be
h^{2,2}=2(22+2 h^{1,1}+2 h^{3,1}- h^{2,1}).
\ee
The Euler number can thus be written as
\be
\chi(X)=6(8+h^{1,1}+h^{3,1}-h^{2,1}).
\ee

The middle cohomology for d even splits into a selfdual 
($*\omega =\omega$) $B_+(X)$ subspace and an 
anti-selfdual ($*\omega =-\omega$) 
subspace $B_-(X)$
$$H^d(X,\IR)=B_+(X)\oplus B_-(X),$$
whose dimensions are determined by the Hirzebruch signature as
\bea(rl) 
\tau(X)&={\rm dim} B_+(X)-{\rm dim} B_-(X)\\ [2 mm] 
&=:\int_X L_{d/2}
\eea
For $K3$ this gives 
\be
\tau(K3)={1\over 3}\int_X p_1=-{2\over 3} 24 
\ee
which leads to the familiar result that 
$H^2(K3,\ZZ)$ is a selfdual lattice of signature $(3,19)$.  
While for general fourfolds we have 
\bea(rl)
\tau(X)&={1\over 45} \int_X (7 p_2-p_1^2)\\ [ 2 mm]
&={\chi \over 3}+32.
\eea

The symmetric inner product 
$(\omega_1,\omega_2)=\int_X \omega_1\wedge * \omega_2$ 
is positive definite on $H^4(X)$ and  $H^4(X,\ZZ)$ is by Poincare 
duality unimodular. The symmetric quadratic form $Q(\omega_1,\omega_2)=
\int_X \omega_1 \wedge \omega_2$ is positive definite on $B_+(X)$ and 
negative on $B_-(X)$. 

\mabs Beside this rough distinction between CY manifolds by  
the Chern classes and the Hodge numbers there exists for three-folds a 
useful finer distinction due to  C.T.C. Wall \cit(wall). 
The statement is that torsion free CY 
threefolds\foot{With $w_2=0$ as it is the case for
CY-threefolds.} are classified up to real diffeomorphism by their 
cohomology groups $H^2(X),H^3(X)$, the 
trilinear coupling $C^0:H\times H\times H\rightarrow \ZZ$, where 
$H$ are classes in $H\in H^2(X)$ i.e. $C^0(H_i,H_k,H_l)=
\int_X H_i\wedge H_k \wedge H_l$ and the evaluation of 
$c_2:H\rightarrow \ZZ$ on  $H\in H^2(X)$, i.e. $c_2(H_i)=
\int_X c_2\wedge H_i$.

\subsection{Construction of the simplest Calabi-Yau spaces}
\label{construction}  
Given the topological condition of the Ricci flatness $c_1(T_X)=0$ one
can readily construct algebraic Calabi-Yau manifolds in 
(weighted) projective spaces, Grassmannian etc. In 
these cases the K\"ahler form
will be inherit from the ambient space $A$ and we only 
have to ensure the vanishing of $c_1$. A weighted projective 
space is defined as 
\bea(rl)
&\IP^n(w_1,\ldots,w_{n+1}):= \{(\vec x)\in 
{\bf C}^{n+1} \setminus (x_1=\ldots = x_{n+1}=0)|\\ [2 mm ] 
&(x_1,\ldots,x_{n+1})\sim (\lambda^{w_1} x_1,\ldots,\lambda^{w_{n+1}} 
x_{n+1}), \lambda\in {\bf C}^*\}
\elea(weightedproj)
For general weights one has various discrete $Z_n$ 
actions on the variables $x_i$, which have to be divided out.   
An in general singular variety can be described in  $\IP^n(\vec w)$ 
by the vanishing locus of $r$  polynomials $p_i(x)=0$, which have to be
quasi-homogeneous 
$$p_i(\lambda^{w_1} x_1,\ldots,\lambda^{w_{n+1}} x_{n+1})=
\lambda^{k_i} p_i(x_1,\ldots x_{n+1})$$ 
of degree $k_i$ and transversal i.e. 
${\rm rank}{\partial p_j\over \partial x_i}=r$ if $p_1=\ldots =p_r=0$ 
and $(x_1,\ldots x_{n+1})\neq (0,\ldots ,0)$. For these manifolds we
will use the short-hand notation $X_{k_1,\ldots,k_r}(w_1,\ldots,w_{n+1})$.
Given such a transversal algebraic embedding\foot{For simplicity we 
assume here first that $X_{k_1,\ldots,k_2}(w_1,\ldots,w_n)$ avoids
the singularities and is smooth, which is not the generic case.} 
one has a decomposition 
of the tangent space $T_A$ of the ambient space into the
tangent space of the manifold $T_X$ and the normal bundle 
${\cal N}$. This is expressed by the following short exact sequence
\be 
0\rightarrow T_X\rightarrow T_A|_X\rightarrow 
{\cal N}|_X\rightarrow 0.
\ele(normal)
In this situation one has for the total Chern classes \cit(Bott)
\be
c(T_A|_X)=c(T_X)\wedge c({\cal N}|_X) . 
\ele(chernfactor) 
By splitting the vector bundles $T_A|_X$ and ${\cal N}|_X$
over $\IP^n(\vec w)$ into line bundles we can write this as 
\be
\prod_{i=1}^{n+1} (1+w_i J)=\sum_{i=0}^{d} c_i(T_X) J^i
\prod_{j=1}^r (1+k_j J,)
\ele(chern)
where $J$ is the pullback of the K\"ahlerform of the ambient 
space. From this equation we have the identity 
\be 
c_1(T_X)=(\sum_{j=1}^{n+1} w_i -\sum_{i=1}^r k_i)J . 
\ele(ceins)
Hence the simplest CY spaces can be defined by the constraints 
$$\begin{array}{rl}  {\rm Torus\, :} &   \sum_{j_i} a_{j_1j_2j_3} 
                x_{j_1} x_{j_2} x_{j_3}=0\\ [2 mm ]  
          K3\, :& \sum_{j_i} a_{j_1j_2j_3j_4} 
                x_{j_1} x_{j_2} x_{j_3} x_{j_4}=0 \\ [ 2 mm]  
         {\rm Quintic\, :} & \sum_{j_i} a_{j_1j_2j_3j_4j_5}
                x_{j_1} x_{j_2} x_{j_3}x_{j_4} x_{j_5}=0
\end{array}
$$
in the ordinary projective spaces  $\IP^2$, $\IP^3$ and 
$\IP^4$ respectively. These polynomials describe
actually families of complex manifolds naively parameterized by the 
$10$, $35$ and  $126$ complex coefficients $a_{j_1\ldots j_n}$ from
which however $3^2$, $4^2$ and $5^2$ can be set to one by the 
${\rm GL}(n+1)$ transformation acting inside the $\IP^n$. This leaves us
with $1$, $19$ and $101$ elements in $H^1_{\bar \p}(X,T_X)$, 
which correspond to complex structure deformations.  
The Lefshetz embedding theorem states that the cohomology
of smooth embeddings is below the middle cohomology inherited from
the ambient space. Therefore degree $n+1$ hypersurfaces in
$\IP^n$ have $h^{p,p}=1$ for $2p < d$ 
and $h^{p,q}=0$ for $p\neq q$, $p+q < d$.     

Let us mention here an elementary technique to calculate the topological
invariants of $X$. From (\ref{chern}) we can calculate $c(T_X)$. 
For example for the quintic 
$$c(T_x)=1+10 J^2-40 J^3\ . $$
If we want to integrate e.g. $c_3(T_X)= - 40 J^3$ over the manifold $X$
to evaluate the Gauss-Bonnet definition of the Euler number: 
$\chi(X)=\int_X c_d(T_X)$ we lift the integral over $X$ to an
integral over the ambient space using the first Chern class of the 
normal bundle ${\cal N}$ i.e.
\be 
\int_X c_d=\int_{A} c_d(T_X)\wedge  \ c_1({\cal N}|_X)\ .
\ele(lift)
The point is that this relates the integral over $X$ to the volume of 
the ambient space, which is normalized e.g. for projective spaces as 
$\int_{\IP^{n_1}\ldots\IP^{n_l}} J_1^{n_1}\ldots J_l^{n_l}=1$.
For the quintic e.g. we get 
$$\int_{X} c_3=\int_{\IP^4} c_3 \wedge 5 J=-200 \int_{\IP^4} J^4=-200.$$ 
Similarly $\int_X c_2\wedge J=50$ and $\int_X J^5=5$. The necessary
information to calculate all characteristic data in Walls theorem
for  arbitrary toric varieties will be provided in appendix E.

If one considers general weighted projective spaces 
$\IP^n(w_1,\ldots,w_{n+1})$ one can 
construct many examples of CY hypersurfaces.
The weighted projective spaces 
$\IP^n(w_1,\ldots,w_{n+1})$ have in general $\ZZ_n$ singularities. 
The criterion $\sum_{i=1}^{n+1} w_i=d$ together with the condition 
of transversality renders the number of possible ambient spaces 
finite and imposes for  ${\rm dim}(X)<4$ restrictions on the weights,
which guarantee that the $\ZZ_n$-singularities can be resolved such that 
the hypersurface in the resolved ambient space has a unique nonvanishing
$(d,0)$-form. We will not go in the general toric machinery, which 
is most useful to establish that fact and to generalize the construction,
a short guide can be found in appendix E and an example for the 
resolution in section (\ref{example}).
After employing Bertini's theorem~\cit(griffithharris)
to derive a criterion for transversality one can classify the
CY hypersurfaces in $\IP(\vec w)$.
There are three tori, 95 $K_3$ surfaces and $7555$  CY 
hypersurfaces in weighted projective spaces \cit(ksks) of this 
type.

\section{$N=2$ String dualities in four dimensions}

Let us start the section with the statement of the conjecture 
for which evidence will be collected, as we go along. 

{\noindent}{\bf Conjecture: } 
{\sl The following $N=2$ string compactifications to four dimensions are 
equivalent: Type IIa theory on CY threefolds $X$,
Type IIb theory on the mirror CY threefolds $\hat X$ and 
heterotic string on $K3\times T^2$.}

\subsection{Mirror Symmetry}
\label{mirrorsymmetry}

The duality between type IIa compactified on $X$ and type 
IIb compactified on $\hat X$ is one application of 
{\sl mirror symmetry}. 

In section \ref{sigmaapproach} we describe a microscopic approach 
to mirror symmetry. Mirror symmetry maps the perturbative sectors 
of the IIa and type IIb theories onto each other. In absence of a
non-perturbative microscopic description of the type II string we can  
therefore still use the perturbative supersymmetric $\sigma$-model 
\cit(wittensusysigma) on $X$ to check part of the conjecture. 

In a macroscopic view we describe in section \ref{macros} the moduli 
spaces of massless fields in the effective low energy Lagrangian of 
the type II theory. $N=2$ space-time supersymmetry restricts the local 
structure of the moduli space and non-renormalization theorems allow 
to calculate certain quantities exactly. 

Ultimately the above equivalence is meant to be true for the full 
non-perturbative theories. In section \ref{branes} we shortly 
discuss the $D$-brane states, which will play 
a r\^ole in understanding some non-perturbative features of the 
theory.

\subsubsection{The microscopic $\sigma$-model approach}
\label{sigmaapproach}

The $\sigma$-model is defined by a map $\phi$ from a Riemann surface 
$\Sigma$ into $X$. The bosonic part of Lagrangian in local coordinates is 
simply 
\bea(rl)
L=&\ds{-{T\over 2}\int_\Sigma d^2 \sigma (h^{\alpha\beta} G_{m n}+
\varepsilon^{\alpha\beta} B_{mn}) \p_\alpha \phi^m \p_\beta  
\phi^n}\\ [ 3 mm] 
  =&\ds{-{T\over 2} \left(\int_\Sigma d^2\sigma ||d \phi||^2  + 
    i \int_\Sigma \phi^*(B)\right)},
\elea(bosonicaction)
where $G_{mn}(\phi)$ is the metric, $B_{mn}(\phi)$ is an antisymmetric 
background field on $X$ and $h^{\alpha\beta}$ is a gauge fixed\footnote{We
might assume that we are in critical dimension so that there are no
anomalies.} metric on the worldsheet. The $N=1$ and $N=2$ supersymmetric 
extensions were discussed in \cit(zumino), see also 
\cit(hklr)\cit(wittenmirror). 
The link between worldsheet properties and the topology of $X$
was pointed out in \cit(wittensusysigma). The simplest quantity one can 
associate to such a $\sigma$-model is the difference between the 
number of bosonic and fermionic states on the worldsheet ${\rm Tr}\ (-1)^F$,
where $(-1)^F$ is defined by requiring that it commutes with the 
bosonic operators, anti commutes with the supersymmetry generators 
$\{ (-)^F, Q^A \}=0$ and has eigenvalues $\pm 1$ on bosonic and 
fermionic states respectively \cit(wessbaggerbook). By the cyclic 
invariance of the trace 
$${\rm Tr}\ \lbrack (-)^F\{Q^A, {\bar Q}_B\} \rbrack
\equiv 0$$ 
and from the supersymmetry algebra in the rest-frame
$$\{ Q^A,{\bar Q}_B \} =2 E \delta^A_B$$ 
one concludes that ${\rm Tr}(-)^F=0$ for every energy level except 
the supersymmetric vacuum, i.e. for $E=0$
(or for the Prasad-Bogomolni-Sommerfield states with $2 E= |Z_i|$ 
$\forall i$, if the supersymmetry algebra is modified by central 
charges $Z_i$). 
Especially for the supersymmetric $\sigma$ model this reduces the 
calculation of ${\rm Tr}(-1)^F$ to the lowest energy configurations 
and these are maps into $X$ which are constant along the spatial 
direction of the world sheet and therefore  in the 
Ramond sector. By this argument the calculation becomes one of 
sypersymmetric quantum mechanics \cit(wittensusysigma) on $X$. 
The ground state operators of the $N=1$ SQM on $X$ are of the type 
$\Psi =b_{i_1,\ldots i_q}(\phi^i) \psi^{* i_1}\ldots \psi^{* i_q}$, 
where the $\psi^{* k}$ are anti-commuting fermion creation 
operators, which carry a cotangent index of $X$, i.e. $1\leq q\leq d$. 
Therefore the $\Psi$ can be identified with elements of $A^q(X)$ 
and, since the Hamiltonian $H={1 \over 2} \{Q,\bar Q \}$ gets 
identified with the $\Delta_d$-Laplacian, ground states of SQM will be 
identified with the {\sl de Rham cohomology\ } of $X$. Especially one has
\cit(wittensusysigma)
$$|\chi(X)|= {\rm Tr}(-)^F.$$

For $X$ k\"ahlerian, the $\sigma$-model has $N=2$ worldsheet 
supersymmetry \cit(zumino)\footnote{$X$ hyperk\"ahlerian leads to 
$N=4$ worldsheet supersymmetry \cit(gf).}. 
Iff $c_1(TX)=0$ the worldsheet theory is superconformal\cit(conformal?) 
with central charge $c=3d$, see \cit(neqtwoscft)\cit(lvw) 
for the $N=2$ superconformal algebra. 

The Ramond ground states $|\Psi \ket$ are {\sl primary}\footnote{Left (right) 
{\sl primary}\  fields are annihilated by all positive modes 
of the operators in the 
left (right) chiral algebra, which is the set of operators with $h\in \ZZ/2$
($\bar h\in \ZZ/2$) whose right (left) part is trivial $h=0$ ($\bar h=0$).
For an introduction into these basic concepts of conformal field theory
see \cit(ginsparg).}
fields, which are annihilated by $G_0^\pm$ and $\bar G_0^\pm$. 
It follows immediately from the $N=2$ algebra that they have conformal 
dimensions $(h,\bar h)=(d/8,d/8)$. 
Ramond ground states of left and right $U(1)$ charge $(q,\bar q)$, can be 
identified with the {\sl Dolbeault cohomology\ } groups 
$H^{d/2-q,d/2+\bar q}(X)$ of $X$ \cit(wittenmirror).
This argument can be followed using the {\sl spectral flow}\                   
of $N=2$ superconformal theories \cit(SS)\cit(lvw). 
It maps $G^\pm_{r}\rightarrow G^\pm_{r\pm\theta}$ and the modes of the energy 
and $U(1)$-charge operators are shifted by
\bea(rl) 
L_n&\rightarrow L_n +  \theta J_n + {d\over 2} \theta^2 \delta_{n,0}, \cr 
J_n&\rightarrow J_n+d \theta \delta_{n,0}.
\elea(specflow) 

For $N=2$ SCFT with $c=3 d$ there are four spectral flow operations with 
$(\theta,\bar \theta)=(\pm {1\over 2},\pm {1\over 2})$, which interpolate 
between the Ramond-Ramond and NS-NS sectors\footnote{The modes of 
$G^\pm=\sum_n G^\pm_{n\pm r} z^{-(n\pm r)-{3\over2}}$ are integer 
in the Ramond and half-integer in the NS sector.}   
of the theory.

\bea(rll)
(c \! \! \! \! \!  &,a)   \cr
& \uparrow ( {1\over 2},-{1\over 2} ) \cr
(c,c)\, { {({1\over 2},{1\over 2} )} \atop {\longleftarrow} } \,  {\rm RR}
\! \! \! \! \! & {\rm vacuum} \ \   { {(-{1\over 2}, -{1\over 2})}\atop 
{\longrightarrow} }\   (a,a)\cr
& \downarrow (-{1\over 2}, {1\over 2} ) \cr
(a\! \! \! \! \! &,c)   
\eea

They map the Ramond-Ramond ground state operators to primary operators
in the NS sector, which form rings under the topological 
fusion algebra \cit(lvw). 
The left/right sectors of these rings are called 
{\bf c}hiral({\bf a}ntichiral) if the operators fulfill the BPS 
condition $2 h=q$ $(2 h=-q)$, i.e. as follows from (\ref{specflow}), 
if they are obtain from the left/right Ramond ground states by the 
$\theta=1/2$ $(\theta=-1/2$) spectral flow \cit(lvw), this implies
that (anti)chiral {\sl primaries}  fulfill $ G^+_{-1/2}|\phi\ket=0$ 
$(G^-_{-1/2}|\phi\ket=0)$ which is often used as definition. 
From  $(G^-_{-n/2})^{\dag}=G^+_{n/2}$, the $N=2$ superconformal algebra  
Sand positivity of the Hilbert space inner product one can easily see that 
the BPS condition for $|\phi\ket$ is equivalent to the later definition; 
$0=\bra \phi|2 h -Q|\phi\ket$ $=\bra \phi|\{G^-_{1/2},G^+_{-1/2}\}|\phi\ket$ 
$=|G^+_{-1/2}|\phi \ket |^2+G^-_{1/2}|\phi \ket|^2$. Similarly from
$0\le \bra \phi|\{G^-_{3/2},G^+_{-3/2}\}|\phi\ket=2 h - 3 Q + {2\over 3} c$
one concludes that $Q\le d=c/3$ for chiral primaries \cit(lvw). 
Application of the spectral flow $(\theta,\bar \theta)=(1,0)$ 
on the vacuum shows that there is a unique states in 
$c=3 d$ theory which satisfies the bound. 
As explained in \cit(gepnertrieste) the spectral flow operation 
correspond to the action of an operator $\varepsilon_a=\exp {i}
\sqrt{c\over 3}a\phi(z)$. Here $\phi(z)$ is a free boson defined 
by the bosonization of the $U(1)$ current $J=i 
\sqrt{c\over 3}\partial_z \phi(z)$. Especially $\varepsilon_{1/2}$ in 
the full critical theory including the space-time part with $c=12$ can be
identified with the space-time supersymmetry operator \cit(gepnertrieste).
Using charge and energy conservation in the operator product expansion 
$\varepsilon_a(z) \Psi_q(w)=(z-w)^{h'-h-a^2 c/6}\Psi'_{q'}=(z-w)^{a q} 
\Psi'{q'}$ one sees that the operator $\varepsilon_{1/2}$ is a local  
fermionic operator, iff the $U(1)$ charges  of all states in the full 
theory are 
odd integers.  Using the spectral flow backwards it follows that the charges 
of the ground states in the Ramond sector of $N=2$ super 
conformal field theories are in the range $-d/2\leq q,\bar  q \leq d/2$ 
and that there are four unique states with $(q,\bar q)=
(\pm d/2, \pm d/2)$.

Depending on the particular 
value of $d$, the worldsheet theory has an extended algebra from the 
chiral states in the $(c,c)$, $(a,c)$ rings; in particular for $d=2$ 
the states of highest dimension in the $c$ and $a$ rings are currents 
$\varepsilon^\pm$ with charges $\pm 2$, which 
extend the $U(1)$ current algebra to a $SU(2)$ affine current algebra 
contained in an $N=4$ superconformal algebra, while for $d=3$; 
$\epsilon^\pm$, 
$J'={1\over 3} J$ and $T={1\over 6}:J^2:$ form a second second $N=2$ 
algebra on the worldsheet and $d=4$ leads to a $W$-algebra structure 
\cit(distnotes).

The ring structure of the images of the Ramond-Ramond ground states 
operators under that flows can be best seen in the  
$N=2$ topological models \cit(eydvv), which are defined
by twisting the stress energy tensor. Depending on whether 
one chose the  $+$ or $-$ twist
\bea(rlrl)
T &\rightarrow \hat T=T+ {1\over 2}\partial_z J & 
\bar T&\rightarrow \widetilde{\bar T}=
\bar T-{1\over 2} \partial_{\bar z} {\bar J}\\ [ 3 mm]
       & \ \ \ \ \hat G^+=H^+         & &\ \ \ \ \tilde G^+= Q^+\cr  
       & \! \! \! \nearrow          & & \! \! \! \nearrow\cr
G^{\pm}&                              & G^{\pm}& \cr
       & \! \!  \! \searrow    & & \! \!  \! \searrow \cr
       & \ \ \ \ \hat G^-=Q^-    & & \ \ \ \ \tilde G^-= H^-\cr
\eea
$G^-$  or $G^+$ becomes a current from which a chiral BRST
operator can be defined as $\hat Q^-=\oint Q^- dz $ or as 
$\hat Q^+=\oint Q^+ dz$. Putting together the chiral half there are up to 
charge conjugation two different topological theories called $A$ and 
$B$ model with BRST operator $Q^{A}=\hat Q^+ + {\widehat {{\bar Q}^-}}$
and $Q^{B}=\hat Q^- + {\widehat {{\bar Q}^-}}$. For the $B$ model, 
which is defined by the $(-,-)$ twist the $(c,c)$ operators become 
local physical operators and for the $A$ model with the 
$(+,-)$ twist the $(a,c)$ operators become local physical 
operators.

Contact with the cohomology of $X$ can  finally be made via 
the topological $N=2$ $\sigma$-model \cit(wittentopolsigma)
\cit(wittenmirror). It is easy to see that the physical operators 
of the $B$-model can be written as ${\cal O}_V$, 
where $V\in A^p(X,\Lambda^q TX)$ \cit(wittenmirror) and $p$ and $q$ are 
identified with the right and left $U(1)$ charges of the $(c,c)$ states. 
Moreover the BRST operator $Q^B$ has the property \cit(wittenmirror)
$$\{Q^B,{\cal O}_V \}=-{\cal O}_{\bar \partial V},$$ 
so that the {\sl local}\footnote{These local physical 
operators ${\cal O}^{(0)}$ correspond to the $(c,c)$ $(a,c)$ 
operators in the SCFT. 
In the topological field theory on can use the descend equations 
$d {\cal O}^(i)=\{ Q,{\cal O}^{(i+1)}\}$ to define in addition 
non-local operators involving integrals over one and two cycles 
on $\Sigma$ \cit(wittenmirror).}  physical operators of the 
$B$-model get identified with the elements in $H^p(X,\Lambda^q TX)$, 
which on a CY manifold are isomorphic to 
$H^{p,d-q}(X,\IC)$, as it follows from contraction with the 
covariant constant $(d,0)$-form and Dolbeaults Theorem. Similarly
the local operators of the $A$-model are ${\cal O}_V$ 
with $V\in A^q(X)$ and $\{ Q^A, {\cal O}_V\}={\cal O}_{dV}$. That is
the local operators of the $A$-model are identified with $H^k(X)$. 

Note that {\sl exactly marginal} $N=2$ operators can be 
constructed from the fields $\psi_{cc}$ and $\psi_{ac}$ 
in the $(c,c)$ and $(a,c)$ rings, which have $(h,\bar h)=({1\over 2}, 
{1\over 2})$ and are identified with $H^{d-1,1}(X)$ and $H^{1,1}(X)$ 
respectively, as $M_{cc}=(G^-_{-1/2} \bar G^-_{-1/2} \psi_{cc})$ and
$M_{ac}=(G^+_{-1/2} {\bar G}^-_{-1/2} \psi_{ac})$. The later fields are neutral 
Virasoro primaries of dimension $(h,\bar h)=(1,1)$, which transform into a 
total derivative under the $N=2$ supersymmetry transformations, 
see e.g. \cit(distnotes). Hence one can add the following terms to the 
action $S_0$ without spoiling the N=2 superconformal invariance
\be 
S(t,z)=S_0 + (t_a \int M_{cc}^a+ z_a \int M_{ac}^a+h.c.)
\ele(perturbed)
For the $\sigma$-model on a CY space the parameter 
$t_a$ and $z_a$ will be identified with the complex structure 
and the complexified K\"ahler structure deformations.

Now there are the following {\sl symmetries}, which do not
change the correlators of the $N=2$ superconformal field theory: 
\begin{itemize}
\item (i) $(q,\bar q)\rightarrow (-q,-\bar q)$
\item (ii) exchange of the left and the right sector and 
\item (iii) $(q,\bar q)\rightarrow (\bar q,-\bar q)$. 
\end{itemize}

If one keeps the identification $(q,\bar q)\leftrightarrow 
H^{d/2-q,d/2+\bar q}(X)$, (i) corresponds to the Hodge $*$-star duality, 
the combination of (i) and (ii) to complex conjugation and hence to the 
symmetries of the homology on $X$ mentioned in section \ref{calabiyau}. 

Symmetry (iii) corresponds to a reflection of the Hodge square on the 
vertical axis, i.e. $H^{p,q}(X)\leftrightarrow H^{p,d-p}(X)$. This is 
{\sl not} a symmetry of the cohomology of $X$. In fact CY 
manifolds  $X$ and $\hat X$ for which $h^{p,q}(X)=h^{p,d-q}(\hat X)$ 
are generally of {\sl different} topological type. Especially for 
$d$ odd $\chi(X)=-\chi(\hat X)$. It was suggested in \cit(dixon)\cit(lvw) 
that this ambiguity in the association of a worldsheet theory to a 
CY target space means actually that there exists a pair of 
CY spaces for which $h^{p,q}(X)=h^{p,d-q}(\hat X)$ such that 
the $\sigma$-model is identical on $X$ and $\hat X$ with the 
r\^ole of the operators of the $(c,c)$ and $(a,c)$ rings exchanged. 
In particular there is no doubt from the conformal field theory 
point of view that the two deformation spaces in (\ref{perturbed}) 
should have a) identical integrability structure; as it can be shown 
using the superconformal Ward-identities they have both special 
K\"ahler structure \cit(distnotes) for $d=3$  and b) they 
will be exchanged by (iii) the relative flip of the $U(1)$ charge.

It is expected that modular invariant $N=2$ SCFT  
have a geometrical interpretation as 
$\sigma$-model on a d-dimensional CY manifold if 
$c=3 d$, $d\in \IN$ and all states have odd integer 
$U(1)$ charges $q$ and $\bar q$. 
E.g. for $d=2$ one can show solely from the above 
requirements\footnote{It would interesting to rederive the index 
theorem (\ref{hirz}) for higher dimensional CY manifolds 
in a similar fashion from the $N=2$ SCFT.}, 
that the degeneracy of the ground states in the Ramond sector 
corresponds to the $K3$ Hodge numbers \cit(ays). In many 
cases the rational $N=2$ SCFT, which corresponds to the 
$\sigma$-model on CY manifolds at a specify point 
in the moduli space, is known \cit(Gepner)\cit(gvw)\cit(fkss)\cit(lr) 
and can be solved exactly. 

The simplest examples of such $c=3 d$ field theories are tensor products of 
minimal $N=2$ SCFTs, in which the charge integrality is achieved by 
an orbifold construction \cit(Gepner). The building blocks are minimal 
$N=2$ superconformal models, which exist for all positive integers $k$ with 
central charge $c={3 k\over k+2}$. The primary fields are labeled by the  
three quantum numbers $l,m,s$ in the range 
\bea(lr)
&0\le l \le k, \ \ 0\le |m-s|\le l\cr
&s=\left\{\matrix{&0,2\quad {\rm NS\ \ Sector }\cr
                  &\pm 1\ \quad {\rm R\ \ Sector}}\right.\cr
&l+m+s=0\ {\rm mod}\ 2
\eea   
and their conformal dimensions and 
charges are given by \cit(gz)\cit(gepnertrieste)
\bea(rl)
h&={l(l+2)-m^2\over 4 (k+2)}+{s^2\over 8}\cr
q&=-{m\over k+2}+{s\over 2}.
\eea
Character functions 
\be
\chi_{l,m,s}(\tau,z,u)=e^{2 \pi i u} 
e^{2 \pi i c\over 24} {\rm Tr}_{{\cal H}_{l,m,s}} e^{2 \pi i J_0 z} 
e^{2 \pi i \tau L_0}
\ee 
of the highest weight representations ${\cal H}_{l,m,s}$ 
belonging to the above primary  fields and 
their transformation under the one loop
modular group are known \cit(gz). 
Beside the obvious left/right symmetric combination by which one can 
form a modular invariant one loop partition function at all values of $k$ 
\be 
Z(\tau ,\bar \tau)=\sum N_{l,\bar l}\delta_{q,\bar q}
\delta_{s,\bar s} \chi_{l,m,s}(\tau){\bar \chi}_{\bar l,\bar m,\bar s}
(\bar \tau)
\ee
with $N_{l, \bar l}=\delta_{l, \bar l}$ there exists a full 
$ADE$ classification for more general possibilities to combine  
$l,\bar l$ in a modular invariant way. 
More precisely one has in addition to the above series of so called 
$A_k$ invariants a $D_k$ series of invariants distinguished for 
$k$ even and $k$ odd and sporadic $E_{6,7,8}$ invariants for $k=10,18,30$  
\cit(ciz),\cit(gepnertrieste). Taking the various combinations of 
products of $ADE$ invariants into account there are 1176 tensor products 
of these models with $c=\sum_{i=1}^r {3 k_i\over k_i+2}=9$ and $4\le r\le 9$ 
\cit(fkss).
The $N=2$ minimal SCFT are conjectured to be the infrared fixed of 
$N=2$ Landau-Ginzburg models with action 
\be
S=\int d^2 z d^4 \theta K(x_i,\bar x_i)+
  \left(\int d^2 z d^2 W(x_i)+h.c.\right)
\ee 
The kinetic terms are irrelevant operators and the infrared limit depends
only on the holomorphic superpotential $W(x_i)$. The $ADE$ classification
of the $N=2$ invariants reflects itself in the classification of 
the superpotentials with no marginal operator, which are mapped 
the classified modality zero or simple singularities\cit(arnold) 
of singularity index $\beta=c/6< 1/2$ see table (\ref{ADEpot}) in
section (\ref{localmirror}).
For the LG discription the $z$ dependent terms can be dropped, as
the correspond to mass terms which do not affect the 
conformal fix point.
For tensor products the LG superpotential is 
$W=\sum_{i=1}^r W_i (x, y)$ and
the connection to CY compactifications can be made in the
most general setting via the gauged Landau-Ginzburg model 
\cit(wittengaugedlg). For $r=4$ factors in the tensor product, with 
three $A$ invariants and one arbitrary invariant 
as well as for $r=5$ factors with five $A$ invariants 
$W$ is a polynomial in five variables 
(note that we can add an irrelevant $y^2$ term 
to $W$) and the connection was discussed before in 
\cit(gvw)\footnote{This approach can be also 
also adapted to more general cases \cit(fkss).}. 
In this case the CY space $X$ is given simply by the 
zero locus of the polynomial $W(x_1,\ldots,x_5)=0$ 
in a four dimensional weighted projective space 
$\IP^4(w_1,\ldots, w_5)$, with suitable weights which make $W$ a 
quasi-homogeneous polynomial of degree $k$, as defined in section 
(\ref{construction}) \cit(gvw). 

Each $A$ type $N=2$ factor theory possess a $Z_{k+2}$ 
symmetry which acts by phase multiplication by 
$\exp 2 \pi i ((m +\bar m)/2)(r/(k+2))$ on the NS-NS states of the
factor theory. The projection onto odd integral $U(1)$ charges 
in the tensor model (times the space-time part) is performed 
by orbifoldizing by the diagonal subgroup $Z_{lcm(k_i+2)}$, which rotates 
simultaneously by the smallest unit $\exp 2 \pi i {1\over k_l+2}$ 
of the cyclic group in each factor theory see \cit(Gepner)
\cit(gepnertrieste)\cit(fkss). In fact one
can construct other orbifolds by modding out other 
subgroups of the $H=\prod_{i=1}^r Z_{k_i+2}/Z_{lcm(k_i+2)}$ 
symmetry, which rotate by $\exp 2 \pi i {r_i\over k_i+2}$ 
in the i'th factor theory. These groups do not introduce 
non-integral $U(1)$ charges in the twisted sector and lead 
therefore to supersymmetric compactifications, iff 
\be 
\sum {r_i\over k_i+2}=0\ {\rm mod} \ 1.
\ele(susycond) 
It can be shown that the maximal subgroup $G_{max}$ of 
$H$ for which all elements fulfill (\ref{susycond}) inverts the 
relative sign of the $U(1)$ charge in the SCFT \cit(gp). On the 
worldsheet this corresponds to the Kramers-Warnier duality, which 
exchanges the order and the disorder operator of the theory.

The group action of $G_{max}$ on the CY space $X$  can be readily 
identified as phase multiplication of the coordinates $x_i$ by 
$\exp 2 \pi i {r_i\over k_i+2}$. Modding out $G_{max}$ in the super 
conformal tensor product field theory with odd integral charges (SCFT) 
and on the CY space $X$ one gets the following diagram.

\bea(rlrl)
&(N=2 \  {\rm SCFT})& \rightarrow & (N=2 \  {\rm SCFT})/G_{max}\\ [ 3 mm]
&\ \ \ \ \  \updownarrow   &    & \ \ \ \ \updownarrow \\ [ 2 mm] 
&  \ \ \ \ X  & \rightarrow & \ \ \hat X= {\widehat {X/G_{max}}} 
\elea(mirrorconstruction)
As explained the upper horizontal arrow correspond 
simply to the orbifold construction in conformal field 
theory; it leads undoubtedly to an identical conformal field theory with 
$(a,c)$ and $(c,c)$ rings exchanged. The lower horizontal arrow 
corresponds likewise to a simple operation in geometry, 
one considers the orbitspace $X/G_{max}$ and the resolves the 
cyclic singularities in $X/G_{max}$ to the smooth space 
$\widehat{X/G_{max}}$. One can check that $h^{p,q}(X)=
h^{3-p,q}(\widehat {X/G_{max}})$ \cit(roan). 
It is more difficult to make the identification indicated by 
the vertical arrow rigorous, especially the heuristic arguments 
about the renormalization group flow in the LG models and the 
path integral arguments \cit(gvw)\cit(wittengaugedlg). Independently 
whether the physical picture (\ref{mirrorconstruction}) is a good
starting point to prove the perturbative part of the Mirror symmetry
conjecture (see \cit(morrisonplesser)), one can use it to construct 
candidate mirror pairs. Reports on the particular aspect of the
construction of mirror pairs can be found in \cit(mconstruction).

\subsubsection{The macroscopic approach via the effective 4d 
supergravity} 
\label{macros}

As we have seen in the previous section mirror,
symmetry states that to every 
CY manifold $X$ there exists a {\sl mirror} CY 
manifold $\hat X$ with $h^{p,d-q}(X)=h^{p,q}(\hat X)$ and 
$\hat {\hat X}=X$ such that after a suitable one to one map between 
moduli parameters and operators all 
correlation functions of type IIa theories on $X$ can be mapped one 
to one to correlation functions of type IIb theory on $\hat X$.  
To learn about the structure  of the moduli spaces 
it is sufficient to use a  Kaluza-Klein like compactification of the 
effective 10d supergravity theory to 4d on a CY manifold, 
which preserves one quarter of the supersymmetry .
Alternatively one can develop the $\sigma$-model point of 
view as it is done in \cit(bs)\cit(distnotes), the main advantage of the
4d point of view is the that space time supersymmetry also constraints the
dilaton modulus. 

The first piece of evidence for mirror symmetry of the type II theories  
is the match of the massless spectrum. In 10d the $SO(8)$-representation 
of the massless modes, which come from 
the left- times right-moving sector of the type II theories is 
\cit(GSW)\cit(PCJ)
\be\begin{array}{lr}
{\rm type \, \, IIa }:  
&({\bf 8}_v \oplus {\bf 8}_s)\otimes ({\bf 8}_v\oplus {\bf 8}_c)\, \cr
{\rm type \, \, IIb }: 
&({\bf 8}_v \oplus {\bf 8}_s)\otimes ({\bf 8}_v\oplus {\bf 8}_s),
\end{array}
\ee
where the vectors comes form the Neveu-Schwarz sectors and the two
kinds of spinors come from the Ramond sectors of the type II worldsheet
theory. In both type IIa and type IIb one gets by decomposing the product 
of the vectors into $SO(8)$ tensors a scalar (the dilaton) $D$, 
the two form antisymmetric tensor field potential $B_{MN}$ and a 
symmetric two form, the metric $G_{MN}$. 
For type IIa the spinors decompose into the a one form 
potential $A_N$ and a three form potential $C_{L,M,N}$. 
For type IIb the Ramond-Ramond fields decompose into a second scalar 
$D'$, a second antisymmetric two form potential $B'_{MN}$ and a 
four-form potential $E^+_{KLMN}$ with seldual field strength. 
Together with the contribution from the mixed sectors the massless 
spectrum of the 10d type IIa and type IIb theory is that of 10d 
nonchiral and chiral supergravity, respectively. The bosonic
components are summarized below.

\begin{center}
\label{tab1}
\begin{tabular}{lrr}
\hline
                   \multicolumn{1}{l}{} 
                 & \multicolumn{1}{c}{NS-NS}
                 & \multicolumn{1}{c}{R-R}    \\
\hline
type IIa: &$ \!\!\! D,B_{KL},G_{KL} $ & $ A_L,  C_{KLM}$ \\
type IIb: &$ \!\!\! D, B_{KL},G_{KL} $ & $ D', 
B'_{KL}, E^+_{KLMN}$ \\
\hline
\end{tabular}
\end{center}

\medskip

Dimensional reduction to 4d links the massless fields of the effective
theory in 4d to the harmonic forms of the internal CY threefold 
$X$ and hence by Hodge theory to the cohomology of $X$. 
More specifically split 
the 10d indices $K,L,\ldots$ into the indices $\kappa,\lambda,\ldots 
= 0,1,2,3$ of vectors (co)tangent to 4d Minkowski-space $M^4$ and 
the indices $k,l,\ldots=1,2,3$, $\bar k,\bar l,\ldots = 1,2,3$ 
(co)tangent to the internal CY threefold $X$. 
Then one sees by splitting the wave equation that the  harmonic part of 
three form potential with index structure $\delta C_{lm\bar l}$ and 
$\delta C_{\bar l\bar m l}$ leads to two massless real scalars $q_1$, 
$\tilde q_1$ in $M^4$ one for every harmonic (2,1)-form 
and one for every (1,2)-form on $X$, the $\delta B_{k\bar k}$ component 
leads to a massless scalar $b^k$ for every harmonic (1,1)-form, while the 
$\delta A_{\mu\kappa l\bar l}$ component gives a vector $A_\mu$ 
in $M^4$ for every harmonic (1,1)-form. 
In addition one gets in four dimensions degrees of freedom from 
gravitational modes, i.e. those independent variations of the metric 
$\delta G$ on $X$, which preserve the Ricci-flatness 
\be
R_{kl}(G+\delta G)=R_{kl}(G)=0. 
\ele(ricciflat) 
In fact two four dimensional real scalars $q_2$, 
$\tilde q_2$ come from each independent pure variation 
$\delta G_{kl}$  and $\delta G_{\bar k \bar l}$ of 
the metric on $X$. Those are in  one to one correspondence with the harmonic 
$(1,2)$- and $(2,1)$-forms, as can be shown from the differential
equation implied by (\ref{ricciflat}), the Lichnerowicz equation, 
and using the contraction with the (anti)holomorphic $(3,0)$ ($(0,3)$)-form 
$\Omega_{ij}^{\bar k}\delta g_{\bar k\bar l}$, see e.g. 
\cit(candelaslectures). 
Furthermore one gets mixed solutions $\delta G_{l\bar l}$ to 
(\ref{ricciflat}), which are in one to one correspondence to the 
harmonic $(1,1)$-forms and contribute each another real 4d scalar $i g^k$, 
which combine with the $b^k$ into a complex scalars $\phi^k=b^k+ig^k$. 
The $\phi^k$ are the lowest component of $h^{1,1}$ $N=2$ vector 
multiplets $\Phi^k$, whose highest component are the vectors 
$A^k_\mu$
\bea(lll)
       & A^k_\mu  &  \cr
\lambda^k&        &  \psi^k \qquad  k=1,\ldots h^{1,1}(X) \cr
         & \phi^k &  
\elea(vectormult)  
The fermionic extension are build using the covariant constant 
spinor $\eta$ on $X$, i.e. they are present iff $X$ is CY 
\cit(CHSW). In this scheme (\ref{vectormult}) $V=(\lambda,A_\mu )$ is an 
$N=1$ vector multiplet, $\Psi=(\phi,\psi)$ is a chiral $N=1$ multiplet and 
the global $SU(2)$ acts horizontally. 

The four real scalars $q^k_1,q^k_2$ and $\tilde q^k_1,\tilde q^k_2$ form two 
complex scalars, which on shell can be interpreted as 
components of $h^{2,1}$ $N=2$ hyper multiplets ${\bf Q}^k$ 
(in fact they combined to a quaternionic quantity)  
\bea(lll)
       & q^k  &  \cr
{\psi_q}^k&        &   \psi_{\tilde q}^k \qquad  k=1,\ldots h^{2,1}(X) \cr
         & \tilde q^k &  
\elea(hypermult)
where $Q=(q,\psi_q)$, $\tilde Q=(\tilde q,\psi_{\tilde q})$ are 
$N=1$ chiral multiplets.  

Beside these fields whose number depend on the cohomology of the particular 
CY space chosen, there is the universal sector of 4d fields, 
which come 
from the $(0,0),(3,0),(0,3)$ and $(3,3)$-forms. The three forms  with index 
structure $\delta A_{klm}$ and $\delta A_{\bar k\bar l \bar m}$ form 
together with $D$ and $\delta B_{\mu,\nu}$ the hyper multiplet of the 4d 
dilaton axion field. While the $\delta G_{\lambda\kappa}$ and 
$\delta A_\mu$ part gives the bosonic degrees of freedom, the graviton and the 
graviphoton, of the $N=2$ gravitational multiplet.  

For the type IIb compactification the r\^ole of the vertical and horizontal
cohomology of $X$ is exactly reversed. From 
$\delta E^+_{\mu k m\bar b}$ 
($\delta E^+_{\mu k \bar m \bar n}$) we get vectors in four 
dimensions, which are completed by the pure gravitational 
deformations $\delta g_{k l}$ and 
$\delta g_{\bar k\bar l}$ to the bosonic degrees of freedom of 
the $h^{2,1}$ vector multiplets in 4d. 
The four bosonic degrees of freedom of 
the $h^{1,1}$ hyper multiplets arise from $\delta B_{l \bar k}$, 
$\delta B'_{k \bar k}$, 
$\delta E_{\mu \nu k \bar l}$ and the gravitational modes 
$\delta G_{k \bar l}$.

The dilaton-axion in the universal sector are from $\delta B'_{\mu,\nu}$, 
$\delta B_{\mu \nu}$, $\delta D$ and $\delta D'$, while the bosonic sectors of 
the gravitational multiplet are from $\delta G_{\mu\nu}$ (graviton) and 
$\delta E^+_{\mu n m l}$ ($\delta E^+_{\mu \bar n \bar m \bar l}$) 
(graviphoton).  

Local $N=2$ supersymmetry of the effective 4d theory restricts the 
structure of the moduli spaces of the theories considerable. 
The scalars of the vector multiplets $V$ 
parameterize a special K\"ahler manifold 
${\cal K}_{\# V}$ of complex dimension $\#V$ \cit(specialkaehler), 
while the scalars of the hyper multiplets $H$ parameterize a quaternionic 
manifold ${\cal Q}_{\# H}$ \cit(quaternionic) of quaternionic 
dimension $\# H$, 
where $\# V$ and  $\# H$ is the number of vector and hyper multiplets 
respectively\footnote{For a recent review of both structures in 4d $N=2$ 
supergravity see \cit(revneq2).}. The effective theory is  that of an 
abelian gauge group $U(1)^{\# V}$. 
For generic values of the moduli there are no light particles (vector or 
hyper multiplets) charged under these $U(1)$'s and in particular 
there are no couplings between the light vectors multiplets and the light 
hyper multiplets at all (\cit(specialkaehler) first reference). 

This can also been argued from the worldsheet point of view similarly as 
in \cit(dkl). From the above one concludes that the moduli 
spaces ${\cal M}^a(X)$ and ${\cal M}^b(X)$ of the type IIa and type IIb 
theory on a CY $X$ and it's mirror $\hat X$ have the structure
\bea(ll)
&{\cal M}^a(X)={\cal K}(X)_{h^{1,1}(X)}\otimes {\cal Q}(X)_{h^{2,1}(X)+1}
\\
&{\cal M}^b(\hat X)={\cal K}(\hat X)_{h^{2,1}(\hat X)}\otimes {\cal Q}(\hat X)_{h^{1,1}(\hat X)+1}
\elea(conmod)
and the mirror symmetry conjecture suggests that they are actually 
identified ${\cal M}^a(X)\simeq {\cal M}^b(\hat X)$. 

\subsubsection{Type II branes}
\label{branes}

From the R-R potentials of the type II theories in the table in 
section \ref{macros} one expects for the type IIa string extended 
R-R ``electric'' sources of spatial dimension $p_e=0,2$ which give 
rise to $2$ and $4$-form field strength which are the curls of the 
$1,3$ form potentials. Furthermore there are the dual 
extended ``magnetic'' sources associated to the dual field strength form.
They are of dimension $p_m=10-4-p_e$, i.e. $p_m=4,6$. 
Likewise for the type IIb theory one has $0,2,4$ form potentials 
leading to $1,3,5$-form field strength, which comes 
from $p_e=-1,1,3$ brane\footnote{The $-1$ is meant to correspond to 
a $D-$-instanton see \cit(green).} sources and the dual magnetic 
sources are of dimension $p_m=3,5,7$. It has been argued
that in addition a $8$ and a $9$ brane exist for the type IIa 
and type IIb theory respectively, compare \cit(polchinski). 
These non-perturbative states, which carry R-R charge were 
identified by Polchinski as $D$-branes 
\cit(polchinski), which can be understood as alternative 
representation of the black $p$-branes \cit(hs). 
$D$-$p$-branes are topological defects along a {\sl dynamical} 
spatial hypersurface $M$ of dimension $p$ in space-time, which is 
defined  by the property that the open string has $p$ Neumann boundary 
conditions tangential to $M$ and $9-p$ Dirichlet boundary 
conditions normal to $M$. An easy but important consequence of this
definition is that $R\rightarrow \alpha' /R$, so called $T$-duality, 
in a compact not-tangential direction to the $p$ brane 
will transform one of the Dirichlet boundary conditions into 
Neumann boundary conditions and transforms therefore the 
$D$-$p$-brane to a $D$-$(p+1)$-brane. 
Similary $T$-duality in a tangential direction transforms the 
$D$-$p$-brane in a $D$-$(p-1)$-brane. This is of course in accordance with 
the long known fact \cit(dlp) that $T$ duality on an odd number of 
compact dimensions reverses the relative chirality of the left- and the 
right-moving ground states and maps therefore the IIa to the type 
IIb theory.

There is a Dirac quantization condition on the R-R charge quanta 
$\mu_p$ of the $D$-$p$-branes defined from 
\be
S={1\over 2}\int F_{p+2}^* F_{p+2}+i\mu_p
\int_{branes} A_{p+1}
\ee
namely
\be
\mu_p\mu_{6-p}=2 \pi n
\ele(diraccond)
with $\mu_p^2=2 \pi (4 \pi^2 \alpha')^{3-p}$ and minimal charge 
quantum $n=1$\cit(PCJ).

Upon compactification on homological nontrivial spaces $X$ of dimension 
$d$, $p$-branes can wrap around $n$ dimensional cycles $n\le p+1$ 
to yield $p-n$ ``dimensional'' objects in $10-d$ dimensions, 
i.e. instantons, solitons, solitonic strings etc. Of particular 
interest are supersymmetric minimal action configurations as they lead
to BPS states in $10-d$ dimensions. 

Let $\phi:\Sigma_{p+1}\rightarrow X$ define the embedding of the 
membrane worldvolume in the target space. For the supersymmetric 
{\sl instantons}\  the conditions boils down to the requirement of maximal 
supersymmetry on the worldvolume \cit(sbb), which is ensured if the 
global 10 d susy of the fermion on the worldvolume 
can be undone by a worldvolume $\kappa$ symmetry. 
That leads to the requirement   
\bea(rl)
P_-\eta&={1\over 2}\biggl(1-{i\over (p+1)!} h^{-1/2}
\varepsilon^{\sigma_1\ldots\sigma_{p+1}} \partial_{\sigma_1}\phi^{n_{p+1}}
\ldots \cr  
& \ \ \ldots \partial_{\sigma_{p+1}}\phi^{m_{p+1}}
\Gamma_{m_1\ldots m_{p+1}}\biggr)\eta=0,
\elea(susy)
where $\eta$ is a covariantly constant 10d spinor and $h$ is the induced
metric on the wordvolume. Submanifolds which fulfill (\ref{susy}) 
are called {\sl supersymmetric cycles}; they are not 
independent elements of the homology of $X$, e.g. for 
$p+1=0$ they are just all points of $X$. 
Supersymmetric two-cycles (\ref{susy}) are holomorphic embedding of 
$\Sigma_2$ in $X$, i.e. $\partial_z\bar \phi=\partial_{\bar z} \phi=0$ 
\cit(sbb). If $X$ is a  Calabi-Yau manifold of dimension $d$ 
one can rephrase (\ref{susy}) for supersymmetric d-cycles as 
$*\phi(\Omega)\propto \omega_d$ and $*\phi(J)=0$, where $\Omega$ is 
the holomorphic $(d,0)$ form $\omega_{d}$ is the volume form on the 
worldvolume, $J$ is the K\"ahlerform on $X$ and $*$ is the Hodge star 
operator on the worldvolume. Such embeddings are also known as 
special Lagrangian submanifolds \cit(gw). As in \cit(sbb) we will 
assume that the single wrapping of higher dimensional objects on the 
supersymmetric cycles will lead to solitonic BPS states. 

The most prominent example for that mechanism is that the wrapping 
of the type IIb threebrane around a supersymmetric three cycle 
in the class $V=m_i A^i+n^k B_k\in H^3(X,\ZZ)$, which vanishes 
as $S^3$ near the 
conifold, leads to an extremal black hole in four dimensions 
\cit(sbb), whose mass is
\be 
M=g_5 e^{K/2}| \int_V \Omega|=g_5 e^{K / 2} |m_i Z^i-n^k F_k|  ,
\ele(blackmass)
where $g_5$ is the five form coupling and $K$ is the K\"ahlerpotential, 
see section (\ref{specialkaehler}) and we expanded $V$ as well as 
$\Omega$ in terms of the basis (\ref{intersectg},\ref{poinc}). The
general form of this central charge in (\ref{blackmass}) term was 
found as the unique K\"ahler and ${\rm SP}(2 h_{21}+2,\ZZ)$ invariant 
expression in \cit(cdfp).
The four dimensional magnetic and electric charges of the black-hole
state can be obtained by integration the associated field strength
over $A^i\times S^2$ or $B_i\times S^2$ respectively, which must 
yield in view of (\ref{diraccond}) integer quantized charges 
$g_5 n^k=\int_{A^k\times S^2} F_5$ and $g_5 m_i=\int_{B_i\times S^2} F_5$.     

In rather generic situations the corresponding dual cycle to the $S^3$ 
cycle $V$ has the topology of a $S^2\times S^1$ and its vanishing gives 
rise to the massless vector multiplet which is needed to complete 
the Seiberg-Witten field theory embedding \cit(klmvw).     

Another important application is that a D-branes in type IIa theory 
wrap\-ped around non-isolated holomorphic curves will lead to 
non-perturbative gauge bosons \cit(km)\cit(kmp)\cit(kkv) which become 
massless if the holomorphic curve vanishes to a curve singularity 
and D-2-branes wrapped around isolated curve give rise \cit(gmp) 
to the dual magnetic monopole states in the sense of \cit(swI).    

Beside the point like states from wrapping $D$-branes also  
instantons can arise when a $D-p$-bane wraps a vanishing 
$p+1$ cycle. These were studied e.g. in \cit(sbb)\cit(ogva). 
More generally if a $D-p$-brane wraps a vanishing $p-n$-cycle 
a tensionless extended object of dimension $n$ arise in the 
compactified theory. The case of  tensionless strings was studied e.g. 
in \cit(gaha)\cit(ganorI)\cit(ganorII)\cit(kmv).

\subsection{ The geometric deformation space and special 
K\"ahler geometry}

In the following we will deal with mainly with the special K\"ahler 
part of the moduli space. Beside for the application to $N=2$ 
Heterotic/TypeII duality we have in mind, the special structure 
was utilized in $(2,2)$ compactifications of the heterotic string 
on CY threefolds, which has $N=1$ supersymmetry and gauge 
group $E_6$. The moduli space of this compactification can be 
obtained at tree level from the moduli space of the type II 
string by setting to zero the Ramond-Ramond fields in the type 
II theories. For instance for the type IIa compactification 
on $X$ this yields 
\be
{\cal M}^{\rm het}(X)={SU(1,1)\over U(1)}\times {\cal K}_{h^{1,1}(X)}
\times {\cal K}_{h^{2,1}(X)}
\ee
The two sorts of moduli parameterize the two - and three 
point couplings between fields in the $\overline {27}$ and $27$ 
representation of $E_6$ respectively. 
This result was derived in \cit(seibergKIII)\cit(specialkaehler)
\cit(candelasetal)\cit(dkl).

As we saw in section (\ref{macros}) the scalars in the special K\"ahler part 
of the moduli space come from the geometric deformations of the CY
metric, which do not spoil Ricci-flatness (\ref{ricciflat}) and from 
the antisymmetric background field
$$
\delta G_{mn},\
\delta G_{\bar m\bar n},\qquad
\delta G_{m\bar n},\
\delta B_{m\bar n}\ . $$ 
It will be useful to introduce a metric on 
the space of metrics 
\be
ds^2= {1\over 2 V}\int_X G^{k\bar m} G^{l, \bar n}
      \biggl[\delta G_{k,l}\delta G_{\bar m \bar n} 
+(\delta G_{k\bar n} \delta G_{l\bar m}+
\delta B_{k\bar n}\delta B_{\bar m l})\biggr] 
\sqrt{G} d^3 z d^3 \bar z
\ele(metricmetric)
This metric will be identified with the special K\"ahler metrics on 
the space-time moduli space of the effective supergravity theory. 
In accord with the expectations from the supergravity Lagrangian it 
is block-diagonal. The first block with the pure deformations, also
known as Peterson-Weil metric \cit(tian), will be identified in type IIa 
with the metric on ``half'' of the quaternionic space-time hyper 
multiplet moduli space, which is special K\"ahler. 
The second block with the mixed deformation and the $B$ field, will 
be identified with special K\"ahler metric  of the  type IIa vector 
moduli space. For the type IIb theory the identification is reversed.

\subsubsection{Special K\"ahler manifolds}
\label{specialkaehler} 
Let us first give a working definition what special K\"ahler manifold
is. Beside the original literature quoted above there are  
recent general reviews \cit(crtp)\cit(abcdffm) and for heterotic/type II 
string duality aspects see especially \cit(lf).

On a complex manifold, here the moduli space ${\cal K}_{\# V}$,  
with any metric one can chose especially an {\sl Hermitian} metric  
for which the pure parts of the metric vanish and 
$\bar {\cal G}_{m\bar n}={\cal G}_{n\bar m}$. 
From the antihermitian tensor $i {\cal G}_{m,\bar n}$ one defines a 
$(1,1)$-form $J=i {\cal G}_{m\bar n} d\phi^m \wedge d \phi^{\bar n}$ in 
coordinates $\phi_m$ $\bar \phi_{\bar m}$ $m,\bar m=1,\ldots \# V$. 
A K\"ahler manifold can be defined by the condition $d J=0$, which is 
nothing then a {\sl local\ } 
integrability condition for the existence of the K\"ahler potential, 
a real function $K(\phi,\bar \phi)$ with the property that 
${\cal G}_{m,\bar n}=\partial_{\phi^m}{\partial}_{\bar \phi^{\bar n}} 
K(\phi,\bar \phi)$. 
We will see in the next section that $J\in H^{1,1}({\cal K})$. 
On a Hodge-manifold by definition $J\in H^2(X,\ZZ)$ \cit(hirzebruch), 
which means that there is a complex line bundle whose Chern class\foot{By the 
famous theorem of Kodaira such manifold admit an complex analytic 
embedding into projective space, see e.g. \cit(hirzebruch).} 
is $c_1(L)=[J]$ \cit(hirzebruch). The latter condition holds for the 
Peterson-Weil metric on the CY moduli space as was shown by 
Tian \cit(tian), which also matches the integrality condition, which was 
previously required by quantum consistency of supergravity \cit(bw). 

A special K\"ahler manifold is Hodge-manifold in which the K\"ahler
potential can itself be derived from {\sl holomorphic} line  bundle over 
${\cal K}$,  called {\sl prepotential} $F(\phi)$ (compare (\ref{rkpot})), 
as follows
\be 
e^{-K}:= i \left( \bar Z^k F_k-Z^k \bar F_k\right),
\ele(kph)
where $Z^k(\phi)$ $k=0,\ldots,\# V$ are special projective coordinates, 
which are multi valued holomorphic functions on ${\cal K}_{\# V}$, 
$F_a:={\p\over \p Z^a} F(Z)$ and $F$ is homogeneous function 
of the $Z^a$ of degree two, i.e. $Z^k F_k =2 F $.
$(Z,\p F)$ is a section of a ${\rm Sp}(2 \#V+2,\IR)\times GL(1,\IC)$ 
bundle, i.e. the transition between adjacent coordinate patches 
$U_i$ and $U_j$ are given by  
\be 
\left(\matrix{Z\cr \p F}\right)_{\{i\}} = e^{f_{ij}}M_{ij}
\left(\matrix{Z\cr \p F}\right)_{\{j\}}\ ,
\ee
with 
$M_{ij}\in {\rm Sp}(2 \# V+2,\IR)$, 
$e^{f_{ij}}\in {\rm GL}(1,\IC)$
subject to the usual cocycle condition. In inhomogeneous coordinates 
$t^a:={Z^a\over Z^0}$ and using the homogeneity of ${\cal F}$ one can
rewrite the K\"ahlerpotential in terms of ${\cal F}$ with $F=-i(t^0)^2
{\cal F}$
\be 
e^{-K}=(t^a-\bar t^a)({\cal F}_a-\bar {\cal F}_a)-2 
({\cal F}+\bar {\cal F})
\ .
\ele(kpi)  
The curvature of a special K\"ahler manifold fullfils in these coordinates 
the constraint 
$$
R_{a{\bar b}c {\bar d}}=
{\cal G}_{a{\bar b}} {\cal G}_{c{\bar d}} +
{\cal G}_{a{\bar d}}  {\cal G}_{c{\bar b}} -
e^{2 K} C_{acm}{\cal G}^{m\bar m}\bar C_{{\bar m}{\bar b}{\bar d}},
$$
with $C_{abm}=\p_{t^a}\p_{t^b}\p_{t^m} {\cal F}$. Depending on the physical 
or mathematical context the $C_{abm}$ are quite differently called: 
Yukawa couplings in the $N=1$ heterotic compactifications, 
magnetic moments in the type II N=2 supergravity,
operator product coefficients or three-point functions 
in context of the conformal or topological field theory on 
the worldsheet and triple intersection numbers from the
point of view of the CY manifold. 

The analog of (\ref{theta}) becomes
\be
{\cal L}= -{1\over 4} g^{-2}_{kl} F^{k\mu\nu} F^l_{\mu\nu} -
{{\theta_{kl}}\over 32 \pi^2} F^{k\mu\nu}  {^*}F^l_{\mu\nu}, 
\ele(thetasugra)
where $k=0,\ldots,\# V$, i.e. $F^0_{\mu\nu}$ is the graviphoton 
field strength,
$g^{-2}_{kl}={i\over 4}({\cal N}_{kl}-{\cal N}_{kl})$, 
$\theta_{kl}=2 \pi^2 ({\cal N}_{kl}-{\cal N}_{kl})$ and 
\be
{\cal N}_{kl}=\bar F_{kl}+2 i {{\rm Im} \ F_{km} {\rm Im}\,F_{ln}Z^m Z^n \over
                              {\rm Im} \ F_{nm} Z^n Z^m}\ .
\ele(bosterms)

\subsubsection{The complexified K\"ahler cone}
\label{complexkaehler}

Let us describe the deformation spaces and start with $\delta G_{m\bar n}$ 
the so called {\sl K\"ahler deformation space\ } which is relatively 
simple. 

Since also the target space $X$ is K\"ahler we have as mentioned 
in the last section a $(1,1)$-form $J=i G_{m,\bar n} dz^m \wedge 
d z^{\bar n}$ with $d J=0$ and hence $ G_{m\bar n}=
\partial_{z^m} \partial_{{\bar z}^{\bar n}} K(z,\bar z)$.
On the other hand as 
\be
\omega ={1 \over d!}
\wedge_{i=1}^d J=i^n\sqrt{g} \wedge_{m=1}^d dz^{m} \wedge dz^{\bar m}
\ele(volumeform) 
the volume form, $J$ {\sl cannot} be exact 
$(J\neq d L)$, as exactness of $J$ would imply that $\omega$ is also exact 
and then by Stokes the volume would be zero $\int_X \omega=0$. That means 
$J\in H^{1,1}(X)$ for CY manifolds actually in $H^{1,1}(X,\ZZ)$.  
From the Licherowicz equation any of the mixed real 
deformations of the metric $i\delta G_{m,\bar n}$ is harmonic and hence 
in $H^{1,1}(X)$.
So that deformation space, called K\"ahler 
deformation space, can be described by the possible real values 
$R_i^2$ in the expansion 
$J=\sum_{i=1}^{h^{1,1}} R^2_i \alpha_i$ for $\alpha_i$ a 
basis of $H^{1,1}(X)$. We might think the $R_i$ 
roughly as sort of ``radii'' which measure 
certain directions in $X$. Of course one does not want 
any volume probed by $J$ to be negative and requires therefore the 
$R_i^2$ are restricted by the conditions
\be 
\int_{S^k}\wedge_{i=1}^k J>0\,,\quad k=1,\ldots n
\ee
for all non-trivial k-cycles $S^k$  on the CY  manifold 
$X$. These inequalities  force the $R_i^2$ to live inside 
the so called {K\"ahler cone}. 

The practical determination of the K\"ahler cone
can be difficult, we collected more literature in appendix E. 
In simple situations the CY manifold 
$X$ is embedded in a in a toric ambient space $Y$ such that the 
all K\"ahler classes of $X$ are pull backs of those on $Y$ and 
the relevant curves in the boundary of the K\"ahler cone of 
$Y$ are also present in $X$. Then the K\"ahler cone of $X$ 
can be determined as a subcone of the one in toric variety\footnote{
Slightly more complicated situations, where some curves are 
missing on $X$ were studied in \cit(bkk),\cit(bkkm).} 
$Y$\cit(hktyI)\cit(hktyII). The determination of the later 
was studied in \cit(torickaehlercone). An easy example of this kind
is the bi-cubic hypersurface $p=\sum c_{ijklmn}x_ix_jx_ky_ly_my_n=0$ 
in $\IP^2\times \IP^2$, with $x_i$ and $y_i$ are homogeneous coordinates
on the first and the second $\IP^2$. In this case $h^{1,1}=2$ and one
has the expansion $J=R_1^2 \alpha_1+ R^2_2 \alpha_2$, 
where $\alpha_i$ are the pull 
back of the K\"ahler form of the first and the second $\IP^2$ and the 
K\"ahlercone is simply the quadrant $R_1^2\ge 0$, 
$R^2_2\ge 0$ in which the volumina of both $\IP^2$ 
are positive. 

The real (1,1) form $B_{m,\bar n}  dz^m\wedge d z^{\bar n}$ is also 
harmonic as a consequence of the equations of motion for the 
antisymmetric tensorfield. As it also suggested by (\ref{bosonicaction})
it is natural to combine
\bea(rl) 
(J+B)&=(i G_{m,\bar n}+B_{m,\bar n}) \ dz^m\wedge d z^{\bar n} \cr 
&=\sum_{i=1}^{h^{1,1}} t_i \alpha_i
\elea(complexifiedkaehler)
and expand it in terms of a fixed basis 
$$\alpha_i \in H^{1,1}(X,\ZZ)$$ 
thereby introducing the  {\sl complex } expansion parameter 
\be 
t_i=i R^2_i+B_i\ .
\ele(parameterck) 

From the discussion of the moduli spaces  of the supersymmetric 
effective theory in section (\ref{macros}) and also from the 
moduli spaces, which deform the $N=2$ superconformal theories in section 
(\ref{sigmaapproach}) it is suggested that this complex structure
is really the natural complex structure which becomes extended 
to the special K\"ahler\footnote{K\"ahler refers here to the K\"ahler 
structure of the moduli space.} structure of the so called {\sl complexied 
K\"ahler\footnote{K\"ahler refers here to the K\"ahler structure on 
$X$.} moduli} space of $X$, which by the mirror hypothesis is identified
with moduli space which parameterize the the deformations the complex 
structure on the mirror manifold $\hat X$.

The most important quantities which dependent on the complexified 
K\"ah\-ler moduli are the two point functions and the three point functions 
between the operators of the topological $A$ model or for that matter 
between the operators of the $(a,c)$ ring of the $N=2$ SCFT. 
Up to some moduli dependent scale factor of the space-time fields, 
which will also be determined, 
they corresponds to the K\"ahler metric of the space-time moduli 
fields and their three-point couplings, known as Yukawa couplings in the 
$N=1$ heterotic compactification and magnetic moments in the $N=2$ 
Type II compactification. 

A non-vanishing three point function on the sphere in the 
topological $A$-model \cit(wittenmirror)\cit(wittentopolsigma)
involves three operators ${\cal O}_{V(p_i)}$ $V\in H^2(X,\ZZ)=H^{1,1}(X,\ZZ)$. 
In order to evaluate it one has to sum in the path integral over all 
instanton sectors\footnote{Such instanton corrections to
string couplings were first discussed in \cit(dsww).}. 
From the classical equation of motion  one learns that $L$ 
(\ref{bosonicaction}) is minimized for the holomorphic maps
$\partial_{\bar z} \phi^i=\partial_z \bar \phi^{\bar \imath}=0$.
The path integral reduces to an integral over the moduli space 
${\cal M }(\phi)$ of certain holomorphic maps whose measure defines 
an intersection number on ${\cal M}(\phi)$ \cit(wittentopolsigma). 

For the case at hand the resulting three-point function 
$C_{abc}$ has a formal expansion as  \cit(wittentopolsigma)
\cit(candelasetal)\cit(am)\cit(wittenmirror)
\be
C_{abc}=A_a\cap A_b \cap A_c +
\sum_{\phi(\IP^1)} n_a n_b n_c
             {e^{2\pi i\int_C\phi^*(J)}\over 
             1-e^{2\pi i\int_C\phi^*(J)}}
\ele(instantonexpansion)

The sum here is over all holomorphic embeddings $\phi(\IP^1)$ into $X$. 
The contribution of such maps will only depend only on  the class of 
$C$, which is determined by the integers $\vec n$ (degrees) 
$n_k=\int_C \phi^*(\alpha_k)=C\cap A_k$, which count how often $C$ 
meets the Poincar\'e dual $A_k$ of $\alpha_k$, note that $\alpha_k$ has 
$\delta$-support on $A_k$. 
Holomorphic maps do contribute only to the path integral if 
$C$ intersects all three $A_k$ or more precisely if the (three) 
points $p_i$ on $\IP^1$ fulfill $\phi(p_i)\subset A_i$ 
\cit(wittentopolsigma). The simplest possibility is that all of 
$\IP^1$ is mapped to a point $P=C$ in $X$ in this case the map contributes 
one, if the point $P$ hits one of the triple intersection points of the divisors, 
which gives rise to the {\sl classical} $A_a\cap A_b \cap A_c$ term. 
Even if $C$ is an single cover\footnote{Which means that the $n_i$ 
have no common nontrivial factor.} of an isolated curve, 
i.e. there are no moduli to move it $C$ in $X$, there are still moduli 
from the reparametrization of $SL(2,\IC)$ of $\IP^1$, 
which are compactified to $\IP^3$. The real dimension of ${\cal M}_{\vec n}$ 
is given by the number of zero modes  $a_{\vec n}$ of those fermions which
become scalars on the world-sheet, i.e. sections of 
$\phi^*(TX)$ ($TX$ real vectorspace over $X$), after the $(+,-)$ twist. 
In the situation at hand each constraint $\phi(p_i)\subset A_i$ 
leads to a linear constraint on this moduli space $\IP^3$ and the 
contribution to the path integral is the number of intersection of 
triples of these hyperplanes in the moduli space $\IP^3$. 
Since $C$ meets $A_i$ generically in $n_i$ points
this gives rise to the combinatorial $n_a n_b n_c$ factor 
in \ref{instantonexpansion}). It was shown in \cit(am) 
(under the restriction that the curves are isolated) that the same 
factor $n_a n_b n_c$ appears in front of the contributions of the 
$k$-cover curves, which have degrees $(k n_1,\ldots,k n_{h^{1,1}})$, hence
the general form of (\ref{instantonexpansion}). 

Unlike in this simple situation, where the constraints $\phi(p_i)
\subset A_i$ kill all the dimension of ${\cal M}(\phi)$, in
general one can end up with a subset $\widetilde {\cal M}(\phi)$ of 
positive real dimension $s$ in ${\cal M}(\phi)$. The dimensions, 
which can be killed by the constraints is the sum of 
the codimensions of the $A_i$. By ghost number conservation this is 
equal to $w_{\vec n}=a_{\vec n}-b_{\vec n}$, where $b_{\vec n}$ are the 
number of zero modes of the fermions which become currents on the 
worldsheet i.e. sections of $K\otimes \phi^*(T^{1,0})$ and $\bar K\otimes 
\phi^*(T^{0,1})$ under the $(+,-)$ twist \cit(wittenmirror). 
By the Riemann-Roch theorem $w_{\vec n}=2 d (1-g)$ for a 
worldsheet of genus $g$. I.e.  $s=dim {\cal M}(\phi_{\vec n})=b_{\vec n}$ 
and if $s>0$ one has to integrate over the Euler class of a 
$s$ dimensional vector bundle over $\widetilde {\cal M}(\phi)$. 
We will call the result of the intersection calculation on ${\cal M}(\phi)$ 
apart from the factor $n_a n_b n_c$ generically the {\sl instanton number\ } 
$N_{\vec n}$. The problem that $w_{\vec n}<{\rm dim}
{\cal M}(\phi_{\vec n})$ can be circumvented by perturbing the 
complex structure on $X$ to a non-integrable almost complex one and consider 
so called pseudo holomorphic embeddings \cit(gromov). 
This has been discussed in the similar context as above in 
\cit(ruantian). Direct mathematical approaches to the calculation of the 
instanton numbers can be found in \cit(km),\cit(giventhal).

After continuation (\ref{instantonexpansion}) to the complexified 
K\"ahler cone one gets
\be
C_{abc}=C^0_{abc}+\sum_{\vec n} 
{N_{\vec n} n_a n_b n_c
\over{ 1-\prod_e q_e^{n_e}}} \prod_e  q_e^{n_e}
\ele(ninst)
with $q_i=\exp(2 \pi i t_i)$. This parameterization is consistent with
the fact that (\ref{bosonicaction}) is unchanged if $B$ is shifted by 
an integer cohomology class.

The importance of the complexification of the K\"ahler cone, as suggested 
by mirror symmetry, can hardly be overestimated. 
Two profound consequences are as follows 

\mabs The three-point functions (\ref{ninst}) are determined, thanks to 
      the special structure of the complexified K\"ahler cone, by the 
      third derivatives of an {\sl holomorphic} section (prepotential) of a 
      line bundle over the moduli space, see section (\ref{complexstructure}).
      Similar as in the Seiberg-Witten case  such holomorphic sections are 
      fixed by finitely many data, in fact by the monodromies of 
      the ``periods'' around the discriminant loci and some boundary 
      values at the discriminant. The ``surprise'' that one can calculate 
      all the world-sheet instantons here is very  similar in nature as 
      the ``surprise'' that one can calculate all space-time instantons 
      in Seiberg-Witten theory.

\mabs The parameter $t_i=iR_i^2 + B_i$ have to be taken seriously as the 
      parameters by which the string theory explores the geometry of $X$.
      Especially the loci of singularities of the theory are determined 
      by {\sl complex  conditions} on the moduli space, i.e. they occur 
      at complex codimension one (at the discriminant locus).
      In contrary to the expectation from classical geometry the theory 
      {\sl cannot be generically} singular at the real codimension one 
      loci at which $X$ is singular as certain $R^2_i$ vanish, see 
      \cit(agm)\cit(agmsd). On the other hand for stability question 
      of the type II solitons real codimension one constraints can play a 
      decisive r\^ole, comp sect \ref{consistencychecks}.

\subsubsection{At the large radius limit} 
\label{largekaehler}
It is obviously very difficult to calculate the instanton sum
(\ref{instantonexpansion}) directly in the $A$ model and despite 
the fact that the counting of the rational curves in algebraic varieties 
is a mathematical subject, which goes back to the nineteenth century,
the amazing symmetries, which allow for recursive description for the 
$N_{\vec n}$ \cit(km),\cit(giventhal) were not suspected before Candelas, 
del la Ossa, Green and Parks had determined {\sl exactly} 
(\ref{ninst}) for the quintic using the mirror symmetry 
hypothesis \cit(candelasetal). Before we discuss this approach we want 
to describe a very important limit in which it is actually easy 
to calculate the quantities on the $A$ model side, 
namely in the limit in which radii all $R_i^2$ are large and 
deep inside the K\"ahler cone of $X$ so that the instanton 
contributions are suppressed as $q_i\rightarrow 0$.

On the harmonic forms in $H^{1,1}(X,\ZZ)$ one can define an 
inner product 
\be
{\cal G}^0_{a\bar b}={1\over {\rm  vol}(X) } 
\int_X \alpha_a\wedge *\alpha_b .
\ele(innerproduct)
For $\sigma\in H^{1,1}(X)$ one has the identity \cit(stromingerI) 
\be
*\sigma=-J\wedge \sigma+{3\over 2} 
{C^0_{\sigma JJ}\over C^0_{JJJ}} J\wedge J,
\ele(stromingereq)
where we abbreviate similarly as before 
$C^0_{\sigma JJ}:=\int_X \sigma\wedge J \wedge J=
A_{\sigma}\cap A_{J}\cap A_J$ in view of (\ref{intersectg}, \ref{poinc}).
Using (\ref{stromingereq}) and the expression for the volume form 
(\ref{volumeform}) and (\ref{complexifiedkaehler},\ref{parameterck}) 
one can write the inner 
product as 
\be 
{\cal G}^0_{a \bar b}= - \partial_{t_a}\partial_{t_{\bar b}} {\rm log}\ 
C^0_{JJJ}.
\ele(limitspecialkahler)
With the definition
\be 
{\cal F}^0=-{1\over 3!} \sum_{a,b,c}C^0_{abc} t_a t_b t_c
\ee
one can moreover express the classical K\"ahler potential $K^0(t, \bar t)$ 
for the metric ${\cal G}_{a \bar b}$ as
\be 
e^{-K^0}= \left[(t_a-\bar t_a)(\cF^0_a-
\bar \cF^0_{\bar a})- 2 (\cF^0 -\bar \cF^0)
\right]
\ele(kpotentiallimit)
and the triple couplings as
\be 
C^0_{abc}=\cF^0_{abc}
\ele(classprepot)
where $\cF_a:=\partial_{t_a}$ etc.
These formulae, valid in the limit of large ``radii'', describe 
precisely the relation between metric and triple couplings in
special K\"ahler geometry in special coordinates.
Sub-leading terms to $\cF^0$ are a priori not determined. 
They will not affect the $C_{abc}$ at the 
large complex structure and if the coefficients are real they 
will not affect the metric either. However even real parameter
will play an physical r\^ole if $\cF$ is analytically continued
to other regions in the moduli space. As it was found in explicit 
calculations \cit(candelasetal)\cit(cdfkm)\cit(hktyII)\cit(cfkm) one has 
$\cF^0\rightarrow \cF^0+ B_{a} t^a + C$,  where $C=i{\zeta(3)\over 
(2 \pi)^3} \int_X c_3$ and real $B_a={1\over 24}\int_X c_2\wedge J_a$.

The world-sheet instanton corrected prepotential on the 
K\"ahler side is expected to converge in a polydisk $|q_i|< q^0_i$
and the general expression obtained from (\ref{ninst}) reads  
\be
\cF=\cF^0+\sum_{n_1,\ldots,n_{h_{11}}\ge 1} N_{\vec n} 
{\rm Li}_3\left(q^{n_1}\cdot \ldots\cdot q^{n_{h_{11}}}\right)\ ,
\ele(fullprepot)
where ${\rm Li}_3=\sum_{k\ge 1} {x^k\over k^3}$. 
This asymptotic behaviour will become crucial for the identification of the 
K\"ahler structure deformation parameter on $X$ with complex 
structure deformation parameter on the mirror and the 
parameterization of the dual heterotic string theories. 

The first part is to identify the matching degeneration of the mirror 
Calabi-Yau space in dependence  of its complex structure. 
The  shift transformation of the periods under 
$t_i\rightarrow t_i+1$, which leaves the physical quantities invariant,
implies on the complex structure side a special degeneration of the 
periods with maximal unipotent monodromy
\cit(candelasetal)\cit(morrisonuni)\cit(batyrev)\cit(hktyI)\cit(agm)\cit(hly), see appendix D 
for the leading  behaviour of the periods on Calabi-Yau complete
intersections in toric varieties 
at this point and (\ref{mirrormap}) for the precise identification 
of the complex parameters with the K\"ahler parameters.
                   
The second part is the identification of the perturbative heterotic 
moduli and especially the heterotic dilaton\cit(kv)\cit(klm)\cit(al) with
the space time moduli of the Calabi-Yau.
This turns out to be rather simple and is 
described in sect. \ref{secperthet}.

\subsubsection{The deformation of the complex structure}
\label{complexstructure}
The pure deformations $\delta g_{\mu\nu}$ and $\delta 
g_{\bar \mu \bar \nu}$  describe the deformations of the 
complex structure. Let $a,b,c,d=1,\ldots, 2 d$ refer to a real 
coordinates $\vec x= (\vec v ,\vec w)$,  $v^m={1\over 2} 
(z^m+\bar z^{\bar m})$,  $w^n={i\over 2}(\bar z^{\bar n}-z^n)$ 
$m,n=1,\ldots, d$ of $X$. As $G_{ab}+\delta G_{ab}$ is still 
a K\"ahler metric close to the original one can find a coordinate 
system in which the pure parts of the new metric vanish. 
Let $x^m\rightarrow x^m + \epsilon^m(x)$ then the variation of 
the metric transforms 
$\delta G_{ab}\rightarrow \delta G_{ab}-{\partial \epsilon^c\over 
\partial x^a}G_{cb}-{\partial \epsilon^c\over \partial x^b}G_{ac}$. 
If $\epsilon^n(z)$ is holomorphic then the pure part of the transformation
can not be removed, or put it differently the new metric cannot be reached 
by a holomorphic coordinate change, i.e. the deformation changes the 
complex structure.

From section \ref{sigmaapproach} we know that the algebra of observables
of the $B$ model is identified with an algebra on 
$\otimes_{p=0^d} H^p(X,\wedge^p T)$. We will now describe the 
calculation of the 2-point and 3-point functions in the topological 
$B$ model, which depend only on the complex structure 
variation of $X$.

As a warm up we start with the case of a CY threefold.
Here we expect to find special K\"ahler structure \cit(tian),
(last reference in \cit(specialkaehler))\cit(comod).

A complex structure on $X$ is fixed by choosing a particular element of 
$H^3(X)$ as the holomorphic (3,0) form $\Omega$.
As in section  (\ref{geompict}) we expand the holomorphic form 
in terms of the topological basis (\ref{intersectg},\ref{poinc}) as 
\be
\Omega=Z^i\alpha_i-F_i\beta^i
\ele(omegaexpandI)
 where 
\be 
Z^i=\int_{A^i}\Omega,\ \ \ \, F_i=\int_{B_i}\Omega
\ele(omegaexpandII)
are periods of $\Omega$. It was shown in
\cit(brgr)\cit(tian) that the  $Z^i$ are local complex projective 
coordinates for the complex structure moduli space in the sense of 
(\ref{kph}), i.e. we have $F_i=F_i(Z)$.
Under a change of complex structure
$\Omega$, which was pure (3,0) to start with,
becomes a mixture of $(3,0)$ and $(2,1)$, i.e.
${\p\over\p z^i}\Omega\in H^{(3,0)}\oplus H^{(2,1)}$. 
In fact as explained 
e.g. in \cit(tian)
${\p\Omega\over\p z^i}=
k_i\Omega+b_i$ where $b_i\in H^{(2,1)}$ is related to elements in
$H^1(M,T_X)$ via $\Omega$
and $k_i$ is a function of the moduli but
independent of the coordinates of $X$. One immediate consequence is
the so called transversality relation 
$\int\Omega\wedge{\p\Omega\over\p Z^i}=0$.
Inserting the expression for $\Omega$ in this equation, one finds
$F_i={1\over 2}{\p\over\p Z^i}(Z^jF_j)$,
or $F_i={\p F \over\p Z^i}$ with
$F={1\over 2}Z^i F_i(Z)$, $F (\mu z)
=\mu^2 {\cal F}(z)$.
{}From ${\p^2\over\p Z^i\p Z^k}\Omega\in H^{(3,0)}\oplus
H^{(2,1)}\oplus H^{(1,2)}$ it immediately follows that
also $\int\Omega\wedge{\p^2\over\p Z^i\p Z^j}\Omega=0$. In fact, this
is already a consequence of the homogeneity of ${\cal F}$.
Finally, ${\p^3\over\p Z^i\p Z^j\p Z^k}\Omega\in H^{(3,0)}\oplus
H^{(2,1)}\oplus H^{(1,2)}\oplus H^{(0,3)}$ and one easily finds
$C_{ijk}=\int\Omega\wedge{\p^3\over\p Z^i\p Z^j\p Z^k}\Omega
={\p^3\over\p Z^i\p Z^j\p Z^k}F
=(Z^0)^2{\p^3\over\p  t_i\p  t_j\p t_k}{\cal F}$ ; here $i,j,k=1,
\dots,h^{2,1}$. It is also easy to see  that in accordance with (\ref{kph})
\be 
K=-\ln i \int\Omega\wedge\bar\Omega\ .
\ele(cskpot) 
If we redefine $\Omega\to{1\over z_0}\Omega$, the periods are
$(1, t_i,{\p\over\p  t_i}{\cal F},2{\cal F}-t_i{\p\over\p  t_i}
{\cal F})$ cf. \cit(hktyII).

One can show that the Yukawa couplings transform homogeneously
under a change of coordinates $t_i\rightarrow \tilde t_i(t)$
and thus $C_{ijk}=\int\Omega\wedge\p_i\p_j\p_k\Omega$ holds in any
coordinate system. In particular in the one given by the coefficients
$a_i$ in front of the monomial deformations of the defining polynomial 
of the CY\foot{See e.g. \cit(comod), why they define local 
inhomogeneous coordinates for the complex structure deformation.} 
as e.g. in (\ref{defpol}). 
On the other side the $C_{ijk}$ can be written as the third 
derivatives of the prepotential only in special coordinates.
To summarize the transformation properties note that $\Omega$ is
fixed only up a gauge transformation $\Omega\rightarrow f(z) \Omega$
with $f(z)$ holomorphic, so $\Omega$ lives in a line bundle $L$ over 
the moduli-space ${\cal K}$ and the above quantities
transform as elements of
\bea(rl) 
C_{ijk}&\in L^2\otimes {\rm Sym}((T^*_{\cal K})^{\otimes 3}), \cr 
 e^{-K}&\in L\otimes \bar L,\ \ F\in L^2\ .
\elea(trans) 

For manifolds of dimension $d$ the  $d$-point  
$C_{i_1,\ldots, i_n}=\int \Omega\wedge \p_{i_1}\ldots \p_{i_n}\Omega$ 
can be easily calculated explicitly from the Picard-Fuchs equations, 
let us say in the coordinates $y_i$ cf. (\ref{example}). 
It is usefull to define
\be 
W^{(k_1,\cdots,k_r)} := \sum_l(Z^l{\bf \p}^{{\bf k}} 
F_l -F_l{\bf \p}^{{\bf k}} Z^l),
\ee
where ${\bf \p}^{{\bf k}}:= \p_{y_1}^{k_1} \ldots \p_{y_r}^{k_r} $.
In this notation, $W^{({\bf k})}$ with $\sum k_i={\rm dim} (X)=d$ 
describes the various types of $d$-point functions and the generalized transversality relations
are
\be
\int_X \Omega \wedge \ {\bf \p}^{{\bf k}} \Omega= 
W^{({\bf k})}\equiv 0 \quad  {\rm for}\  \sum k_i<n \ .
\ele(vanish)
If we now write the Picard-Fuchs differential operators in the form
${\cal L}_a=\sum_{\bf k} f_a^{({\bf k})}{\bf \p}^{\bf k}$ 
then we immediately obtain the relation $\sum_{\bf k} 
f_a^{({\bf k})}W^{({\bf k})}=0.$
Further relations among the $W^{({\bf k})}$ must be obtained in general 
from operators ${\bf \p}^{\bf k}{\cal L}_a$. 
If the system of PF differential equations is complete, 
these equations are sufficient to  derive linear relations among 
the Yukawa couplings and their derivatives, which can be integrated 
to give the Yukawa couplings up to an overall normalization.
See \cit(candelasetal) for the simplest example. For more 
details we refer to \cit(hktyI). It follows from the general
theory of the singular loci of systems of differential 
operators\cit(yoshida) that the denominators of these $d$-point 
functions correspond to components of the singular loci.

Next we turn to a general discussion for the  calculation 
of the basic two- and three-point functions for general CY 
$d$-folds. 
Let $\pi:{\cal X}\ra S$ be a complex structure deformation family 
whose generic fiber is a CY $d$-fold $X_s$. One writes now 
the three-point is a cubic form on 
$H^p(X_z,\wedge^p T)$. Put $\cB_s=\oplus H^p(X_s,\wedge^pT)$ defined by 
\be 
C(a,b,c)=\int\Omega(a\wedge b\wedge c)\wedge\Omega
\ele(bcoupling)
where $\Omega(a\wedge b\wedge c)$ is the contraction along the 
tangent direction producing an $d$-form on $X_z$.

We shall first fix a base point $0\in S$, a topological
base of homology cycles and the dual base $\gamma_a^{(p)}$
on $H^d(X_0)$ with the property that 
$\bra\gamma_a^{(p)},\gamma_b^{(q)}\ket=0$ 
for $p+q\geq d$. For fixed $p$, the label $a$ in $\gamma^{(p)}_a$
takes $h^{d-p,p}(X_0)$ different values.
Due to mirror symmetry such a base will be the image of a 
base on $\cA$ under $\phi_0$. In fact in practice, there is usually a 
canonical choice of such a base on the A-model side.

There is a filtration of holomorphic vector bundles over $S$: 
$F_{(0)}\subset F_{(1)}\subset\cdots\subset F_{(d)}$,
where the fiber over $s\in S$ of $F_{(k)}$ is given by the vector space
$\oplus_{p=0}^{k} H^p(X_s,\wedge^p T)$. 
We now provide a set of frames for the these bundles. We shall
express these frames as linear combinations in the base $\gamma_a^{(p)}$
with holomorphically varying coefficients.
We shall see that these coefficients completely determine the
three-point function $C$. For each $k$, let $\{\alpha^{(0)}:=
\Omega,\alpha^{(1)}_a,..,\alpha^{(k)}_b\}$
be a frame of $F_{(k)}$ having the following upper-triangular
property with respect to the $\gamma^{(p)}_a$:
\be 
\alpha^{(k)}_a=\gamma^{(k)}_a+\sum_{p>k}g^{(p)c}_a\gamma^{(p)}_c.
\ele(alphaEQ)
(The $g^{(p)}$ actually depends on $k$, which we have suppressed
in the notation above.)
These frames can be obtained by row reduction on a given arbitrary base
of sections. (See \cit(gmp).) Note that for $k=0$
the coefficients $g^{(p)}$ are exactly the periods 
of the above given homology cycles.
These periods are solutions to the Picard-Fuchs equations (in an
appropriate gauge). We will give explicit formulas later for these
periods for CY complete intersections in a toric variety. Note that
in $\alpha^{(0)}$ the coefficients $t_a:=g^{(1)}_a$ are regarded
as local coordinates on $S$. These are the so-called flat coordinates.
In these coordinates the Gauss Manin connection $\nabla_{a}$ becomes $\partial_{t_a}$, 
and the three-point functions  of type $(1,k,d-k-1)$ is given by 
\be 
C^{(1,k,d-k-1)}_{a,b,c}=
\int_X \alpha^{(d-k-1)}_a\wedge \partial_{t_a} \alpha^{(k)}_b 
=:\bra \partial_{t_a}\alpha^{(k)}_b,\alpha^{(d-k-1)}_c\ket.
\ele(CubicForm)

Using the upper-triangular property of the $\alpha^{(k)}_a$ and
the topological basis $\gamma^{(k)}$, it
is easy to show that
\be
\eta_{ab}^{(k)}:=\bra\alpha^{(k)}_a,
\alpha^{(d-k)}_b\ket=
\bra\gamma^{(k)}_a,\gamma^{(d-k)}_b\ket.
\ele(InnerProduct)
In particular these matrix coefficients are independent of $t$. 
Furthermore we claim that
\be\partial_{t_a}\alpha^{(k)}_b=
C^{(1,k,d-k-1)}_{a,b,c}\eta_{(d-k-1)}^{cd}\ \alpha^{(k+1)}_d.
\ele(Transversality)
By Griffith's transversality, we have
$\partial_{t_a}\alpha^{(k)}_b\in F_{(k+1)}=Span\{\alpha^{(0)},..,
\alpha^{(k+1)}_a\}$. But because of the 
upper triangular form of $\alpha^{(k)}_b$, 
$\partial_{t_a}\alpha^{(k)}_b$ has
zero component along $\gamma^{(0)},..,\gamma^{(k)}_a$. Thus
it can be expressed as a linear combination (with holomorphically
varying coefficients) of the $\alpha^{(k+1)}_b$. To determine the
coefficients, we take its inner product with $\alpha^{(d-k-1)}_c$
and apply eqns (\ref{CubicForm},\ref{InnerProduct}). The 
claim above then follows.

To summarize, our strategy for computing the A-model three-point-func\-tion 
$Q$ on $X$ by mirror symmetry is as follows. Actually we will only do it for
a Frobenius subalgebra $\cA$ (see below) of the A-model algebra. First
we fix a topological basis on $\cA$ (In the case
of toric hypersurfaces, this basis
will come from toric geometry). We define
our isomorphism $\phi_s$ so that it sends this basis to the
holomorphically varying basis $\alpha^{(k)}_a$ of the B-model
with $1\mapsto\alpha^{(0)}$.
Then  we shall use eqns (\ref{CubicForm},\ref{InnerProduct},
\ref{Transversality})) as our crucial ingredients for computing the
B-model three-point functions $C$ explicitly. The actual computation
will be subject of appendix D.

\subsection{Heterotic-Type II String-duality}
\label{stringduality}

$K3\times T^2$ break ${1\over 2}$ and the CY threefold ${3\over 4}$ 
of the supersymmetry generators of the ten dimensional theory.
Therefore the type II string theory has $N=4$ or $N=2$ 
supersymmetry, when compactified on $K_3\times T^2$ or on CY
threefolds. Similarly the heterotic string has $N=2$ or $N=1$ when 
compactified on $K_3\times T^2$ or on CY threefolds. Candidate
dual pairs appear of $N=2$ theories appear in\cit(kv)\cit(fhsv). 
Evidence that the $N=2$ theories are perturbatively equivalent was 
first given in\cit(kv)\cit(klm)\cit(kltt), while first 
non-perturbative properties where tested in\cit(kklmv).

\subsubsection{Perturbative heterotic prepotential and $K3$-fibrations}
\label{secperthet}

As  $K3\times T^2$ gives rise to a $N=2$ supergravity theory the 
general macroscopic structure is  as in (\ref{macros})  
In particular the vector moduli space is special K\"ahler and 
governed by a holomorphic prepotential and despite the fact that
we have local supersymmetry the discussion parallels in many aspects 
the discussion in section (\ref{defaction}). 

Like in (\ref{finfinit}) the perturbative prepotential has, because
of corresponding renormalization theorems \cit(lf), only the classical 
three-level term $\cF^0$ and the string one-loop term. Because 
of the special r\^ole of the dilaton $S$ in the vector multiplet moduli 
space we separate the fields $t^a$ $a=1,\ldots, \# V$, in $S$ and
$T^a$, where the latter are the scalars of neutral space-time moduli.   
As the dilaton arises in the universal sector it does not couple 
to any other of the non-universal $T^a$, in particular 
${K^0}=-\log (S+\bar S)+ K(T,\bar T)$. That implies 
\cit(fvp)  
\be
\cF_{pert}=-S (\eta_{a b} T^a T^b)+{\cF}_{1-loop}\ . 
\ele(fhetpert)
Here $\eta_{a,b}={\rm diag}(1,-1,\ldots,-1)$ and we have suppressed 
charged vector multi\-plets\foot{They would contribute with 
$S(\delta_{ij} Q^iQ^j)$ to ${\cal F}_{pert}$.}. What will become important
for us is the fact that  the tree-level coupling to any 
non-abelian gauge factor is given simply by
\be
g^{-2}={\rm Re }\ S\ .
\ele(gaugecouplingI)
The space-time instanton effects, i.e. the nontrivial self-dual gauge field
configurations in (\ref{thetasugra}), brake the freedom 
of shifting the dilaton by an arbitrary real constant\cit(lf) to integer
shifts $S\rightarrow S+{i n\over 4 \pi}$.
The non-perturbative prepotential close to a perturbative limit must 
therefore be of the form
\be 
\cF= \cF_{pert}+\cF_{np}(e^{-8 \pi^2 S},T^a).
\ele(fullhetprop)
Comparison with (\ref{fullprepot}) shows immediately 
that, while we have the discrete shift symmetry on all of the CY 
K\"ahler moduli at the large radius limit, the fact that one modulus 
must couple {\sl only} linearly in the classical term singles out the
one which is to be identified with the dilaton. Since the classical 
terms of the CY prepotential are fixed by the intersections of 
divisor classes, it means that there is one divisor class say $D_S$ 
with $D_S \cap D_S=0$. That implies that the CY is a 
fibration $F\rightarrow  X \rightarrow \IP^1$, where $D_S$ is to be 
identified with the class of the fiber $F$. Such a fibration is defined  
by a projection map $\pi: X\rightarrow B=\IP^1$ such that the 
pre-image  of the {\sl generic} point in $B$  $F_p=\pi^{-1}(p)$ is a smooth 
manifold. However at codimensions $1$ over the base $F_p$ is 
allowed to  degenerate.
Trivially the class of the fiber has the property $F\cap F=0$ since two 
divisors $F$ can only intersects on the base, but points do not intersect 
generically\foot{In general of course  $\cap_{i=1}^r F=0$ for 
$r > {\rm dim}(B)$.} in $\IP^1$. Conversely if  one has a numerically 
effective divisor class $F$ in $X$ with $F\cap F=0$ one can project along the 
$F$ and $X$  admits a fibration with fiber $F$. For $X$ to be CY 
$c_1(T_F)=0$, so the fiber above can only be $K3$ or $T^4$. It was argued\cit(klm) that in pairs of dual type II/heterotic strings\cit(kv), 
which admit a perturbative heterotic limit, the CY must be $K3$-fibration.
The two form $\sigma_S$ dual to $F$ has support on the base so that
the geometrical parameter $t_S$ measures the complexified `size' of 
the base $\IP^1$, i.e its imaginary part measures the size of 
$\IP^1$ and the real part the flux of the antisymmetric $B$-field over 
it. Comparing now eq. (\ref{fullhetprop}) with eqs. 
(\ref{classprepot},\ref{fullprepot}) we learn that we should identify 
\be 
t_S= 4 \pi i S \ . 
\ele(ind) 
The higher derivative couplings $g_n$ between the curvature tensor 
and the gravi\-pho\-ton field strength $G$ i.e. 
\be 
{\cal L}=g_n^{-2} R^2 G^{2 n-2}
\ele(higherderivative) 
\cit(bcov)\cit(agnt) are governed at least at tree-level by the so 
called topological n-loop partition functions \cit(bcov) 
${\cal F}_n$ as $g_n^{-2}={\rm Re} (\cF_n)$, see \cit(lf) for a review. 
Especially  it was shown in \cit(bcov) that $\cF_1\propto \sum_i t^i 
\int_X c_2\wedge \sigma_i+O(q_i)$ at the large radius limit. 
This distinguishes between the $F=T^4$ with $\int_X c_2 \wedge \alpha_S=0$ 
and $F=K3$ with $\int_X c_2\wedge \alpha_S=24$ and as argued in \cit(al)
the generic situation is $\cF_1\propto S$ hence $F=K3$.
    
The above  statement does by no means imply that type II compactifications  
on CY manifolds, which are no $K3$-fibrations do not have heterotic 
duals. In particular if we associate a heterotic string on 
$K3\times T^2$ to a type II compactification on a 
$K3$ fibration\foot{Which is, by the way, not unique as there exist CY 
which admit  several (even infinitely) many possible 
projection maps with $F=K3$.}, one can study transition to non $K3$ 
fibered CY manifold, well defined for the type II as argued by Strominger 
for conifolds transitions and by \cit(km)\cit(bkkm)\cit(kmv) for other 
transitions, there are no indications that one looses the correspondence 
to the heterotic string. There is  just no the perturbative limit in 
this branch of the parameter space of the heterotic string.   

\subsubsection{Spectra of the heterotic string on $K3\times T^2$}

In order to get some concrete examples let us next discuss the 
spectrum for the heterotic string on $K_3\times T_2$, following Kachru 
and Vafa \cit(kv). In the heterotic case the generic 
unbroken gauge group of the eight dimensional theory will
be ${\cal G}=E_8\times E_8\times G_{T^2}$ where the last part is 
a rank two gauge group from the $T^2$; generically an $U(1)^2$, 
but enhanced at special values in the moduli space of the torus.
Instead of considering only the standard embedding of the 
$SU(2)$ holonomy into the gauge group, we like to allow the more general
situation, where one takes stable gauge bundles with gauge 
group $H_a$ over $K3$ and embeds their connection into ${\cal G}$ 
to break ${\cal G}\supset \otimes_a H_a$ to the maximal commutant with the 
$\otimes_a  H_a$ subgroups. 
To yield a vacuum configuration these gauge bundles have 
to fulfill the constraints 
\bea(rl)
\sum_a h_a:&=\sum_a \int c_2(V_a)=\int c_2(T_{K_3})=24, \\ [ 3 mm]
   c_1(V_a)&=0,
\elea(cherngauge)
where $h_a$ is called the instanton number of $H_a$.
The dimension of the moduli space of the gauge bundle $H_a$ is 
given by 
\be
{\rm dim}_{\IR}({\cal M}_a)=4\ h_a {\rm cox}(H_a)-
4\ {\rm dim} (H_a)\ ,
\ele(dimha)
where cox$(H_a)$ is the dual Coxeter number of the group $H_a$.
Furthermore we need the number of fields transforming
in the matter representations of unbroken gauge group $G$, 
${\cal G}\supset \otimes_{a=1}^n H_a \times G $. 
Decomposing ${\rm adj}\ ({\cal G})=\sum_i (R_i,M_i)$ one has
from the index theorem that the number of fields in the $M_i$
representation is  
\be 
N_{M_i}={1\over 2}\int_{K_3} c_2(V) \
{\rm index}(R_i)- {\rm dim} (R_i)\ .
\ele(numbermi)

\mabs 
If we embed just the holonomy group $H=SU(2)$ into one $E_8$ factor,  
$E_8\supset SU(2)\times E_7\ $, we get, since 
$h=\int_{K_3} c_2(V)=\int_{K_3} c_2(T_{K_3})=24$ from (\ref{dimha}): 
$48-3$ quaternionic scalars of hyper multiplets. 
From the universal gravitational sector we get $h^{1,1}(K3)=20$
further scalars. The number of $\underline {56}$ is  
$N_{\underline {56}}=10$ by (\ref{numbermi}). One can use the latter
fields to Higgs the $E_7$ gauge group completely. This gives
rise to $(56\cdot 10-133)$ further neutral scalars. To summarize, 
the number of hypermultipletts is $45+20+427=492$ and the number of
vector multiplets is ${\rm rank}( E_8\times U(1)^2)=10$ plus the 
dilaton\footnote{One other vector the graviphoton sits in the 
gravitational multiplet, but does not correspond to a modulus}, 
i.e. $(\# H,\# V)=(492,11)$. 

\mabs 
Similarly if we take two 
$SU(2)$ gauge bundles with 
$h_1=12$ and $h_2=12$ and embed them
symmetrically into the two $E_8$-factors we get 
from (\ref{dimha})  $2\cdot(24-3)$ and from (\ref{numbermi}) and 
complete higgsing:
$2\cdot(4\cdot 56-133)$ plus 20 hyper multiplets and three 
vector multiplets; i.e. $(\# M ,\# V)=(244,3)$. This model may be 
called $(STU)$ model, because it contains the dilaton  $S$ and the 
K\"ahler and the complex modulus of the $T^2$  in $K3\times T^2$, 
which were called previously $T$ and $U$ .

\mabs 
For a last example take three copies of 
$SU(2)$  gauge bundles with $h_1=h_2=10$ and $h_3=4$ and embed them into
the three factors of ${\cal G}$ where $G_{T^2}=SU(2)\times U(1)$ 
this yields $2\cdot(20-3)+(8-3)+20+ 2(3\cdot 56-133)=129$
hypermultipletts and only two vector multiplets, the reason is that 
we had to fix one modulus, say the complex one, of the torus to the 
$G_{T^2}=SU(2)\times U(1)$ enhancement point, 
i.e. $(\# H,\# V)=(129,2)$. So we may call this the $(ST)$ model.

The moduli space of these theories is again governed 
by the $N=2$ supersymmetry and has therefore the basic structure
\be 
{\cal M}^{het}={\cal K}_{\# V } \otimes 
{\cal Q}_{\# H}\ .
\ele(hetmod)
Comparing that with (\ref{conmod}) we see that the most naive 
macroscopic requirement on potential dual IIa compactifications for the 
three heterotic string theories discussed above, is that the 
CY manifolds  should have $(h_{1,1},h_{2,1})=(11,491)$, 
$(3,243)$ and $(2,128)$. Indeed such CY manifolds exist
in the lists of\cit(ksks), namely the $K3$ fibrations 
$X_{84}(1,1,12,28,42)$, $X_{24}(1,1,2,8,12)$ and 
$X_{12}(1,1,2,2,6)$. As is turns out by a closer analysis of these
potential pairs this identification \cit(kv) is almost correct. 
The rectification is that the second model with ${\rm SU}(2)^2$ 
instanton numbers $(h_1=12,h_2=12)$ has to be identified with a 
closely related $K3$ fibration, which is is most easily characterized 
as an elliptic fibration over $\IP^1\times \IP^1=:F_0$. 
The $X_{24}(1,1,2,8,12)$ CY, which can also be viewed as 
elliptic fibration over the Hirzebruch surface $F_2$, corresponds 
actually to the heterotic model with ${\rm SU}(2)^2$ instanton numbers 
$(h_1=10,h_2=14)$ \cit(mvI). As the weak coupling behavior is the same 
as in the $(STU)$ model we call this $(STU)''$ model. 
In general it has been shown using $F$-theory compactification 
to six dimensions that $(h_1=12-k,h_2=12+k)$ heterotic strings correspond 
to elliptic fibrations over $F_k$ Hirzebruch surfaces \cit(mvII). 

The conjectured identification between the heterotic vector multiplet moduli 
space on $K3\times T2$ and the K\"ahler moduli space of 
CY spaces, implies a wealth of strange strong coupling 
physics for the heterotic
string. In particular it is  known that the CY moduli-spaces 
are connected by transitions, at least \cit(bkk)\cit(connect) 
for the wide class of examples in \cit(ksks) and complete intersections 
\cit(intersect) in which the dimension of the K\"ahler moduli space 
ranges between $1-491$. The above conjecture would be incomplete if it
would not be possible to follow this transitions on the heterotic side. The
dimensions of the vector moduli space and so the maximal rank of the 
gauge group is bounded in the perturbative description of 
the heterotic string due to the simple fact that a vertex operator for 
abelian gauge boson contributes with $1$ to the central charge of the 
Virasoro algebra and hence ${\rm rank}(G)+1\le 22+1$, where the $1$ comes 
from the dilaton modulus and the (-)22 from the ghost sector of the bosonic
string. In other words, there should be an enormous non-perturbative
gauge-symmetry enhancement possible on the 
heterotic side, which increases the rank of the gauge group 
to at least $491$. One known mechanism to obtain higher rank gauge groups 
is by small instantons as discussed in \cit(wittensi). 
The fitting effect on the Type II side comes from  
degenerate $K3$-fibers as was analyzed by in\cit(ag), 
corresponding transition where studied in\cit(lsty). 
For a review on heterotic/typeII duality in six and also in 
four dimensions and a more complete list of references to this
subject see\cit(aspinwallkiii).

\section{The $(ST)$ model, a concrete example}
\label{example}

We want to now to exemplify all the formal concepts about mirror 
symmetry, moduli spaces and type II/heterotic duality that we discussed in 
the last sections with a simple $K3$-fibration Calabi-Yau manifold. 
This $K3$-fibration is dual to the heterotic string $(ST)$ model, defined
in the last section.

The manifold is given by a degree $k=12$ 
hypersurface constraint in a weighted $\IP^4(1,1,2,2,6)$. 
According to (\ref{chern}) the first Chern class vanishes 
$c_1(T_X)=0$. We can represent this 
manifold as the zero of
\be 
p_0=a_1 x_1^{12}+a_2 x_2^{12}+a_3 x_3^6+a_4 x_4^6+a_5 x_5^{2}\ ,
\ele(defpol)
which is clearly transversal. The subscript `0' refers to the fact 
that we can perturb the polynomial by monomial perturbations which are 
also of degree 12. It is not difficult to find all possible, up to 
weighted homogeneous coordinate transformations, 126 monomial 
independent perturbations of degree 12. 
The coefficients of these monomials in the general 12th order 
polynomial are coordinates in the complex structure moduli space. 
This CY manifold does however admit 128 complex structure
deformations, two of which have no algebraic description in terms of a 
monomial perturbation of the defining polynomial\foot{One can find 
related representations in which all deformations are geometric 
see e.g. \cit(bkk)} $p_0$. This model thus has $h^{2,1}=128$.
The two non-geometric representations will be interpreted in section 
(\ref{strongcoupling}). To count the $h^{1,1}$ forms, we would like
to count the dual homologically inequivalent sub manifolds of 
codimension one in $X$, called divisors. As it turns out these 
come all in this simple example from restrictions of divisors classes 
of the ambient space $\IP^4(1,1,2,2,6)$. These divisors classes are not yet 
visible in the parameterization used in (\ref{defpol}). The reason is,
that we are working with a singular model. As we already mentioned 
weighted projective spaces have $Z_N$ singularities due to nontrivial 
(common) factors in the weights, here a $Z_2$ hyperplane $H$ given 
by $x_1=x_2=0$ and a $Z_6$ singular point $P$ given by 
$x_1= \ldots =x_4=0$. The constraint $p_0=0$ meets $H$ but 
not $P$, so we will need at least a representation of 
$\IP(\vec w)$ in which the $Z_2$ singularity is resolved. 
That can be achieved following \cit(batcox) by introducing more 
coordinates\foot{It is convenient but not necessary to also add $x_0$ 
at this point.} $(x_0;x_1,\ldots, x_6)\in\IC^7$ and 
more equivalence relations
\bea(rl)
(x_0;x_1,\ldots x_r) &\sim (
\lambda_{(k)}^{l^{(k)}_0} x_0;\lambda_{(k)}^{l^{(k)}_1} 
x_1, \ldots, \lambda_{(k)}^{l^{(k)}_r} x_r) \\[ 2 mm] 
{\rm with}& \left\{\matrix{
l^{(1)}&=(-6;0,0,1,1,3,\phantom{-}1)\cr
l^{(2)}&=(\phantom{-} 0;1,1,0,0,0,-2)}\right.  
\elea(equiv) 
and $\lambda_{(i)}\in \IC^*$. Similar as the locus $x_1=\ldots =x_{n+1}=0$
in the definition for the $\IP^n(\vec w)$ one has also here some forbidden 
loci, the technical terminus is Stanley-Reisner ideal, which have
to be subtracted from $\IC^7$, so that the $\IC^*$-actions are well 
defined. Here it is $x_3=\ldots= x_6=0$ and $x_1=x_2=0$. 
In fact dropping the $x_0$ coordinate (\ref{equiv}) together with the
Stanley-Reisner ideal defines the toric variety of the partly 
resolved $\IP^4(1,1,2,2,6)$, the associated polyhedron is the convex hull
of 
$\nu^{(1)}=(1,0,0,0)$, 
$\nu^{(2)}=(-1,-2,-2,-6)$
$\nu^{(3)}=(0,1,0,0)$, 
$\nu^{(4)}=(0,0,1,0)$, 
$\nu^{(5)}=(0,0,0,1)$, 
 and $\nu^{(6)}=(0,-1,-1,-3)$ compare 
appendix E. This partial resolution of $\IP(1,1,2,2,6)$ fits the general
description of a resolution given below (\ref{blowup}) and the map 
$\pi$ is just the identity outside the exeptional locus $x_6=0$. 
Namely if $x_6\neq 0$ we can use one $\IC^*$ action to set it 
to $x_6=1$ and the remaining $\IC^*$ action which respects this 
(gauge) choice: $l=(1,1,2,2,6,0)$ is the one of the $\IP(1,1,2,2,6)$.      
 
We now write the CY manifold as zero locus of the ``proper transform'' 
of the polynomial (\ref{defpol}) 
\be 
p_0=x_0(x_6^6(x_1^{12}+x_2^{12})+x_3^6+x_4^6+x_5^2),
\ele(batcoxpol)
which is invariant under (\ref{equiv}) and restricts to (\ref{defpol}) if
we set the new coordinates $x_0$ and $x_6$ to 
$1$. Looking at (\ref{batcoxpol}) it 
is nicely visible that there is a {\sl new} divisor at $x_6=0$, 
which is a ruled surface over the curve $C$ defined by
$x_3^6+x_4^6+x_5^2=0$, with fiber $\IP^1$. The $(x_1:x_2)$ can be viewed 
as the homogeneous coordinates of that $\IP^1$.  
Beside this divisor there are the {\sl old} divisors at 
$x_i=0$ $i=1,\ldots,5$, from which we obtain one additional 
independent divisor class (\ref{reldivisor}), so that one has two 
divisor classes and hence $h^{1,1}=2$. 
Also visible is the $K3$ fibration structure: if we fix a point in 
the $\IP^1$ (\ref{batcoxpol}) defines a hypersurface of degree $6$ 
in $\IP^3(1,1,1,3)$ denoted $X_6(1,1,1,3)$, which is a $K3$ according 
to the criterion (\ref{chern}). 

The mirror will be defined in this example\foot{This method of constructing the mirror 
can very significantly be generalized  by the reflexive polyhedra construction 
pioneered by Batyrev \cit(batyrev)\cit(hktyI)\cit(cok).} by orbifolding 
(\ref{defpol}) w.r.t. $G_{max}$ of (\ref{mirrorconstruction}). 
$G_{max}$ has the following three generators 
$g_1:(x_1,x_2,x_3,x_4,x_5)\mapsto (x_1 \mu_1, x_2 \mu_1^{11},x_3,x_4,x_5)$
$g_2:(x_1,x_2,x_3,x_4,x_5)\mapsto (x_1 \mu_2^2, x_2 ,x_3\mu_2^{10},x_4,x_5)$
$g_3:(x_1,x_2,x_3,x_4,x_5)\mapsto (x_1 \mu_3^{2}, x_2,x_3,x_4\mu^{10},x_5)$
with $\mu_i$ 12th unit roots $\mu_i^{12}=1$.
Of the $126$ possible monomial perturbations only two  survive the
orbifolding by the discrete phase symmetry, which we now display 
\be
p^*=p_0+a_0 x_1 x_2 x_3 x_4 x_5+a_6 (x_4 x_5)^6
\ele(polmirror)
The mirror can again be expressed as the vanishing locus the transverse 
polynomial $p^*$ in an embedding space with the same weights as for the
original CY. To show that the manifold $X^*$ so defined also 
has $h^{1,1}(X^*)=128$ takes more effort.   
One has to resolve the singularities introduced by the 
orbifolding and count the divisors
that have to be introduced in the process of resolution of the 
singularities. There exists a completly
systematic approach using toric geometry how to do this \cit(batyrev).

\mabs {\sl The Picard-Fuchs equations and its solutions:} 
As in the example in appendix \ref{c} the  manifold (\ref{polmirror}) 
is redundantly parameterized; here we have a $(\IC^*)^5$ action on the 
$a_i$. With some prescience one chooses the following 
invariant parameters $y_k=(-1)^{l^{(k)}_0}\prod_{i=0}^6 a_i^{l^{(k)}_i}$.
The point here is that, since the $l^{(k)}$ are actually the Mori-cone 
edges, the $y_k$ are not only invariant but actually the parameterization  
whose origin $y_i=0$ is the point of maximal unipotent monodromy, 
see appendix D,E . As a consequence the period vector
takes the form (\ref{sol}). The Picard-Fuchs equations can derived
in this case as in appendix \ref{c}, using the scaling symmetries alone.
The relevant operators identities on the integrals over the holomorphic 
$(3,0)$ form (\ref{periodI})  are 
\be
\prod_{l^{(k)}_i>0} \left( \p \over \p a_i\right)^{l^{(k)}_i}= 
\prod_{l^{(k)}_i<0} \left( \p \over \p a_i\right)^{-l^{(k)}_i}.
\ele(rpf) 
Rewriting that in the $\tilde x=y_1$ and $\tilde y=y_2$ variables  
we get, after factorizing the six'th order operator from $l^{(1)}$
to a third order one, the following operators\cit(hktyI) 
\bea(rl)
{\cal L}_1&=\ttx^2(\ttx-2\tty)- 8 \tilde x(6\ttx+5)(6\ttx+3)(6\ttx+1)\cr
{\cal L}_2&=\tty^2-\tilde y(2\tty-\ttx+1)(2\tty-\ttx)
\elea(tspf)
The topological triple intersection numbers can be calculated using 
(\ref{topring}) (up to a normalization see \cit(hktyII)\cit(hly), for 
further explanations)  or classical intersection theory (see appendix E)
\cit(fultonoda)\cit(schubert) to be
\be
C^0_{111}=4, \quad C^0_{112}=2,
\ele(intersect)
where the `2' refers to the dilaton whose divisor class is 
the fiber. The dual curve is the $\IP^1$ base of the $K3$-fibration 
and in particular by the identification (\ref{ind}) 
\be 
t_S:=t_2:=4 \pi i S
\ele(dilatonmodulus) 
the dilaton modulus is identified with the `size' of the base of 
the $K3$-fibration. The modulus $t_T:=t_1$ controls the size of a curve in 
the $K3$-fiber. The classical couplings specify the solutions 
according to (\ref{sol})
as   
$\Pi^{(0)}=S_0$,
$\Pi_1^{(1)}=l_1 S_0+S_1$,
$\Pi_2^{(1)}=l_2 S_0+S_2$,
$\Pi_1^{(2)}=4\ (l_1^2 S_0/2+l_1 S_1+S_{1 1}) +
             2\ (l_1 l_2 S_0+l_1 S_2+l_2 S_1+S_{12})$,
$\Pi_2^{(2)}=2\ (l_1^2 S_0/2+l_1 S_1+S_{1 1})$,
$\Pi^{(3)}=4(l_1^3 S_0/6 + l_1^2 S_1 /2+\ldots)+
2 (l_1^2 l_2 S_0/2+ l_1^2 S_2/2 +l_1l_2 S_1+\ldots) $,

Moreover we calculate  (see appendix E)
\be 
\int_X c_2 \alpha_1=24, \quad  
\int_X c_2 \alpha_2=52 \ .
\ee

\mabs {Singularities :}
The manifold $p^*=0$ is transverse for generic values of the moduli. It, 
however, fails to be transverse if one of following discriminant  vanishes 
\be
\Delta_{c}=(1-x)^2- x^2 y, \quad \Delta_{s}=(1-4 y)
\ele(stdis) 
where we have rescaled the coordinates $x=1728 \tilde x$ and 
$y=4 \tilde  y$. See last paragraph of appendix C for hints 
how to calculate (\ref{stdis}) for (\ref{polmirror}).

The nature of this singularities is a follows: 
$\Delta_c=0$ is a conifold locus in the moduli space. For these 
values of the moduli the manifold $X^*$ develops an isolated singularity 
called node, or ordinary double point. It is characterized 
by the fact that $p=0$ and $d p=0$, but already the matrix of second 
derivatives is non-degenerate. As such it is the most harmless 
failure of transversality which is possible. The leading terms of a 
multi Taylor expansion around the singular point in the CY can be 
brought to the form   
$$\sum_{i=1}^4 \zeta^2_i=0\ .$$
This is a cone with the singularity at the appex which coincides with the orign. 
In order to analyse its base one intersects the cone with a real seven sphere 
$\sum_{i=1}^4 |\zeta_i|^2=2 r^2$ following \cit(candelasetal)\cit(intersect). 
The intersection is characterized by the
equations $\vec x \cdot \vec x = r^2$, $\vec y \cdot \vec y = r^2$ and 
$\vec x \cdot \vec y = 0$, where $\zeta_k=x_k+i y_k$. For each point 
on the $x$-$S^3$ defined by the first equation,  the last two equations 
describe a hyperplane intersecting with the $y$-$S^3$ giving thereby a 
$S^2$. As there are no non-trivial fibrations $F$: $S^2\rightarrow 
F\rightarrow S^3$ the base of the cone is actually $S^3\times S^2$.   
As a consequence one can desingularize the appex of the cone by replacing 
it with an $S^3$ or an $S^2$.       

In fact the cycle $V$ vanishes as $S^3$, when we approach the conifold point. 
Vice versa there is always the possiblity of resolving the node by the $S^3$ 
by just deforming the complex structure away from $\Delta_c=0$. 

For the CY to develop nodes we must fix some complex structure moduli. 
On the other hand one can resolve the nodes also by a so called small 
resolution in which the node is resolved by the $S^2$ \cit(lefshetz)\cit(intersect). 
Roughly speaking the size of the $S^2$ can become a 
new K\"ahler modulus. In this process one can hence drastically change
the topology in a transition which decreases $b_3={\rm dim}(H^3)$ and 
increases $b_2={\rm dim}(H^2)$. However after the small resolution the 
new manifold $\hat X$ is not necessarily K\"ahler. 
E.g. in our case at we cannot make the transition, without loosing 
the K\"ahler property. Doing it nevertheless might in fact lead to 
an interesting mechanism for supersymmetry breaking \cit(candelasetal).
It was analyzed e.g. in\cit(hiwe), when a configuration of nodes 
can be small resolved so that the new  smooth manifold is still 
projective algebraic, which implies that it is K\"ahler.
In this situation neither all $S^3$s vanishing at different nodes 
nor all the blown up $S^2$s are homologically inequivalent in $X$ and 
$\hat X$ respectively. Remarkably, as it was demonstrated in\cit(gms) 
on the type IIb side, the changes in the Hodge numbers in these 
``allowed'' transitions fit perfectly the physical picture of the 
Higgs effect by giving vevs to the massless hypermultiplets (black
holes)nothing more then a Higgs effect !

The nature of the second singularity for the manifold $X$ is most 
easily deciphered by
noting that the second differential equation ${\cal L}_2|_{x=0}$ can be 
integrated in terms of elementary functions  
\be
t_S={1 \over 2 \pi i} \log \left({1- 2 y - 2\sqrt{1-4 y}}\over 2 y\right) \ .
\ele(ele)
The ``area''  $t_S$ becomes zero at $y=1/4$, this correspond 
physically to strongly coupled heterotic string theory. 
Since $t_S$ resolves the $\ZZ_2$ singular curve we get a non-isolated
singularity all along $C$.

As an aside: Shrinkings of isolated $\IP^1$'s were discussed  in \cit(agm) 
in the context of flops. In this case after blowing down the $\IP^1$ one can blow up
a topologically different one. This can lead to mild topology changes, mild in the 
sense that  the Hodge numbers will not change. Such changes are called birational 
transformation in the mathematical literature. 

In the case of the non-isolated $\IP^1$ one finds a far more 
drastically topology changing transition possible at the singularity, which 
also changes the Witten index (Euler number)\cit(km)\cit(gmp).  
Thanks to our new understanding of non-perturbative string physics, this
transition is physically perfectly smooth, in fact as in the conifold case 
it is nothing more then the Higgs effect, but with an additional 
enhancement of an $SU(2)$ group. Compare \cit(strominger) and (\ref{strongcoupling}).

In addition to these singularities we see, most easily 
from the differential equations, that there are further regular 
singularities at $x=0$ and $y=0$ meeting normally at the large complex 
structure point. Other singularities can found similarly by transforming 
(\ref{tspf}) into other coordinate patches. They are at $x^{-1}=0$ $y^{-1}=0$ 
and $x y^{-1}=0$, comp.\cit(cdfkm). 
In \figref{stmod} we show schematical drawing of the singular loci in 
the moduli space. It is analogous to the situation in the 
$X_{8}(1,1,2,2,2)$ model discussed in great detail in \cit(cdfkm). 
The only difference is that the variables\cit(hktyI) we use here 
automatically resolve the large complex structure point to 
divisors with normal crossing. 

\figinsert{stmod}
{Singularities in the vector moduli space of the $ST$ model.}
{2.5truein}{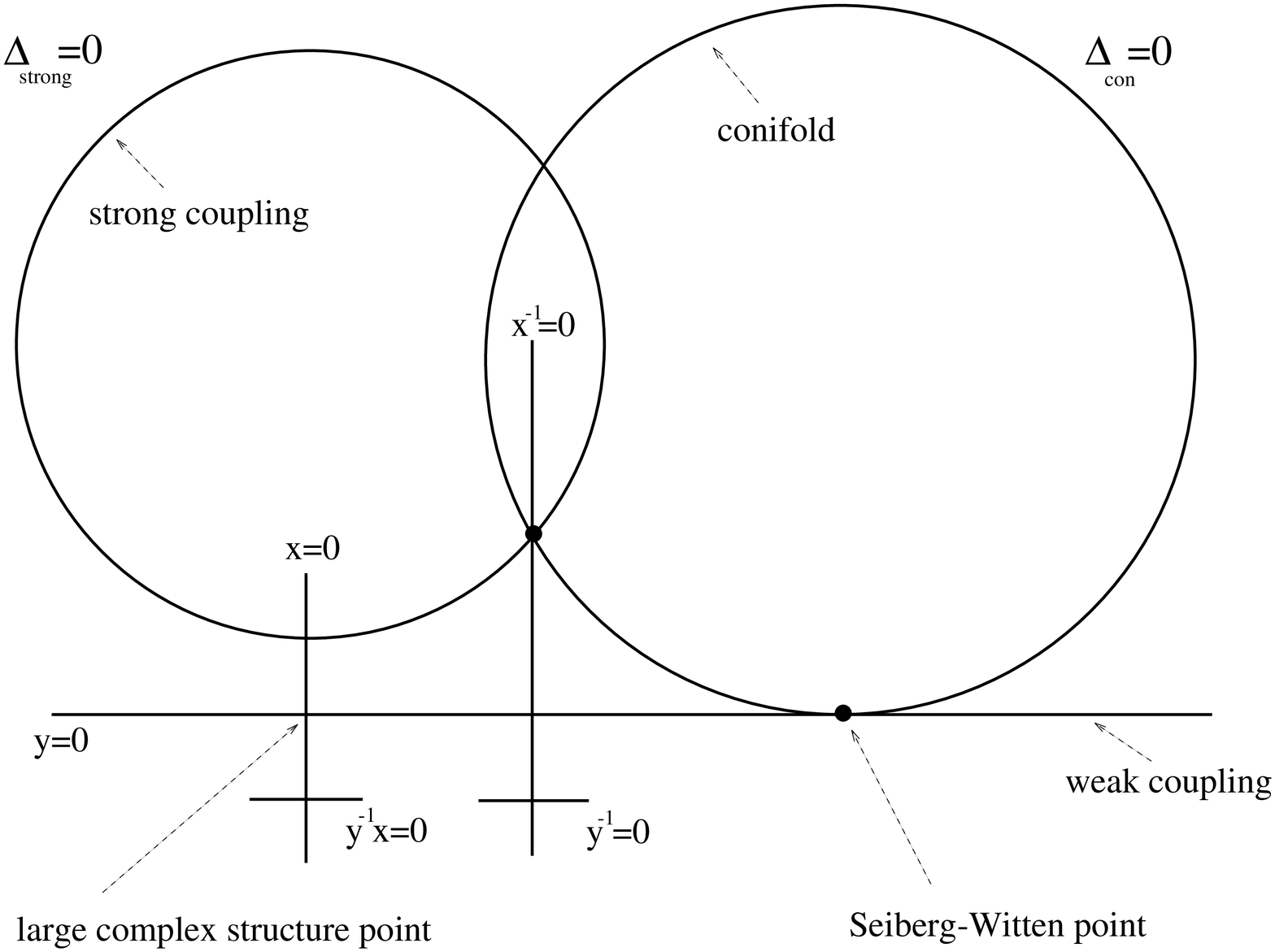}

\subsection{Weak coupling tests or modular functions again}
 
According to our identification of the dilaton (\ref{ind}
\ref{dilatonmodulus}) weak coupling 
is at $y=0$. Let us check the weak coupling structure of the $ST$ model in
particular the one loop corrected  heterotic gauge coupling.
This appears in the second derivative of the prepotential (\ref{fhetpert})
\be
\cF^{het}_{TT}\sim \tilde S + \log (j(T)-j(i)),
\ele(gauge)
where $\tilde S$ is the so called tree-level dilaton. The relation between this 
tree-level dilaton and the geometrical dilaton (\ref{ind}) is defined by formula
(\ref{dilrel}). As the $T$ modulus is on the heterotic string side the space 
time-modulus of the torus and we have space-time modular modular invariance the
corresponding one-loop function must be some modular invariant of the corresponding 
$SL(2,\ZZ)$ action on $T$. That fact and some knowledge about the asymptotic enables 
the explicit calculation of this contribution to (\ref{gauge})  \cit(oneloophet).

What is the reason for the appearance of modular functions on the type 
II side ?  Taking a look back at the way
we derived the differential equations we will recognize 
(\ref{equiv}) as the fundamental data. We may interpret them as follows: 
Take the exponents of the monomials in the polynomial (\ref{polmirror}) 
as integer coordinates for points in $\ZZ^5$, the convex hull of them is 
called the Newton polytope \cit(arnold)\cit(fultonoda). 
Because of the quasihomogenity of (\ref{polmirror}) they lie actually 
in a 4 d hyperplane.  The $l^{(k)}$ are just coefficients of linear 
relations between the points. Especially the $l^{(1)}$ is a linear 
relation between points in the $K3$ Newton polyhedron of the mirror 
of $X_{6}(1,1,1,3)$ and in the limit $\tilde y\rightarrow 0$ the first
operator in (\ref{tspf}) approach the PF equation for the $K3$ and the
periods  
$\Pi^{(0)}$, 
$\Pi_1^{(1)}$ and  
$\Pi_2^{(2)}$ approach the
periods $\varpi_0,\varpi_1,\varpi_2$ of the $K3$. Unlike as in the odd 
case we get for a even complex dimensional manifold  from (\ref{vanish})  
by (\ref{poinc}) an algebraic 
relation for the periods\footnote{For $K3$ we get this algebraic relation 
from (\ref{vanish}). For CY threefolds we
get differential relations from which special geometry follows. For
the CY fourfolds we get a mixture of differential and algebraic 
relations, which ensure among other things the associativity of the
correlation functions of the topological B-model, comp. appendix D 
\cit(gmp).}, 
here for a one modulus $K3$ family $\varpi_0\varpi_2=\varpi_1^2$. In
particular if we normalize the periods such that $\tilde \varpi_0=1$ 
and $\tilde \varpi_1=t:=\varpi_1/\varpi_0$ then it must be that 
$\tilde \varpi_2=t^2$ and the Picard-Fuchs operator looks in the
$t$ coordinates simply $\tilde {\cal L}=\p_t^3$. By 
rescaling $\varpi $ we can write any third order differential equation 
like e.g. ${{\cal L}_1|}_{\tilde y=0} \varpi =0$ in the form\foot{For 
reference write it first in the form $\varpi '''+3p\varpi''+3q\varpi'+
r \varpi=0$. Then 
rescale $\varpi =y\exp(-\int p d \tilde x)$ which does not affect $t$. That 
gives $Q={3\over 4}(q-p^2-p')$ and $R=r-3pq+2p^3-p''$.} 
$y'''+4 Q y' + R y=0$. As it  was observed in \cit(forsyth) 
this can be further transformed by 
\be 
\{ t , \tilde x \}=2 Q\ ,
\ele(unifI)
$y{\dd t \over \dd \tilde x}=:u(t)$ and $I\left(\dd t \over \dd \tilde x\right)^3=
R-2 Q$
into the form $\tilde {\cal L}u= (\p_t^3+I)u=0$. Here $I$ is an 
invariant of the equation, which cannot transformed to zero 
by a change of the dependent or independent variable, so for $K3$ 
PF equations it must be zero from the outset. On the other hand
(\ref{unifI}) determines the mirror map. In particular for 
${\cal L}_1|_{\tilde y=0} \varpi =0$ one finds 
$$
Q={{1-1968 \tilde x +  2654208 {\tilde x}^2} \over 
{4 {\tilde x}^2(-1+1728 \tilde x)^2}}\ . 
$$ 
Equating that with (\ref{qj}) we get a so called consumerability 
relation of $J$ with $\tilde x$, with shows in this case simply
\be 
\tilde  x(t)={1\over 1728 J(t)} \ {\rm or}\  x(t)={1728\over j(t)}\ .
\ele(jmap) 
This and other consumerability relations have been observed in 
\cit(ly) for various cases of one parameter families of $K3$. 
In particular they show that the inverse mirror map $x(t)$ of $K3$ surfaces 
is a Hauptmodul of various subgroups of ${\rm SL}(2,\IR)$ related
to subgroups of $SL(2,\ZZ)$ by adding Atkin-Lehner involutions see \cit(ly)\cit(cn).

To check the heterotic one-loop correction (\ref{gauge}) we have to
calculate the Type II prepotential. This can be done following 
\cit(hktyII)\cit(hktyI)\cit(cdfkm). We will take the route of first 
computing the three-point functions on $X^*$ by the method discussed in 
(\ref{complexstructure}). One finds
\bea(rl)
C_{xxx} &=\ds{\! 4\over (1728)^3\Delta_c x^3}, \ \ \  
C_{xxy}  =\ds{{2-2x\over (1728)^2 4\Delta_c  x^2 y}},\cr
C_{xyy}&=\ds{{2x-1\over 4^2 1728 \Delta_c\Delta_{s} x y }},\ \ \
C_{yyy} =\ds{{1-x+ y-3  x y\over
4^3 2\Delta_c \Delta_s^2 y^2}}\ .
\eea
We have already identified from the classical terms the special 
inhomogeneous  large radius coordinates $t_i$, defined concretely in 
(\ref{mirrormap}), as the relevant ones for the
comparison with the heterotic string prepotential. To transform the couplings  
to these coordinates we use (\ref{trans}) and get
$$
C_{t_it_jt_k} = \sum_{lmn}{1\over (\Pi^{(0)})^2} C_{x_lx_mx_n}
{\p x_l\over \p t_i}  
{\p x_n\over \p t_j}  
{\p x_m\over \p t_k}\ .
$$
Because of the $K3$ fibration structure the couplings $K_{TSS}=K_{SSS}=0$
must vanish in the $y\rightarrow 0$ limit. To be more specific the limit 
is defined by $q_{\tilde S}=e^{-8 \pi^2 \tilde S}\rightarrow 0$, where the relation 
between the tree-level $\tilde S$ and the geometrical dilaton $S$ is given by 
\be 
y=q_{\tilde S} f(q_1)+O(q_2),
\ele(dilrel)
with $t_s:= t_2=4 \pi i S,t_T:=t_1$.

The non-vanishing couplings in this limit are in the 
$T$, $\tilde S$ coordinates with $C_{t_it_jt_k}=\cF_{ijk}$
\bea(rl)
C_{TTT}&\propto {\ds {(\p_T j(T))^3\over E_4(T) j(T) (j(T)-j(i))^2}}\\ [ 3 mm]
C_{TT\tilde S}&\propto {\ds {(\p_T j(T))^2\over E_4(T) j(T) (j(T)-j(i))}}\ ,
\elea(coup)
where we used  $j(i)=1728$ and $(\Pi^{(0)})^2(x(T))=E_4(T)$.

Applying the identity  $(\p_T j(T))^2 \propto E_4(T) j(T) (j(T)-j(i))$  
we see that this matches exactly the one-loop correction 
\bea(rl)
C_{TTT}&\approx {\ds {\p_T j(T) \over (j(T)-j(i))}} \\ [ 3 mm]
C_{TT\tilde S}&\approx 1
\eea
from the heterotic string \cit(kv) ! In the view 
of (\ref{ninst}) one might be tempted to consider 
(\ref{coup}) as a direct relation between modular 
functions and worldsheet instanton numbers, but due 
to the non-geometrical choice of the variable $\tilde S$ 
the coefficients in (\ref{coup}) do not represent instanton 
contributions.  Modular functions in the instanton expansion
occur however if the Calabi-Yau contains del Pezzo divisors 
\cit(cfkm)\cit(kmv).     

Similar, in their complexity even more 
striking matchings, can be observed is the $(STU)''$-model\footnote{These
perturbative results in the $(STU)''$ apply also to the $(STU)$ 
model.}\cit(klm). 
In fact on the heterotic side the one-loop contribution could only
be determined in leading order $(T-U)$ i.e. the CY 
calculation \cit(hktyI) is a simpler method to get the one-loop result.
The main virtue of the type II formulation is of course that on the 
CY the dilaton modulus of the heterotic string is exactly 
treated as the spacetime moduli.  The calculation on the CY  
gives exact non-perturbative values for the gauge couplings
the BPS masses and the couplings (\ref{higherderivative}). Comparing
the latter with  the perturbative heterotic string was subject of \cit(kltt).
Just like in the Seiberg-Witten theory monodromies on the CY 
moduli space are exact non-perturbative symmetries of the theory. 
Such exact symmetries were discussed completly\cit(kklmv) for the $(ST)$ 
and in part\cit(klm) for the $(STU)''$ model.

\subsection{Deriving the Seiberg-Witten Theory from the Type II string:}
\label{swtII}
The perturbative gauge symmetry enhancement on the heterotic side is  
described by the $T$ dependent momenta and windings energies for the 
heterotic string on the torus. Generically the gauge group is broken to 
$U(1)$ by the stringy Higgs effect but for  $T=i$ the $W^\pm$ gauge 
of a ${\rm SU}(2)$ become massless, see \cit(lt) for a review. 
By (\ref{jmap}) this locus is mapped to $x=1$.

To decouple the string effects and the gravitational effects  
we want to take $M_{string}={1\over \sqrt{\alpha'}}\rightarrow \infty$ 
and $M_{planck}\rightarrow \infty$  to recover the Seiberg-Witten 
${\rm SU} (2)$ field theory. Because of the asymptotic freedom 
the bare coupling 
constant of the $SU(2)$ theory must go to zero if the string scale 
is pushed to infinity. We must take therefore $y\rightarrow 0$ or said
differently the volume ${\rm Im}\, t_S\sim {4\pi \over g^2}$ 
of the base $\IP^1$ to infinity.

Taking both arguments together one finds that the region in the moduli 
space where we expect the Seiberg-Witten theory is is near 
$\Delta_c \cap W$, with $W:=\{y=0\}$ is the weak coupling divisor.  

With a little insight in the nature of type II  non-perturbative
states we have not to refer to the heterotic side. Finding
the correct locus in the $x,y$ plane is naturally 
a type IIb question. We expect that the hypermultiplet, which becomes 
massless at the conifold \cit(strominger) of of $X^*$, has to be
identified with  the  magnetic monopole of the Seiberg-Witten theory 
in the field theory limit. So again we are forced to look at the 
intersection between $y=0$ and $\Delta_{con}=0$.

In the type IIa theory, where the $K3$ fibration is the valid 
picture, the light  $W^\pm$ gauge bosons come from to differently 
oriented two-branes 
wrapping around a non-isolated vanishing holomorphic 
curve. Let us assume for the moment we know the this 
``curve'' and its ``area'' $t_{W^\pm}$. 
According to the interpretation of the $W^\pm$ as a single wrapping 
state of a two-brane we expect $M_{W^\pm}/M_{string}\propto |t_{W^\pm}|$, 
i.e. to keep a finite $W^\pm$ mass we have to send the ``area'' to zero, 
when pushing the string scale to infinity. 
The limits of $t_S$ and $t_{W^\pm}$ are related by the running of the 
coupling constant, which is in the weak coupling region given by the 
one--loop  $\beta$-function 
\be
{8 \pi^2\over g^2}= \kappa\log \left( M_{W^\pm}\over \Lambda  \right) \ .
\ele(limitI)
in other words 
\be
\exp({2 \pi i t_S}) \sim \left(\Lambda \over  M_{W^\pm}\right)^{\kappa}
\ele(limitII)     
where $\kappa$ is th from the $\beta$-function (\ref{kappa}) and by 
the multiplet of anomalies it is related to the way a spacetime 
instanton of instanton
number $n$ is weighted in (\ref{nonpert}), e.g. in pure 
${\rm SU}(2)$ by $\exp({ 2 \pi i n t_s})=(\Lambda^4/a^4)^n$. 
As $a \propto M_{W^\pm}$ is proportional to the ``area'' of the 
holomorphic curve we have the simple double scaling limit
\bea(rl)
y\sim \exp (2 \pi i t_S)& \sim \epsilon^4 \Lambda^4 \\ [ 3 mm]
t_{W^\pm} & \sim \epsilon \ a \ .
\elea(limit) 
We have yet not given the precise relation between 
$t_{W^\pm}$ and the $x,y$ coordinates near the Seiberg-Witten 
point. At this point one could use (\ref{jmap}) and refer 
to string/string duality \cit(clm).  On the other hand that 
information follows also from the resolution process to which we turn 
now. The naive question which arises is how can we stay near the 
Seiberg-Witten point and still get a parameter, 
which plays the r\^ole of the $u$  modulus of the Seiberg-Witten 
theory ? The most naive
idea to introduce the direction in which approach this point as 
parameter see \figref{blowi} is almost the correct answer. This is what
physicst would simply call  a double scaling limit. Here we have to 
repeat this procedure two times. Let us explain this important limit 
in some detail.  

\mabs {\sl Resolution of the Seiberg-Witten point to the 
Seiberg-Witten plane: } 

We discuss the resolution process with an example which encorporates the
situation we are interested in. This example
and some introduction into the general theory of monodial or quadratic 
transformations, commonly called blow ups,  
can be found in\cit(laufer)\cit(hartshorne). The book
of Laufer deals in concrete terms
with the desingularisation of the $ADE$ surface singularities 
which become relevant in the next chapter.  

The example is the cusp defined by the affine equation in 
$M=\IC^2$, 
\be 
\Delta=b^2-a^3=0\ ,
\ele(cusp)
which is singular at $a=b=0$. The general 
idea is to introduce more variables and more (quadratic) relations 
so that the singularity becomes weaker, i.e. in the new variables 
the first non vanishing derivatives at the singular locus of 
$\Delta$ are of lower order.  
This process is not unique, but as we know from Hironakas 
work it can always be chosen such that we end up with a situation with only 
normal crossing divisors \cit(hironaka). 
In our example the first step is to  introduce $c,d$ subject to 
\be 
a c = b d \ .
\ele(blowup) 
As we want to have the direction $(c/d)$ at 
$a=b=0$ as the new coordinate, $(c:d)$ must be homogeneous 
coordinates of a $\IP^1$, i.e. $(c,d)\sim (\lambda c,\lambda d)$ 
$\lambda\in \IC^*$ and the locus $c=d=0$ is excluded.  
Eq. (\ref{blowup}) defines a holomorphic 
one to one map $\pi^{-1}$ from the variety $\IC^2\setminus \{ \vec 0\}$ 
in the $(a,b)$ parameterization to the one $\hat M$ in the $(a,b,c:d)$ 
parametrisation. But at $a=b=0$ the new $\IP^1$ parametrized by 
the $(c:d)$ becomes unconstrained, see \figref{blowi}. 

That indicates the general setting for concept of the
resolution of a complex manifold.

\mabs For the resolution we search a smooth\foot{In practice the
resolution process is typically divided in several steps, 
so that $\hat M$ might be not smooth, but just less singular then $M$.} 
manifold $\hat M$ and a map $\pi:\hat M\rightarrow M$, 
such that $\pi^{-1}:(M\setminus S)\rightarrow (\hat M\setminus E)$ 
is a holomorphic one to one map outside the singular set $S \in M$. 
The set $E=\pi^{-1}(S)$ is called the exceptional divisor, 
here a $\IP^1$.

The singular divisor $\{ \Delta=0 \}\in M$ is modified after a finite 
iteration of these non-unique processes into a set of regular 
divisors with normal crossing. 

In fact in our case after the first blow up we still have a tangency between 
$\Delta$ and $\IP^1$ also visible in \figref{blowi}, 
so we have to iterate the procedure.

\figinsert{blowi}
{The first step in the resolution of the cusp. The righthand side
shows a neighborhood of the cusp singularity. The radial lines
indicate the directions at $a=b=0$. The exceptional $\IP^1$ is the horizontal
line on the lefthand side and each point on this $\IP^1$ corresponds
to a particular direction.   As it is nicely
visible in this picture the singularity of the cusp is smoothed by 
introducing the direction at the singular point as the new 
coordinate of the new {\sl exceptional} $W:=\IP^1$.}
{1.2truein}{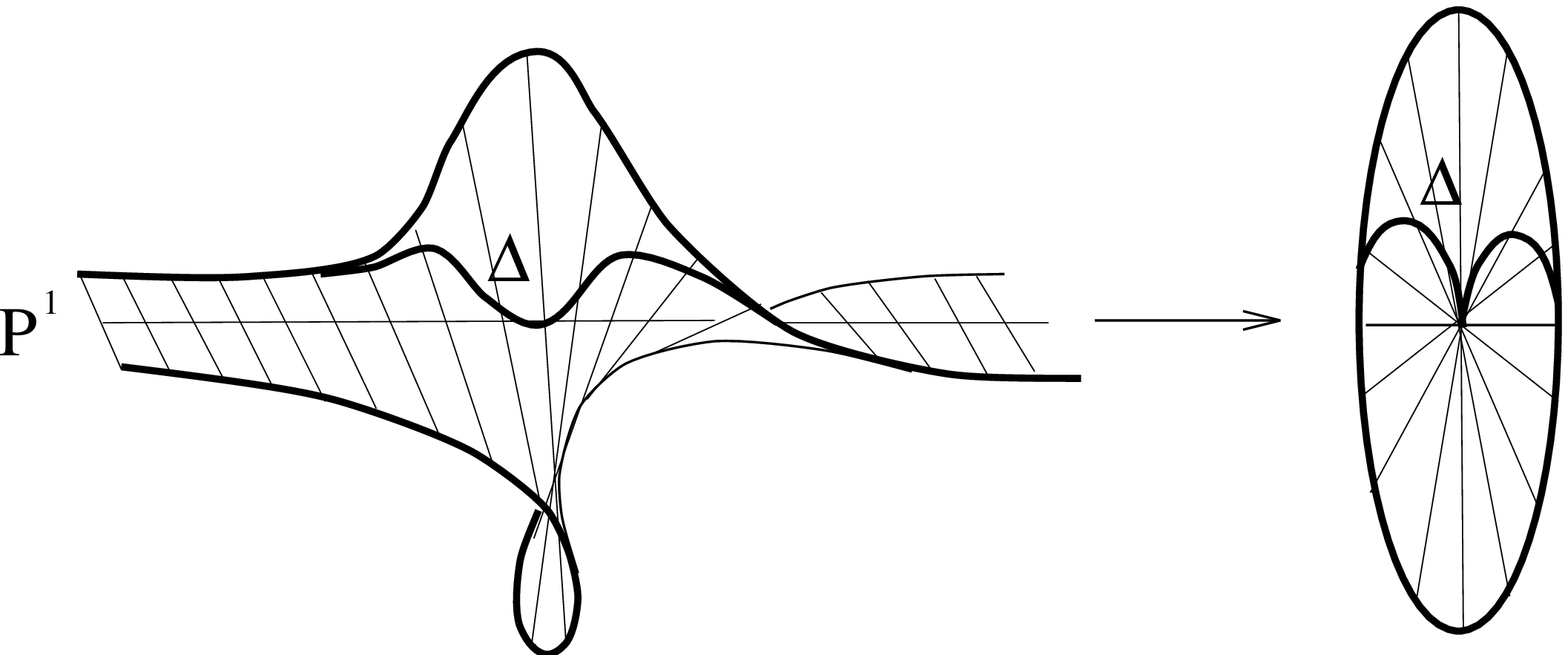}

In each step of the blow procedure we must check the structure of the
singularity in all coordinate patches of the $\IP^1$, i.e. here 
in the $c=1$ chart with  $a=bd$ and in the $d=1$ chart with $b=ac$. 
In the present case the singularity is in the $d=1$ chart and we 
continue the blow ups in this chart. 
In  practice one may keep  track of the variables and choices 
of charts in form of a table~1.  
In the $a,c$ coordinates $\Delta$ looks like 
$\Delta=a^2(c^2-a)$ and at this point we make contact with leading pieces 
of the relevant components of the CY discriminant 
$y^2 \cdot ((1-x)^2-y)\sim y^2\cdot \Delta_c$ in that region, by identifying
$(x-1)= c$ and $y=a$. The next steps in the blow up procedure appear in
\figref{blowii}. In the last step against the arrow direction, which
indicates $\pi$, we have the desired result all divisors are normal 
to each other.

\figinsert{blowii}
{The full blow up process.}
{1.8truein}{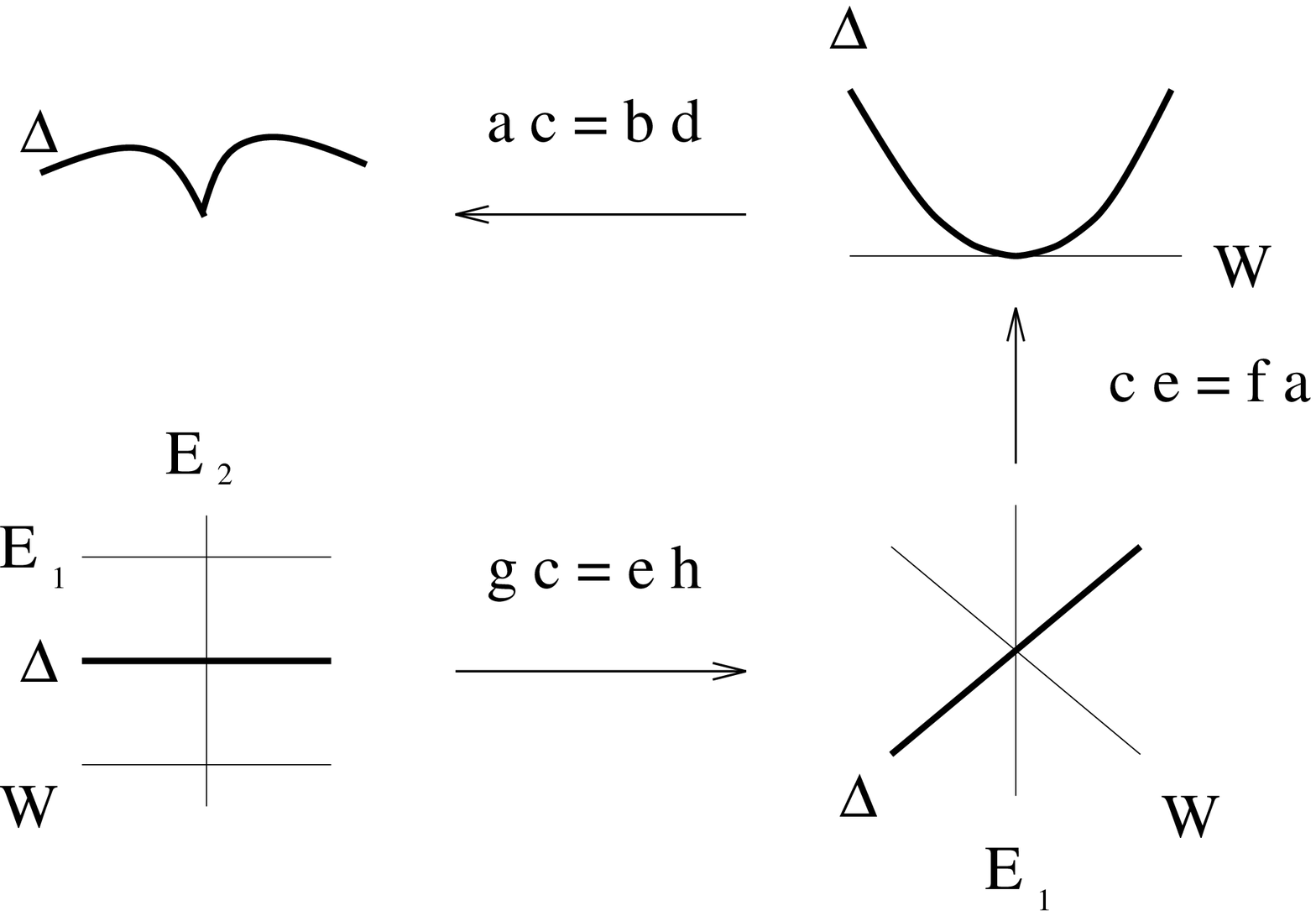}

\begin{table}
\begin{center}\label{coordinates}
\begin{tabular}{|l|ll|ll|ll|ll|}\hline
          &   a&b            &            c&d  &             e&f  &   g&h     
\\ \hline
$ \Delta$ &   a&$a^{3\over2}$&$a^{1\over 2}$&1  & $a^{1\over 2}$&1 &   1&1  \\
$W  $ &   0&0            &            c&1  &             0&1  &  0&1  \\
$E_1$     &   0&0            &            0&1  &             e&1  &  1&0  \\   
$E_2$     &   0&0            &            0&1  &             0&1  &  g&h  \\
\hline  
\end{tabular}
\end{center}
\caption{Coordinates introduced in the blow up procedure.}
\end{table}

The coordinates for the different normal crossings in the last 
blow up of \figref{blowii} must be such that they move along the 
corresponding divisor, and can be read off from the table.
At $W \cap E_2$ we can use $g={a\over c^2} = {y \over (x-1)^2}$
and $c=(x-1)$ at $E_2\cap \Delta_c$ one must go away from $(1:1)$ 
along $E_2$, e.g. using $h-1={(x-1)^2\over y}-1$ and along $\Delta_c$ 
$c=(x-1)$ is a good variable. Finally at $E_2\cap E_1$, i.e. 
at the other tip of the 
$E_2$, we use $h={1\over g}={(x-1)^2\over y}$ and along $E_1$ 
$e={a\over c}= {y\over (x-1)}$.   
In view of the identification $y=\epsilon^4 \Lambda^4$ we must chose
\be 
(x-1)=\epsilon^2 u=:\epsilon^2\Lambda^2\sqrt{\alpha},
\ele(identI)
where $u$  is the Seiberg-Witten variable, to keep the quotients
finite and the dimensions correct. At the various normal crossing the 
coordinates are then  
$$\begin{array}{rl} 
W\cap E_2 &:\ds{\  ({ 1 \over \alpha},\epsilon \sqrt{\alpha}),\ \ 
\Delta_c \cap E_2:\  (\alpha-1 ,\epsilon \sqrt{\alpha} )},\\ [ 3 mm]
E_1 \cap E_2 &:\ds{\  (\alpha, {\epsilon\over \sqrt{\alpha}})}
\end{array}$$
and we identify the exceptional $\IP^1$ called $E_2$ with the Seiberg-Witten 
$\IP^1$ in \figref{mod}.  This matches perfectly the physical 
requirement that $y$ and $(1-x)^2$ have to be small at the same time and
identifies in leading order $t_{W^\pm}$ with $\sqrt{1-x}$. 
  
The relation between $\epsilon$ and $\alpha'$ is defined by 
$y=(\alpha')^2 e^{-\hat S} \Lambda^4$ so $\epsilon=\sqrt{\alpha'} e^{-\hat S/4}$.
In particular we can now solve the Picard-Fuchs equations in the 
prescribed variables and get the following local form of the
six solutions e.g. at $W\cap E_2$
\bea(rlrl)
&1+\cO(\epsilon^4 u^2), &   & \epsilon^2 u +\cO(\epsilon^4 u^2),\cr
&\sqrt{\alpha'}a(\alpha)(1+\cO(\epsilon^2 u)),& &-S (1+\cO(\epsilon^4 u^2)),\cr
&\epsilon^2 u S(1+\cO(\epsilon^4 u^2)),& & \sqrt{\alpha'}
a_D(\alpha)(1+\cO(\epsilon^2 u)).
\elea(leadingterms)
Especially the occurrence of the Seiberg-Witten periods $a_D(\alpha)$ and 
$a(\alpha)$ (\ref{ada}) can be easily established to all orders by 
analyzing the local form of the Picard-Fuchs operators.
Near $W\cap E_2$ $x_1=y/(x-1)^2$ and 
$x_2=(x-1)$ are good local variables.
To compare the differential operator with (\ref{periodeqii}) 
$\sqrt{\alpha'}=\sqrt{x_1} x_2^{1\over 4}$ has to be commuted with the
operators ${\cal L}_i$, i.e. $\sqrt{\alpha'} \tilde {\cal L}_i f={\cal L}_i 
\sqrt{\alpha'} f $ before taking the limit $x_2\rightarrow 0$. 
It is then easy to establish that $\tilde {\cal L}_1(x_1,x_2)$ acts 
in the limit trivially on the relevant periods while 
$\tilde {\cal L}_2(x_1,x_2)$ can be identified precisely with 
(\ref{periodeqii}). That establishes the fact that the non-perturbative 
type II string reproduces exactly the Seiberg-Witten result !  

Of course the explicit solutions determine the exact non-perturbative 
gravitational corrections to that result. It is an interesting question 
which properties of this corrections depend on the specific CY manifold 
and which are universal. 

Further properties of this model, especially the full non-perturbative 
monodromies, were worked out in \cit(kklmv). One can establish the fact that one
has sub monodromies $\Gamma$ with $\Gamma\in {\rm SP}(6, \ZZ)$ acting 
as $\tilde \Gamma$ on the corresponding periods to define the 
Riemann-Hilbert problems 

a.) with $\tilde \Gamma\sim SL(2,\ZZ)$, which explains the occurrence of  
the $j$-function in the weak coupling limit and

b.) with $\tilde \Gamma\sim \Gamma_0(4)$, which is responsible for the
occurrence of the Seiberg-Witten functions.

Beside that the whole non-perturbative structure of the effective 
supergravity action is encoded in the periods of the CY and
we will use it to investigate the strong coupling behaviour of the
theory. 

\subsection{The strong coupling gauge symmetry enhancement and extremal
transitions.}
\label{strongcoupling}

By the two highly non-trivial checks we might have gained enough 
confidence in our the identification of the complex moduli space of 
$X_{12}(1,1,2,2,6)$ with the moduli space of the non-perturbative 
heterotic string that we go now to explore genuine strong coupling 
behaviour of the $(ST)$-model.  From the type IIa theory point of 
view the understanding of the theory at the strong coupling 
singularity $y=1/4$ is easier then at 
the Seiberg-Witten point. 
The reason is that the realization of the 
supersymmetric vanishing cycle is geometrical simpler. 
As we have pointed out in (\ref{ele}) the holomorphically embedded 
base $\IP^1$ shrinks down to a point with vanishing $B$-field 
over the genus $2$ curve $C$ $z_3^6+z_4^6+Z_5^2=0$. 
What makes the situation clear cut is that the vanishing of this 
holomorpic curve occurs at the boundary of the K\"ahler cone.
What we expect from the non-isolated vanishing $\IP^1$ is
a ${\rm SU}(2)$ gauge symmetry enhancement, where the $W^\pm$ bosons
come from wrapping the the type IIa brane around the non-isolated 
$\IP^1$.

The question is what is the precise field especially the corresponding 
matter content? One rough tool to address this question 
is the topological index. Let $\cal V$ be the period which
vanishes at a component of the discriminant. As was pointed out in
\cit(vafat1) the contribution of that period to the 
singular behaviour of the topological one-loop partition 
function \cit(bcov) is 
\be
F_1= - {b \over 12} \log {\cal V} \bar {\cal V}\ ,
\ele(topi) 
where $ b= \#V-\#H$ is
the difference between the massless solitonic vector and hyper multiplets. 
In particular the observation that $b=-1$ at
all conifolds \cit(hktyII) was interpreted in \cit(vafat1) as confirmation 
of the picture that {\sl one}\foot{The argument leading to (\ref{topi}) 
comes from a one-loop amplitude and the normalisation of $b$ 
depends on the precise normalization of the coupling. 
We chose it in (\ref{topi}) to fit the ${\rm SU}(2)$ conventions.}
massless black hole appears at the conifold as suggested in 
\cit(strominger). This was further checked from the terms 
(\ref{higherderivative})\cit(agnt), especially second reference.

Now at $y=1/4$ the index $b$ was determined in 
\cit(km)\cit(kmp) to be $b=-{1\cdot 2}$, i.e we have a surplus of two hyper
multiplets and the massless $W^\pm$ vector boson cannot be the full story.
We will see in fact that there will be four additional light hypermultiplets
completing the two neutral hyper multiplets, which are associated to
the non-geometric deformation to two adjoints of ${\rm SU}(2)$. 

One way to argue is from the transition between this manifold, 
through the strong coupling singularity to a manifold, which is 
defined as a complete intersection of two polynomials\foot{The
instantons of this manifold were calculated first in \cit(kt).}
of degrees $(6,2)$ in $\IP^5(1,1,1,1,1,3)$ \cit(km)\cit(kmp). 
The Hodge numbers change from $h_{1,1}=2$ and  $h_{2,1}=128$ to 
$h_{1,1}=1$ and $h_{2,1}=129$. Let us try to understand that as a Higgs
mechanism and assume we have $g={|b|\over 2}+1$ new 
hypermultiplets in the adjoint of the ${\rm SU}(2)$. 
The scalar potential (spelled out e.g. 
in \cit(kmp)) shows that we can give a vacuum expectation
value to one of them. That breaks the ${\rm SU}(2)$ completly 
and reduces the abelian vector multiplets by one: 
$h_{1,1}\rightarrow h_{1,1}-1$. Since the ${\rm SU}(2)$ is broken, 
the off diagonal parts of the new hypermultiplets in the adjoint 
become neutral, i.e. the surplus of neutral hypermultiplets is 
$2 g-3=|b|-1$ so the expected change of the Hodge number is 
$h_{2,1}\rightarrow h_{2,1}+|b|-1$.
That behaviour was indeed observed for the transition in question as well
as for various transitions of {\sl analogous} type\cit(km)\cit(kmp). 
Near the transition points the simple
factorization (\ref{conmod}) fails. That is not a big surprise due
to the presence of charged massless states. 
The easiest transition, without enhancement of the gauge group, 
is the one at the conifold which was discussed by \cit(strominger)\cit(gms).   
      
For  a more direct way to obtain the matter content consider, as  in \cit(kmp),
the volume of the curve as very large against the rest of the CY manifold.
This is possible since at $y=1/4$ the volume $t_{W^\pm}=0$ is zero 
independently of the value of $x$ and in particular we can choose $x$ such 
that $C$ becomes very large. 
Now the compactification on the part with the vanishing $\IP^1$
leads in six dimensions to an $N=(1,1)$ theory which has 
a vector, a complex scalar and two fermions, all in the adjoint of 
${\rm SU}(2)$. The charged parts stem from the $\IP^1$-wrapping 
modes of the D-2-branes. 

The six dimensional theory has a global R symmetry 
${\rm SU}^{(1)}(2)\times {\rm SU}^{(2)}(2)$. If one compactifies the 
six dimensional theory on $C\times \IR^4$ the six dimensional 
Lorentz ${\rm SO}(6)$ group splits into ${\rm SO}(4)\times {\rm U}(1)$ 
and the representations of the four dimensional bosonic fields 
are collected below ($SO(4)\sim {\rm SU}(2)\times {\rm SU}(2)$) 
$$
\begin{array}{rrclclcl}
      & SO(4)&\times &U(1)&\times & {\rm SU}^{(1)}(2)&\times &{\rm
SU}^{(2)}(2) \\ [ 2 mm]
V_\mu &(2,2) &       & 0  &       &   1              &       & 1 \\ [ 2 mm] 
V_{++}&(1,1) &       & 1  &       &   1              &       & 1 \\ [ 2 mm]
V_{--}&(1,1) &       & -1 &       &   1              &       & 1 \\ [ 2 mm]
\phi  &(1,1) &       & 0  &       &   2              &       & 2 
\end{array}
$$  
Normally supersymmetry is broken upon compactification on $C$. To 
obtain an $N=2$ supersymmetry in four dimensions one has consider 
an exotic embedding of action of the Lorentz group generator $J$ of 
the $U(1)$ on $C$ into the $R$ symmetry group \cit(wittentwist). 
The unbroken supercharges have to be scalars under that action. 
The so called twisted $J_T$ was found in \cit(kmp) to be related to 
the standard $J_S$ by $J_T=J_N-J_3^{(1)}-J_3^{(2)}$. 
By  the same argument as in the 
CY case, section (\ref{macros}), massless states are linked to the 
cohomology of $C$. From their charges under $J_T$ one sees that 
the bosonic part of a vector multiplet comes from $h_{0,0}=1$ while 
the bosonic part of hypermultiplet comes from $h_{1,0}=h_{0,1}=g$. 
Let us summarize the situation with a picture \figref{p1}. 

If $g>1$ the theories are not asymptotically free, but they can still 
be consistently defined when embedded into the type II theory. The 
$g=1$ case leads to a $N=4$ spectrum and a conformal theory. 
It is realized e.g. in the $(STU)''$ model \cit(km)\cit(kmp)\cit(bkkm).

The generalisation to other $ADE$ groups is more or less straightforward. 
$A_n$ was discussed in terms of toric diagrams in \cit(km)\cit(kmp).
In fact the singularity in the compactification space to six 
dimensions can be described locally as in (\ref{ADEpot}). 
The light gauge bosons come from the wrapping modes of the two branes
around the  $ADE$ sphere-tree, which is fibred over the 
holomorphic curve and the irreducible divisors $D_j$ are ruled 
surfaces over $C$. They will have an intersection form with the generic
fibre which is the negative of Cartan-Matrix of the $ADE$ group. 
The corresponding periods on the type IIb side exhibit as 
monodromy the Weyl-group of the $ADE$ algebra \cit(km).
Further generalisations to non-simply laced 
groups can be achieved in this context by considering an additional 
outer automorphism\cit(slowody) (twists) on the singularity as 
in\cit(ag)\cit(bikmsv),see\cit(ims) for a discussion in a five 
dimensional $M$-theorie compactification . 

A simple D-brane picture for the non-compact $A_n$ case was previously 
presented by \cit(bsv). The $D$-brane approach can be generalized 
to the $D_n$ series, by introducing orientifold planes. The $E$-cases 
turn out to be less accessible using $D$-branes.

More general ``Higgs'' transitions involving $k$ matter multiplets 
in  the fundamental can be obtained, when in addition to
the divisors from the $ADE$ sphere-tree a conic bundles 
over $C$ with $k$ singular line pairs as exceptional fibers 
degenerate \cit(bkkm).

Let us end this section with a small overview what can happen
if one approaches a codimension one wall in the K\"ahler cone
of a CY threefold (compare \cit(mvII)\cit(ims)): 

\mabs An isolated curve can collapse to zero volume. Physically that 
leads to a $U(1)$ enhancement of the gauge group and  if flat 
directions to higgs exist to the type IIa perspective of the 
conifold transition as described by \cit(strominger)\cit(gms). 

\mabs A curve in a ruled surface can collapse 
leaving behind a curve singularity, which lead as we have just 
discussed to an ${\rm SU}(2)$ gauge symmetry enhancement. The
possibility to higgs by matter in the adjoint leads to a so called
extremal transition.

\mabs A conical bundle can collapse, this can lead to matter in the 
fundamental and the corresponding Higgs transitions were 
discussed in \cit(bkkm). \

\mabs A del Pezzo surface $B_d$ $d=0,\ldots,8$ 
can be contracted \cit(mvII), which in six dimensions correspond 
on the heterotic side corresponds to an $E_d$ instanton shrinking 
to zero size and in the $M$-theory picture to a tensionless string 
\cit(horavawitten)\cit(gaha), whose properties can be inferred by compactifying 
further on a $S^1$ \cit(ganorI)\cit(kmv). Four dimensional 
interpretations of that situation where studied in \cit(ganorII)
\cit(lmw).
   
We understand some combinations of these contractions at
higher codimensions in the K\"ahlermoduli space, as for instance 
the $ADE$ enhancements. However higher codimension degeneration
will exhibit genuine new types of singularities \cit(reid), whose 
physical interpretation is not investigated yet.

\figinsert{p1}
{An Hirzebruch-Jung sphere tree with $A_r$ (more generally $ADE$) 
intersection fibred over a genus $g$ Riemann surface inside the CY
threefold will lead to an $SU(r+1)$ ($ADE$) gauge group with $g$ matter 
multiplets in the adjoint. The non-abelian gauge boson become 
massless if the $\IP^1$s in the fiber shrink, i.e. the irreducible
components of the divisor, which are ruled surfaces over $C$, shrink to a 
singular curve. At this boundary points in the moduli space the 
CY admits a ``Higgs'' transition changing the Hodge numbers 
by $h_{1,1}\rightarrow h_{1,1}-r$ and $h_{2,1}\rightarrow 
(2g-2)\left((r-1)\atop 2\right)-r$. If one has in 
addition a conic bundle which splits in k-points over $C$ into line pairs
one gets in addition k light  matter multiplets in the fundamental 
representation of ${\rm SU}(r+1)$, if this bundle is also contracted.}
{2.8truein}{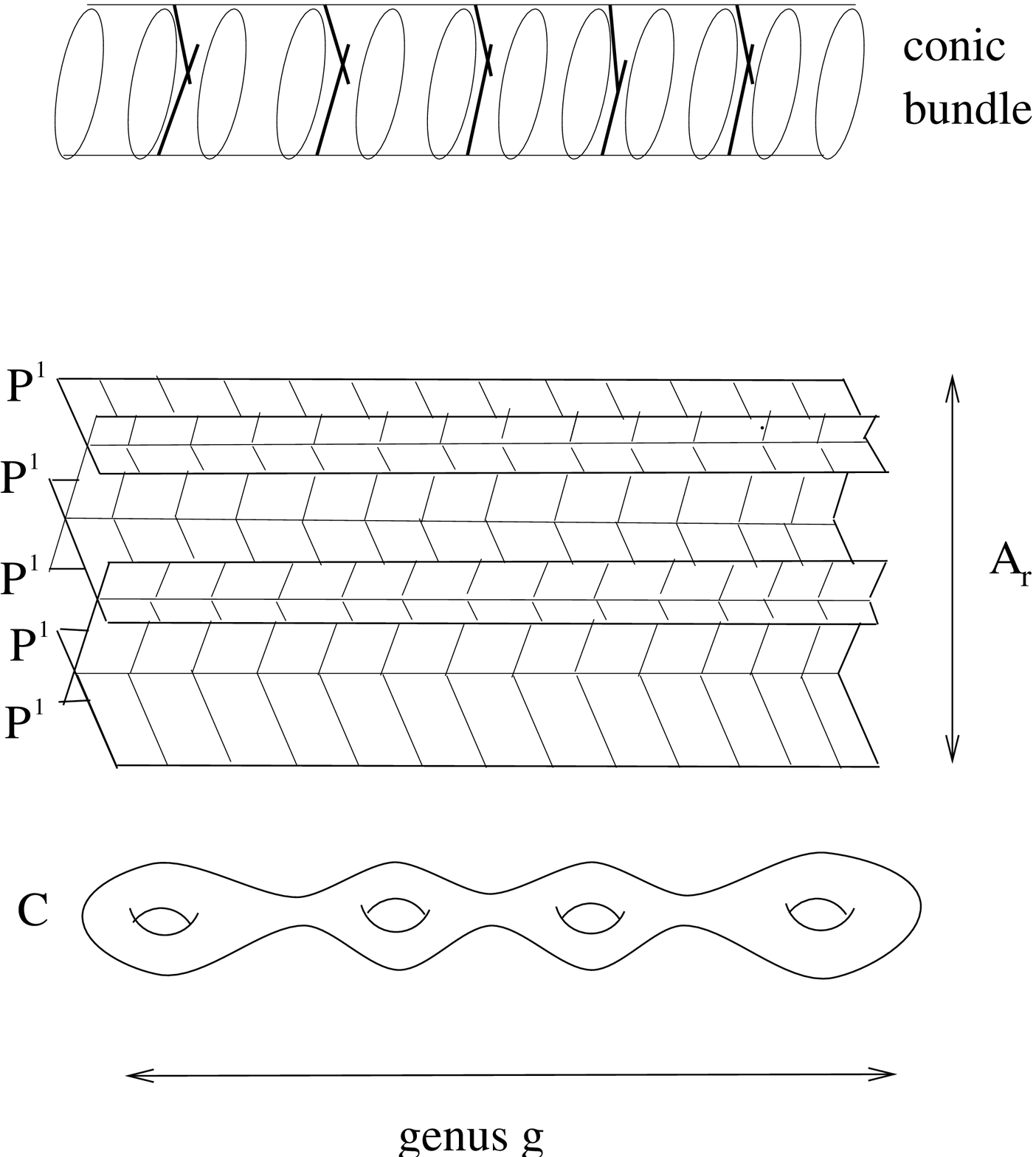}

\section{Local Mirror symmetry}
\label{localmirror}

We have seen in the last section that the gauge theory and the 
matter content can be obtained from the local singularity 
structure of the Calabi-Yau manifold. The situation can be analyzed 
in five dimensions by compactifying $M$-theory on the Calabi-Yau 
threefold\cit(ims), which clears the picture from worldsheet instanton 
corrections. However thanks to mirror symmetry the 
worldsheet instantons are very well under control and they are 
a crucial ingredient for the 4 dimensional non-perturbative gauge dynamics.  
Recently remarkable progress has been made to give a strict mathematical 
proof for the part of the mirror conjecture that we will
need here \cit(giventhal), namely the relation of the worldsheet instanton 
sums to the solutions of the differential equations that physicists 
have conjecturally provided. 
Because of (\ref{conmod}) type IIa space-time instantons 
corrections to the K\"ahler moduli space of the 
type IIa theory are absent. This gives us the possibility \cit(kkv) to
``proof'' the Seiberg-Witten result from our present understanding 
of some basic non-perturbative features of the type IIa theory.

\subsection{Space-time versus Word-sheet Instantons:}

Let us start with the the six dimensional picture as in the
last section to relate the worldsheet instantons of type IIa to 
space-time instantons. Here we have\cit(kkv) from a $K3$ compactifiaction
of the type IIa string \cit(wittenvafa)
\be
\dd * ({\rm exp}(-2 \phi) H = {\rm tr} R\wedge R - 
{\rm tr} F\wedge F\ ,
\ele(eqn) 
where $H$ is the 3-form field strength and $\phi$ is the dilaton. 
In our applications this is fibred over the base $\IP^1$. 
To count a 4d space-time instanton number $n$ we want to integrate  
${\rm tr}F\wedge F$ term over euclidian space-time.
As we are interested in the $M_{str}\rightarrow 0$ limit we 
may ignore the ${\rm tr} R\wedge R$ term integrate partially and 
relate that to $\int_{S^3}*({\rm exp}(-2 \phi)H=n$. But that is the 
wrapping number of a worldsheet instanton which wraps n-times 
the base  $\IP^1$ at the point $x$ in the uncompactified 4d space-time.  
I.e. (\ref{eqn}) associates point-like spacetime gauge instanton with 
instanton number $n$ to configurations of worldsheet 
instantons wrapped n-times around the base.

\subsection{Landau-Ginzburg description of the local A-model}

As the enhanced gauge symmetry or other interesting physics 
arises when divisors shrink in the Calabi-Yau space we are basically 
interested in a classification of three dimensional singularities, 
which can be resolved by adding exeptional divisors without 
changing the canonical class.  The canonical class of the blown up 
manifold is $K(\hat \Xi)=K(\Xi)+\sum a_i E_i$, where $E_i$ are the exceptional
divisors. Depending on how much the canonical class changes in the
resolution process the singularities are said to have a 
\cit(reid) terminal: $a_i>0$, canonical: $a_i\ge 0$ or crepant resolution:
$a_i=0$. In the last case the singularity is also called Gorenstein 
singularity  and these are the most interesting ones, if we want to
end up with the same amount of supersymmetry after compactification 
of a supersymmetric theory on the resolved and the unresolved space.
 
In two dimension the Gorenstein singularities are classified 
see\cit(slowody)\cit(laufer) for  reviews.
One can either describe such a singularity by the quotient $\IC^2/G_F$ 
of $\IC^2$ with respect to a finite subgroup $G_F\in {\rm SL}(2,\IC)$ 
or as a hypersurface singularity. In two dimensions these two descriptions
are equivalent and have a beautiful $ADE$ classification. The 
non-compact spaces $\IC/G_F$ are known as ALE spaces the simplest one 
with $G_F={\rm diag}(-1,-1)$ being the Eguchi-Hansen space 
$\cO_{\IP^1}(-2)$ and unlike in the compact case ($K_3$) the 
metric on them can be studied explicitly\cit(hklr)\cit(kronheimer).

In the table below we show the classification and the correspondence. 
The way the index $k$ appears seems slighty odd, but it is put this 
way to highlight an other beautiful connection namely the one to 
the minimal rational $N=2$ superconformal  field theories at level $k$ of 
sect.(\ref{sigmaapproach}). 
\begin{table}
\begin{center}
\begin{tabular}{|l|l|l|l|l|}\hline
 Group& Level& HS-singularity& $G_F$  & order$(G_F)$ \\ \hline
$A_{k+1}  $      &$k\in \IN^+      $   &$ W=x^{k+2}+yz$ & 
$\ZZ_{r+2}$& k+2 \\  
$D_{{k\over 2}+2}$ &${k\over 2} \in \IN^+$ & $W=x^{k+2\over2 } + x y^2+z^2$
&${\bf D}_{k\over 2 } $& 2k \\ 
$E_6$, & $k=10$ &$ W=x^3 + y^4+z^2$ & {\bf T} & 24  \\ 
$E_7$, & $k=18$ &$ W=x^3+ x y^3+z^2$& {\bf O} & 48 \\ 
$E_8$, & $k=30$ &$ W=x^3 + y^5 +z^2$& {\bf I} & 120\\ 
\hline  
\end{tabular}
\end{center}
\caption{$ADE$ classification of rational double points and Kleinian groups.}
\label{ADEpot}
\end{table}
Here ${{\bf D}_n}$ is dihihedral group, and ${\bf T}$, ${\bf O}$ and ${\bf I}$
are discrete space goups leaving the  tetrahedron the octahedron and
the icosahedron invariant.

In three dimensions we have a classification of the 
discrete subgroups $G_F\in {\rm SL}(3,\IC)$ \cit(blichfeld)  and S. Roan 
constructs\cit(roanI) crepant resolutions for all $\IC^3/G_F$. 
All these cases (A)-(J) can be analysed physically, e.g. 
within the (A) case of Blichfeldt  the choice 
$G_F={\rm diag }(\alpha,\alpha,\alpha^{2r})$ with $\alpha^{2r+2}=1$ 
leads to  $A_r$ gauge groups in four dimension. Differently then in 
two dimensions the quotient singularities will not be equivalent 
with hypersurface singularities.

Since we are mainly interested in asymptotic free gauge groups without
matter in the adjoint we consider reducible configurations of divisors 
$S$ whose irreducible components $C_i$ are ruled surfaces as in 
the last section but now over $\IP^1$. 
To get pure Yang-Mills theory we furthermore first assume that there 
are no exceptional fibers in the ruled surfaces. 
As we already metioned $G_F={\rm diag }(\alpha,\alpha,\alpha^{2r})$
leads to the $A_r$ case.

The description of the local geometry on the type IIa side is given 
by a non-compact Calabi-Yau threefold $\Xi$, which contains the 
configuration $S$ and the non-compact direction is given by the
canonical linebundle of $S$, i.e. the total space is 
$\Xi=\cO_S(K_S)$ and by the adjunction formula it has 
vanishing first Chern class. 
In the simplest case of ${\rm SU}(2)$ we can choose for $S$  
one of the ruled surfaces $F_0=\IP^1 \times \IP^1$, $F_1$ or $F_2$, 
the difference between them will become irrelevant in the rigid field 
theory limit. We will focus on the $F_2$ case, which can be compactified 
e.g. to the $(STU)''$ hypersurface $X_{24}(1,1,2,8,12)$. Our 
advantage is however that our arguments are locally, and can
applied whenever such a surface becomes small inside a not 
necessarily compact CY threefold. In fact as we see in \figref{matter2},
we can immediatly generalize to situations with arbitrary rank of the gauge
group. In case of very high rank we cannot not expect in general 
to find a compactification to a Calabi-Yau threefold. In the noncompact 
case $c_1(\Xi)=0$ does not necessarily
imply the existence of a ricciflat metric, which become the standard 
flat metric at infinity. 

The local situation can be rephrased in terms of  a $N=2$ 
gauged Landau-Ginzburg model with abelian gauge group $U(1)^n$ \cit(wittenglg). 
The defining data are the charges of the 
$n+3$-fields $l^{(k)}=(q_0^{(k)};q_1,\ldots,q_{n+2}^{(k)})$. 
Non-anomalous $R$-symmetry implies that the charges must fulfill
$\sum_{i=0}^{n+2}q_i^{(k)}=0$, which,  morally an equivalent of 
(\ref{chern}), ensures trivialiy of the canonical bundle of 
$\Xi$. The space-time geometry we are interested in 
is actually the moduli space of that theory. 
So we must analyse the  zero locus of the 
scalar potential. The charge vectors for the model are 
\bea(rl)
l^{(1)}&=(\phantom{-}0; 1,1,-2,0)\cr
l^{(2)}&=(-2;0,0,\phantom{-}1,1)\ .
\elea(lls)

Since  we have no $D$ terms  the scalar potential is given by\cit(wittenglg) 
\be  
U={e_1^2\over 2}(|x_1|^2+|x_2|^2-2 |x_3|^3-r_1)^2+
{e_2^2\over 2}(|x_3|^2+|x_4|^2-2 |x_0|^3-r_2)^2
\ee
If $r_1$ and $r_2$ are positive, which is an equivalent way of saying
that we are inside the K\"ahlercone of the geometrical phase of
our model, then we cannot have $x_1=x_2=0$ or $x_3=x_4=0$. 
As usual we denote the loci $x_i=0$ as
divisors $\tilde D_i$ in our case $\tilde D_i$ are non-compact divisors
in $\Xi$. $\tilde D_0$ on the other hand is easily recognized as the 
Hirzebruch surface $F_2$. The scaling relations specified by 
(\ref{lls}) act analogous to (\ref{equiv}) on the $(x_0,\ldots,x_4)$ 
and define for $x_0=0$ the $F_2$ surface. Physically they corresponds to the $U(1)^2$ 
gauge freedom and the possibility to rescale the parameters 
in the potential $r_1,r_2\in \IR^+$. 
Moreover we have the correct excluded loci, or Stanley-Reisner ideals, 
to match the $F_2$ description, compare\cit(mvI) and appendix E. 

There are useful mneneotechnic diagrams for this kind 
of manifolds  called toric diagrams, 
which makes it easy to visualize the homological dependencies 
between the divisors \cit(fultonoda) as linear dependencies of points. 
We will give review some basic facts about this subject in appendix E. 
Using that it is easy to see that the 
compact divisors inside the Hirzebruch surface $F_2$  
$\tilde D_i=D_i\cap D_0$ are the class of the fiber 
$\tilde F=\tilde D_1= \tilde D_2$, a section $\tilde S= \tilde D_3$ 
and a disjoint section $\tilde S'=\tilde D_4$, which
generate the cohomology of $F_2$ modulo a relation $\tilde S'=2 \tilde F
+\tilde S$. 
Using the formalism of appendix E we readily calculate $\tilde S^2=-2$ 
$\tilde F^2=0$, $\tilde F \tilde S=1$ $\tilde F \tilde H=1$. 
The divisor $D_0$ is 
the restriction to $F_2$ as a section of the canonical line bundle. 
Using $K_{F_2}=-c_1(T_{F_2})$ and (\ref{chernI},\ref{reldivisor}) 
we get $K_{F_2}=-(2\tilde S + 4 \tilde F)$.

\figinsert{makematter}
{In the upper part of the picture we show the Hirzebruch-surfaces 
$F_0$, $F_1$, $F_2$. If a CY contains such a ruled 
surface, which can be contracted, we get pure ${\rm SU}(2)$. 
Local mirror symmetry converts non-compact CY space $\cO_{F_n}(K_{F_n})$ 
into the Seiberg-Witten curve!  
Blowing up the $F_n$ once as shown in the lower part 
yields ${\rm SU} (2)$ with one matter multiplet, again local mirror 
converts this into the ${\rm SU}(2)$ curve with matter.}
{3.2truein}{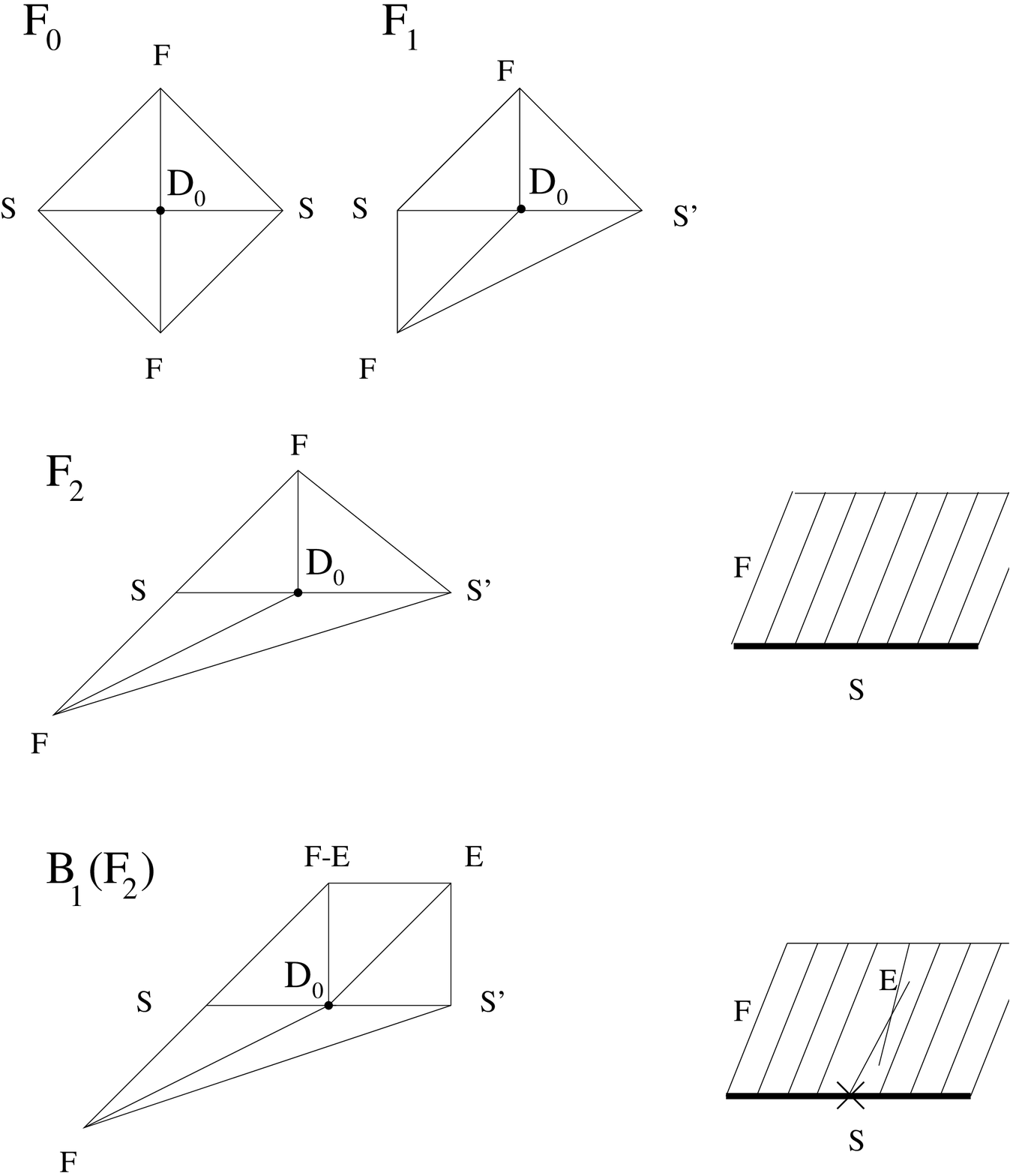}

Note that in cases of compact toric CY manifolds $X$ such as the 
$X_{24}(1,1,2,8,12)$ hypersurface the $F_2$ polyhedron appears  
as face of the defining dual reflexive polyhedron $\Delta(F_2)$, 
i.e. the  non-compact affine CY manifold arises by ``forgetting'' 
about the rest of the complete fan $\Sigma(\Delta^*)$. The noncompact 
Calabi-Yau manifold is defined by the fan $\Xi_{\Delta}$ in 
$\IR^3$ in \figref{fan}, comp. appendix E remark 
{\bf i)}.  The contraction of 
divisors to a singular variety with Gorenstein singularities 
appendix E remark {\bf iii}), corresponds to deleting the solidly 
drawn points in the interior of $\Delta$ and leads to enhanced 
gauge symmetry.

\figinsert{fan}
{The fan $\Xi_\Delta$ drawn for $\Delta=\Delta(F_2)$. It lies 
in a one dimension higher space $N'_{\IR}$ than $\Delta$, 
and is defined as the set of points hit by rays from  
the origin $\{0\}\in N'_{\IR}$ through $
\bar \Delta$, where $\bar \Delta$ is the convex hull of the points 
$\bar \nu^{(i)}=(1,\nu^{(i)})$, which lie on a hyperplane of distance
$1$ from the origin in $N'_{\IR}$. $\Xi_\Delta$ inherits its subdivison 
in a fan from the triangulation of $\Delta$. In this setting $\Delta$
is often called the trace of $\Xi_\Delta$.}
{1.4truein}{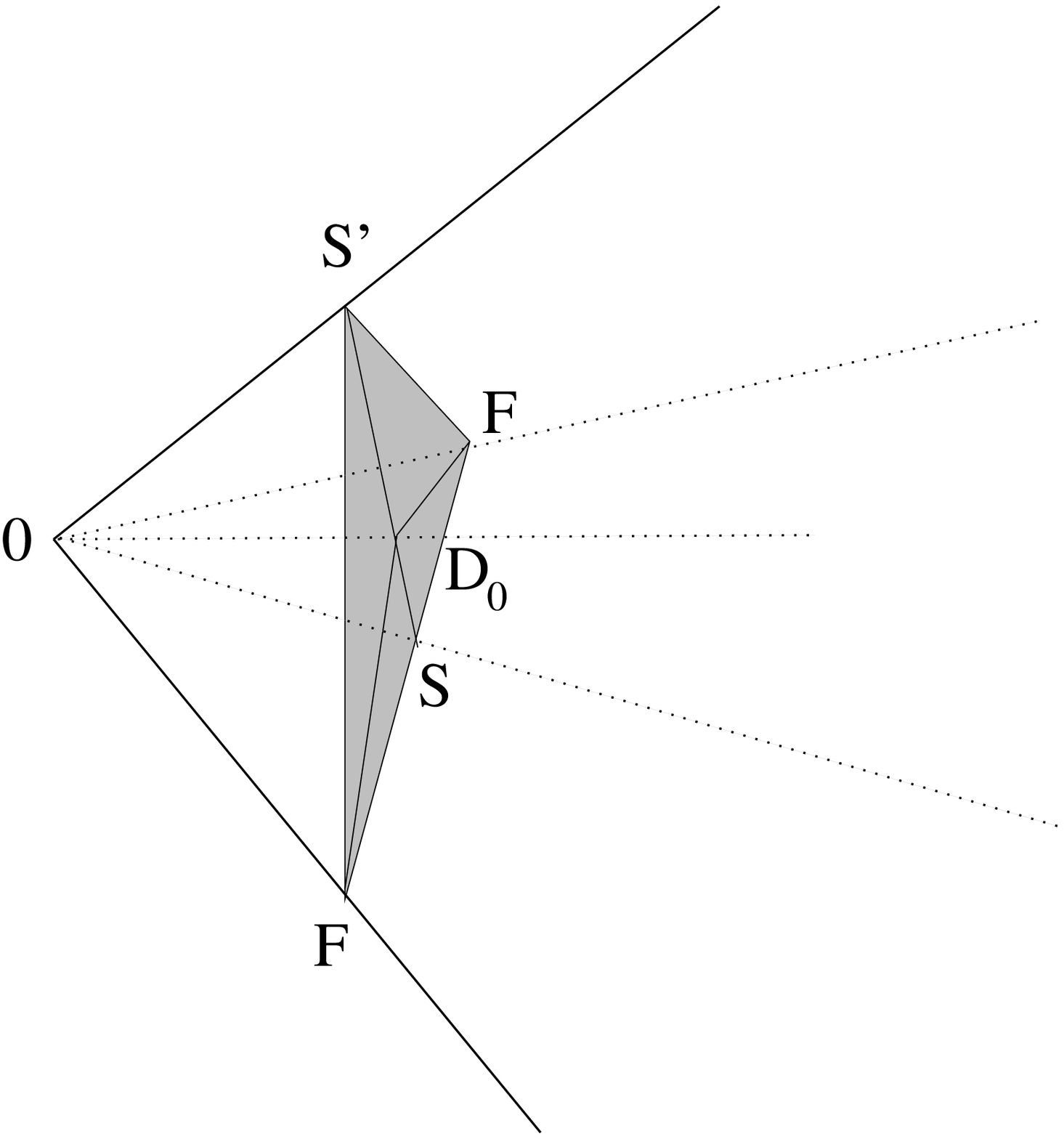}

\subsection{ Seiberg-Witten curves from the local B-model}

We use now the construction of \cit(batyrev) to assign a local mirror
description to the local $A$ model geometry. This construction assigns new 
variables $y_i$ $i=1,\ldots r$ to every field in the 
gauged $\sigma$-model, which are subject to relations defined from the 
$U(1)^k$ charge vectors $l^{(i)}$ $i=1,\ldots,k$ as
\be
\prod_{l^{(i)}_n> 0} y_n^{l^{(i)}_n}=\prod_{l^{(i)}_n< 0} y_n^{-l^{(i)}_n}\ 
\ele(yrelations)
where we left open the dimension for the moment. The mirror manifold 
$\hat \Xi$ is given by the $r-k-1$ dimensional manifold \cit(batyrev)
\be 
P=\sum_{i=1}^r a_i y_i(t_i)=0\ , 
\ele(mp)  
where we solved the relations (\ref{yrelations}) in terms of $r-k$ 
variables $s_i$ which we will projectivize. This defines an 
$r-k-2$-dimensional
manifold, which encodes the local geometry of the $A$ model and
turns out to be in the limit discussed in the previous section the 
Seiberg-Witten curve of the associated field-theory!
In addition to the curve we have to provide a meromorphic differential 
$\Lambda$ whose periods fulfill the Picard-Fuchs equations, which 
is associated to the quantum-cohomology of the $A$-model. These are 
derived from the charge vectors as in (\ref{rpf}) and are identically
fulfilled by any period over the form
\be 
\Lambda=-{\rm res}\left(\log (P) {d s_1\over s_1}\wedge \ldots\wedge 
{d s_{r-k-2}\over s_{r-k-2}}\right)\ .
\ele(Lambda)
In order to convert (\ref{rpf}) into a differential equation involving 
only on the invariant parameters $z_i=(-1)^{l^{(i)}_0}
\prod_n a_n^{l^{(i)}_k}$ we must show that $\int_C \Lambda$ 
depends only on the $z_i$. This is essentially the case, in fact
it is easy to see that $\int_C \Lambda$ transforms at worst by a 
constant shift under the projective $\IC^*$ action on the $s_i$ as 
well as under the $k$ $\IC^*$-star actions on the $a_i$ defined 
by the $l^{(i)}$. That is consistent with the fact that the differential
equation has the constant as solutions, as has to be expected for 
the Picard-Fuchs equation for a meromorphic differential with
non-vanishing residue. 

Let us convert $\Lambda$ for the relevant case $r-k=3$ into a 
differential in a patch of the projective coordinates $s_1,s_2,s_3$. As 
usually $C$ is a cycle on the $1$-complex dimensional manifold and
and $\gamma$ is a cycle around $P=0$ enforcing the residue 
\bea(rl) 
&
-\ds{\int_C \int_\gamma {\rm log}(P) {\dd s_1\over s_1}\wedge 
{\dd s_2\over s_2}=-\int_C \int_\gamma {\rm log}(P) 
\dd {\rm log}(s_1)\wedge {\dd s_2\over s_2}}=\\ [ 3 mm]
&
\ds{\int_C \int_\gamma {\rm log}(s_1) {\dd P\over P} \wedge {\dd s_2\over s_2}
= \int_C {\rm log}(s_1){\dd s_2\over s_2}} =:\int_C \lambda\ .
\elea(convert)

For $F_2$ the relations (\ref{yrelations}) are 
$y_1 y_2=y_3^2$ and $y_3y_4=y_0^2$ and will be identically 
solved by identifying $(y_0;y_1,y_2,y_3,y_4)=(st;sz,s/z,s,st^2)$. 
Projectivizing $s=1$ gives 
\be 
P=a_1 z + a_2 {1\over z} +a_3+a_0 t + a_4 t^2=0 \ ,
\ele(swc)   
which depends de facto only on  the good coordinates 
$z_S:=z_1={a_1 a_2\over a_3^2}$  and $z_F:=z_2={a_3 a_4\over a_0^2}$ and
we might use the $(\IC^*)^2$-action to set $a_1=a_4=1$ and  
in the following. The differential $\Lambda$ becomes by (\ref{convert})
\be 
\lambda={\rm log}(t){\dd z \over z}\ .
\ele(lambda) 
What remains at this point is to implement the double scaling 
limit discussed in (\ref{limit}), i.e. 
$z_S\sim \epsilon^4 \Lambda^4$, $t_{W^\pm}=t_2\sim \epsilon a$. The
corresponding scaling can be achieved by setting 
$a_3=\epsilon^{-2}$ and $a_2=\Lambda^4$. 
Netxt we bring (\ref{swc}) in the standard form (\ref{martinecform}); in order
to get rid of the next to leading term in $t$ we define 
\be
t=:(\sqrt{2} x- {a_0\over 2})
\ele(idII) 
that converts (\ref{swc}) into
\be
P=z+{\Lambda^4\over z}+2\, ( x^2 -u)=0
\ele(swcl)
with $u:=-{1\over 2} \left(a_3- {a_0^2 \over 4}\right)$. As $u$ is required 
to be finite we must identify  $2\epsilon^2 u:=-\left(1-{1\over 4 z_2}\right)$,
which gives precisely the definition of the Seiberg-Witten curve 
in the physical limit ! Inserting (\ref{idII}) into (\ref{lambda}) shows
that the leading pieces of the periods can go with
$$\int {\rm log}(x-\cO(\epsilon)){\dd z \over z}=-\log(\cO(\epsilon))
\int{\dd z\over z}+\epsilon \int x { \dd z \over z}\ . $$
In fact the residue  around the first term reproduces periods
which go with $(1 \ {\rm or}\ S) +\ldots$ and the residue around 
the second term reproduces periods, which go with  $
\sqrt{\alpha'} (a \ {\rm or} \  a_D)$ in agreement with (\ref{leadingterms}). 

\subsection{Including matter}

The inclusion of matter on the type IIa side is very simple. Basically
we have to introduce in the ruled surface over the $\IP^1$ 
a singular fiber which splits into two $\IP^1$s as indicated in 
figure \figref{makematter}. This is the situation explained more
generally in \cit(katzvafa). Similar as in (\ref{strongcoupling}) 
one may start the consideration in six dimensions with a 
configuration of vanishing cycles, which give rise to a gauge group 
$G$ of rank $r+1$ with a vector, a complex scalar and two fermions, 
all in the adjoint. Let us consider a decomposition of $G$  
into $H\times U(1)\subset G$. After fibering that configuration, 
the position on the base can be viewed in the right geometric setting 
as the scalar vev in the $U(1)$ direction of the Cartan subalgebra 
of $G$, which breakes the group generically to $H$ except at the point 
where the scalar vev is zero. 
That leaves generically a gauge group $H$ with matter 
from the decomposition of the complex scalar (and the fermionic 
completion) into $H\times U(1)$. Here we have simply $SU(2)\times 
U(1)\subset SU(3)$ and expect one matter hypermultiplet in 
the fundamental representation.

The toric description of the situation is depicted in \figref{makematter}
and the toric data are
$$
\begin{array}{rrrrrrr}
K    &F&F-E&E&S&S'&\cr
D_0  &D_1&D_2&D_3&D_4&D_5&\cr
\left(\matrix{0\cr 0}\right)&
\left(\matrix{-2\cr -1}\right)&
\left(\matrix{0\cr 1}\right)&
\left(\matrix{1\cr 1}\right)&
\left(\matrix{-1\cr 0}\right)&
\left(\matrix{1\cr 0}\right)&
\end{array}
$$
The corresponding charges of the $U(1)^3$ gauged Landau-Ginzburg model 
can be obtained from the calculation of the Mori-vectors 
\bea(rrl)
B:& l^{(1)}&=(
\phantom{-} 0;
\phantom{-} 1, 
\phantom{-} 1, 
\phantom{-}0,
-2,
\phantom{-}0)\cr
F-E:&l^{(2)}&=(
- 1;
\phantom{-}0, 
-1,
\phantom{-}1,
\phantom{-}1,
\phantom{-} 0)
\cr
E:&l^{(3)}&=
(-1;
\phantom{-}0,
\phantom{-}1,
-1,
\phantom{-}0,
\phantom{-}1)\cr
\eea

The relations (\ref{yrelations}) are fulfilled by introducing
$(s t,z s,s/z,t/z,s,t^2)$ which leads after projectivisation to
the constraint
$$
P=a_1 z+a_2{1\over z}+a_3{t\over z} + a_4 + a_0 t + a_5 t^2=0\ .
$$
Replacing again $t\rightarrow \sqrt{2} x -{1\over 2} a_0$ 
and substituting $z\rightarrow y -(x^2-u)$ we convert that precisely into 
the form (\ref{masscurves}) 
$$
y^2=(x^2-u)^2-\Lambda^3 (x+m)
$$
with the parameters
\be
u=-{1\over 2}(a_4-{1\over 4} a_0^2), \quad
\Lambda^3=\sqrt{2} a_3,\quad
\Lambda^3 m=(a_2-{1\over 2} a_3 a_0)\ ,
\ee
In particular from (\ref{limitI}) with (\ref{kappa}) $\kappa=3$ we
know that $z_B\sim \epsilon^3 \Lambda^3$ and hence
$( (4 z_E z_{F-E})^{-1}-1)\sim 2 \epsilon u^2$ and 
$(z_E-{1\over 2})\sim \Lambda^3 m \epsilon$, which is perfectly 
consistent with the picture of growing base, a mass generation 
of the gauge boson from wrapping the $D$-2-brane around 
the non-isolated curve and a mass generation of the hypermultiplet 
from wrapping the isolated curve. 

\subsection{Other Gauge groups}
 
The toric diagrams for the generalizations to $A_n$ groups are shown 
in \figref{matter2}. Using this toric representation it is
simple to calculate, using the 
the description given in appendix E, the following intersections

\be
F C_i C_j =\cases{-2 &  if  $i=j$\cr 
                  \phantom{-}1&  if  $|i-j|=1$\cr
                  \phantom{-} 0& otherwise}\ .
\ele(ani)
For $A_n$ more matter in the fundamental representation can be easily 
added by further blowing up the same toric diagram compare\cit(ims). 

By exactly  same construction as above, 
this reproduces the Seiberg-Witten curves, with more matter in the 
fundamental representation.

\figinsert{matter2}
{In the left  part of the picture we show the trace of a three  
dimensional fan $\Xi_\Delta$ with apex at the origin with the 
hyperplane $H$ at distance one from the origin, see figure 15 
for the definition of the trace. This toric diagram corresponds 
to a configuration of ruled surfaces over $\IP^1$ indicated on the left. 
If all components $C_i$ shrink simultaneously (or partly) we get  
Gorenstein singularities in the non-compact Calabi-Yau and 
pure $A_n$ (or a subgroup) as gauge theory.  
The right part shows the modification by a $\IP^1$ blow up, 
which leads to one matter multiplet in the fundamental 
representation.}{2.2truein}{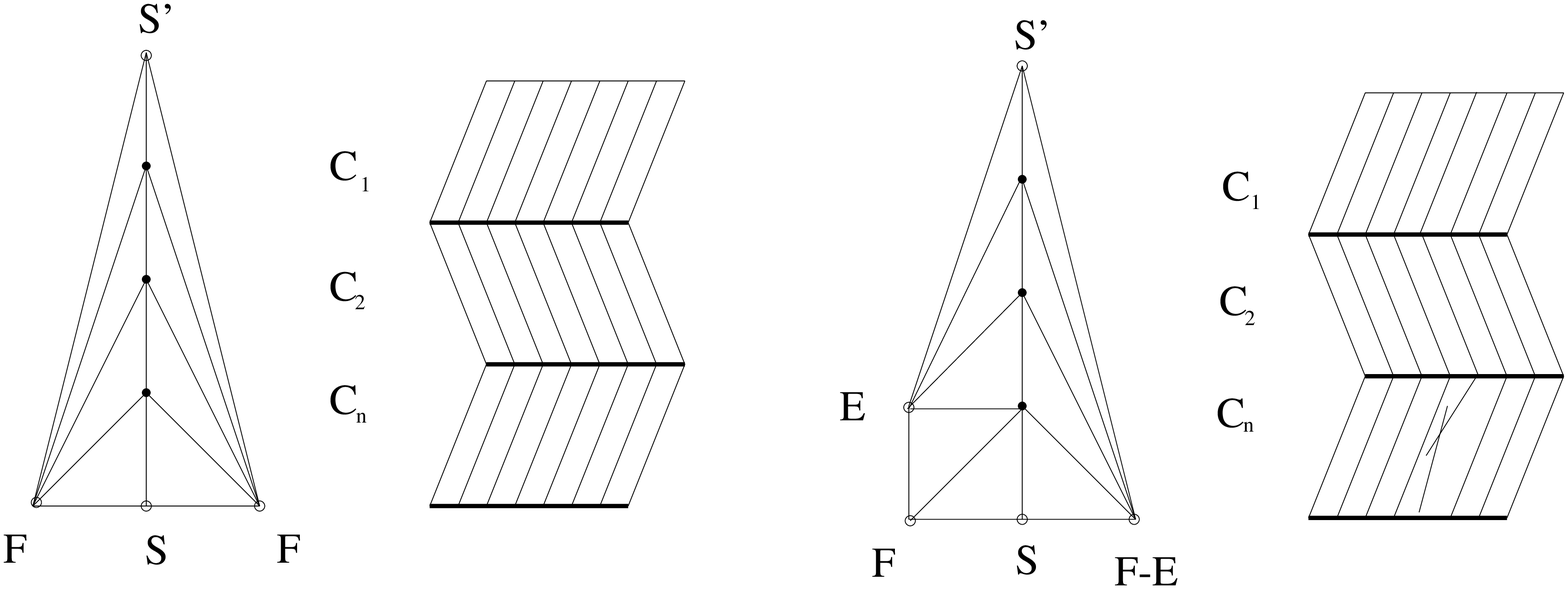}

\mabs {\sl Field theory from worldsheet instantons:}
Let us finally comment on the detailed field theoretic 
interpretation, which this picture assigns to certain 
worldsheet instantons. We consider the pure $SU(2)$ case and 
denote by  $N_{n,m}$ the instanton number of a worldsheet 
instanton, which wraps $m$-times around the base and $n$-times 
around the fiber of the Hirzebruch surface $F_n$.  
The K\"ahlermoduli of the fiber and the base are $t_W^\pm$ and $t_S$ respectively. 

We find $N_{0,1}=-2$, $N_{0,i}=0$, $\forall i>1$. What is the
field theoretic interpretation of this worldsheet instanton 
contribution ?  Using (\ref{limit}) and (\ref{fhetpert}) we can
express the gauge coupling as  
$$-i\tau={4 \pi \over g^2}+{\theta\over 2 \pi i}=
\partial_{t_W^\pm}^2 \cF =
\sum_{n=0}^\infty \sum_{m=1 \atop k=1}^\infty N_{n,m} m^2
{q_S^{nk} q_{W^\pm}^{mk}\over k}.$$
In the perturbative limit we have to consider only the contributions 
with $n=0$ and can sum over $k$ to obtain 
$\partial_{t_W^\pm}^2 \cF_{pert}=-2 \sum_{k=1}^\infty {q_{W^\pm}^k\over
k}=2\, \log (1-q_{W^\pm})$. In the limit (\ref{limit}) 
$q_{W^\pm}\sim 1-\epsilon a$ and so  $-i \tau_{pert}=2 \log(a)+const. + 
\cO(\epsilon)$. In other words the instanton, which wraps only the 
base produces exactly the one-loop contribution of the field theory.
It is easy to see that one gets the general perturbative 
one-loop part (\ref{sunpert}) if one  replaces the $\IP^1$ by a 
$ADE$ Hirzebruch-Jung sphere three. 

More generally it was observed in \cit(kkv) that the spacetime
instantons $\cF_n$ in the expansion (\ref{finfinit}), which are
made explicit in the table above (\ref{calg}), describe nothing
else then the growth of the worldsheet instantons. More precisely 
if we parameterize the growth by $N_{n,m}=\gamma_n m^{4n-3}$ then
${\cal F}_n={2^{3(3n-1)}\over (4 n-3)!}\gamma_n$.

\section*{Acknowledgments}
It is a pleasure to thank Per Berglund,
Shinobu Hosono, Sheldon Katz, Bong Lian,  Peter Mayr, Wolfgang Lerche, 
Shir-Shyr Roan,  Rolf Schimmrigk, Stefan Theisen, Cumrun Vafa, Niclas
Warner, Shimon Yankielowicz and Shing-Tung Yau for 
collaborations on these subjects. Also I would 
like to thank Eric Zaslow for discussions and
Micha\l\ Spalinski for reading parts of the manuscript.

\begin{appendix}

\section{BPS multiplets}

Let us briefly remind the reader about a basic fact 
from the representation theory of extended supersymmetry
algebras\cit(wessbaggerbook). If $N>1$ the general 
supersymmetry algebra 
\bea(rl)
\{Q^I_\alpha,Q_{\dot \beta\, J}\}&=
2 \sigma^\mu_{\alpha\beta}P_\mu \delta^I_J\\
\{Q^I_\alpha,Q^J_\beta\}&=
2\sqrt{2} \varepsilon_{\alpha\beta}Z^{IJ}\\
\{Q_{\dot \alpha I},Q_{\dot \beta J}\}&=2\sqrt{2} 
\varepsilon_{\dot \alpha \dot \beta}Z^*_{IJ}
\eea
allows in general for a nonvanishing 
central extension $Z^{IJ}$, $I,J=1,\ldots ,N$, which is antisymmetric
in $IJ$. For $N$ even one can skew-diagonalize 
$Z=\varepsilon \otimes {\rm diag}(Z_1,\ldots Z_{N\over 2})$
by a basis transformation. Defining new generators 
$^k\delta_\alpha^\pm$,  $k=1,\ldots,{N\over2}$
\be  
{^k}\delta^\pm_\alpha ={1\over 2}\left(
Q^{2k-1}_\alpha \pm\varepsilon_{\alpha\beta}{(Q^{2k}_\beta)}^{\dag}
\right)
\ele(newgen)
we can write the supersymmetry algebra in the restframe ($P_0=M$) 
in the form
\be
\{
{{^k}\delta^\pm_\alpha,({^k}\delta_\beta^\pm)}^{\dag}
\}
=
\delta_{\alpha\beta}(M\pm \sqrt{2} Z_k) \ ,
\ele(algconv)
where all other anticommutators vanish.
As the physical Hilbert space norm must be positive definit
$\bra \phi| \{\ldots \}|\phi\ket\ge 0$ one has 
$M\ge \sqrt{2} |Z_k|$. If none of these inequalities is
saturated one gets  $2^{2N}$ states, which can be built 
by application of the creation and annihilation operators
on a ``vacuum'' state, which allows for a representation
with highest possible spin difference. Those 
representations are called `long'. 
If $r$ of these inequalities are saturated one gets in
the same way only $2^{2N-r}$ states and the corresponding
representations are called short, ultrashort etc.  
In particular for $N=2$ the short multiplet is the BPS-multiplet 
and, as mentioned, those states can be viewed 
as topological subsector of the theory.

\section{Some properties of the fundamental region of discrete 
subgroups in ${\rm PSL}(2,\IR)$}
\label{a}
The upper half plane $\IH$ is parameterized by $\tau=x+i y$ with 
$x\in \IR$, $y\in \IR_0^+$. ${\rm PSL}(2,\IR)$ acts on $\IH$ by $\tau 
\mapsto {A \tau + B\over C \tau +D}$ with  $M=
\left(\matrix{A&B\cr C&D}\right)\in {\rm SL}(2,\IR)$, i.e. 
$A,B,C,D\in \IR$ and $AD-BC=1$.
As explained e.g. in \cit(cunning)\cit(lehner) the fundamental 
region $F$ in $\IH$ of a discrete subgroup 
$\Gamma\subset {\sl SL}(2,\IR)$ will be a polygon bounded arcs 
(including the ones of infinite radius), which are 
perpendicular to the real axis $E$. The reader will check that 
(\ref{frac}) maps indeed the interior of these arcs to the exterior.
All monodromy matrices (\ref{matinfinity}) and (\ref{matmonopole}) 
have $|{\rm Tr}(M)|=2$, such elements of ${\rm SL}(2, \IR)$ are known as 
{\sl parabolic} elements and conjugated to shifts.  The 
significance of ${\rm Tr}(M)^2-4$ is that it is the discriminant of 
the fixpoint equation $C\tau^2+(D-A)\tau-B=0$ of the ${\rm PLS}(2,\IR)$ 
action. So if $C\neq 0$ the fixpoint of a parabolic element is on $E$.
If $C=0$ the parabolic element is a shift $\tau\mapsto \tau+m$ with 
fixpoint $\tau=i\infty$. If $|{\rm Tr} (M)|<2$ ($|{\rm Tr}(M)|>2$) the 
element is called elliptic (hyperbolic). An elliptic element $M$ for which 
$\rho$, as in $\rho+\rho^{-1}={\rm Tr}(M)^2-2$, is an n'th root of unity 
corresponds to a transformation of order $n$ 
($M^n=\pm 1$) in ${\rm PSL}(2,\IR)$. 
For parabolic elements one defines formally $n=\infty$. 
It is easy to check that in ${\rm PSL}(2,\ZZ)$ one has only 
$n=2,3$ or $\infty$. 
The arcs bounding $F$ contain fixed points $P_\gamma$ of elliptic or parabolic 
elements in $\Gamma$ and the sum of the inner angles at the equivalent 
fixed points on the boundary of $F$ is $2 \pi/n$. This is clear from
the fact that $n$ equivalent regions will meet at the fixpoint of order 
$n$ and since (\ref{frac}) is angle preserving each of these regions 
occupy the same angle at $P_\gamma$. In hyperbolic geometry the
metric is $\dd s={|\dd \tau|\over y}$ and the area differential 
$\dd^2\sigma={\dd x \dd y\over y^2}$. 
The ${\rm PSL}(2,\IR)$ action on $\IH$ obviously preserves
distances and areas in hyperbolic geometry. 
The sides of in the boundary of the fundamental region are pairwise
identified\footnote{If we make this identification we get 
the Riemann surface on which $\tau(u)$ becomes single valued. Its
genus is given by $g=(2+n-c-1)/2$, where $c$ is the number of 
inequivalent vertices. Luckily we will encounter only the genus 
zero case, which is by far simpler then the general case.} 
and the area of the normal polygons with $2n$ sides is easily 
calculated to be
\be 
A=2\pi(n-1-\sum_{fp}{1\over n_i})\ .
\ele(area) 
E.g.the standard fundamental region of ${\rm SL}(2,\ZZ)$, whose boundary 
contains the parabolic fixpoint of $T$: $i\infty$ the $Z_2$ fixpoint of 
$S$: $i$ and the cycle of $Z_3$ fixpoints: of $(TS)$: 
$\exp {2\pi i\over 6}$ and of $(ST)$: $\exp {2 \pi i \over 3}$ is 
$A_0=2\pi(2-1-1/3-1/2)={\pi\over 3}$. 
For subgroups $\Gamma$ of ${\rm SL}(2,\ZZ)$ the area will clearly be a 
multiple $\mu$ of the the area $A_0$, which is known as the index of 
$\Gamma$. Subgroups of small index are classified \cit(kleinfricke).

{\sl Isometric circles\ } of an element $M\in \Gamma$ are defined 
by $|C z+ D|^2=1$. Their radius is $1/|C|$ and their center is at $
({\rm Im}(\tau)=0,{\rm Re}(\tau)=C D/|C|^2)$. 
If $\Gamma$ has a translation $\tau\rightarrow \tau+b$ 
the construction of the fundamental region is especially  simple. 
One draws a vertical strip of width $b$ called $R_\infty$. 
The fundamental region is the union of $R_\infty$ with the exterior 
of every isometric cycle. As we have only parabolic elements to our
disposal in our application the isometric cycles have to fit exactly 
in the strip $R_\infty$ as in \figref{fundamental}.

\section {Weighted projective form of the $\Gamma_0(2)$ curve, 
Picard-Fuchs equations and discriminants} 
\label{c}
The elliptic curve ${\cal E}$ which gives rise to $\Gamma(2)$ or to 
$\Gamma_0(2)$ can be represented in different forms. We will first 
chose a representation, which will prove useful later when we discuss 
CY manifolds and describe it by the zero locus of the 
quasi-homogeneous degree $k=4$ polynomial 
\be 
p={a_1 x_1^4}+{a_2 x_2^4} + 
{a_3 x_3^2}-{a_0} x_1 x_2 x_3 \ .
\ele(poli)
in the weighted projective space $\IP^2(1,1,2)$. See section 
\ref{construction} for the definitions. Clearly this manifold is one
complex dimensional and from (\ref{ceins}) with $k=4$, $w_1=w_2=1$ and 
$w_3=2$ it has vanishing first Chern class $c_1=0$, hence it is a
torus\footnote{The reader will find later much more
down to earth arguments why that is a torus.}. 

\mabs {\sl Picard-Fuchs equations from scaling symmetries.}
The parameterization in (\ref{poli}) is redundant as an elliptic 
curve will have only one independent parameter in the 
defining polynomial, which deforms the complex structure. In fact
we have three $\IC^*=(\IC\setminus \{0\})$ actions on the parameters as
$a_i\mapsto \lambda^{k/w_i} a_i$, $x_i\mapsto \lambda^{-1} x_i$ and 
$a_0\mapsto \lambda a_0$ for  $i=1,2,3$  leave (\ref{poli}) invariant.
We might therefore introduce later $z={a_1 a_2 a_3^2 \over a_0^4}$ as 
invariant parameter. The period integral can be defined by the
residue expression 
\be 
\tilde \varpi_i= \oint_{C_i} {1\over 2 \pi i} 
\oint_{\Gamma_\epsilon} {a_0 \dd \mu \over p}  \ . 
\ele(periodI)
Here $\dd \mu=\sum_{i=1}^3 (-1)^j w_i x_i  \dd x_1\ldots 
\widehat {\dd x_i}\ldots \dd x_3$ and the hat means omission. 
$\Gamma_{\epsilon}$ is a small circle looping around $p=0$ and 
$C_i\in H^1({\cal E},\ZZ)$ is an element in the integral homology 
of ${\cal E}$. 

Note that in $\tilde \varpi$ we put an $a_0$, which is 
essential to keep the $(\IC^*)^3$ invariance, which we will use now 
to derive the Picard-Fuchs equation. After the derivation we will 
use the more conventional form of period integral 
$\varpi={1\over a_0}\tilde \varpi$ . First note, using 
$\vartheta_{i}:=a_i\p_{a_i}$, $[a_i,\vartheta_{i}]=-a_i$, 
$\theta_z:=z\p_z$ and  $\vartheta_{i} f(z:=\prod_{k=1}^n a_k^{l_k})=
l_i \theta_z f(z)$ the following identities 
$$\begin{array}{rl}
0=& a_1a_2a_3^2 a_0( \p_{1}\p_{2}\p_{3}^2-\p_{0}^4){\dd \mu \over p}\\ [ 2 mm] 
 =&a_0
    \left(\vartheta_{1}\vartheta_{2}\vartheta_{3}(\vartheta_{3}-1)-
        z \prod_{i=0}^3(\vartheta_{0}-i)\right)
       {\dd \mu \over p} \\ [ 2 mm]
 =&\left( \vartheta_{1}\vartheta_{2}\vartheta_{3}(\vartheta_{3}-1)-
         z \prod_{i=1}^4(\vartheta_{0}-i)\right){a_0\dd \mu \over p}\\ [ 2 mm]
 =&  (2 \theta_z^3 (2 \theta_z+1)-z \prod_{i=1}^4(4 \theta_z-i))
  {a_0\dd \mu \over p}\\    [2 mm]
 =&\theta_z (4 \theta_z- 2)[\theta_z^2- 
     z (4\theta_z+1)(4 \theta_z+3)]
       {a_0\dd \mu \over p}\\ [ 2 mm]
 =&\theta_z (4 \theta_z- 2){\tilde {\cal L}}{a_0\dd \mu \over p}
\end{array}
$$
From the last expression it is clear that $\theta_z 
(4\theta_z-2){\cal L}\tilde\varpi=0$. From the four solutions to this 
equation only the two which fulfill ${\tilde {\cal L}}\tilde\varpi=0$ 
have the right asymptotic at $z=0$ to be periods of ${\cal E}$. 
It is straightforward to see that this system  after the variable 
substitution $z=(64 u^2)^{-1}$ is  equivalent to ${\cal L} \varpi=0$ 
with ${\cal L}$ as in (\ref{periodeqi}).

The fact that two elliptic curves have the same Picard-Fuchs equation
does not quite imply that they belong to the same parameterization 
family, which has a unique $\Gamma\in SL(2,\ZZ)$ and a 
unique $\tau(u)$.  The ratios of the period 
integrals over the generating elements of 
$H^1({\cal E},\ZZ)$ have also to agree. This is for example not
the case for curves, which are only isogeneous, here $\tau(u)$ 
differs by an integer factor. 

We can complete the check that (\ref{poli}) is a $\Gamma(2)$ curve 
by calculating the $j$-invariant. We may use first an invariance 
transformation of the $\IP^2(1,1,2)$ $x_1\mapsto \sqrt{i}x_1$, 
$x_2\mapsto \sqrt{i}x_2$ and $x_3\mapsto x_2+\sqrt{2 u}x_1x_2$ and go 
to inhomogeneous coordinates $x_2=1$, $x_1=x$, $x_3=y$ so that the 
constraint $p=0$ looks like
\be 
y^2\equiv x^4+2 u x^2+1\ .
\ele(ellii)
This is further transformed to the Weierstrass form (\ref{weierstrass}) 
(comp. footnote \ref{jfoot}) with $3 g_2=3+u^2$ and $27 g_3=9 u(1-u^3)$ 
so the $j$-invariant is 
\be 
j={(3+u^2)^3\over 27(1-u^2)^2}.
\ele(gamiij)
Comparing that with (\ref{jfunct}) solves the inversion problem 
for the triangle functions and by comparing it with the asymptotic 
of $\tau(u)$  (\ref{ellii}) is established as $\Gamma(2)$ curve. 
From (\ref{gamiij}) we also see that 
$u$ branches sixfold over $j$, which is another way to see that 
$\Gamma(2)$ is of index six in $SL(2,\ZZ)$. If we do not introduce 
the double covering variable $u$ but just rescale $z=64\tilde z$, then we 
get $\Gamma_0(2)$ of index $3$ and the hypergeometric system 
(\ref{paramidentification}) with (note that we exchanged $0$ 
and $\infty$) $\alpha_\infty=1/2$, $\alpha_0=0$ and $\alpha_1=0$, 
i.e the system $(1/4,3/4,1)$. 

In (\ref{periodI}) with $a_0=1$ we might use (\ref{ellii}) and  perform the 
integration over the loop $\Gamma_\epsilon$ in the $y$-plane. This
leads to the integral
\be
\varpi_i=\oint_{C_i} {{\rm d}x \over {{\rm d} p\over 
{\rm d} y}|_{p=0}}= 
\oint_{C_i} \omega\ .
\ele(periodii)
with the holomorphic $(1,0)$-form $\omega={\dd x \over y}$ as 
derived in sec \ref{algtor}.

\mabs {\sl Discriminant and Picard-Fuchs equation for general Hyperelliptic
           curves:}

The residue expression (\ref{periodI}) is well defined under the 
equivalence relation in $\IP^n(\vec w)$ only for $c_1=0$ manifolds 
(\ref{ceins}).  The above symmetry considerations are a powerful 
tool and often sufficient to derive the Picard-Fuchs equations
for $K3$ and higher dimensional CY, see \cit(hktyI),\cit(hktyII) and 
references therein for additional techniques. 
For higher genus Riemann surfaces $c_1<0$ and we will need a little 
more algebra to derive the Picard-Fuchs equations. Let
\bea(rl) 
f&=a_0 x^m+a_1 x^{m-1}+\ldots + a_m \\ [ 2 mm]
g&=b_0 x^n+b_1 x^{n-1}+\ldots +b_n  \\ [ 2 mm]
\eea
polynomials of degree $n$ and $m$. The resultant\foot{The word is actually
derived as a short form of the phrase ``result of elimination''. What we 
want to eliminate are powers of $x$.} $R(f,g)$ is defined
as the determinant $|M|$ of the $(n+m)\times (n+m)$ matrix see e.g.
\cit(langalg)

\be
\matrix{n\!\!\! 
&\left\{\phantom{\matrix{a_1\cr a_2 \cr a_3 \cr a_4}}\right. \cr      
m \!\!\!&\left\{\phantom{\matrix{a_1\cr a_2 \cr a_3 \cr a_4}}\right.}      
\!\!\!\!\!\!\!\!\!\!\!
\left|\matrix{ a_0&  a_1& \ldots & a_m    & 0 &   &\ldots  & 0 \cr
                 0&  a_0&\ldots  & a_{m-1}&a_m& 0 &\ldots  & 0\cr
                  &     &        &  \ldots&   &   &        & \cr
                  &  0   & \ldots &0        &a_0&a_1& \ldots & a_m \cr 
                 b_0&  b_1& \ldots & b_n    & 0 &   &\ldots  & 0 \cr
                 0&  b_0&\ldots  & b_{m-1}&b_n& 0 &\ldots  & 0\cr
                  &     &        &  \ldots&   &   &        & \cr
                  & 0   & \ldots & 0        &b_0&b_1& \ldots & b_n 
}\right|
\ele(detdisc)
It is clear that this determinant vanishes only if $f,g$ have
a common root or $a_0=b_0=0$. Now define $\vec m=(x^{m-1},\ldots, 1)^t$
and $\vec n=(x^{n-1},\ldots, 1)^t$. Obviously 
$(f\cdot \vec m, g\cdot \vec n)^t=M\cdot (\vec m,\vec n)^t$. 
Cramers rule applied 
to the last entry $1$ in $(\vec m,\vec n)^t$ gives
$ R(f,g)=|M|=|M_{n+m}|$, which
implies that $ R(f,g)=a(x)f(x)+b(x)g(x)$ where $a(x)$ ($b(x)$) 
is of degree $n-1$ ($m-1$). 

This is useful for the derivation of the Picard-Fuchs equation 
for the period $\varpi=\oint \omega$ of the hyperelliptic Riemann 
surface $y^2=p(x,u_i)$ with holomorphic differential 
$\omega={\dd x\over y}$. The discriminant is given by the 
resultant of $p=0$ and $p':={\dd \over \dd x}p=0$ i.e. 
$\Delta( u_i)=R(p,p')$. As we just shown $\Delta=a p+ b p'$. 
We want find differential relations of the form 
${\cal L}(u_i) {\dd x\over y}={\p h \over \p x}\dd x$. 
Derivatives on $\omega$ w.r.t. the 
moduli $u_i$  produce ${\phi(x,u_i)\dd x\over y^n}$ and we have 
to relate this terms up to exact terms to ${\dd x\over y}$. To reduce the  
degree of $y$ in the denominator one uses the following algorithm.
By partial integration we have up to exact terms 
$${\phi(x)\over y^n}={1\over \Delta} {a\phi+{2\over n-2} (b \phi)'
\over y^{n-2}}.$$
This substitution increases the  powers of $x$ in the numerator. They
have to be lowered by expressing the highest power in $x$  
in terms of lower ones in terms of $p$ or $p'$ and lower powers. 
In the later case a partial integration must follow. 
Combining these steps the desired relations ${\cal L}(u_i) \varpi=0$ can 
be derived. 

For the calculation of the discriminante of the weighted projective 
Calabi-Yau hypersurfaces it is certainly not a practical way to use 
the generical alogarithm (\ref{detdisc}) iteratively on $p=0$ and 
${\dd \over \dd x_i}p=0$. Direct elimination of the coordinates $x_i$ 
using the symmetries leads much quicker to the result, whose complexity
grows however rapidly with the number of moduli. Note also
that to find all possible components of the discriminate one has to 
test $p=0$ and ${\dd \over \dd x_i}p=0$ on all strata, i.e. for all
allowed combinations of $x_{i_1}=\ldots x_{i_k}=0$. The component of the
discriminante, which is calculated for all $x_i\neq 0$ is called 
principal part. For (\ref{polmirror}) it is $\Delta_{c}$, while  
for $x_3=x_4=x_5=0$ we get the independent component of the 
discriminante $\Delta_s$. For complete intersections examples see 
\cit(hktyII).   

\section{Instantons corrected triple intersections for the practioner}

In this appendix we want to summarize how to calculate the periods 
the mirror map and the triple intersections on general toric 
CY $d$-folds following largely \cit(hktyI)\cit(hktyII)\cit(gmp)
\cit(mayr) and especially \cit(klry). The solution for the periods 
is given by eq. (\ref{sol}) the mirror map is defined in (\ref{mirrormap}) 
and (\ref{soldr},\ref{cop}) give the basic instanton corrected triple
coupling. The reconstruction of the other ones, for $d>4$ a problem, 
is described in section (\ref{reconstructionI}). There are two conceptional 
straightforward but technically involved problems which must be solved
before applying these formulas. The calculation of the classical 
intersection numbers, is described in appendix D, for the practical 
application there is a program \cit(schubert). 
The second is the construction of the K\"ahler
cone respectively its dual the  Mori-cone. It is described in 
\cit(torickaehlercone) and a problem which arises in this context 
namely the triangulation of polyhedra can be solved with the program
\cit(puntos).

\subsection{Frobenius algebras}

To obtain all $k$-point functions we introduce some basic 
notions of Frobenius algebras.  In this section, all vector 
spaces are finite dimensional.
A Frobenius algebra is a commutative graded algebra
$A=\oplus_{i=0}^dA_{(i)}$, generated by $A_{(1)}$, has $A_{(0)}=\C\cdot 1$,
and a nondegenerate degree $n$ bilinear symmetric invariant pairing
$\bra,\ket:A\times A\ra\C$. Note that because we require
generation by $A_{(1)}$, this notion is slightly stronger than
the usual notion of a Frobenius algebra. We give some well-known
examples from geometry. Let $\bP$ be a complete toric variety,
and $A^*(\bP)$ be its Chow ring. Then $A^*(\bP)\otimes\C$ is
a Frobenius algebra. The pairing here is the Poincar\'e pairing.
If $X$ is a hypersurface in $\bP$, then it can be shown that the
ring
\be
\tilde A^*(X):=Im(A^*(\bP)\ra A^*(X))=A^*(\bP)/Ann([X])
\ee
tensored with $\C$
is a Frobenius algebra. More generally, if $A$ is a Frobenius algebra,
and $x\in A_{(1)}$ is a nonzero element, then
$\tilde A:=A/Ann(x)$ is a Frobenius algebra with the
induced pairing $\bra a+Ann(x),b+Ann(x)\ket:=\bra a,b\cdot x\ket$
having degree $d-1$.

Let $V_1,V_2,V_3$ be vector spaces, and $C:V_1\otimes V_2\otimes V_3\ra\C$
be a three-point function. It is call $V_1$-nondegenerate if that $C_{(a,b,c)}=0$
for all $b,c$ implies that $a=0$. Similar notion of $V_i$-nondegeneracy
applies. We call the form nondegenerate if it is $V_i$-nondegenerate for all
$i$. Now suppose $C$ is $V_3$-nondegenerate. Then we have the following
invertibility property. Let $D:V_3^*\otimes V_4\ra\C$ be any bilinear
form. Then the knowledge of the 3-form
$E_{(a,b,d)}:=C_{(a,b,c_i)}D_{(\gamma^i,d)}$ ($\{c_i\},\{\gamma^i\}$ being dual bases),
allows us to determine $D$ completely. In fact, there exists (in general
not unique) a 3-form $F$ such that
$D_{(\gamma,d)}=F_{(\gamma,\alpha^i,\beta^j)}E_{(a_i,b_j,d)}$. 
This is just the statement that the $V_3$-nondegenerate three-point function
$C$ defines an onto map $V_1\otimes V_2\ra V_3^*$, hence choosing
a section gives us a left inverse $F$ to this map.

We now return to a Frobenius algebra $A$.
it determines a collection of three point functions
$C^{(ijk)}: A_{(i)}\otimes A_{(j)}\otimes A_{(k)}\ra\C$ with
$i,j,k\geq0, i+j+k=d$.
These three-point functions  are $A_{(i)}$-nondegenerate
whenever either $j=1$  or $k=1$ because $A_{(1)}\cdot A_{(i)}=A_{(i+1)}$.

\subsection{Reconstruction}
\label{reconstructionI}
Let $A=\oplus_{i=0}^d A_{(i)}$ be a graded space with $A_{(0)}=\C$
and equipped with a degree $d$ nondegenerate symmetric bilinear form
$\eta$. Suppose we are given three-point function:
$C^{(ijk)}:A_{(i)}\otimes A_{(j)}\otimes A_{(k)}\ra\C$,
$i,j,k\geq0$ with the following properties:
\begin{itemize}
\item (a) (Degree) $C^{(ijk)}=0$ unless $i+j+k=d$.
\item (b)  (Unit) $C^{(0ij)}_{(1,b,c)}=\eta^{(i)}_{b,c}$.
\item (c) (Nondegeneracy) $C^{(1ij)}$ is nondegenerate in the second slot.
\item (d)  (Symmetry) For any permutation $\sigma$ of 3 letters,
$C^{(ijk)}_{(a,b,c)}=C^{\sigma(ijk)}_{\sigma(a,b,c)}$.
\item (e)  (Associativity)
\end{itemize}
\be
 C^{(i,j,d-i-j)}_{(a,b,c_p)}\eta_{(d-i-j)}^{pq}C^{(i+j,k,d-i-j-k)}_
{(d_q,e,f)}=
C^{(i,k,d-i-k)}_{(a,e,c_p')}\eta_{(d-i-k)}^{pq}C^{(i+k,j,d-i-j-k)}_
{(d_j',b,f)}
\ee
where the $c$ and the $d$ are bases of the appropriate spaces.

Then $A$ is a Frobenius algebra with the product
\be
a\cdot b=C_{(a,b,c_p)}\eta^{pq}d_q.
\ee
The rules above are known as fusion rules.
One can also build a $k$-form by fusing together 2- and 3-forms.
The associativity law says that there will often be many ways
to build a given $k$-form. Similarly the 3-forms are not
independent. We claim that the forms of type $(i,j,d-i-j)$ for
$i,j>1$ are determined by the those of type $(1,r,d-r-1)$.
To see this without loss of generality, we can assume $1<d-i-j\leq i,j$.
Now by the associativity law above with $k=d-i-j-1$ and the invertibility
property of $C^{(i+j,k,d-i-j-k)}=C^{(i+j,k,1)}$, it follows that
$C^{(i,j,d-i-j)}$ are determined  in terms of forms of type
$(i,d-i-j-1,j+1)$ and $(i+k,j,1)$. By the symmetry property,
$(i,d-i-j-1,j+1)$ is equivalent to $(i,j+1,d-i-j-1)$.
Thus we have reduced the value of $d-i-j$ by 1. 
By induction, we see that all $(i,j,d-i-j)$ can be expressed
in terms of those of type $(1,r,d-r-1)$. In terms of the algebra
$A$ itself, an alternative way to state the result is that
all the products $A_{(i)}\otimes A_{(j)}\ra A_{(i+j)}$ is
determined by those of the form $A_{(1)}\otimes A_{(r)}\ra
A_{(r+1)}$ because $A$ is generated by $A_{(1)}$ and that
\be 
(a_1\cdots a_i)(a_{i+1}\cdots a_{i+j})
=a_1(a_2\cdots a_{i+j}).
\ee

\subsection{Application}

Let $X$ be a CY $d$-fold, and let $\cA$ be the corresponding
 Frobenius subalgebra
of $\oplus_{p=0}^d H^p(X,\wedge^pT^*)$. Suppose mirror symmetry holds:
there is a mirror family $X^*$ whose B-model algebra coincides with
the A-model algebra of $X$. We shall now compute the Frobenius
subalgebra $\cB$ of the B-model algebra corresponding to $\cA$.
{}From our general discussion of Frobenius algebras, it is enough
to compute the three-point functions $C$ of types $(1,r,n-r-1)$ which come
with $\cB$. Once we have a period expansion in the topological base 
(\ref{alphaEQ})these can be 
easily obtained using eqns (\ref{CubicForm},\ref{InnerProduct},
\ref{Transversality}). 
To obtain the coefficients in (\ref{alphaEQ}) we will use the 
fact \cit(hktyI) that the universal structure of the solution of 
the Picard-Fuchs equation on $X^*$ at the large radius point mirrors the 
primitive part of the vertical cohomology of $X$ and the 
leading structure of logarithm enables us to associate  
this solutions with the expansion of the periods in a 
topological base. This leads to a direct generalization of the formulas of 
\cit(hktyII) to some correlation functions on $d$-folds. 

More precisely there are $h^{r,r}_{prim}(X)$ solutions $0\le r\le d$ 
with leading degree $r$ in the $\log(z_i)$, which have the form 
\be
\tilde \Pi^{(r)}_k=\sum_{\Pi} 
{^0C^{d-r,1\ldots 1}_{k,i_1,\ldots,i_r}} 
\bigl({1\over r!} l_{i_1} \ldots l_{i_r} S_0 
+ {1\over (r-1)!} l_{i_1} \ldots l_{i_{r-1}} S_{i_r} +
\ldots + S_{i_1,\ldots,i_r}\bigr),
\ele(sol)
here we defined $l_i:=\log(z_i)$ and the $S_{i_1,\ldots i_r}$ are 
holomorphic series in the $z_i$, whose explicit form are given 
in section (\ref{explicite}). $\Pi$ means permutation over distinct 
indices see below Eq. (\ref{intersect}) for an example. The map to an specific 
element of the cohomology $H^{d-r,d-r}$ of $X$ can be made precise by 
noting that the $^0C^{d-r,1\ldots 1}_{k,i_1,\ldots i_r}$ are 
given by the classical intersection of that specific element with the
 intersection of divisors $J_{i_1}\cdot\ldots\cdot J_{i_r}$. 
We discuss the primitive part of the (co)homology generated by 
$J_1\ldots J_{h^{1,1}}$ only and by
Poincare duality, this data fix the element in $H^{d-r,d-r}$ completely.

As mentioned above the covariant derivative $\nabla_{a}$ in \cit(gmp) 
becomes the ordinary derivative in the flat complexified K\"ahler 
structure coordinates $t_k$.
The coordinate change from the natural complex structure coordinates 
$z_a$ to the $t_k$ variables is given by the mirror map 
\be 
t_k={\tilde \Pi^{(1)}_k(z_i) \over \tilde \Pi^{(0)}(z_i)}=
\log(z_k)+{S_k\over S_0}\ .
\ele(mirrormap)
If we substitute this coordinate transformation in the normalized periods 
$\Pi_i^{(r)}={\tilde \Pi_i^{(r)}\over \tilde \Pi^{(0)}}$ 
some simplifications occur as the first sub-leading terms in the 
$t_i$ cancel out:       
\be
\Pi^{(r)}_k=\sum_{\Pi} 
{^0C^{d-r,1\ldots 1}_{k,i_1,\ldots,i_r}} 
\bigl({1\over r!} t_{i_1} \ldots t_{i_r}+     
{1\over (r-2)!}  t_{i_1} \ldots t_{i_{r-2}} \hat S_{i_{r-1}} \hat S_{i_r}+
\ldots + \hat S_{i_1,\ldots,i_r}\bigr).
\ele(soldr)
Now we notice from the monodromy around $z_i=0$ ($t_i\rightarrow t_i+1$) 
that the periods $\Pi_k^{(r)}$ correspond to a expansion 
of $\alpha^{(0)}=\Omega$ in terms of the topological 
basis\footnote{This is actually only true up to the addition of 
solutions with sub-leading logarithms, which however does not
affect the holomorphic couplings discussed below. It will 
affect however the non-holomorphic Weil-Peterson metric.} 
$\gamma_{(r)}^k$ of (\ref{alphaEQ}) 
$\alpha^{(0)}=\sum_{k,r} \Pi^{(r)}_k {\gamma}_{(r)}^k$.

The coupling $C^{(1,1,d-2)}_{a,b,c}:H^{1,1}\times 
H^{1,1}\times H^{d-2,d-2}\rightarrow \IC$ is especially 
simple to obtain. Applying (\ref{Transversality}) in the case $k=0$ we have 
$\partial_{t_a}\alpha^{(0)}=
\alpha^{(1)}_a$. This determines $\alpha^{(1)}_a$, hence all its coefficients.
Now using (\ref{CubicForm}) for $k=1$, (\ref{alphaEQ})  for $k=1,d-2$,
`and the fact that
$\bra\gamma^{(k)}_a,\gamma^{(l)}_b\ket=0$ for $k+l>d$, we see
that
\be
C^{(1,1,d-2)}_{a,b,c}=\partial_{t_a}g^{(2)d}_b\eta^{(2)}_{dc}=
\partial_{t_b}\partial_{t_b}\Pi^{(2)}_c\ ,
\ele(cop)
where the $g^{(2)}$ are the coefficients of the $\gamma^{(2)}$ in the 
$\alpha^{(1)}$. Note that the last equation follows from the fact that 
$\Pi^{(r)}_a$ is an expansion in the dual base $\gamma^a_{(r)}$ 
and that the associativity of the classical parts in (\ref{sol}) is manifest.
Eqs. \ref{soldr} \ref{sol} are direct generalizations 
of eqs. (4.9) and (4.18) to the $d$-fold case. 
For $H^{1,1}$ we have always a canonical choice 
of the basis say $J_1\ldots J_{h^{1,1}}$, as there is a canonical basis 
for the tangent space of the moduli space corresponding to elements 
$H^{d-1,1}(X^*)$, which is mapped by the monomial divisor mirror map 
to $H^{1,1}(X)$ and (\ref{cop}) reduces for $d=3$ to the 
expressions given in \cit(hktyI). 
For $d>3$ there is a priori no canonical choice for the 
basis of $H^{d-2,d-2}$. However toric geometry can be used as in 
\cit(hly) to show that the graded ring 
\be 
{\cal R}=\IC[\theta_1,\ldots ,\theta_{h^{1,1}}]/{\cal J},
\ele(topring) 
where ${\cal J}$ is the ideal generated by the leading $\theta$-terms of 
Picard-Fuchs equations, gives, by the identification 
$\theta_i\rightarrow J_i$, a presentation of the primitive part of 
$H^{*,*}$. Because of Poincare duality it is of course sufficient to pick
a basis of half of $H^{*,*}$ and as mentioned above the choice 
of the basis in $H^{1,1}$ is canonical. It was shown in 
\cit(hktyI)\cit(hktyII)\cit(hly) that any element of 
$\cal R$ can be mapped to a 
solution (\ref{sol}) , i.e.  the 
$^0C^{d-r,1\ldots 1}_{i_1,\ldots,i_r}$ are 
determined by the principal part of the Picard-Fuchs equation. 
This can be viewed as a proof of mirror symmetry at the level
of the classical intersections, which readily generalizes to $d$-folds.

Now proceed by induction. Suppose we know (the coefficients of)
the $\alpha_{(i)}$ 
and the three-point functions of types $(1,i,n-i-1)$
for $i=0,1,..,k$. Then by the invertibility property of
a three-point function of type $(1,k,n-k-1)$ in a Frobenius algebra,
we can solve for the $\alpha_{(k+1)}$ using (\ref{Transversality}).
Thus the $\alpha^{(k+1)}$ are determined.
By (\ref{alphaEQ}), we can write $\partial_{t_a}\alpha^{(k+1)}_b=
\partial_{t_a}g^{(k+2)d}_b\gamma^{(k+2)}_d+\cdots$ (which is
now known),
arguing as before using (\ref{CubicForm}) with $k$ replaced by $k+1$, 
and using the inner product property of the $\gamma$, we find that
$C^{(1,k+1,n-k-2)}_{abc}=
\partial_{t_a} g^{(k+2)d}_b\eta^{(k+2,n-k-2)}_{dc}$.
Thus the three-point functions of type $(1,k+1,n-k-2)$ is also determined.
This shows that all three-point functions 
of type $(1,k,n-k-1)$ for $k=1,2,..,n-1$
can be expressed in terms of the coefficients of $\alpha_{(0)}$ alone.

\subsection{Explicit expressions for periods in toric varieties}   
\label{explicite}

Following \cit(hktyI)\cit(hktyII) we can determine the holomorphic series  
$S_{i_1,\ldots,i_r}$ from the generators of the Mori cone.
Consider a CY $d$-fold defined as complete 
intersection with $p$ polynomial constraints in a toric 
variety of dimension $d+p$. The generators of the Mori cone will be
of the form
$$l^{(i)}=(\hat l^{(i)}_0,\ldots,\hat l^{(i)}_{p-1};
l^{(i)}_1,\ldots,l^{(i)}_q),$$ 
where $q=d+p+h^{d-1,1}$. The series $S_{i_1,\ldots,i_r}$ are
obtained by the Frobenius method from the coefficients of the 
holomorphic function $\omega(\vec z,\vec \rho)$
$$\begin{array}{rl} 
\omega(z,\vec \rho)&=\ds{\sum c(\vec n,\vec \rho)
\prod_{j=1}^{h^{1,d-1}} z_j^{n_j+\rho_j}}\\ [ 3 mm]
c(\vec n,\vec \rho)&=\ds{ { \prod_{k=1}^p \Gamma(1-\sum_{i=1}^{h^{1,d-1}} 
\hat l^{(i)}_k(n_i+\rho_i))\over
 \prod_{k=1}^q \Gamma(1+\sum_{i=1}^{h^{1,d-1}} 
 l^{(i)}_k (n_i+\rho_i))}}\\ [ 4 mm]
S_{i_1,\ldots,i_r}&=\partial_{\rho_{i_1}}\ldots\partial_{\rho_{i_r}}
\omega(\vec z,\vec \rho)|_{\vec \rho=\vec 0}
\end{array}
$$     
Notably with leading behavior 
$S_0=1+\ldots$, $S_i=z_i+\ldots$. 

This gives the explicit expansion of $C^{(d-2,1,1)}_{A,b,c}
=^0C^{(d-2,1,1)}_{A,b,c}+\cO(q_i)$, with $q_i=e^{t_i}$. The latter has 
a conjectured interpretation as being the counting  function 
for invariants of maps from the two sphere into $X$. 

\section{Toric geometry in a nutshell:} 

In this appendix we want to review some basic facts about toric 
geometry, which were used in lectures. Even a basic introduction 
into toric geometry would require much more space, so the following 
is merely intended to list these facts and to give a guide to the
mathematical literature or sources phycists might find useful. 
From the mathematical reviews\cit(danilov)\cit(fultonoda) the 
book of Fulton might be easiest to read for physicists and an useful 
recent introduction in the subject 
motivated from physics can be found in 
\cit(greenerev).    

For illustration we work with the $F_2$ example see \figref{makematter} and
remind the reader for convenience  of our identification of the toric
divisors 
\bea(rrrrrr)
K    &F&F&S&S'&\cr
D_0  &D_1&D_2&D_3&D_4&\cr
\left(\matrix{0\cr 0}\right)&
\left(\matrix{-2\cr -1}\right)&
\left(\matrix{0\cr 1}\right)&
\left(\matrix{-1\cr 0}\right)&
\left(\matrix{1\cr 0}\right)\ ,
\eea 
where we dropped the tildes. To every toric divisor 
$D_i$ $i=1,\ldots,k$ we associate a point
$\nu^{(i)}$ in a $n$ dimensional lattice $N\sim \ZZ^n$ and a  
variable $x_i$, here $D_1\sim (-2,-1)$  etc. 
The convex hull of the points defines a $n$ dimensional 
polyhedron $\Delta$ in $N_{\IR}\sim \IR^n$ see 
\figref{makematter} and we call the origin $\nu^{(0)}=(\vec 0)$.
An additional data is the choice of a triangulation\foot{For $F_2$
there is nothing to choose, but for $B_1(F_2)$ there is a choice.} 
into $n$ dimensional simplices.  
It defines the Stanley-Reisner ideal which is generated by the 
intersection of those $D_i=\{ x_i=0\}$ whose associated points 
do not share a common simplex\foot{E.g. for $F_2$ $D_1 D_2:=D_1\cap D_2
=\emptyset$, i.e. the locus $x_1=x_2=0$ has to be excluded and similarly 
$D_3\cap D_4=\emptyset$ in accordance to our Stanley-Reisner ideal 
in sect. (\ref{localmirror}).}. 
The latter condition has to be tested for simplices of any 
dimension yielding sets of indices of points $\{J\}$, 
which are not on a common simplex and the full Stanley-Reisner ideal ${\cal SR}$ is 
generated by
\be 
\prod_{\{J\}} D_{j_1}\cap \ldots \cap D_{j_{|J|}}, \  \forall J \ .
\ele(sr) 

We consider the $x_i$ as variables parameterizing $\IC^k$ and define
the $k-s$ dimensional toric variety following\cit(cox) 
(see also \cit(batcox)) as 
$\IC^k\setminus {\cal SR}$ modulo the $s$ equivalence relations 
\be
(x_1,\ldots x_k) \sim (\lambda_{(i)}^{l^{(i)}_1} x_1, 
\ldots, \lambda_{(i)}^{l^{(i)}_k} x_k)
\ele(toricdeff) 
with $\lambda_{(i)}\in \IC^*$ and the $l^{(i)}$ $i=1,\ldots,k$  
are an integral basis for the linear relations 
$\sum_{i=1}^k \nu^{(i)} l_i^{(p)}=0$ between the points 
in $\Delta$. 

This definition applies to smooth toric varities. 
If there are singularities a discrete torsion group acting on $x_i$ has 
also be divided out. E.g in the $\IP(\vec w)$ examples these are 
the $Z_n$ actions.

The  tricky part is the definition of
those generators $l^{(k)}$, called edges of the 
Mori-cone\cit(torickaehlercone)\cit(fultonoda),
which is defined by the secondary cone\foot{A cone for a given triangulation. 
If various triangulations are considered these cones together 
form the so called secondary fan.}  
of strictly convex piecewise linear function on the fan 
$\Xi_\Delta$ \figref{fan}. The K\"ahler cone is dual to the
cone defined by the convex piecewise linear functions on $\Xi_{\Delta}$ 
modulo the smooth functions on $\Xi_\Delta$. Pragmatical procedures, 
how to determine the Mori-cone and the K\"ahler cone for toric 
varieties be found in \cit(agm)\cit(hktyI)\cit(bkk)\cit(hly)\cit(greenerev) 
also\cit(batyrev) discusses the problem. Often the K\"ahler cone
restricts simply to the K\"ahlercone of the (CY) hypersurface 
in the toric variety, which may be the main object of interest.  
The discussion of 
the CY K\"ahler cone, when this is not the case can be 
found in\cit(bkk)\cit(bkkm).

If the triangulation is a star triangulation, i.e. every $n$ dimensional 
simplex $s_j$ has the origin $\nu^{(0)}$ as vertex, then we can 
associate a  fan $\Sigma(\Delta)$ to the polyhedron $\Delta$ by viewing 
the vertices $\nu^{(i_j)}\neq \nu^{(0)}$ of the simplex $s_j$ as vectors 
spanning the edges of a cone $\sigma_j$ from $\nu^{(0)}$, which is 
defined as $\sigma_j=\{ \sum_{i_j} r_{i_j} \nu^{(i_j)}|r_{i_j}\in \IR^+\}$.  
The fan  $\Sigma(\Delta)=\cup_i \sigma_i$ 
is the collection of all cones $\sigma_i$ (and its
faces) and it is easy to obtain from $\Sigma$ a representation of 
the manifold in terms of charts and transition functions\cit(danilov)
\cit(fultonoda). 

\noindent {\bf i)} The manifold is compact if the fan $\Sigma$ is 
complete, i.e. it covers\foot{Which is obviously the case for the 
fan $\Sigma(\Delta)$.} all of $N_{\IR}$, while an incomplete fan describes 
an non-compact affine toric variety, see\cit(fultonoda) for a 
proof of that statement. For more instructive examples, especially $\IP^2$,
whose polyhedron is the hull of $(1,0)$, $(0,1)$, $(-1,-1)$, see
\cit(greenerev). 

\noindent {\bf ii)} If all cones are spanned with positive coefficients 
by a subset of an integral basis for the lattice $N$, the manifold is 
smooth, again \cit(fultonoda) can be consulted for the proof. 

\noindent {\bf iii)} Consider the cone $\sigma$. If the lattice points
on each edge, which are nearest to the origin, lie all in a hyperplane 
$H$, which is in distance one from the origin, i.e  it exists
a $m$ in the dual space $M_{\IR}={\rm Hom}(N_{\IR},\IR)$ to $N_{\IR}$ such 
that $H=\{x\in N_{\IR} | \bra x,m\ket=1\}$, and there are no lattice 
points $x\in \sigma$ with $0<\bra m,x\ket <1$, then the affine
toric variety defined by $\sigma$ has only {\sl canonical Gorenstein 
singularities}.

\mabs {\sl The $\IP^2(1,1,2)$ example}:
For instance, as we check most easily by comparing the definition of 
the toric variety (\ref{weightedproj}) with (\ref{toricdeff}), 
$\IP^2(1,1,2)$ can be defined by the polyhedron, which is the convex 
hull of the the points $\nu^{(1)}=(-2,-1)$, 
$\nu^{(2)}=(0,1)$, $\nu^{(3)}=(1,0)$. It has simplices  
$s'_1=\{\nu^{(0)},\nu^{(1)},\nu^{(2)}\}$,
$s_3=\{\nu^{(0)},\nu^{(2)},\nu^{(4)}\}$,
$s_4=\{\nu^{(0)},\nu^{(1)},\nu^{(4)}\}$. Condition {\bf ii} is not
fulfilled for the cone $\sigma'_1$. There is a canonical
Gorenstein $\ZZ_2$-singularity in the chart associated to $\sigma'_1$.
The lattice $N$ modulo the lattice $N'$, which is spanned by the edges of
$\sigma_1'$, defines the torsion group, which is the $\ZZ_2$ in the 
case at hand. It acts on 
the normalbundle to the fixed locus of $(x_1,x_2,x_3)\mapsto 
(\mu x_1, \mu x_2,\mu^2 x_3)$ $\mu^2=1$ i.e. $(x_1=0,x_2=0,x_3=1)$, 
by $(z_1,z_2)\mapsto (-z_1,-z_2)$. Generally the order of the discrete
group  is the volume of unit cell in $N'$ divided by the unit cell of 
$N\cap \sigma$. Similarly if $r$ weights $w_{i_1}\ldots w_{i_r}$ of a 
$\IP^n(\vec w)$ have a common factor $n$ one gets a $n+1-r$ 
dimensional singular cone and  $\ZZ_n$ action on the normal bundel 
to the stratum $x_{i_1}=\ldots = x_{i_r}=0$.  For fuller explanation
of weighted projective spaces we refer to \cit(dolgachev). 
Generally singularities can be resolved by adding
points and making a finer subdivision into cones, such that 
property {\bf ii} holds for all of them. 
For $\IP^2(1,1,2)$ we achieve that by adding the point 
$\nu^{(3)}\in\sigma'_1$ and splitting $s'_1$ into 
$s_1=\{\nu^{(0)},\nu^{(1)},\nu^{(3)}\}$,
$s_2=\{\nu^{(0)},\nu^{(2)},\nu^{(3)}\}$, i.e. the non-singular 
resolution of $\IP^2(1,1,2)$ is $F_2$.

What makes smooth toric varieties so easy to deal with is the fact that 
linear relations of the type $\sum_{i}\bra m,\nu^{(i)}\ket D_i=0$ with 
$m\in M_{\IR}$, which may be written in coordinates as  
\be 
\sum_{i=1}^k \nu^{(k)}_i D_k=0\ , 
\ele(reldivisor)
are homological relation  between the divisors classes, e.g. for 
$F_2$: $D_1=D_2= F$ and $S'=S+2 F$. 
The Chern class of the toric variety is simply
\be
c(T_{A})=\prod_{i=1}^k (1+D_i)\ , 
\ele(chernI)  
the Chern classes $c_i(T_A)$ are terms, which are homogeneous
in the $D_i$. The evaluation of intersection 
defined in $c_d(T_A)$ gives by Gauss-Bonnet theorem 
the Eulernumber $\chi(A)$ of $A$. On the other hand $\chi(A)$ 
can be expressed by the sum of all simplices of maximal 
dimension \cit(fultonoda)
\be 
\chi(A)=\# {\rm d-simplices}\ ,
\ele(eul)
here 4. Eqs. (\ref{sr},\ref{reldivisor},\ref{chernI},
\ref{eul}) are strong enough to calculate by a, in general very 
involved but otherwise straigthforward, algebraic manipulation all
intersections $D_{i_1}\cdots D_{i_d}$ and in particular using (\ref{chernI})
one can calculate the evaluation of the Chern classes on the divisors. 

Special hypersurfaces $X$ can be described in $A$ by  
combinations of toric divisors $L=\sum_i a_i D_i$. 
The first Chern class on the normal 
bundle is just 
\be 
c_1({\cal N } )=\sum_{i=1}^k a_i D_i 
\ele(cerntoric)
Refering back to section (\ref{calabiyau}) we see by 
(\ref{normal},\ref{chernfactor},\ref{chern},\ref{chernI}) that the choice 
$H=\sum_{i=1}^k D_i$ leads to  $c_1(T_X)=0$ for the singular variety 
defined by $H$.
It was shown by Batyrev that the requirement of transversality of the
constraint $H=0$ and the requirement that only canonical 
Gorenstein singularities appear on $H$, which is necessary to get 
$c_1(T_X^{smooth})=0$, leads to a combinatorial condition for the 
polyhedron $\Delta$ called reflexivity, by which 
a natural dual polyhedron $\Delta^*$ can be defined. Moreover the divisor 
$H^*=\sum_{i=1}^{k^*} D_i^*$ in the toric variety defined by $\Delta^*$ 
describes a Calabi-Yau manifold, which has the mirror 
cohomology\cit(batyrev)! 

Using (\ref{normal},\ref{chernI},
\ref{cerntoric}) all topological data of $X$ can be ``straightforwardly'' 
calculated. However as the combinatoric and the algebra can get quickly 
quite involved, programs such as Schubert \cit(schubert) were developed. 

In non-generic situations on the other hand there exist very simple 
formulas, e.g. for two-dimensional spaces the intersections are very 
easy to determine. $D_i D_j=1$ if $D_j$ and $D_j$ share a common simplex
and zero otherwise.
The self-intersection of $D_i^2$ is determined  
by the linear equation $\nu^{(i-1)}+D_i^2 \nu^{(i)} 
+\nu^{(i+1)}=0$ where $\nu_i$ are labeled e.g. clockwise around 
$\Delta$. I.e. $\tilde F^2=0$, $\tilde S^2=-2$ etc.  

Similarly if we consider the smooth non-compact variety defined by
n-dimensional fan as e.g. $\Xi_\Delta$ then the intersections are 
\be 
D_{i_1}\ldots D_{i_n}=\cases{1 & if $\nu_{i_1}\ldots \nu_{i_n}$ 
span a cone\cr
                   0 & otherwise}
\ele(inter)
and using (\ref{reldivisor}) we ge immediatly
(\ref{ani}). Also for certain selfintersections simple formulas 
can be formulated \cit(hktyI).

\end{appendix}

\end{document}